\documentclass[amsmath,amssymb,aps,prx,reprint]{revtex4-2} 
\usepackage{amsmath,amssymb}
\usepackage[english]{babel}
\usepackage{ bbold }
\usepackage{upgreek}
\usepackage{dsfont}
\usepackage{graphicx}
\usepackage{comment}
\usepackage{braket}
\usepackage{colortbl}
\usepackage{blindtext}
\usepackage[hidelinks]{hyperref}
\usepackage{booktabs}
\usepackage{enumitem}

\newcommand{\bls}[1]{\mathbf{#1}} 
\newcommand{\blg}[1]{\boldsymbol{#1}} 
\newcommand{\bsf}[1]{\mathbf{#1}}

\begin{abstract}
    Levitating charged particles in ultra-high vacuum provides a preeminent platform for quantum information processing, for quantum-enhanced force and torque sensing, for probing physics beyond the standard model, and for high-mass tests of the quantum superposition principle. Existing setups range from single atomic ions, to ion chains and crystals, to charged molecules and nanoparticles. Future technological applications of such quantum systems will be crucially affected by fluctuating electric fields emanating from nearby electrodes, which interact with the levitated particles' monopole and higher charge moments. In this article, we provide a theoretical toolbox for describing how the rotational and translational quantum dynamics of charged nano- to microscale objects is affected by near metallic and dielectric surfaces, as characterized by their macroscopic dielectric response. The resulting quantum master equations describe the coherent surface-particle interaction, due to image charges and Casimir-Polder potentials, as well as surface-induced decoherence and heating, with the experimentally observed frequency and distance scaling. We explicitly evaluate the master equations for typical charge distributions and types of motion, thereby providing the tools required for describing and mitigating surface-induced decoherence in a variety of experiments with charged objects.
\end{abstract}

\begin{document}

\title{Surface-induced decoherence and heating of charged particles}

\author{Lukas Martinetz}
\author{Klaus Hornberger}
\author{Benjamin A. Stickler}
\affiliation{University of Duisburg-Essen, Lotharstra\ss e 1, 47058 Duisburg, Germany}

\maketitle

\tableofcontents

\section{Introduction}

\subsection{Quantum technologies with charged particles}

Electrically charged, levitated objects are at the core of numerous cutting-edge quantum experiments and technologies. Atomic ions are amongst the most promising platforms for quantum computation and simulation \cite{monroe2021,de2021materials}; trapped ion crystals act as ultra-precise sensors  \cite{gilmore2021quantum,cetina2022quantum} and enable studies of novel quantum phases of matter \cite{zhang2017observation1,kyprianidis2021observation}; charged molecules will enable encoding of quantum information \cite{andre2006coherent,rabl2006hybrid,rabl2007molecular,xiang2013hybrid,albert2020robust,sawant2020ultracold,campbell2020dipole,gregory2021robust}, hold the prospects of state-controlled quantum chemistry \cite{krems2008cold,ospelkaus2010quantum,quemener2012ultracold,balakrishnan2016perspective,hu2021nuclear,hirzler2022observation}, and allow probing the fundamental laws of nature \cite{zelevinsky2008precision,hudson2011improved,demille2017probing,safronova2018search,urban2019coherent,budker2022millicharged}; electrically trapped microscale particles \cite{gonzalez2021levitodynamics} can be cooled via embedded paramagnetic spins \cite{perdriat2021spin} or via electric feedback cooling \cite{goldwater2018levitated,magrini2020optimal,tebbenjohanns2021quantum}, preparing future test of quantum foundations \cite{arndt2014a,millen2020quantum,martinetz2020,carney2021trapped}.

Unlike atomic ions, molecules and nanoparticles possess higher charge moments so that the trapping fields can exert torques. Already the free quantum rotation dynamics of rigid bodies are intrinsically non-harmonic, rendering them attractive for quantum applications \cite{lemeshko2013manipulation,koch2019quantum,stickler2021}. For instance, the rotation states of molecular ions have been proposed for quantum information processing \cite{demille2002quantum,andre2006coherent,rabl2006hybrid,xiang2013hybrid,hudson2018dipolar,sawant2020ultracold,campbell2020dipole}, promising significant advantages in comparison to harmonic degrees of freedom due to the special topology of rotations in three dimensions \cite{albert2020robust}.
At the same time, experiments with levitated nanorotors \cite{hoang2016,kuhn2017a,kuhn2017b,rashid2018precession,reimann2018ghz,ahn2018optically,ahn2020ultrasensitive,bang2020,schafer2020,vanderlaan2020observation,jin20216} reach millikelvin rotational temperatures \cite{delord2020spincoupling,bang2020,vanderlaan2020observation}, pointing the path towards ultra-precise torque sensors \cite{hoang2016,kuhn2017b,rashid2018precession,delord2018,ahn2018optically,ahn2020ultrasensitive,schafer2020,jin20216}, the search for new physics \cite{moore2021searching}, and tests of the quantum superposition principle with massive objects \cite{rusconi2017quantum,stickler2018probing,ma2020quantum,delord2020spincoupling,kaltenbaek2022maqro}.

Controlling atomic, molecular, and nanoscale ions with electrodes and superconducting circuits \cite{tian2004interfacing,sorensen2004capacitive,martinetz2021electric,an2021coupling} brings them close to metallic and dielectric surfaces. Thermally fluctuating charges and currents in the surface then unavoidably  induce decoherence and motional heating of the trapped particle \cite{negretti2011quantum,xiang2013hybrid,kurizki2015quantum,bruzewicz2019trapped,brown2021materials}. Given that building a quantum sensing device or a quantum computer will likely require chip-based miniaturization \cite{bruzewicz2019trapped}, the role of surface induced heating and decoherence can be expected to become worse in future devices, presenting a potential roadblock. This calls for a universal theoretical framework for describing charge-induced decoherence and heating close to metallic and dielectric surfaces, which allows incorporating microscopic and heuristic models, and which points the way towards noise mitigation techniques.

\subsection{Surface-induced heating and decoherence}

The experimental characterization of surface noise and decoherence relies on measuring heating rates and coherence times of trapped ions near the surface  \cite{wineland1998experimental,leibfried2003quantum,hite2013surface,monroe2013scaling,brownnutt2015ion,bruzewicz2019trapped,brown2021materials,de2021materials} and of interfering free electron beams travelling above a plane \cite{sonnentag2007measurement,beierle2018experimental,kerker2020quantum}. While different setups can exhibit distinct noise spectral densities, most experiments seem to agree on a characteristic inverse frequency scaling and on a proportionality to the fourth power of inverse distance \cite{brown2021materials}. That said, the microscopic origin of surface noise is still debated \cite{brownnutt2015ion,bruzewicz2019trapped,de2021materials,brown2021materials}. Theoretical models range from the damping of conduction electrons \cite{anglin1997deconstructing}, the dissipative formation of image charges \cite{machnikowski2006theory}, the excitation of surface plasmons \cite{howie2011mechanisms}, to the fluctuations of surface-bound \cite{safavi2011microscopic,safavi2013influence} or diffusing \cite{kim2017electric} adatoms, to patch potentials \cite{sandoghdar1992direct,turchette2000heating,dubessy2009electric,low2011finite}, and to thermally activated two-level adsorbates \cite{noel2019electric}. In an approach that does not require specifying a microscopic mechanism for surface-induced heating and decoherence, one may characterize the surface material solely by means of its dielectric response function \cite{wylie1984quantum,buhmann2008surface,scheel2012path,kumph2016electric,teller2021heating}. 

This mechanism-independent approach can be formalized by using the framework of macroscopic quantum electrodynamics \cite{buhmann1,buhmann2}, which describes the quantum and thermal noise of the quantized electromagnetic field in terms of fluctuating bound charges and currents. This framework expresses the observable forces on the particle through empirically accessible response functions, while providing a versatile, i.e.\ non-material specific, description. Macroscopic quantum electrodynamics can reproduce the classical field fluctuations found above lossy dielectric layers \cite{kumph2016electric,teller2021heating,holz2021electric}, including the experimentally observed magnitude, distance, and frequency scaling \cite{brown2021materials}. This supports that the noise originates from a thin surface layer. In addition, macroscopic quantum electrodynamics has been employed to quantitatively describe the decoherence of fast electrons \cite{scheel2012path,kerker2020quantum}, of spins \cite{skagerstam2006spin}, and of the center-of-mass motion of magnetic \cite{pino2018chip} and polarizable \cite{sinha2020quantum} spheres above surfaces.

Future quantum technologies, which utilize molecules and nanoparticles levitating and rotating in micro-traps near electrodes and micro-structured surfaces, will require a comprehensive toolbox to predict surface-induced heating and decoherence for arbitrary charge distributions and surface configurations. This article provides this toolbox in the form of Lindblad master equations for the ro-translational quantum dynamics of arbitrarily charged particles in front of general dielectrics.

\subsection{This work}

This article derives quantum master equations for charged rigid bodies above general metallic and dielectric surfaces. These master equations contain a coherent particle-surface potential, accounting for the formation of image charges and the impact of Casimir-Polder interactions, and a dissipator, describing surface-induced decoherence and heating. We express the resulting translational and rotational heating and decoherence rates in terms of the particle charge distribution, the surface geometry and layer structure, and the surface dielectric response and surface temperature. Based on the general theory, we provide specific formulas for typical charge distributions, surface dielectric functions, and types of motion. This includes the oscillation of monopoles, the libration of dipoles, and the rotation of quadrupoles above metallic and superconducting surface covered by thin dielectric layers, whose noise spectrum can decohere the particle motion either resonantly or off-resonantly.

The chosen framework facilitates a pragmatic description of surface-induced decoherence and heating in terms of macroscopic dielectric response functions. The latter are accessible both by {\it in situ} and {\it ex situ} measurements as well as by microscopic ab-initio calculations \cite{botti2007time,ping2013electronic} and molecular-dynamics simulations \cite{foulon2022omega}. By disentangling the effects of the surface geometry from the influence of the surface material, the derived master equations can be used for the design of surfaces with favourable noise properties given a specific charge distribution and ro-translational motion, providing a bridge to device engineering and materials design \cite{brown2021materials}. We expect the here presented toolbox to be relevant for a variety of state-of-the art setups ranging from trapped ion quantum computers \cite{brown2021materials}, to hybrid quantum devices with charged atoms or molecules \cite{xiang2013hybrid}, to nanoparticles in electric and optical traps for fundamental tests of physics \cite{stickler2021}.

The article is structured as follows: Section~\ref{sec:summary} summarizes the theoretical framework of surface-induced decoherence and presents the resulting master equations in their most generic form. The master equations are then specified in Sects.~\ref{sec:specific_charge_distributions} and \ref{sec:specific_dielectrics} for typical charge distributions, for metallic, dielectric, and superconducting surfaces, and for planar and layered surface geometries. In Sect.~\ref{Sec:decoherenceAndHeating} we discuss criteria how the presented toolbox can be utilized, and we illustrate this with two examples. Finally, Sect.~\ref{sec:conclusion} discusses possible future research directions and provides our conclusions. The technical details of the derivation are presented in Apps.~\ref{sec:derivation} to \ref{sec:quasi_static}. Appendix  \ref{app:ThomsonScattering} discusses the ro-translational decoherence due to elastic photon scattering (Thomson scattering).

\begin{figure}
\centering
\includegraphics[width = 0.48\textwidth]{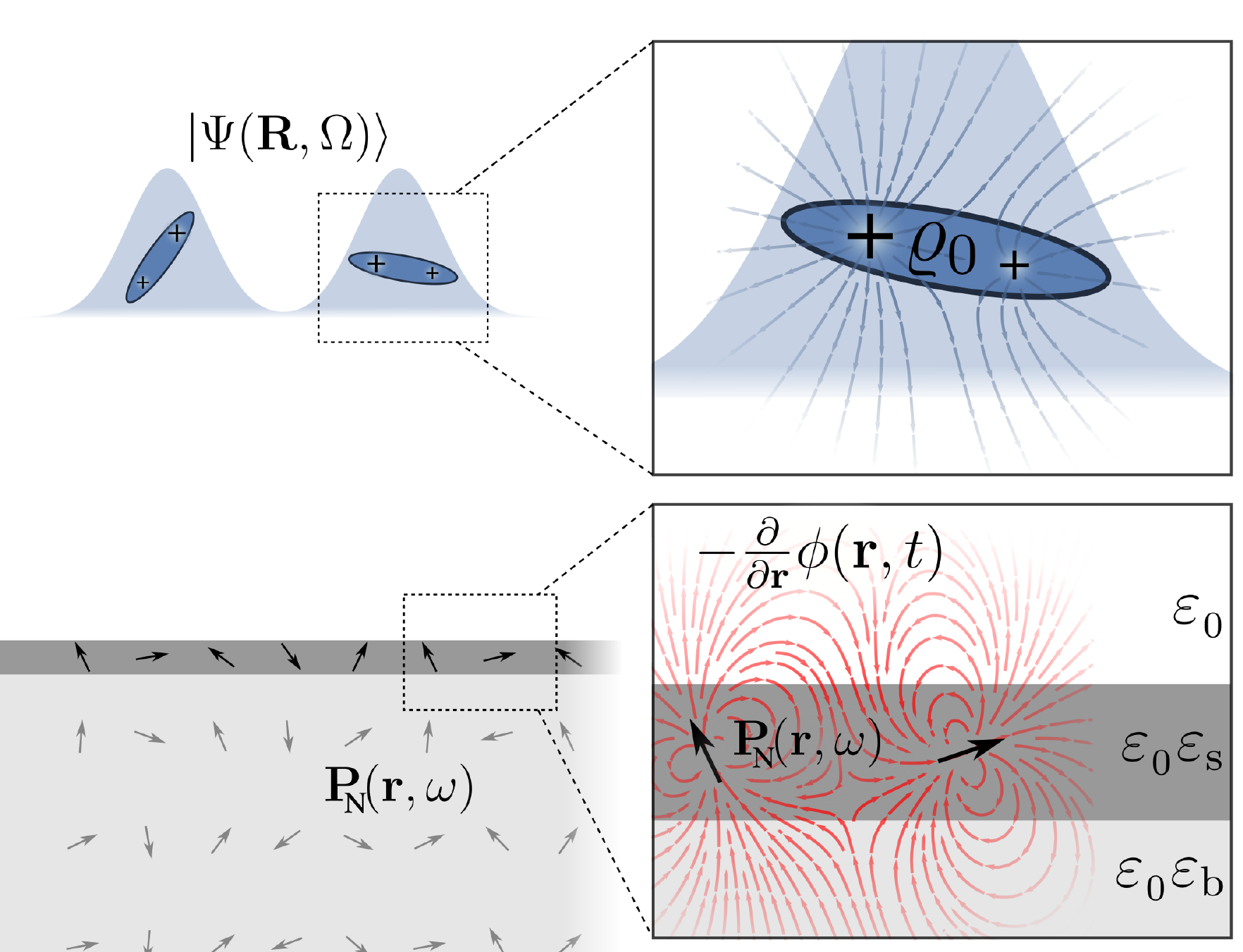}
\caption{A charged particle (blue) with charge distribution $\varrho_0$ evolving in a spatio-orientational superposition state $\ket{\Psi(\bls{R},\Omega)}$ will suffer from motional decoherence due to electric fields (red lines) originating from a fluctuating polarization (black arrows) in nearby metallic or dielectric surfaces. The fluctuation strength of the polarization at a position $\bls{r}$ depends on the dielectric material response $\varepsilon_0\varepsilon_{\rm r}(\bls{r},\omega)$ and the temperature $T$. For small distances, the electric field propagates quasistatically from each fluctuation to the particle, as can be described by a scalar potential $\phi(\bls{r},t)$. A layered dielectric consisting of a surface layer with $\varepsilon_0\varepsilon_{\rm s}(\omega)$ and a bulk dielectric with $\varepsilon_0\varepsilon_{\rm b}(\omega)$ can explain the noise levels and scalings measured in ion traps \cite{kumph2016electric}. }\label{figure1}
\end{figure}

\begin{center}
\begin{table*}
\caption{Selection of quantum master equations for typical charge distributions and types of motion, see Sec.~\ref{sec:specific_charge_distributions}. The prefactors in line {\it a.}, {\it b.}, and {\it c.} are special cases of the spatio-orientational decoherence rate \eqref{eq:DecoRateSlowParticle}.}
\bgroup
\begin{tabular}{c l l}
 & Illustration &  \multicolumn{1}{c}{Dissipator}  \\ \toprule
 \toprule
{\it a.} & \begin{minipage}{.16\textwidth}
      \includegraphics[width=30mm, height=20mm]{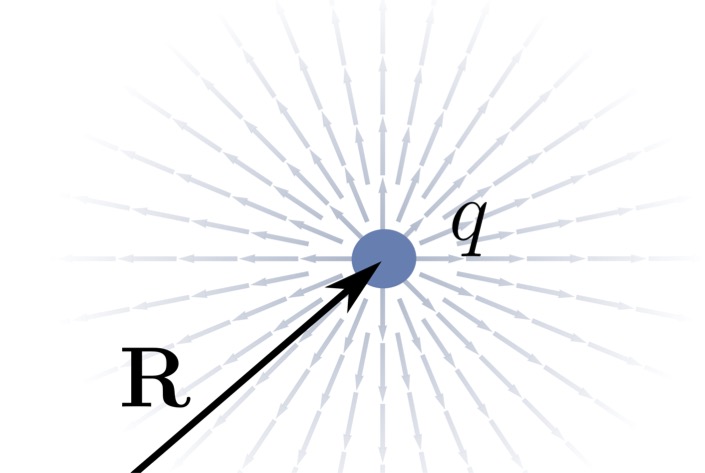}
    \end{minipage} & $\begin{aligned}&\langle {\bf R},\Omega | {\cal L} \rho|{\bf R}',\Omega'\rangle = -\frac{q^2}{\hbar} \left[h(\bls{R},\bls{R})+h(\bls{R}',\bls{R}')-2h(\bls{R},\bls{R}')\right]\langle {\bf R},\Omega | \rho|{\bf R}',\Omega'\rangle\\
    &\text{ with}\qquad h(\bls{R},\bls{R}')=-\lim_{\omega\downarrow 0}n(\omega){\rm Im}\left[g(\bls{R},\bls{R}',\omega)\right]\end{aligned}$ \\ \midrule

{\it b.}& \begin{minipage}{.16\textwidth}
      \includegraphics[width=30mm, height=20mm]{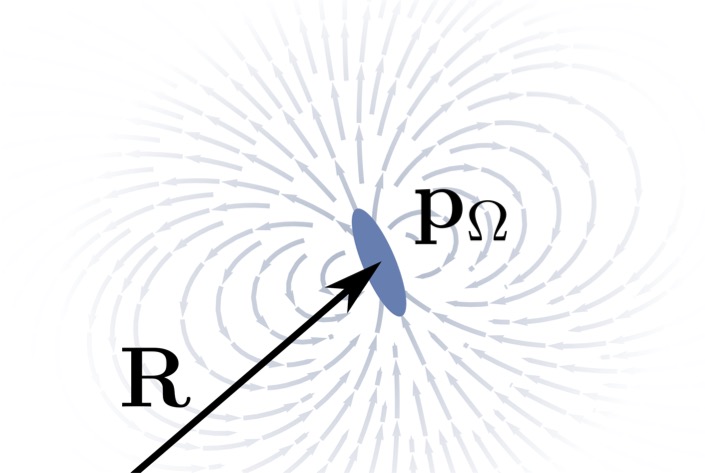}
    \end{minipage} & $\begin{aligned}&\langle {\bf R},\Omega | {\cal L} \rho|{\bf R}',\Omega'\rangle =-\frac{1}{\hbar} \left[h^{(2)}_{\Omega\Omega}(\bls{R},\bls{R})+h^{(2)}_{\Omega'\Omega'}(\bls{R}',\bls{R}')-2h^{(2)}_{\Omega\Omega'}(\bls{R},\bls{R}')\right]\langle {\bf R},\Omega | \rho|{\bf R}',\Omega'\rangle\\
    &\text{ with}\qquad h^{(2)}_{\Omega\Omega'}(\bls{R},\bls{R}')=\left(\bls{p}_\Omega\cdot\frac{\partial}{\partial \bls{R}}\right)\left(\bls{p}_{\Omega'}\cdot\frac{\partial}{\partial \bls{R}'}\right)h(\bls{R},\bls{R}')\end{aligned}$ \\
\midrule

{\it c.} & \begin{minipage}{.16\textwidth}
      \includegraphics[width=30mm, height=20mm]{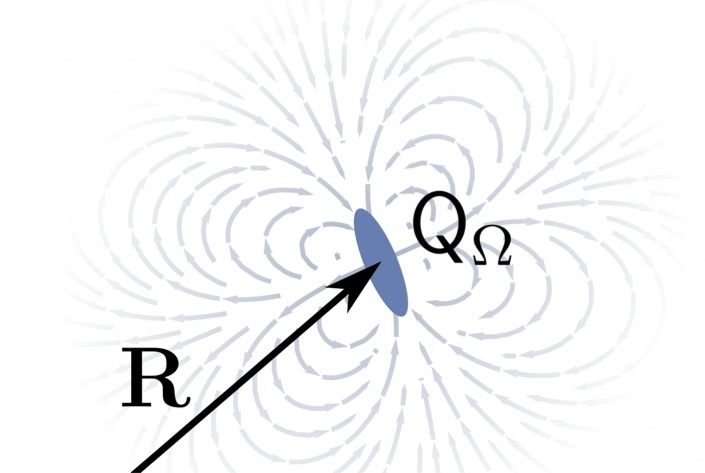}
    \end{minipage} & $\begin{aligned}&\langle {\bf R},\Omega | {\cal L} \rho|{\bf R}',\Omega'\rangle =-\frac{1}{\hbar}\left[h^{(4)}_{\Omega\Omega}(\bls{R},\bls{R}) +h^{(4)}_{\Omega'\Omega'}(\bls{R}',\bls{R}')-2h^{(4)}_{\Omega\Omega'}(\bls{R},\bls{R}') \right]\langle {\bf R},\Omega |  \rho|{\bf R}',\Omega'\rangle\\
    &\text{ with}\qquad h^{(4)}_{\Omega\Omega'}(\bls{R},\bls{R}')=\frac{1}{36}\left(\frac{\partial}{\partial \bls{R}}\cdot \mathsf{Q}_\Omega\frac{\partial}{\partial \bls{R}}\right)\left(\frac{\partial}{\partial \bls{R}'}\cdot \mathsf{Q}_{\Omega'}\frac{\partial}{\partial \bls{R}'}\right)h(\bls{R},\bls{R}')\end{aligned}$  \\
\midrule

{\it d.} & \begin{minipage}{.16\textwidth}
      \includegraphics[width=30mm, height=20mm]{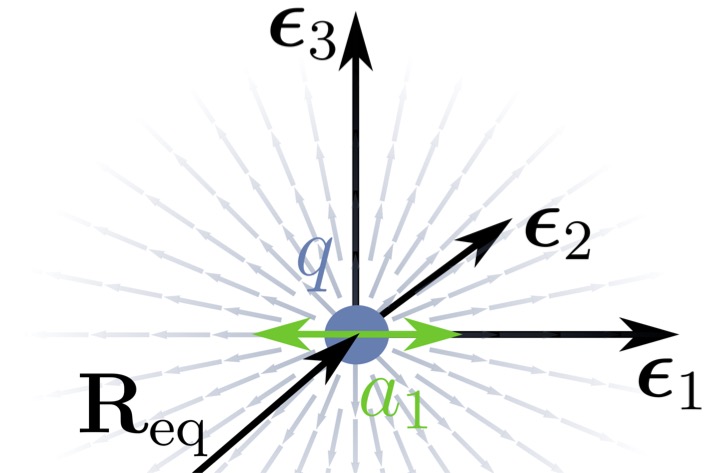}
    \end{minipage} & $\begin{aligned}&\mathcal{L}\rho=\sum_{k=1,2,3}\frac{q^2}{m\omega_k}h_k(\bls{R_{\rm eq}})\left[\left[n(\omega_k)+1\right]\left(a_k\rho a^\dagger_k-\frac{1}{2}\left\{a^\dagger_k a_k,\rho\right\}\right)+n(\omega_k)\left(a^\dagger_k\rho a_k-\frac{1}{2}\left\{a_k a^\dagger_k,\rho\right\}\right)\right]\\
    &\text{ with}\qquad h_k(\bls{R_{\rm eq}})=-\left(\blg{\epsilon}_k\cdot\left.\frac{\partial}{\partial \bls{r}}\right)\left(\blg{\epsilon}_k\cdot \frac{\partial}{\partial \bls{r}'}\right){\rm Im}\left[g(\bls{r},\bls{r}',\omega_k)\right]\right|_{\bls{r}=\bls{r}'=\bls{R}_{\rm eq}}\end{aligned}$  \\
\midrule

{\it e.} & \begin{minipage}{.16\textwidth}
      \includegraphics[width=30mm, height=20mm]{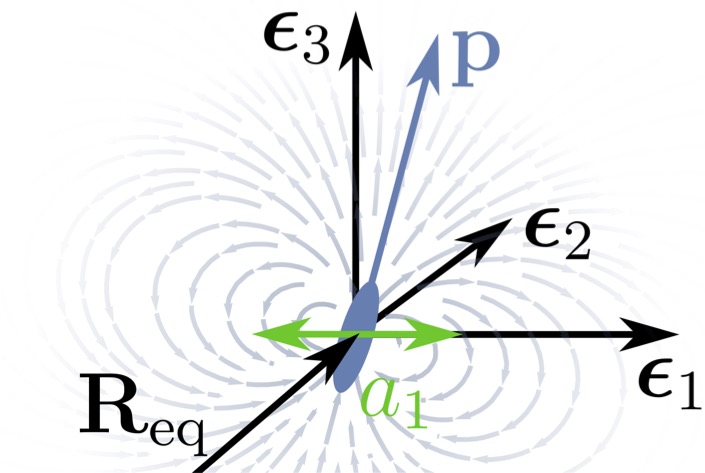}
    \end{minipage} & $\begin{aligned}&\mathcal{L}\rho=\sum_{k=1,2,3}\frac{1}{m\omega_k} h_k(\bls{R}_{\rm eq})\left[\left[n(\omega_k)+1\right]\left(a_k\rho a^\dagger_k-\frac{1}{2}\left\{a^\dagger_k a_k,\rho\right\}\right)+n(\omega_k)\left(a^\dagger_k\rho a_k-\frac{1}{2}\left\{a_k a^\dagger_k,\rho\right\}\right)\right]\\
    &\text{ with}\qquad h_k(\bls{R}_{\rm eq})=-\left(\blg{\epsilon}_k\cdot\frac{\partial}{\partial\bls{r}}\right)\left.\left(\blg{\epsilon}_k\cdot\frac{\partial}{\partial\bls{r}'}\right)\left(\bls{p}\cdot\frac{\partial}{\partial\bls{r}}\right)\left(\bls{p}\cdot\frac{\partial}{\partial\bls{r}'}\right){\rm Im}\left[g(\bls{r},\bls{r}',\omega_k)\right]\right|_{\bls{r}=\bls{r}'=\bls{R}_{\rm eq}}\end{aligned}$  \\
\midrule

{\it f.} & \begin{minipage}{.16\textwidth}
      \includegraphics[width=30mm, height=20mm]{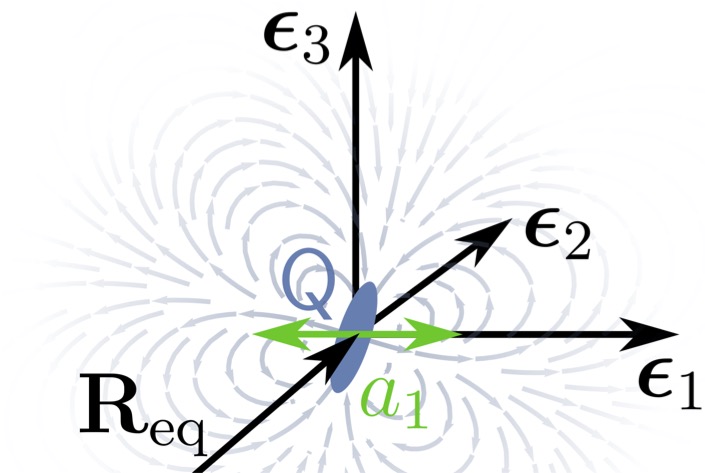}
    \end{minipage} & $\begin{aligned}&\mathcal{L}\rho=\sum_{k=1,2,3}\frac{1}{36m\omega_k} h_k(\bls{R}_{\rm eq})\left[\left[n(\omega_k)+1\right]\left(a_k\rho a^\dagger_k-\frac{1}{2}\left\{a^\dagger_k a_k,\rho\right\}\right)+n(\omega_k)\left(a^\dagger_k\rho a_k-\frac{1}{2}\left\{a_k a^\dagger_k,\rho\right\}\right)\right]\\
    &\text{ with}\qquad h_k(\bls{R}_{\rm eq})=-\left(\blg{\epsilon}_k\cdot\frac{\partial}{\partial\bls{r}}\right)\left.\left(\blg{\epsilon}_k\cdot\frac{\partial}{\partial\bls{r}'}\right)\left(\frac{\partial}{\partial\bls{r}}\cdot\mathsf{Q}\frac{\partial}{\partial\bls{r}}\right)\left(\frac{\partial}{\partial\bls{r}'}\cdot\mathsf{Q}\frac{\partial}{\partial\bls{r}'}\right){\rm Im}\left[g(\bls{r},\bls{r}',\omega_k)\right]\right|_{\bls{r}=\bls{r}'=\bls{R}_{\rm eq}}\end{aligned}$  \\
\midrule

{\it g.} & \begin{minipage}{.16\textwidth}
      \includegraphics[width=30mm, height=20mm]{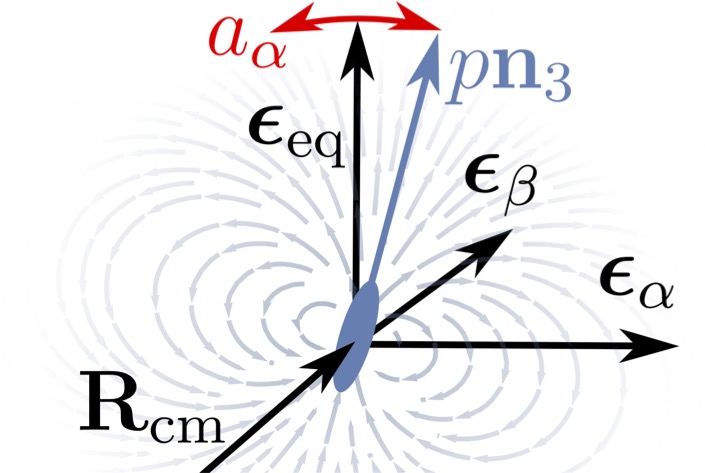}
    \end{minipage} & $\begin{aligned}&\mathcal{L}\rho=\sum_{\ell = \alpha,\beta}\frac{p^2}{I\omega_\ell}h_\ell(\bls{R_{\rm cm}})\left[\left[n(\omega_\ell)+1\right]\left(a_\ell\rho a^\dagger_\ell-\frac{1}{2}\left\{a^\dagger_\ell a_\ell,\rho\right\}\right)+n(\omega_\ell)\left(a^\dagger_\ell\rho a_\ell-\frac{1}{2}\left\{a_\ell a^\dagger_\ell,\rho\right\}\right)\right]\\
    &\text{ with}\qquad h_\ell(\bls{R_{\rm cm}})=-\left(\blg{\epsilon}_\ell\cdot\left.\frac{\partial}{\partial \bls{r}}\right)\left(\blg{\epsilon}_\ell\cdot \frac{\partial}{\partial \bls{r}'}\right){\rm Im}\left[g(\bls{r},\bls{r}',\omega_\ell)\right]\right|_{\bls{r}=\bls{r}'=\bls{R}_{\rm cm}}\end{aligned}$  \\
\midrule

{\it h.} & \begin{minipage}{.16\textwidth}
      \includegraphics[width=30mm, height=20mm]{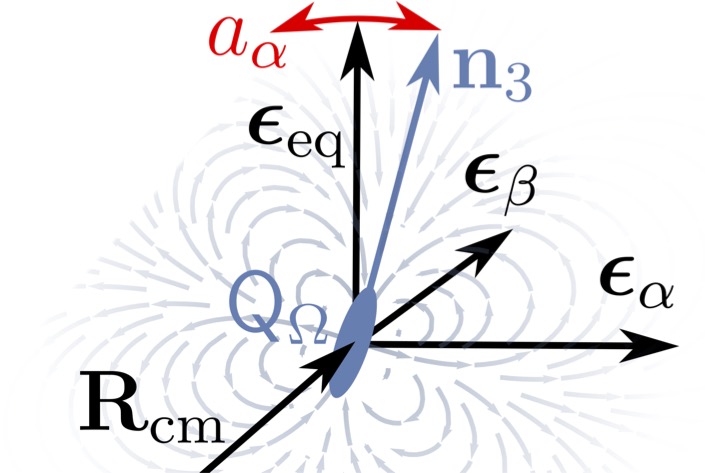}
    \end{minipage} & $\begin{aligned}&\mathcal{L}\rho=\sum_{\ell=\alpha,\beta}\frac{Q_{33}^2}{4I\omega_\ell} h_\ell(\bls{R}_{\rm cm})\left[\left[n(\omega_\ell)+1\right]\left(a_\ell\rho a^\dagger_\ell-\frac{1}{2}\left\{a^\dagger_\ell a_\ell,\rho\right\}\right)+n(\omega_\ell)\left(a^\dagger_\ell\rho a_\ell-\frac{1}{2}\left\{a_\ell a^\dagger_\ell,\rho\right\}\right)\right]\\
    &\text{ with}\qquad h_\ell(\bls{R}_{\rm cm})=-\left(\blg{\epsilon}_\ell\cdot\frac{\partial}{\partial\bls{r}}\right)\left.\left(\blg{\epsilon}_\ell\cdot\frac{\partial}{\partial\bls{r}'}\right)\left(\blg{\epsilon}_{\rm eq}\cdot\frac{\partial}{\partial\bls{r}}\right)\left(\blg{\epsilon}_{\rm eq}\cdot\frac{\partial}{\partial\bls{r}'}\right){\rm Im}\left[g(\bls{r},\bls{r}',\omega_\ell)\right]\right|_{\bls{r}=\bls{r}'=\bls{R}_{\rm cm}}\end{aligned}$  \\
\midrule

{\it i.} & \begin{minipage}{.16\textwidth}
      \includegraphics[width=30mm, height=20mm]{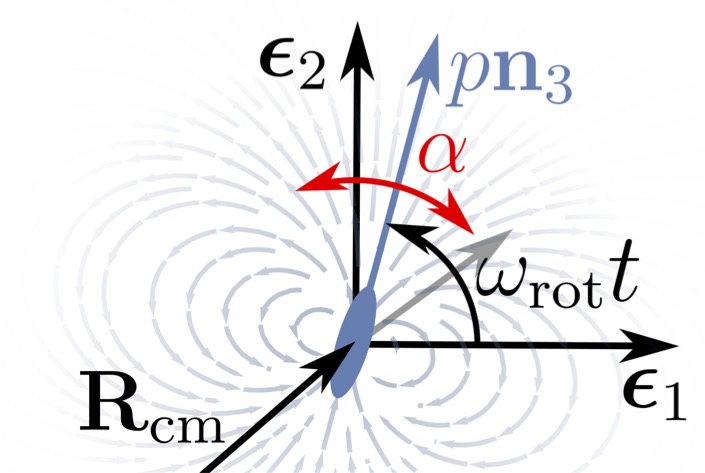}
    \end{minipage} & $\begin{aligned}&\mathcal{L}\rho=\sum_{\ell=1,2}\frac{p^2}{2\hbar}h_\ell(\bls{R_{\rm cm},\omega_{\rm rot}})\left[\left[n(\omega_{\rm rot})+1\right]\left(e^{-i\alpha}\rho e^{i\alpha}-\rho\right)+n(\omega_{\rm rot})\left(e^{i\alpha}\rho e^{-i\alpha}-\rho\right)\right]\\
    &\text{ with}\qquad h_\ell(\bls{R_{\rm cm},\omega_{\rm rot}})=-\left(\blg{\epsilon}_\ell\cdot\left.\frac{\partial}{\partial \bls{r}}\right)\left(\blg{\epsilon}_\ell\cdot \frac{\partial}{\partial \bls{r}'}\right){\rm Im}\left[g(\bls{r},\bls{r}',\omega_{\rm rot})\right]\right|_{\bls{r}=\bls{r}'=\bls{R}_{\rm cm}}\end{aligned}$  \\
\midrule

{\it j.} & \begin{minipage}{.16\textwidth}
      \includegraphics[width=30mm, height=20mm]{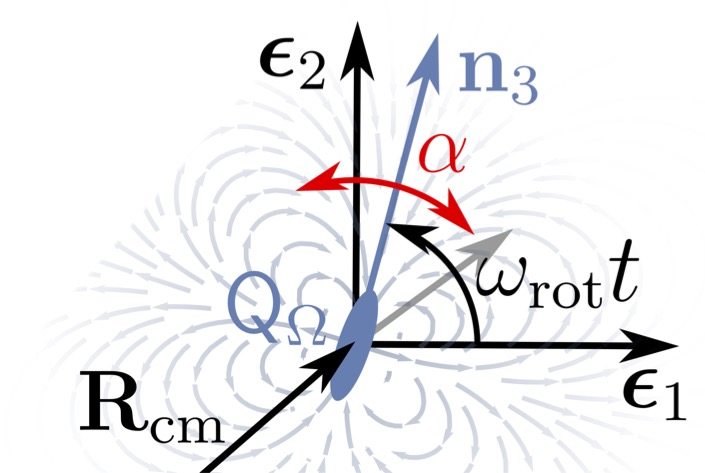}
    \end{minipage} & $\begin{aligned}&\mathcal{L}\rho=\frac{Q_{33}^2}{128\hbar}h_{12}(\bls{R}_{\rm cm})\left\{\left[n(\omega_{\rm rot})+1\right]\left(e^{-2i\alpha}\rho e^{2i\alpha}-\rho\right)+n(\omega_{\rm rot})\left(e^{2i\alpha}\rho e^{-2i\alpha}-\rho\right)\right\}\\
    &\text{ with}\qquad h_{12}(\bls{R}_{\rm cm})=-\left[\left(\blg{\epsilon}_1-i\blg{\epsilon}_2\right)\cdot\frac{\partial}{\partial\bls{r}}\right]^2\left.\left[\left(\blg{\epsilon}_1+i\blg{\epsilon}_2\right)\cdot\frac{\partial}{\partial\bls{r}'}\right]^2{\rm Im}\left[g(\bls{r},\bls{r}',\omega_{\rm rot})\right]\right|_{\bls{r}=\bls{r}'=\bls{R}_{\rm cm}}\end{aligned}$  \\
\bottomrule
\bottomrule
\end{tabular}
\egroup
\label{tab:MasterEquations}
\end{table*}
\end{center}

\section{Summary of main results}\label{sec:summary}

\subsection{Particle-surface coupling}

We consider the motion of a particle of mass $m$ and with moments of inertia $I_{1}$, $I_{2}$, $I_{3}$ moving in vacuum at center-of-mass position ${\bf R}$. The particle orientation $\Omega$ is specified by the rotation tensor $\mathsf{R}_\Omega$, relating the space-fixed axes $\bls{e}_i$ to the body-fixed principal axes  $\bls{n}_i=\mathsf{R}_\Omega\bls{e}_i$, so that the inertia tensor is ${\rm I}_\Omega=\sum_i I_i\bls{n}_i\otimes\bls{n}_i$. The body carries a rigid charge distribution $\varrho(\bls{r})=\varrho_0[\mathsf{R}_\Omega^{\rm T}(\bls{r}-\bsf{R})]$, where $\varrho_0$ is the charge distribution at the reference position ${\bf R} = 0$ and orientation ${\sf R}_\Omega = {\mathbb 1}$.

The particle interacts with the fluctuating electro-magnetic field emanating from a nearby surface, see Fig.~\ref{figure1}. This field is due to thermal polarization currents in the surface material and can be expressed via the scalar and vector gauge potential $\phi(\bls{r},t)$ and $\bsf{A}(\bls{r},t)$. The resulting particle dynamics are described by the minimal coupling Hamiltonian
\begin{align}\label{eq:couplingHamiltonian}
    H&=\frac{1}{2m}\left[\bsf{P}-\int d^3r\,\varrho_0(\bls{r})\bsf{A}\left(\bsf{R}+\mathsf{R}_\Omega\bls{r},t\right)\right]^2\nonumber\\
    &+\frac{1}{2}\left[\bsf{J}-\int d^3r\,\varrho_0(\bls{r})\left(\mathsf{R}_\Omega\bls{r}\right)\times\bsf{A}\left(\bsf{R}+\mathsf{R}_\Omega\bls{r},t\right)\right]\nonumber\\
    &\cdot {\rm I}_\Omega^{-1}\left[\bsf{J}-\int d^3r\,\varrho_0(\bls{r})\left(\mathsf{R}_\Omega\bls{r}\right)\times\bsf{A}\left(\bsf{R}+\mathsf{R}_\Omega\bls{r},t\right)\right]\nonumber\\
    &+\int d^3r\,\varrho_0(\bls{r})\phi\left(\bsf{R}+\mathsf{R}_\Omega\bls{r},t\right).
\end{align}
Here, $\bsf{P}$ and $\bsf{J}$ are the canonical linear and angular momentum vectors. The canonical angular momentum vector is understood in terms of the canonical momentum coordinates $p_\Omega$, conjugate to the orientation coordinates $\Omega$, see App.~\ref{sec:RotorFieldCoupling}. (It is not identical to the kinetic angular momentum due to gauge coupling to the transverse field.) 

Promoting the gauge fields to Heisenberg-picture operators, their dynamics can be described within the framework of {\it macroscopic quantum electrodynamics} \cite{buhmann1,scheel2008macroscopic}. This theory relates the gauge fields at position $\bls{r}$ to the  quantized polarization currents $-i \omega \bls{P}_{\rm N}(\bls{r}',\omega)$ at position $\bls{r}'$ and frequency $\omega$, which fluctuate in the dielectric with complex-valued permittivity $\varepsilon_{\rm r}(\bls{r}',\omega)$ (App.~\ref{sec:macroscopicQED}). The polarization noise field is modelled by
\begin{align} \label{eq:pn}
    \bls{P}_{\rm N}(\bls{r},\omega)=i\sqrt{4\pi\hbar\varepsilon_0{\rm Im}\left[\varepsilon_{\rm r}(\bls{r},\omega)\right]}\bls{f}_{\rm e}(\bls{r},\omega),
\end{align}
with bosonic annihilation operators $\bls{f}_{\rm e}(\bls{r},\omega)$, from now on dubbed as polarization oscillators. They satisfy the standard (dyadic) canonical commutation relations \cite{buhmann1}, e.g.,
\begin{align}
    &[\bls{f}_{\rm e}(\bls{r},\omega),\bls{f}_{\rm e}^\dagger(\bls{r'},\omega')]=\mathbb{1}\delta(\bls{r}-\bls{r'})\delta(\omega-\omega').
\end{align}
In the following, we assume that the oscillators are always in a thermal state characterized by the surface temperature $T$, so that
\begin{align}
\langle \bls{f}^\dagger_{\rm e}(\bls{r},\omega)\otimes\bls{f}_{\rm e}&(\bls{r}',\omega')\rangle=n(\omega)\delta(\bls{r}-\bls{r}')\delta(\omega-\omega')\mathbb{1},
\end{align}
with the Bose-Einstein occupation factor $n(\omega)=1/[{\rm exp}(\hbar\omega/k_{\rm B}T)-1]$. 

The polarization noise field \eqref{eq:pn} induces an electromagnetic field at the particle position, as given by Maxwell's equations. In most experimental setups, this field is well approximated by its longitudinal part, and therefore described by a quasistatic scalar Green function (App.~\ref{sec:quasi_static}) 
\begin{equation}\label{eq:DGLscalarGreenfunction}
    \frac{\partial}{\partial \bls{r}}\cdot\left[\varepsilon_0\varepsilon_{\rm r}(\bls{r},\omega)\frac{\partial}{\partial \bls{r}}g(\bls{r},\bls{r}',\omega)\right]=-\delta(\bls{r}-\bls{r}').
\end{equation}
The Green function $g(\bls{r},\bls{r}',\omega)$ determines the electrostatic potential at $\bls{r}$ of a charge oscillating in magnitude with frequency $\omega$ at position $\bls{r}'$ in presence of a dielectric. This Green function fulfills the relation
\begin{align}\label{eq:relationScalarGreenfunction}
    {\rm Im}\left[g(\bls{r},\bls{r}',\omega)\right]=&-\varepsilon_0\int d^3s \,{\rm Im}\left[\varepsilon_{\rm r}(\bls{s},\omega)\right]  \left[\frac{\partial}{\partial \bls{s}}g(\bls{r},\bls{s},\omega)\right]\nonumber\\
    &\cdot \left[\frac{\partial}{\partial \bls{s}}g^*(\bls{s},\bls{r}',\omega)\right],
\end{align}
which will be useful in the following. App.~\ref{sec:quasi_static} derives this relation together with further properties of $g(\bls{r},\bls{r}',\omega)$.

In the quasistatic approximation \eqref{eq:quasistaticGDerivation} the vector potential vanishes in Coulomb gauge so that the particle-surface interaction is determined by the operator-valued scalar potential
\begin{align}\label{eq:phiviag}
    \phi(\bls{r},t)=&i\sqrt{\frac{\varepsilon_0\hbar}{\pi}}\int d^3r'\,\int_0^\infty d\omega\, \sqrt{{\rm Im}\left[\varepsilon_{\rm r}(\bls{r'},\omega)\right]}e^{-i\omega t}\nonumber\\
    &\times\bls{f}_{\rm e}(\bls{r}',\omega) \cdot\frac{\partial}{\partial\bls{r}'}g(\bls{r},\bls{r}',\omega)+{\rm h.c.},
\end{align}
where ${\rm Im}[\varepsilon_{\rm r}({\bf r},\omega)]\geq0$ for positive frequencies \cite{buhmann1}. The hermitian Heisenberg-picture operator \eqref{eq:phiviag} acts on the Hilbert space of the polarization oscillators.

Having specified how the particle-field coupling Hamiltonian relates to the surface noise, one can obtain the reduced  quantum dynamics of the particle by tracing out the field dynamics \cite{breuer2002}. For realistic experimental situations, the weak-coupling Born-Markov approximation is justified (Sec.~\ref{sec:adequacy}). The resulting quantum master equation depends on the relation of the involved  timescales. We will start with the {\it slow-particle limit} before discussing the case of {\it resonant motion}.

\begin{center}
\begin{table*}
\caption{The resonant surface-fluctuation kernel for particles in front of a homogeneous planar surface (top) and a planar surface covered by a dielectric layer (bottom). For the latter case we assume the particle to be at a large distance from the surface, $\bls{r}\cdot\bls{e}_3,\bls{r}'\cdot\bls{e}_3\ll d_{\rm s}$, as described in Sec.~\ref{sec:specific_dielectrics}.  In both expressions $\bls{r}$ and $\bls{r}'$ are evaluated in vacuum, and the mirror tensor $\mathsf{M}=\mathbb{1}-2\bls{e}_3\otimes\bls{e}_3$ serves to reflect a vector at the surface plane.}
\bgroup
\begin{tabular}{l l}
  Illustration &  \multicolumn{1}{c}{Surface-fluctuation kernel}  \\ \toprule
 \toprule
  \begin{minipage}{.16\textwidth}
      \includegraphics[width=30mm, height=30mm]{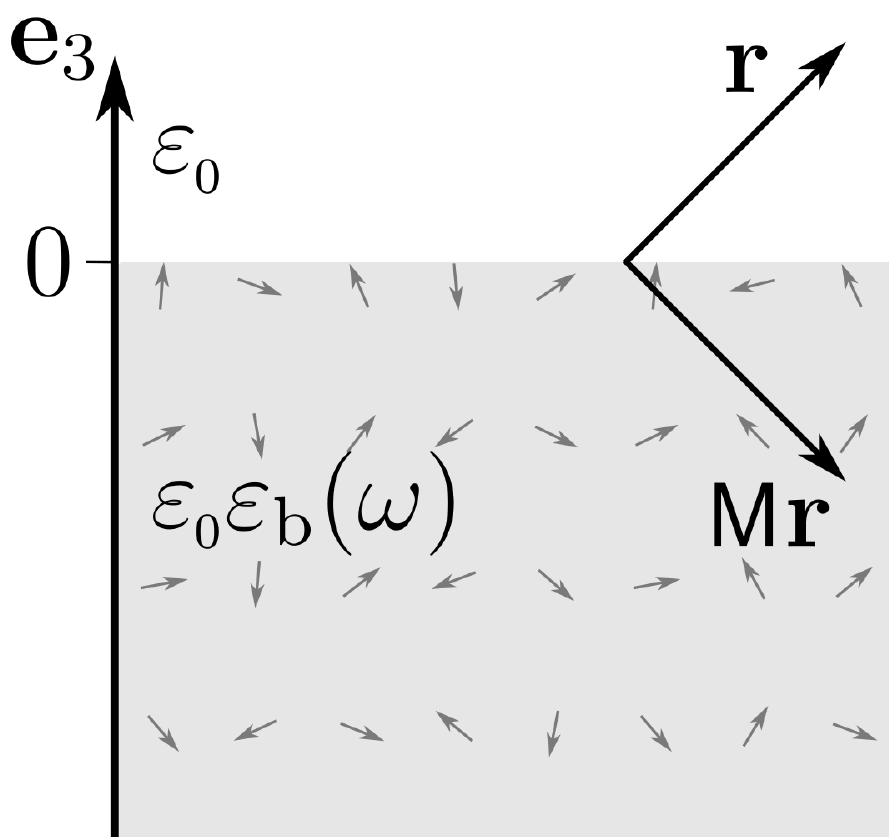}
    \end{minipage} & $\begin{aligned}&{\rm Im}\left[g(\bls{r},\bls{r}',\omega)\right]=-\frac{2}{4\pi\varepsilon_0}\frac{{\rm Im\left[\varepsilon_{\rm b}(\omega)\right]}}{|\varepsilon_{\rm b}(\omega)+1|^2}\frac{1}{|\bls{r}-\mathsf{M}\bls{r}'|}
    \end{aligned}$  \\
\midrule
 \begin{minipage}{.16\textwidth}
      \includegraphics[width=30mm, height=30mm]{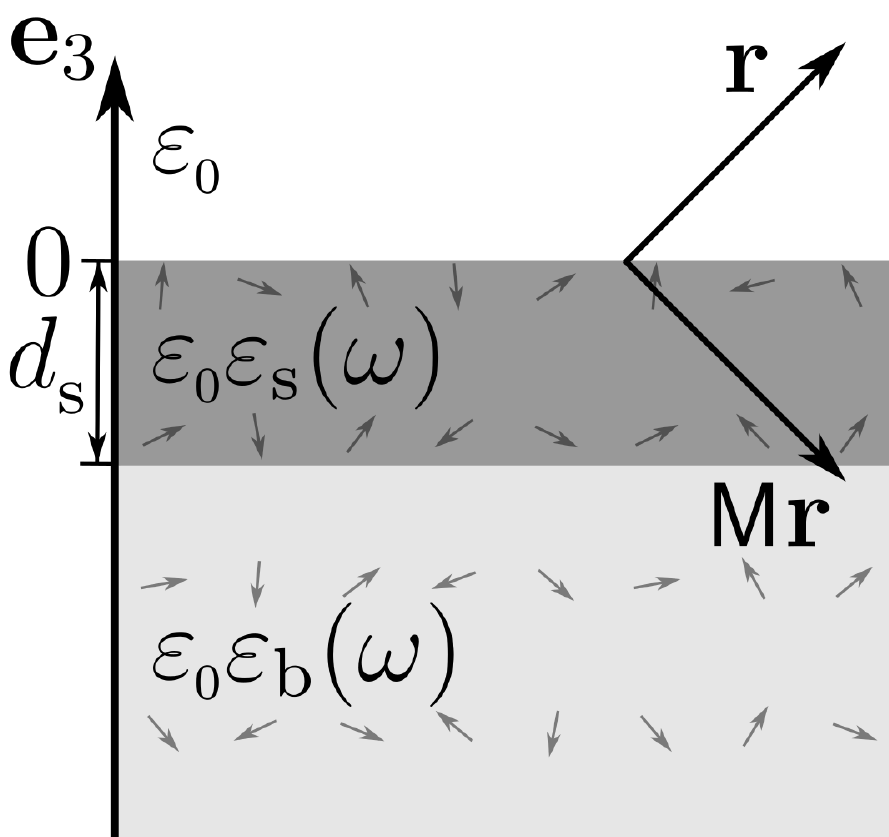}
    \end{minipage} & $\begin{aligned}
     {\rm Im}\left[g(\bls{r},\bls{r}',\omega)\right]=& \frac{2d_{\rm s}}{4\pi\varepsilon_0}
    {\rm Im}\left[\frac{\varepsilon_{\rm s}^2(\omega)-\varepsilon_{\rm b}^2(\omega)}{\varepsilon_{\rm s}(\omega)\left[\varepsilon_{\rm b}(\omega)+1\right]^2}\right]\bls{e}_3\cdot\frac{\partial}{\partial \bls{r}}\frac{1}{|\bls{r}-\mathsf{M}\bls{r}'|}\\
    &-\frac{2}{4\pi\varepsilon_0}\frac{{\rm Im}\left[\varepsilon_{\rm b}(\omega)\right]}{|\varepsilon_{\rm b}(\omega)+1|^2}\frac{1}{|\bls{r}-\mathsf{M}\bls{r}'|}
    \end{aligned}$  \\
\bottomrule
\bottomrule
\end{tabular}
\egroup
\label{tab:dielectrics}
\end{table*}
\end{center}

\subsection{Slow particle motion}

In the slow-particle limit, the surface-free particle dynamics due to the Hamiltonian $H_0$ and the surface-induced particle dynamics are much slower than the surface-fluctuation correlation time. As shown in App.~\ref{sec:derivationSlowParticle}, the resulting master equation 
\begin{align}\label{eq:MasterEquationUnspecifiedSlowParticle}
    \frac{\partial \rho}{\partial t}=-\frac{i}{\hbar}\left[ H_0 + H_{\rm si},\rho\right]+\mathcal{L}\rho
\end{align}
takes Lindblad form
\begin{align}
\label{eq:Lsp}
    \mathcal{L}\rho=&\frac{\varepsilon_0}{\hbar}\lim_{\omega\downarrow 0}\frac{2k_{\rm B}T}{\hbar \omega}\int d^3r\,\frac{{\rm Im}[\varepsilon_{\rm r}(\bls{r},\omega)]}{|\varepsilon_{\rm r}(\bls{r},\omega)|^2}\Big(\bls{L}_{\rm e}(\bls{r})\cdot \rho \bls{L}^\dagger_{\rm e}(\bls{r})\nonumber\\
    &-\frac{1}{2}\left\{\bls{L}^\dagger_{\rm e}(\bls{r})\cdot \bls{L}_{\rm e}(\bls{r}),\rho\right\}\Big)
\end{align}
with Lindblad operators
\begin{align}\label{eq:lbo}
    \bls{L}_{\rm e}(\bls{r})&=\int d^3s \,\varrho_0\!\left[\mathsf{R}_\Omega^{\rm T}\left(\bls{s}-\bls{R}\right)\right] \lim_{\omega\downarrow 0}\varepsilon_{\rm r}(\bls{r},\omega)\frac{\partial}{\partial \bls{r}} g(\bls{r},\bls{s},\omega).
\end{align}
The limit is taken from above because ${\rm Im}[\varepsilon_r({\bf r},\omega)]$ can exhibit a discontinuity at $\omega = 0$ due to the Schwarz reflection principle. The surface-interaction Hamiltonian $H_{\rm si}$ describes the coherent interaction with image charges in the dielectric,
\begin{align}
H_{\rm si} =& \frac{1}{2}\int d^3rd^3r'\,g(\bls{r},\bls{r}',0)\varrho_0\!\left[\mathsf{R}_\Omega^{\rm T}(\bls{r}-\bls{R})\right]\nonumber \\
&\times \varrho_0\!\left[\mathsf{R}_\Omega^{\rm T}(\bls{r}'-\bls{R})\right].
\end{align}

Equations \eqref{eq:MasterEquationUnspecifiedSlowParticle}--\eqref{eq:lbo} describe ro-translational decoherence and heating in terms of (i) the dielectric properties of the surface material, (ii) the surface geometry, and (iii) the particle charge distribution. As one expects, only the dielectric regions of space, where ${\rm Im}[\varepsilon_{\rm r}({\bf r},\omega)]>0$, contribute to the integral in \eqref{eq:Lsp}. Note that the form of Eq.~\eqref{eq:lbo} implies that the Lindblad operators correspond to the (negative) displacement field induced by the particle charge distribution located at position ${\bf R}$ and orientation $\Omega$.

Using relation \eqref{eq:relationScalarGreenfunction}, one can disentangle the contributions of charge distribution and surface properties by showing that (App.~\ref{sec:derivationSlowParticle})
\begin{align}\label{eq:MasterEquationSlowParticleIntro}
    \mathcal{L}\rho=&\frac{2}{\hbar}\int d^3rd^3r'\,h(\bls{r},\bls{r}')\nonumber\\
    &\times\bigg(\varrho_0\!\left[\mathsf{R}_\Omega^{\rm T}\left(\bls{r}-\bsf{R}\right)\right]\rho\varrho_0\!\left[\mathsf{R}_\Omega^{\rm T}\left(\bls{r}'-\bsf{R}\right)\right]\nonumber\\
    &-\frac{1}{2}\left\{\varrho_0\!\left[\mathsf{R}_\Omega^{\rm T}\left(\bls{r}'-\bsf{R}\right)\right]\varrho_0\!\left[\mathsf{R}_\Omega^{\rm T}\left(\bls{r}-\bsf{R}\right)\right],\rho\right\}\bigg).
\end{align}
Here, the surface geometry and response function enter through the {\it surface-fluctuation kernel}
\begin{equation}\label{eq:12}
    h(\bls{r},\bls{r}')=-\lim_{\omega\downarrow 0}n(\omega){\rm Im}\left[g(\bls{r},\bls{r}',\omega)\right].
\end{equation}

The particle charge distribution takes the role of the Lindblad operators, leading to a localization in position and orientation. The decoherence rate in this basis, defined via 
\begin{align}\label{eq:definingEquationDecoherenceRate}
    \bra{\bls{R},\Omega}&\mathcal{L}\rho\ket{\bls{R}',\Omega'}\nonumber\\
    &=-\Gamma_{\Omega\Omega'}(\bls{R},\bls{R}')\bra{\bls{R},\Omega}\rho\ket{\bls{R}',\Omega'},
\end{align}
takes the form
\begin{align}\label{eq:DecoRateSlowParticle}
    \Gamma_{\Omega\,\Omega'}(\bls{R},\bls{R}')=&\frac{1}{\hbar}\int d^3r d^3r'\,h(\bls{r},\bls{r}')\Big(\varrho_0\!\left[\mathsf{R}_{\Omega}^{\rm T}(\bls{r}-\bls{R})\right]\nonumber\\
    &-\varrho_0\!\left[\mathsf{R}_{\Omega'}^{\rm T}(\bls{r}-\bls{R}')\right]\Big)\Big(\varrho_0\!\left[\mathsf{R}_{\Omega}^{\rm T}(\bls{r}'-\bls{R})\right]\nonumber\\
    &-\varrho_0\!\left[\mathsf{R}_{\Omega'}^{\rm T}(\bls{r}'-\bls{R}')\right]\Big),
\end{align}
where we used $g(\bls{r},\bls{r}',\omega)=g(\bls{r}',\bls{r},\omega)$ (see App.~\ref{sec:symmetryOfGreenFunction}).

The decoherence rate is non-negative, as follows from \eqref{eq:relationScalarGreenfunction}, it vanishes for ${\bf R} = {\bf R}'$ and $\Omega = \Omega'$, and remains bounded for large lateral superposition sizes. It thus describes how superposition states of the charged particle gradually lose coherence due to their interactions with surface charge fluctuations. 

The Lindblad operators are specified in Sec.~\ref{sec:specific_charge_distributions} for common charge distributions, while the form of the surface-fluctuation kernel (\ref{eq:12}) is discussed in Sec.~\ref{sec:specific_dielectrics} for typical surface geometries and generic material responses. Table \ref{tab:MasterEquations} summarizes the master equations for particles characterized by their first multipole moments and for typical types of particle motion. Table \ref{tab:dielectrics} gives the surface-fluctuation kernel for a half-space and a layered surface geometry.

\subsection{Resonant motion}

The resonant limit applies if both the free particle motion and the decay of surface-fluctuation correlations are much faster than the surface-induced particle dynamics. The resulting master equation can be stated by diagonalizing the surface-free particle dynamics,
\begin{eqnarray} \label{eq:h0}
    H_0&=&\frac{\bsf{P}^2}{2m}+\frac{1}{2}\bsf{J}\cdot {\rm I}_\Omega^{-1}\bsf{J}+V_{\rm ext}(\bls{R},\Omega)\nonumber
    \\ &=& \sum_n E_n |\Psi_n\rangle \langle \Psi_n|,
\end{eqnarray}
including a possible external trapping potential $V_{\rm ext}$. This allows expanding the particle charge density by means of the operators
\begin{align}\label{eq:eigenoperatorsExplicitly1}
    c_\ell(\bls{r})=&\sum_{\substack{n,m \\E_m-E_n=\hbar\omega_\ell>0}}c_{nm}(\bls{r})\ket{\Psi_n}\bra{\Psi_m},
\end{align}
with coefficients
\begin{equation}\label{eq:definitionCnm}
    c_{nm}(\bls{r})=\bra{\Psi_n}\varrho_0\!\left[\mathsf{R}_\Omega^{\rm T}(\bls{r}-\bls{R})\right]\ket{\Psi_m},
\end{equation}  
where $\omega_\ell$ labels the level spacings of $H_0$.

A rotating wave approximation of the interaction Hamiltonian with respect to the particle dynamics yields a dissipator of Lindblad form  (App.~\ref{sec:DerivationResonantLimit}),
\begin{align}\label{eq:LindbladDissipatorResonantLimitLindbladForm}
    &\mathcal{L}\rho=\frac{\varepsilon_0}{\hbar}\sum_{\omega_\ell>0}\int d^3r\,\frac{{\rm Im}\left[\varepsilon_{\rm r}(\bls{r},\omega_\ell)\right]}{|\varepsilon_{\rm r}(\bls{r},\omega_\ell)|^2}\nonumber\\
    &\times\left[ \left[n(\omega_\ell)+1\right]\left(\bsf{L}^{\rm e}_\ell(\bls{r})\cdot\rho{\bsf{L}_\ell^{\rm e}}^\dagger(\bls{r})-\frac{1}{2}\left\{{\bsf{L}_\ell^{\rm e}}^\dagger(\bls{r})\cdot\bsf{L}^{\rm e}_\ell(\bls{r}),\rho\right\}\right)\right.\nonumber\\
    &+\left.n(\omega_\ell)\left({\bsf{L}_\ell^{\rm e}}^\dagger(\bls{r})\cdot\rho\bsf{L}^{\rm e}_\ell(\bls{r})-\frac{1}{2}\left\{\bsf{L}^{\rm e}_\ell(\bls{r})\cdot{\bsf{L}_\ell^{\rm e}}^\dagger(\bls{r}),\rho\right\}\right)\right],
\end{align}
with
\begin{align}
    \bsf{L}^{\rm e}_\ell(\bls{r})=\sqrt{2}\int d^3s\, c_\ell(\bls{s}) \varepsilon_{\rm r}(\bls{r},\omega_\ell)\frac{\partial}{\partial\bls{r}}g(\bls{r},\bls{s},\omega_\ell).
\end{align}
The Lindblad operators thus describe the displacement field smeared out over the field-free motion of the charged particle.

Only  the dielectric regions contribute to the integral in the master equation \eqref{eq:LindbladDissipatorResonantLimitLindbladForm}. The latter describes resonant transitions between the particle motion \eqref{eq:h0} and the dielectric, as weighted by the energy loss function ${\rm Im}[\varepsilon({\bf r},\omega_\ell)]/|\varepsilon({\bf r},\omega_\ell)|^2$. The first term in \eqref{eq:LindbladDissipatorResonantLimitLindbladForm} accounts for spontaneous and stimulated de-excitation of the particle motion; the second term accounts for surface-induced excitations of the particle dynamics. This leads to heating, damping, and thermalization of the particle motion with the surface temperature.

The associated surface-interaction Hamiltonian (Lamb shift) reads in the zero-temperature limit
\begin{align}\label{eq:LambShiftResonantQuasistaticTzero}
    H_{\rm si}=&\frac{1}{2}\int d^3rd^3r'\,\sum_{\omega_\ell>0}\left({\rm Re}\left[g(\bls{r},\bls{r}',\omega_\ell)\right]-\frac{1}{4\pi\varepsilon_0}\frac{1}{|\bls{r}-\bls{r}'|}\right)\nonumber\\
    &\times\left\{c_\ell^\dagger(\bls{r}),c_\ell(\bls{r}')\right\}+\int d^3rd^3r'\,\sum_{\omega_\ell>0}\frac{\omega_\ell}{\pi}\mathsf{P}\int_0^\infty d\omega\,\nonumber\\
    &\times\frac{{\rm Im}\left[g(\bls{r},\bls{r}',\omega)\right]}{\omega^2-\omega_\ell^2}\left[c_\ell^\dagger(\bls{r}),c_\ell(\bls{r}')\right]
    .
\end{align}
It contains two contributions: The first accounts for the interaction of the particle charge distribution with the particle-induced surface polarization -- i.e. the image-charge potential (see App.~\ref{sec:DerivationResonantLimit}). The second contribution describes modifications of the particle dynamics induced by zero-point fluctuations in the surface, similar to the Casimir-Polder potential. 

One can again disentangle the contribution of particle and surface properties  by using Eq.~\eqref{eq:relationScalarGreenfunction}. This yields
\begin{align}\label{eq:dissipatorQuasistatic}
    \mathcal{L}\rho=&-\frac{2}{\hbar}\int d^3rd^3r'\,\sum_{\omega_\ell>0}{\rm Im}\left[g(\bls{r},\bls{r}',\omega_\ell)\right]\nonumber\\
    &\times\left[\left[n(\omega_\ell)+1\right]\left({c}_\ell(\bls{r}')\rho{c}^\dagger_\ell(\bls{r})-\frac{1}{2}\left\{{c}^\dagger_\ell(\bls{r}){c}_\ell(\bls{r}'),\rho\right\}\right)\right.\nonumber\\
    &+\left. n(\omega_\ell)\left(c_\ell^\dagger(\bls{r})\rho{c}_\ell(\bls{r}')-\frac{1}{2}\left\{{c}_\ell(\bls{r}'){c}^\dagger_\ell(\bls{r}),\rho\right\}\right)\right],
\end{align}
where the surface properties now enter through the imaginary part of the Green function in \eqref{eq:DGLscalarGreenfunction}, evaluated on resonance with the particle motion. The Lindblad operators $c_\ell({\bf r})$ are fully specified by the particle charge distribution $\varrho_0$ and the field-free Hamiltonian $H_0$, implying that the integrands in \eqref{eq:dissipatorQuasistatic} are localized within the smeared-out charge distribution.

The form of these  Lindblad  operators  is discussed in Sec.~\ref{sec:specific_charge_distributions} for typical charge distributions. The {\it resonant surface-fluctuation kernel} Im$[g(\bls{r},\bls{r}',\omega_\ell)]$  of typical surface geometries and generic material responses is then specified in Sec. \ref{sec:specific_dielectrics}, and Sec.~\ref{Sec:decoherenceAndHeating} discusses the resulting heating and decoherence rates for state-of-the-art setups. Master equations for typical charge distributions and types of motion are illustrated in Tab.~\ref{tab:MasterEquations}. The resonant surface-fluctuation kernel is given in Tab.~\ref{tab:dielectrics} for a half-space and a layered surface geometry.

\section{Lindblad operators for charge multipoles}\label{sec:specific_charge_distributions}

In the master equations \eqref{eq:MasterEquationSlowParticleIntro} and \eqref{eq:dissipatorQuasistatic}  the particle properties, entering through the Lindblad opertors, are separated from the dielectric surface, entering through the surface-fluctuation kernel. 
In the slow particle limit, only the instantaneous charge distribution appears, while  the resonant limit blurs the charge  over the fast particle motion, as described by the operators ${c}_\ell(\bls{s})$. This section gives the Lindblad operators for typical experimental setups. The resulting master equations are provided in Tab.~\ref{tab:MasterEquations}.

\subsection{Slow particle motion}

{\it a.\ Slow monopole---}The Lindblad operator of a point monopole with total charge $q$ at position $\bls{r}$ is independent of the particle orientation,
\begin{equation}\label{eq:c0OperatorSlowMonopole}
    \varrho_0[\mathsf{R}_\Omega^{\rm T}(\bls{r}-\bsf{R})]=q\delta\!\left(\bls{r}-\bls{R}\right).
\end{equation}

{\it b.\ Slow dipole---}A point particle with a body-fixed dipole moment $\bls{p}_0$, but no other multipoles, is characterized by $\varrho_0(\bls{r})=-\bls{p}_0\cdot\partial _\bls{r}\delta(\bls{r})$, yielding the Lindblad operator
\begin{equation}\label{eq:c0OperatorSlowDipole}
    \varrho_0[\mathsf{R}_\Omega^{\rm T}(\bls{r}-\bsf{R})]=-\left(\mathsf{R}_\Omega\bls{p}_0\right)\cdot\frac{\partial}{\partial \bls{r}}\delta\!\left(\bls{r}-\bls{R}\right),
\end{equation}
which involves the orientation state-dependent dipole vector $\bls{p}_\Omega=\mathsf{R}_\Omega\bls{p}_0$. 
After inserting \eqref{eq:c0OperatorSlowDipole} into the master equation \eqref{eq:MasterEquationSlowParticleIntro}, the  spatial derivatives of the $\delta$-distributions can be shifted to $h(\bls{r},\bls{r}')$ by partial integration. 

{\it c.\ Slow quadrupole---}A point particle with a body fixed quadrupole tensor $\mathsf{Q}_0$, but no other multipoles, is characterized by $\varrho_0(\bls{r})=\partial_\bls{r}\cdot\mathsf{Q}_0\partial_\bls{r}\delta(\bls{r})/6$, leading to the Lindblad operators
\begin{equation}\label{eq:c0OperatorSlowQuadrupole}
    \varrho_0[\mathsf{R}_\Omega^{\rm T}(\bls{r}-\bsf{R})]=\frac{1}{6}\frac{\partial}{\partial \bls{r}}\cdot\mathsf{R}_\Omega\mathsf{Q}_0\mathsf{R}_\Omega^{\rm T}\frac{\partial}{\partial \bls{r}}\delta\!\left(\bls{r}-\bls{R}\right).
\end{equation}
Similar to the case of a slow dipole, the Liouvillan \eqref{eq:MasterEquationSlowParticleIntro} depends on the orientation state-dependent quadrupole tensor $\mathsf{Q}_\Omega=\mathsf{R}_\Omega\mathsf{Q}_0\mathsf{R}_\Omega^{\rm T}$ and the Hessian of $h(\bls{r},\bls{r}')$ at the particle position.

{\it General charge distribution---}The dissipator for a particle with a charge distribution characterized by several multipole moments is not simply obtained by adding the multipole dissipators. The Lindblad operator in (\ref{eq:MasterEquationSlowParticleIntro}) is given by the operator-valued charge distribution, implying that the master equation contains cross-terms involving different multipole moments. 

\subsection{Resonant motion}

{\it d.\ Oscillating monopole---}The quantum dynamics of an oscillating monopole at position $\bls{R}=\bls{R}_{\rm eq}$ can be decomposed into three bosonic modes with operators $a_\ell$, frequencies $\omega_\ell$, and oscillation directions $\blg{\epsilon}_\ell$, 
\begin{equation}\label{eq:RvectorResonantOScillatingMonopole}
    U_0^\dagger(t)\bls{R}U_0(t)=\bls{R}_{\rm eq}+\sum_{\ell={1}}^3\sqrt{\frac{\hbar}{2m\omega_\ell}}\left(a_\ell e^{-i\omega_\ell t}+{\rm h.c.}\right)\blg{\epsilon}_\ell.
\end{equation}
If the frequencies are sufficiently well separated, so that the rotating wave approximation applies, and for  amplitudes sufficiently small, so that field can be approximated locally, the Lindblad operators take on the form (App.~\ref{sec:DerivationResonantLimit})
\begin{equation}\label{eq:resonantMonopolecl}
    c_\ell(\bls{r})=-q\sqrt{\frac{\hbar}{2m\omega_\ell}}\blg{\epsilon}_\ell\cdot\frac{\partial}{\partial\bls{r}}\delta(\bls{r}-\bls{R}_{\rm eq})a_\ell.
\end{equation}

{\it e.\ Oscillating dipole---}The corresponding calculation for an oriented dipole with time-independent dipole vector $\bls{p}$ whose center of mass is oscillating, yields the Lindblad operators 
\begin{align}
    c_\ell(\bls{r})=&\sqrt{\frac{\hbar}{2m\omega_\ell}}\left(\bls{p}\cdot\frac{\partial}{\partial \bls{r}}\right)\left(\blg{\epsilon}_\ell\cdot\frac{\partial}{\partial \bls{r}}\right)\delta(\bls{r}-\bls{R}_{\rm eq})a_\ell.
\end{align}

{\it f.\ Oscillating quadrupole---}Likewise, for an oscillating oriented quadrupole with time-independent quadrupole tensor ${\mathsf Q}$, one obtains
\begin{align}
    c_\ell(\bls{r})=&-\frac{1}{6}\sqrt{\frac{\hbar}{2m\omega_\ell}}\!\left(\frac{\partial}{\partial \bls{r}}\cdot\mathsf{Q}\frac{\partial}{\partial \bls{r}}\right)\!\left(\blg{\epsilon}_\ell\cdot\frac{\partial}{\partial \bls{r}}\right)\!\delta(\bls{r}-\bls{R}_{\rm eq})a_\ell.
\end{align}

{\it g.\ Librating dipole---}A linear rotor with moment of inertia $I$ and dipole operator $\bls{p}_\Omega=p\bls{n}_3$ parallel to the symmetry axis $\bls{n}_3$ is assumed to librate at $\bls{R}_{\rm cm}$ in the two directions $\blg{\epsilon}_\alpha$, $\blg{\epsilon}_\beta$ perpendicular to its equilibrium orientation $\blg{\epsilon}_{\rm eq}$. Denoting the  frequencies of the libration modes by $\omega_\alpha$, $\omega_\beta$ and  the bosonic operators  by $a_\alpha$, $a_\beta$, the rotor symmetry axis evolves as
\begin{align}\label{eq:libratingAxis}
    U_0^\dagger(t)\bls{n}_3U_0(t)=\blg{\epsilon_{\rm eq}}+\sum_{\ell=\alpha,\beta}\sqrt{\frac{\hbar}{2I\omega_\ell}}\left(a_\ell e^{-i\omega_\ell t}+{\rm h.c.}\right)\blg{\epsilon}_\ell.
\end{align}
For sufficiently well-separated libration frequencies, the Lindblad operators take the form
\begin{align}
    c_\ell(\bls{r})=-p\sqrt{\frac{\hbar}{2I\omega_\ell}}\blg{\epsilon}_\ell\cdot\frac{\partial}{\partial\bls{r}}\delta(\bls{r}-\bls{R}_{\rm cm})a_\ell.
\end{align}

{\it h.\ Librating quadrupole---}Likewise, if the linear rotor is a quadrupole $\mathsf{Q}_\Omega=Q_{33}\left[3\bls{n}_3\otimes\bls{n}_3-\mathbb{1}\right]/2$ with symmetry axis $\bls{n}_3$, librating as in \eqref{eq:libratingAxis}, the Lindblad operators are given by
\begin{align}
    c_\ell(\bls{r})=&\frac{Q_{33}}{2}\sqrt{\frac{\hbar}{2I\omega_\ell}}\left(\blg{\epsilon}_{\rm eq}\cdot\frac{\partial}{\partial \bls{r}}\right)\left(\blg{\epsilon}_\ell\cdot\frac{\partial}{\partial \bls{r}}\right)\delta(\bls{r}-\bls{R}_{\rm cm})a_\ell.
\end{align}

{\it i.\ Rotating planar dipole---}A linear rotor located at $\bls{R}_{\rm cm}$ is assumed to rotate in a plane spanned by $\blg{\epsilon}_1$ and $\blg{\epsilon}_2$. For sufficiently sharp angular momentum $p_\alpha$ the symmetry axis $\bls{n}_3$ rotates approximately at a frequency $\omega_{\rm rot}=p_\alpha/I$ determined by the moment of inertia $I$,
\begin{align}
   U_0^\dagger(t)\bls{n}_3U_0(t) \simeq& \frac{1}{2}e^{-i\omega_{\rm rot} t}e^{-i\alpha}(\blg{\epsilon}_1+i\blg{\epsilon}_2)+{\rm h.c}.,\label{eq:rotatingAxis}
\end{align}
with Euler-angle operator $\alpha$. For a dipolar charge distribution $\bls{p}_\Omega=p\bls{n}_3$ with symmetry axis $\bls{n}_3$ the Lindblad operators read as
\begin{equation}
    c_{\rm rot}(\bls{r})=-\frac{p}{2}(\blg{\epsilon}_1+i\blg{\epsilon}_2)\cdot\frac{\partial}{\partial\bls{r}}\delta(\bls{r}-\bls{R}_{\rm cm})e^{-i\alpha}.
\end{equation}

{\it j.\ Rotating planar quadrupole---}Likewise, for a  quadrupolar charge distribution with quadrupole tensor $\mathsf{Q}_\Omega=Q_{33}\left[3\bls{n}_3\otimes\bls{n}_3-\mathbb{1}\right]/2$ rotating as in  \eqref{eq:rotatingAxis} one obtains the Lindblad operators
\begin{align}
    c_{\rm rot}(\bls{r})=&\frac{Q_{33}}{16} \left[(\blg{\epsilon}_1+i\blg{\epsilon}_2)\cdot\frac{\partial}{\partial\bls{r}}\right]^2\delta(\bls{r}-\bls{R}_{\rm cm})e^{-2i\alpha}.
\end{align}

{\it k.\ Freely rotating dipole---}Finally, we derive the Lindblad operators 
of a freely rotating linear rotor at a position $\bls{R}_{\rm cm}$ with the dipole operator specified by the Euler angle operators $\alpha$ and $\beta$,
\begin{align}
    p\bls{n}_3=p\left(\sin\beta\cos\alpha\blg{\epsilon}_1+\sin\beta\sin\alpha\blg{\epsilon}_2+\cos\beta\blg{\epsilon}_3\right).
\end{align}
The energy eigenstates  entering \eqref{eq:eigenoperatorsExplicitly1} and \eqref{eq:definitionCnm} are given by the spherical harmonics $\braket{\alpha,\beta|\ell,m}=Y_{\ell m}(\beta,\alpha)$, with associated energies $E_\ell=\hbar^2\ell(\ell+1)/2I$. Expanding the dipole operator in the angular momentum basis 
\begin{align}\label{eq:definitionPllmm}
    \bra{\ell,m}&p\bls{n}_3\ket{\ell'm'}=p\sum_{i=1}^3n^{(i)}_{\ell m\ell'm'}\blg{\epsilon}_i
\end{align}
yields
\begin{subequations}\label{eq:plmlm}
\begin{align}
    n^{(1)}_{\ell m \ell' m'}=&(-1)^{m}\frac{1}{\sqrt{2}}\sqrt{(2\ell+1)(2\ell'+1)}\left(\begin{array}{rrr}\ell & \ell' & 1 \\ 0 & 0 & 0 \\\end{array}\right)\nonumber\\
    &\times\left[\left(\begin{array}{rrr}\ell & \ell' & 1 \\ -m & m' & -1 \\\end{array}\right)-\left(\begin{array}{rrr}\ell & \ell' & 1 \\ -m & m' & 1 \\\end{array}\right)\right],\\
    n^{(2)}_{\ell m \ell' m'}=&(-1)^{m}\frac{i}{\sqrt{2}}\sqrt{(2\ell+1)(2\ell'+1)}\left(\begin{array}{rrr}\ell & \ell' & 1 \\ 0 & 0 & 0 \\\end{array}\right)\nonumber\\
    &\times\left[\left(\begin{array}{rrr}\ell & \ell' & 1 \\ -m & m' & -1 \\\end{array}\right)+\left(\begin{array}{rrr}\ell & \ell' & 1 \\ -m & m' & 1 \\\end{array}\right)\right],\\
    n^{(3)}_{\ell m \ell' m'}=&(-1)^{m}\sqrt{(2\ell+1)(2\ell'+1)}\left(\begin{array}{rrr}\ell & \ell' & 1 \\ 0 & 0 & 0 \\\end{array}\right)\nonumber\\
    &\times\left(\begin{array}{rrr}\ell & \ell' & 1 \\ -m & m' & 0 \\\end{array}\right).
\end{align}
\end{subequations}
The expansion coefficients \eqref{eq:definitionCnm} thus depend on four quantum numbers,
\begin{align}
    c_{\ell m \ell' m'}(\bls{r})=&\bra{\ell,m}\varrho_0\!\left[\mathsf{R}_\Omega^{\rm T}(\bls{r}-\bls{R}_{\rm cm})\right]\ket{\ell'm'}\nonumber\\
    =&-p\frac{\partial}{\partial\bls{r}}\delta(\bls{r}-\bls{R}_{\rm cm})\cdot\sum_{i=1}^3n^{(i)}_{\ell m\ell'm'}\blg{\epsilon}_i,
\end{align}
which are nonzero only for $|\ell'-\ell|=1$ and $m-m'=\pm 1,0$ due to the  Wigner 3-j symbols in (\ref{eq:plmlm}). Since \eqref{eq:eigenoperatorsExplicitly1} is restricted to positive energy spacings, only the Lindblad operators
\begin{equation}\label{eq:clFreeRotation}
    c_\ell(\bls{r})=\sum_{m,m'}c_{\ell\,m\,\ell+1\,m'}\ket{\ell,m}\bra{\ell+1,m'},
\end{equation}
with frequencies $\omega_\ell : =(E_{\ell'}-E_\ell)/\hbar=\hbar(\ell+1)/I$ contribute. We will use the operators \eqref{eq:clFreeRotation} in Sec.~\ref{sec:decoherenceHybridQuantumDevices} to simulate the rotational decoherence of a dipole above a superconducting surface covered with a dielectric layer.

{\it General charge distribution---}The dissipator for  a charge distribution characterized by several multipole moments does in general not decompose into a sum of pure multipole dissipators. However, this sum may be a good approximation if the characteristic frequencies associated with the motion of the distinct multipole components are sufficiently well separated, so that all cross-terms effectively average to zero.

\section{Surface-fluctuation kernel}\label{sec:specific_dielectrics}

Having specified the Lindblad operators, we now discuss the surface-fluctuation kernel, which depends on the geometry of the surface and its dielectric response,
as characterized by the relative permittivity $\varepsilon_{\rm r}({\bf r},\omega)$. The latter is subject to the Kramers-Kronig relations, implying that high frequency resonances described by peaks in the imaginary part of $\varepsilon_{\rm r}({\bf r},\omega)$ contribute to the low frequency behaviour of the real part (Fig.~\ref{figure2}). As we will see below, even weak resonances due to surface-adsorbed molecules and imperfections can strongly affect the energy loss function at the typical trapping frequencies of ions and nanoparticles.

A simple model for the dielectric response is given by a sum of Drude-Lorentz oscillators \cite{jackson1999}
\begin{equation}\label{eq:genDL}
    \varepsilon_{\rm DL}(\omega)=1+\sum_{n\geq 0} \frac{f_n \omega_n^2}{\omega_n^2-\omega^2-i\gamma_n\omega}.
\end{equation}
The imaginary part exhibits peaks at the resonance frequencies $\omega_n$, whose relative weights are determined by the (real-valued) coefficients $f_n$, and whose widths are determined by the damping rates $\gamma_n$. The special case of a single resonance at $\omega_{\rm n}=\omega_{\rm r}$ with $f_n = \omega_{\rm p}^2/\omega_r^2$ yields the Drude-Lorentz model for a single oscillator
\begin{equation} \label{eq:DrudeLorentzDielectric}
    \varepsilon_{\rm sDL}(\omega)=1+\frac{\omega_{\rm p}^2}{\omega_{\rm r}^2-\omega^2-i\gamma\omega}.
\end{equation}
In the limit that the resonance frequency vanishes, this is referred to as the \emph{Drude metal},
\begin{equation}\label{eq:responseDrudeMetal}
    \varepsilon_{\rm m}(\omega)=1-\frac{\omega_{\rm pm}^2}{\omega^2+i\gamma_{\rm m}\omega}.
\end{equation}

We will also discuss decoherence close to a superconducting surface, with critical temperature $T_{\rm c}$, at temperature $T<T_{\rm c}$. Its permittivity can be written in the two-fluid model as \cite{london1935electromagnetic,hohenester2007spin}
\begin{equation}\label{eq:responseSuperconductor}
    \varepsilon_{\rm sc}(\omega)=1-\frac{\omega_{\rm pm}^2}{\omega^2+i\gamma_{\rm m}\omega}\left(\frac{T}{T_{\rm c}}\right)^4-\frac{1}{\omega^2\Lambda_0\varepsilon_0}\left[1-\left(\frac{T}{T_{\rm c}}\right)^4\right],
\end{equation}
where $\Lambda_0$ is a material-specific constant. For temperatures $T$ approaching $T_{\rm c}$ from below, this superconductor turns into a normal metal. At much smaller temperatures it behaves like an ideal electron gas without damping.

\subsection{Homogeneous half-space}

We consider a particle levitated in vacuum at $\bls{r}\cdot\bls{e}_3>0$ above a semi-infinite dielectric half-space at $\bls{r}\cdot\bls{e}_3<0$ with homogeneous permittivity $\varepsilon_{\rm b}(\omega)$. The Green function (\ref{eq:DGLscalarGreenfunction}) for this configuration can be found in App.~\ref{sec:derivationGHalfSpace}. For the special case that $\bls{r}$ and $\bls{r}'$ are both in vacuum, as appearing in the master equations \eqref{eq:MasterEquationSlowParticleIntro} and \eqref{eq:dissipatorQuasistatic}, it is given in Tab.~\ref{tab:dielectrics}. The imaginary part of the Green function decays with the inverse distance between the location ${\bf r}$ and the mirror image of ${\bf r}'$ at the location ${\sf M}{\bf r}'$, with the mirror tensor ${\sf M} = \mathbb{1} - 2{\bf e}_3 \otimes {\bf e}_3$. In the special case of the Drude-Lorentz model \eqref{eq:genDL} one obtains
\begin{align}\label{eq:fFunctLorentzDrude}
 h(\bls{r},\bls{r}')=\frac{1}{4\pi\varepsilon_0}\frac{2 k_{\rm B}T}{\hbar}\frac{1}{|\bls{r}-\mathsf{M}\bls{r}'|}\frac{1}{[\varepsilon_{\rm DL}(0)+1]^2}\sum_{n\ge 0}\frac{f_n\gamma_n}{\omega_n^2},
\end{align}
and for the Drude-metal half-space
\begin{align}\label{eq:fFunctDrudeMetal}
 h(\bls{r},\bls{r}')=\frac{1}{4\pi\varepsilon_0}\frac{2k_{\rm B}T}{\hbar}\frac{1}{|\bls{r}-\mathsf{M}\bls{r}'|}\frac{\gamma_{\rm m} }{\omega_{\rm pm}^2}.
\end{align}
We will use this surface-fluctuation kernel to calculate below the decoherence rate of electron beams near a metallic surface.

For a superconductor, the surface-fluctuation kernel follows from \eqref{eq:responseSuperconductor}. In the static limit one obtains $h({\bf r},{\bf r}') = 0$ since $n(\omega){\rm Im}\left [2/\left[\varepsilon_{\rm sc}(\omega)+1\right]\right]$ vanishes. This means that there is no charge-induced decoherence of a slowly moving charge distribution in front of a pristine superconductor.

\subsection{Half-space covered by thin surface layer}

In typical experimental situations, the surface may be covered by a thin dielectric surface layer. The presence of this layer can strongly affect the charge-induced decoherence and heating, even though its impact on the overall dielectric response of the surface is rather weak. Denoting the layer permittivity by $\varepsilon_{\rm s}(\omega)$ and its thickness by $d_{\rm s}$, one finds the imaginary part of the Green function for ${\bf r}$ and ${\bf r}'$ in vacuum
 \begin{align}\label{eq:ImGreenFunctionLayer}
     {\rm Im}&\left[g(\bls{r},\bls{r}',\omega)\right]\nonumber\\
     =&\frac{1}{4\pi\varepsilon_0}\int_0^\infty dk\, {\rm Im}\left(\frac{\xi_{\rm b}(\omega) e^{-2kd_{\rm s}}-\xi_{\rm v}(\omega)}{1-\xi_{\rm v}(\omega)\xi_{\rm b}(\omega)e^{-2kd_{\rm s}}}\right) \nonumber\\&\times e^{-k(\bls{r}+\bls{r}')\cdot\bls{e}_3}J_0\left(k\sqrt{\left[(\bls{r}-\bls{r}')\cdot\bls{e}_1\right]^2+\left[(\bls{r}-\bls{r}')\cdot\bls{e}_2\right]^2}\right)
 \end{align}
where 
\begin{align}
\xi_{\rm b}(\omega)&=\frac{\varepsilon_{\rm s}(\omega)-\varepsilon_{\rm b}(\omega)}{\varepsilon_{\rm s}(\omega)+\varepsilon_{\rm b}(\omega)},\\
\xi_{\rm v}(\omega)&=\frac{\varepsilon_{\rm s}(\omega)-1}{\varepsilon_{\rm s}(\omega)+1}.
\end{align}
Here, $J_0(\cdot)$ denotes the zeroth order Bessel function. The full Green function is calculated in App.~\ref{sec:derivationGLayer}.

For large distances, $\bls{r}\cdot \bls{e}_3\gg d_{\rm s}$ and $\bls{r}'\cdot \bls{e}_3\gg d_{\rm s}$, Eq.~\eqref{eq:ImGreenFunctionLayer} can be expanded to first order in $kd_{\rm s}\ll 1$, yielding the expression given in Tab.~\ref{tab:dielectrics}. In case of a low-permittivity dielectric layer on a metallic (or superconducting) bulk, $|\varepsilon_{\rm b}(\omega)|\gg |\varepsilon_{\rm s}(\omega)| \gtrsim 1$, the metal (or superconductor) effectively acts as a mirror. One then obtains
\begin{align}\label{eq:thinLayerReflectingMetal}
    {\rm Im}&\left[g(\bls{r},\bls{r}',\omega)\right]=-\frac{2}{4\pi\varepsilon_0}\frac{{\rm Im}\left[\varepsilon_{\rm b}(\omega)\right]}{|\varepsilon_{\rm b}(\omega)|^2}\frac{1}{|\bls{r}-\mathsf{M}\bls{r}'|}\nonumber\\
    &+\frac{2d_{\rm s}}{4\pi\varepsilon_0}\frac{{\rm Im}\left[\varepsilon_{\rm s}(\omega)\right]}{|\varepsilon_{\rm s}(\omega)|^2}\bls{e}_3\cdot\frac{\partial}{\partial \bls{r}}\frac{1}{|\bls{r}-\mathsf{M}\bls{r}'|}.
\end{align}
For a dielectric layer characterized by the surface permittivity $\varepsilon_{\rm s}(\omega)=\varepsilon_{\rm DL}(\omega)$ on top of a Drude metal, $\varepsilon_{\rm b}(\omega)=\varepsilon_{\rm m}(\omega)$ the general expression in Tab.~\ref{tab:dielectrics} yields
\begin{align}\label{eq:diffLimitLayer}
 h(\bls{r},\bls{r}')=&\frac{1}{4\pi\varepsilon_0}\frac{2 k_{\rm B}T}{\hbar}\frac{\gamma_{\rm m}}{\omega_{\rm pm}^2}\frac{1}{|\bls{r}-\mathsf{M}\bls{r}'|}\nonumber\\
 &-\frac{d_{\rm s}}{4\pi\varepsilon_0}\frac{2k_{\rm B}T}{\hbar}\left[\sum_{n\ge 0}\frac{f_n\gamma_n}{\omega_n^2}-\frac{2\gamma_{\rm m}\varepsilon_{\rm DL}(0)}{\omega_{\rm pm}^2}\right]\nonumber\\
 &\times\frac{1}{\varepsilon^2_{\rm DL}(0)}\bls{e}_3\cdot\frac{\partial}{\partial \bls{r}}\frac{1}{|\bls{r}-\mathsf{M}\bls{r}'|}.
\end{align}
Here $\varepsilon_{\rm DL}(0)=1+\sum_{n\ge 0} f_n$ is the static value of the surface response.

Replacing the metal by a superconductor, $\varepsilon_{\rm b}(\omega)=\varepsilon_{\rm sc}(\omega)$, one obtains
\begin{align}\label{eq:genDLhfunctionLayer}
 h(\bls{r},\bls{r}')=&-\frac{d_{\rm s}}{4\pi\varepsilon_0}\frac{2k_{\rm B}T}{\hbar}\frac{1}{\varepsilon^2_{\rm DL}(0)}\sum_{n\ge 0}\frac{f_n\gamma_n}{\omega_n^2}\bls{e}_3\cdot\frac{\partial}{\partial \bls{r}}\frac{1}{|\bls{r}-\mathsf{M}\bls{r}'|},
\end{align}
which is equivalent to \eqref{eq:diffLimitLayer} for negligible damping $\gamma_{\rm m}=0$ in the bulk. Like in \eqref{eq:thinLayerReflectingMetal}, the superconductor acts like a mirror for the fluctuations generated in the dielectric layer. This approximation
is justified if the surface layer is thinner than the London penetration depth, which is the case for typical superconducting materials.

\subsection{Electric-field power spectral density}

The surface noise kernel also characterizes the electric field fluctuations observed above thermal surfaces. As we shall see below, the resulting electric field power spectral density (PSD) is closely related to the decoherence and heating rates of electric monopoles and dipoles.

The PSD at position $\bls{R}_{\rm 0}$ is defined as the Fourier transform of the autocorrelation function,
\begin{equation}
    \mathsf{S}_{EE}(\bls{R}_0,\omega)=\frac{1}{2}\int_{-\infty}^\infty d\tau \braket{\{\bls{E}(\bls{R}_0,t),\bls{E}(\bls{R}_0,t+\tau)\}}e^{i\omega\tau}.
\end{equation}
where ${\bf E}({\bf r},t)$ denotes the electric field operator, described by the scalar potential \eqref{eq:phiviag}, and $\{\cdot,\cdot\}$ is the dyadic anti-commutator. For a thermal distribution one finds in the quasistatic limit
\begin{align}\label{eq:52}
    \mathsf{S}_{EE}(\bls{R}_0,\omega)=&-2\hbar \left [n(\omega) + \frac{1}{2} \right ]\nonumber\\
    &\times\frac{\partial}{\partial\bls{r}}\otimes \left.\frac{\partial}{\partial\bls{r}'}{\rm Im}\left[g(\bls{r},\bls{r}',\omega)\right]\right|_{\bls{r}=\bls{r}'=\bls{R}_0},
\end{align}
which reduces for $\omega = 0$ to
\begin{align}
    \mathsf{S}_{EE}(\bls{R}_0,\omega=0)=2\hbar\frac{\partial}{\partial \bls{r}}\otimes\left.\frac{\partial}{\partial \bls{r}'}h(\bls{r},\bls{r}')\right|_{\bls{r}=\bls{r}'=\bls{R}_0}.
\end{align}

For instance, a thin dielectric layer, as described by Eq.~\eqref{eq:thinLayerReflectingMetal}, gives rise to
\begin{align}\label{eq:electricFieldNoiseThinLayer}
    &\mathsf{S}_{EE}(\bls{R}_0,\omega)=\left [ n(\omega) + \frac{1}{2}\right ]\frac{\hbar}{4\pi\varepsilon_0}
    \left(\frac{{\rm Im}\left[\varepsilon_{\rm b}(\omega)\right]}{|\varepsilon_{\rm b}(\omega)|^2}\frac{1}{2|\bls{R}_0\cdot\bls{e}_3|^3}\right.\nonumber\\
    &\left.+\frac{{\rm Im}\left[\varepsilon_{\rm s}(\omega)\right]}{|\varepsilon_{\rm s}(\omega)|^2}\frac{3d_{\rm s}}{4|\bls{R}_0\cdot\bls{e}_3|^4}\right)\left(\mathbb{1}+\bls{e}_3\otimes\bls{e}_3\right).
\end{align}
In the large temperature limit $k_{\rm B} T/\hbar \omega\gg1$, this is equivalent to the expression given in \cite{kumph2016electric}. On the other hand, the Drude-metal half-space with \eqref{eq:responseDrudeMetal} yields in the limit of small frequencies 
\begin{equation}
    \mathsf{S}_{EE}(\bls{R}_0,0)=\frac{k_{\rm B}T}{4\pi\varepsilon_0}\frac{\gamma_{\rm m} }{2\omega_{\rm pm}^2}\frac{1}{|\bls{R}_0\cdot\bls{e}_3|^3}\left(\mathbb{1}+\bls{e}_3\otimes\bls{e}_3\right),
\end{equation}
which is consistent with the calculations in \cite{kumph2016electric, henkel1999loss}\footnote{Note however that the quasistatic approximation of the electric field emanating from the metal ceases to be valid for distances larger than the skin depth $\delta$, with $\delta^2=c^2\gamma_{\rm m}/\omega\omega_{\rm pm}^2$ \cite{kumph2016electric, henkel1999loss}.}. As a final example, a Drude-Lorentz layer on a superconductor with \eqref{eq:genDLhfunctionLayer} yields
\begin{align}\label{eq:fieldPSDLayerSlowParticle}
    \mathsf{S}_{EE}(\bls{R}_0,0)=&\frac{k_{\rm B}T}{4\pi\varepsilon_0}\frac{3d_{\rm s}}{4|\bls{R}_0\cdot\bls{e}_3|^4}\left(\mathbb{1}+\bls{e}_3\otimes\bls{e}_3\right)\nonumber\\
    &\times\frac{1}{\varepsilon^2_{\rm DL}(0)}\sum_{n\ge 0}\frac{f_n\gamma_n}{\omega_n^2}.
\end{align}

These examples illustrate that the presence of a thin dielectric layer induces the experimentally observed  scaling
\begin{equation}
    \mathsf{S}_{EE}(\bls{R}_0,\omega)\propto\frac{1}{d^4}
\end{equation}
with $d=\bls{R}_0\cdot\bls{e}_3$ the distance to the surface. Importantly, this contribution from the layer dominates for typical electrode and surface layer materials \cite{kumph2016electric}. We will see in the following section that this distance behavior translates to nanoparticle decoherence and heating rates.   

\begin{figure}
\centering
\includegraphics[width = \columnwidth]{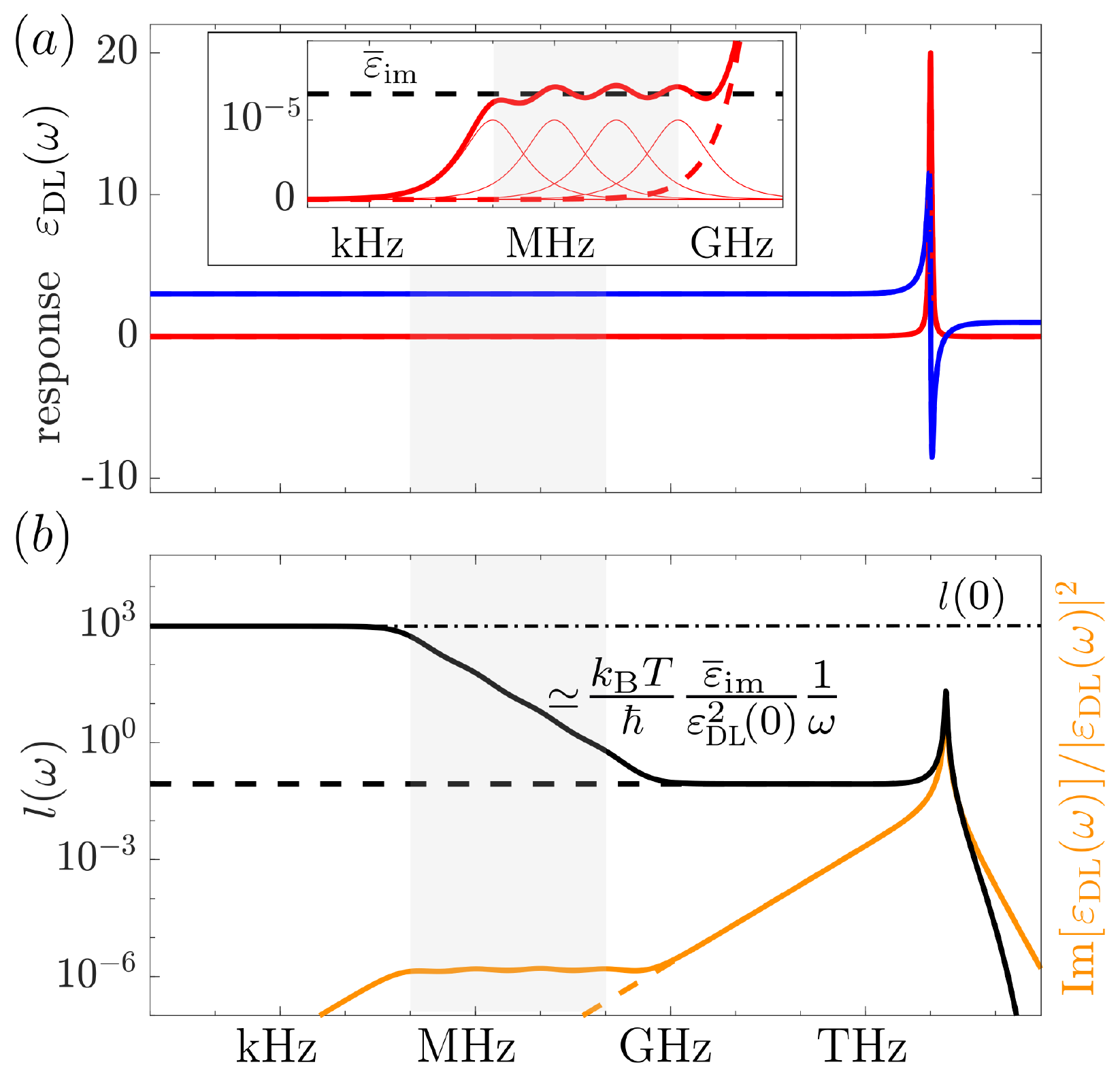}
\caption{Panel (a) shows the real (blue) and imaginary (red) part of the dielectric permittivity \eqref{eq:genDL}, characterized by a single strong THz resonance ($\omega_5=10^{13}{\rm s}^{-1}$, $\gamma_5=10^{12}{\rm s}^{-1}$, $f_5=2$) and weak low-frequency resonances (thin red lines), as illustrated in the inset. The weak resonance form a broad region (grey area) of approximately constant imaginary part. Panel (b) shows the energy loss function ${\rm Im}[\varepsilon_{\rm DL}(\omega)]/|\varepsilon_{\rm DL}(\omega)|^2$ (orange) and its thermal version $l(\omega)$ (black) in presence (solid) and absence (dashed) of the low-frequency resonances. For small frequencies,  the low-frequency peaks determine $l(\omega)$ despite their small weights, and therefore dominate the surface-induced decoherence, while the energy loss is negligible. In the region of nearly flat imaginary part (grey area), $l(\omega)$ exhibits the characteristic $1/\omega$-scaling. We used $T=300\,{\rm K}$, $f_n=2\times 10^{-5}$, $\omega_n = 10^{6+n}{\rm s}^{-1}$, and $\gamma_n = 10^{8+n} {\rm s}^{-1}$ for $n=1,...4$.} \label{figure2}
\end{figure}

\subsection{Relevance of low-frequency surface excitations}

Inserting the surface-fluctuation kernel into the master equation \eqref{eq:dissipatorQuasistatic}, for instance using Eq.~\eqref{eq:thinLayerReflectingMetal}, demonstrates that the magnitude of surface-induced decoherence is determined by the energy loss function ${\rm Im}[\varepsilon_{\rm r} (\omega)]/|\varepsilon_{\rm r} (\omega)]|^2$ and by its thermal version
\begin{align}\label{eq:thermenloss}
   l(\omega) = n(\omega)\frac{{\rm Im}\left [ \varepsilon_{\rm r}(\omega)\right]}{|\varepsilon_{\rm r}(\omega)|^2}.
\end{align}
The latter dominates the former for small frequencies (see Fig.~\ref{figure2}b), and thus determines the high-temperature frequency behaviour of the electric field PSD \eqref{eq:electricFieldNoiseThinLayer}.

For instance, for a Drude-Lorentz dielectric \eqref{eq:genDL} (Fig.~\ref{figure2}a), the absolute value in the denominator {of \eqref{eq:thermenloss}} can be approximated by its static value  {$\varepsilon_{\rm DL}(0)$}, so that the function plateaus at
\begin{equation}\label{eq:low_freq_deco_genDL}
    l(0)= \frac{k_{\rm B} T}{\hbar \varepsilon_{\rm DL}^2(0)} \sum_{n\geq 0} \frac{f_n \gamma_n}{\omega_n^2},
\end{equation}
for frequencies much less than $\omega_n,\omega_n^2/\gamma_n$. Typically, high-frequency electronic transitions exhibit significantly larger weights $f_n$ and thus dominate the static value of the dielectric permittivity. However, Eq.~\eqref{eq:thermenloss} can be enhanced by broad low-frequency contributions to the spectrum, each contributing with relative strength $f_n\gamma_n/\omega_n^2$. Such weak and broad absorption bands could for instance be due to weakly bound surface adsorbates \cite{brown2021materials,foulon2022omega}.

Moreover, let us assume that these low lying transitions give rise to a broad plateau in the absorption spectrum with mean imaginary dielectric response $\overline{\varepsilon}_{\rm im}$ (see Fig.~\ref{figure2}a). It then follows that the function $l(\omega)$ exhibits the experimentally observed $1/\omega$ frequency scaling (Fig.~\ref{figure2}b). Specifically, one obtains
\begin{equation}
    l(\omega) \simeq \frac{k_{\rm B} T}{\hbar}\frac{\overline{\varepsilon}_{\rm im}}{ \varepsilon_{\rm DL}^2(0)} \frac{1}{\omega}.
\end{equation}

This illustrates the decisive role played by the low-frequency dielectric response of the electrode surface in ion traps. As we shall see below, this frequency scaling carries over to the surface-induced decoherence and heating rates. We note that non-flat combinations of low-lying resonances can give rise to other power-law scalings \cite{jonscher1977universal}.

\section{How to use the toolbox}\label{Sec:decoherenceAndHeating}

Having discussed how the particle charge distribution and the surface properties enter the master equations \eqref{eq:MasterEquationSlowParticleIntro} and \eqref{eq:dissipatorQuasistatic}, we now explain how one can identify the suitable master equation for a given experimental setting. This will be followed by two examples and a short discussion of relevant limiting cases, in which the master equations simplify to  known results.

\subsection{Finding the right master equation}

To identify the appropriate master equation for a given physical situation requires comparing the involved timescales. These  are determined by the setup, as described by the form and location of the particle charge-distribution, its free dynamics, as well as the surface geometry, dielectric response, and temperature. Given this information, one proceeds as follows:

\begin{enumerate}[itemsep = -0.2mm]
    \item Specify the body-fixed charge distribution of the particle, e.g.\ in terms of its multipole moments, and estimate the characteristic timescales of the particle dynamics in absence of the surface.
    \item Specify the surface response function, e.g.\ by experimental measurements or by a suitable model such as Eq.~\eqref{eq:genDL}.
    \item Check that electric-field retardation is negligible and calculate the quasistatic Green function \eqref{eq:DGLscalarGreenfunction}. If the surface is clean/covered by a thin dielectric layer, one can use the first/second row in Tab.~\ref{tab:dielectrics}.
    \item Estimate the resulting decay rate of surface-fluctuation correlations via Eqs.\,\eqref{eq:dissipationKernel}--\eqref{eq:spectralDensityDefinition}.
    \item[5.1] If the particle dynamics are much slower than the surface-fluctuation correlation time, use the slow-particle master equation \eqref{eq:MasterEquationSlowParticleIntro}.
    \begin{itemize}[itemsep = -0.2mm]
        \item[{\tiny$\blacksquare$}] If the particle charge distribution is characterized exclusively by its monopole, dipole, or quadrupole moment, one can use the master equations \emph{a}., \emph{b}., or \emph{c}. in Tab.~\ref{tab:MasterEquations}.
        \item[{\tiny$\blacksquare$}] Otherwise, insert the charge distribution into \eqref{eq:MasterEquationSlowParticleIntro}.
    \end{itemize}
    \item[5.2] If both the free particle dynamics and the decay of surface-fluctuation correlations are much faster than the surface-induced system dynamics, use the resonant master equation \eqref{eq:dissipatorQuasistatic}.
        \begin{itemize}[itemsep = -0.2mm]
        \item[{\tiny$\blacksquare$}] If the particle centre-of-mass is oscillating and its charge distribution is characterized solely by its monopole, dipole, or quadrupole moment, one can use the master equations \emph{d}., \emph{e}., or \emph{f}. in Tab.~\ref{tab:MasterEquations}.
        \item[{\tiny$\blacksquare$}] If the particle orientation is librating and its charge distribution is characterized solely by its dipole or quadrupole moment, one can use the master equations \emph{g}. or \emph{h}. in Tab.~\ref{tab:MasterEquations}.
        \item[{\tiny$\blacksquare$}] If the particle is rotating in a plane with approximately constant angular velocity and its charge distribution is characterized only by its dipole or quadrupole moment, one can use \emph{i}. or \emph{j}. in Tab\,\ref{tab:MasterEquations}.
        \item[{\tiny$\blacksquare$}] If the particle is simultaneously oscillating and librating/rotating or for a general charge distribution, calculate the operators \eqref{eq:eigenoperatorsExplicitly1} and insert them into \eqref{eq:dissipatorQuasistatic}.
    \end{itemize}
\end{enumerate}

It is worth noting that for a flat surface-fluctuation spectrum, $l(\omega_\ell) \simeq l(0)$, the resonant master equation describes the same dynamics as the slow-particle master equation if  a rotating wave approximation with respect to the system's surface-free dynamics is justified.

The next subsection shows that the Markov and quasistatic approximations are justified for typical state-of-the-art setups, ranging from trapped ions, to molecules and nanoparticles. We will then illustrate how to use the master equations in two examples, (i) a slowly rotating two-ion Coulomb crystal and (ii) a freely rotating polar molecule, both  close to a surface covered by a thin dielectric layer. Finally, this section closes by discussing limiting situations in which the derived master equations reduce to known results.

\subsection{Experimental adequacy of the description}
\label{sec:adequacy}
{\it Quasistatic approximation---}The quasistatic approximation applies if any retardation in the wave-propagation from the source in the surface to the particle can be neglected. The experimentally observed distance scaling $\propto d^{-4}$ of the electric field noise \cite{hite2013surface,monroe2013scaling,brownnutt2015ion,bruzewicz2019trapped,brown2021materials} indicates that the predominant noise sources are longitudinal fields originating from a thin surface layer. Retardation effects can be usually neglected given that the time it takes a photon to travel through the surface layer of thickness $d_{\rm s}$ to the particle at distance $d$ is much smaller than the characteristic timescale $1/\omega_0$ of the particle motion in the absence of the surface,  $(\sqrt{|\varepsilon_{\rm s}|}d_{\rm s} +d) \omega_0/c \ll 1$. Since $|\varepsilon_{\rm s}|\lesssim40$ and $d_{\rm s}\lesssim 5\,{\rm nm}$ for typical surface contaminants \cite{kumph2016electric}, retardation is negligible for frequencies up to $\omega_0=30\,{\rm GHz}$ at $d=1\,{\rm mm}$, as applies to typical setups with atoms \cite{bruzewicz2019trapped,de2021materials,brown2021materials}, molecules \cite{andre2006coherent,rabl2006hybrid,rabl2007molecular}, and nanoparticles \cite{delord2017b,nagornykh2017optical,urban2019coherent, bykov2019,tebbenjohanns2019cold,conangla2019optimal,martinetz2020}. In the absence of a thin surface layer, the quasistatic approximation is still justified as long as the skin depth $\delta$ is greater than the trapping height. For a Drude metal, one finds $\delta^2 =\gamma_{\rm m}c^2/\omega_0\omega_{\rm pm}^2$, yielding   $\delta\approx4.3\,{\rm mm}$ for gold at kHz frequencies and  $\delta \approx 140\,\mu$m at MHz frequencies.

{\it Born-Markov approximation---}The weak-coupling Born-Markov approximation is justified if the temporal correlations in the surface fluctuations are irrelevant for the particle dynamics. Rigorously, this requires the surface-induced particle relaxation to take place on a timescale much greater than that of temporal correlations. The surface-noise correlation timescale is dominated either by the thermal correlation time $\hbar/k_{\rm B}T$ or by the width of the spectral density \eqref{eq:spectralDensityDefinition}. For temperatures above $100\,{\rm mK}$ the thermal correlation time falls below $100\,{\rm ps}$, while the correlation timescale due to e.g.\ the THz-resonance in Fig.~\ref{figure2} reaches $1\,{\rm ps}$ at most. Similar timescales are obtained for typical metals, though low-lying overdamped surface resonances may dominate ($\gamma_n/\omega_{\rm n}^2\approx 10\,\mu$s for Fig.~\ref{figure2}). These timescales can be compared to the particle-relaxation time, which lies in state-of-art setups in the range of up to milli-seconds. In addition, the rigorous Born-Markov approximation requires the surface-free particle dynamics  to be either much slower than the decay of surface correlations (slow-particle limit) or much faster than particle-relaxation timescale (resonant limit). For instance, a rotating nanoparticle at $100\,{\rm kHz}$ is described by the slow-particle limit given the above surface properties, while the resonant limit applies for an oscillating ion at MHz frequencies. We emphasize that even if the Born-Markov approximation is not justified formally, it often yields an adequate description given that non-Markovian modifications of the dynamics are often marginal \cite{breuer2016colloquium}.

\subsection{Example 1: Two-ion Coulomb crystal}
We consider two equally charged ions with $q=e$, rotating with constant angular velocity $\omega_{\rm rot}$ at a constant interparticle distance $d=5\,\mu$m around their center of mass $\bls{R}\cdot\bls{e}_3=100\,{\rm \mu m}$ above a flat surface with surface normal ${\bf e}_3$, motivated by the experiment of Ref.~\cite{urban2019coherent}. Choosing the reference charge distribution such that the interparticle axis is aligned with the surface normal, one has
\begin{align}
    \varrho_0(\bls{r})=&q\delta\left(\bls{r}-\frac{d}{2}\bls{e}_3\right)+q\delta\left(\bls{r}+\frac{d}{2}\bls{e}_3\right).
\end{align}
It is  characterized not only by the monopole and quadrupole moment, but also exhibits higher charge moments. The direction of the interparticle charge axis at a given crystal orientation $\Omega$ is given by $\bls{n}_3=\cos\alpha\sin\beta\bls{e}_1+\sin\alpha\sin\beta\bls{e}_2+\cos\beta\bls{e}_3$.

\begin{figure}
\centering
\includegraphics[width = 0.35\textwidth]{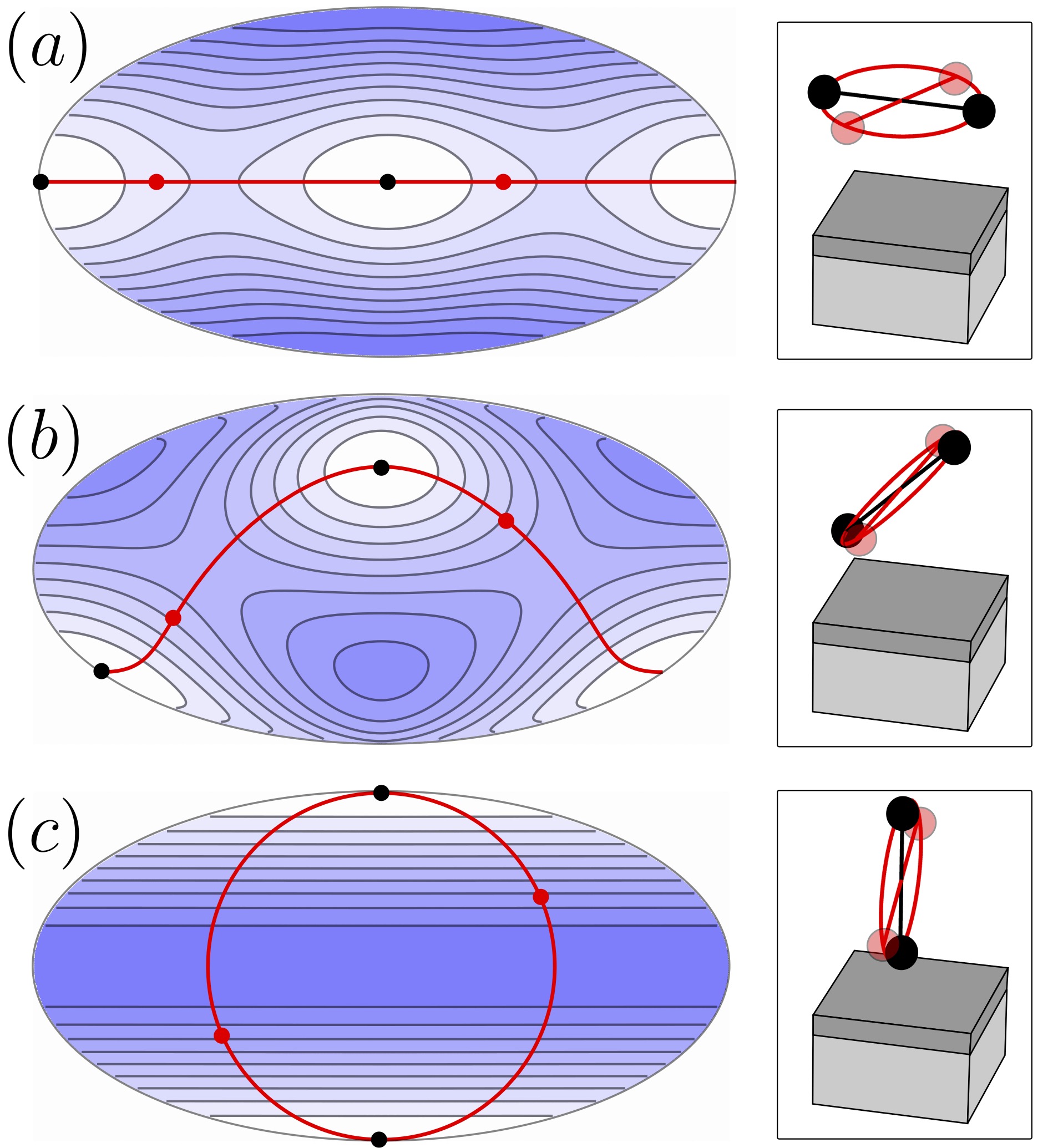}
\caption{Rotational decoherence rate of a two-ion Coulomb crystal \cite{urban2019coherent} rotating around a fixed center-of-mass position near a dielectric layer deposited on bulk gold. (a) The contour plot in Mollweide projection displays the rate \eqref{eq:DecoRateSlowParticle} for a superposition between two orientations with one parallel to the surface (black dumbell and dots). The plot is normalized to the maximum decoherence rate of $\Gamma_{\rm max}=3.7\,{\rm kHz}$ (blue). Great-circle trajectories of the second branch are shown in red. In panel (b) and (c) the fixed orientation (black dumbell) is set to 45$^\circ$ and 90$^\circ$ to the surface plane, respectively.}\label{figure5}
\end{figure}

For a flat surface layer of thickness $d_{\rm s}=5\,{\rm nm}$ at $T=300\,{\rm K}$ with the exemplary response of Fig.~\ref{figure2}, exhibiting spectrally flat and weak features in the kHz to GHz range on a bulk metal, the quasistatic approximation is justified. This is because (i) retardation between the surface and the particle is negligible on the timescale of the particle motion and (ii) because even the weak features with $\overline{\varepsilon}_{\rm im} \approx 1\times 10^{-5}$ dominate the total noise, i.e.\ noise from the gold bulk as well as blackbody radiation from infinitely distant surfaces, at the particle frequencies and its small distance above the surface. In addition, the surface-fluctuation correlations \eqref{eq:dissipationKernel} and \eqref{eq:noiseKernel} decay on a timescale that is determined by the spectral widths of $l(\omega)$ and ${\rm Im}[\varepsilon(\omega)]/|\varepsilon(\omega)|^2$. In Fig.~\ref{figure2} the width of $l(\omega)$ is approximately given by $\omega_2^2/\gamma_2=10^5{\rm s}^{-1}$.  The width of the energy loss function is approximately given by the width of the underdamped high frequency peak of $\gamma_5=10^{12}{\rm s}^{-1}$. The Markov approximation and Eq.~\eqref{eq:MasterEquationSlowParticleIntro} are justified since  for rotation frequencies $\omega_{\rm rot}\lesssim 100$\,kHz the crystal moves (and relaxes) much slower than those surface correlation timescales.

The resulting orientational decoherence rate \eqref{eq:DecoRateSlowParticle} of the crystal with the surface-fluctuation kernel \eqref{eq:genDLhfunctionLayer} is depicted in Fig.~\ref{figure5}. We display the decoherence of a superposition between a fixed orientation (black dumbell and dots) and all other possible orientations. The rate as a function of the varying orientation is depicted in a contour plot in Mollweide projection. In superposition experiments, such as in \cite{urban2019coherent}, the varying orientation typically rotates on great circles (red).

\subsection{Example 2: Rotating polar molecule}\label{sec:decoherenceHybridQuantumDevices}

Next, we consider a point dipole in a superposition of angular momentum states $|\ell,m\rangle$ (quantization axis $\blg{\epsilon}_3$) of the form $\ket{\Psi}=\left(\ket{2,0}+\ket{1,0}\right)/\sqrt{2}$ above a superconducting surface at the temperature $T=100\,{\rm mK}$ \cite{andre2006coherent,xiang2013hybrid}. Following Ref.~\cite{andre2006coherent}, we assume a dipole moment of $p=4.36\,{\rm D}$ and $\omega_0=\hbar/I=2\pi\times 5.5\,{\rm GHz}$, implying that $n(\omega_0)\approx 0.07 \ll 1$. At such low temperatures no energy states outside $\ell \in \{0,1,2\}$ get populated since $n(\omega_\ell)\simeq 0$ for all $\ell$ with $\omega_\ell=(\ell+1)\hbar/I$. Thus, the only operators needed to specify the dissipator \eqref{eq:dissipatorQuasistatic} are
\begin{align}\label{eq:c1RotationalDecoherence}
    c_\ell(\bls{r})=-p\sum_{i=1}^3n_{\ell,i}\blg{\epsilon}_i\cdot\frac{\partial}{\partial\bls{r}}\delta(\bls{r}-\bls{R}_{\rm cm}).
\end{align}
Here we defined the operators
\begin{subequations}
\begin{align}
    n_{0,i}=&\left(n^{(i)}_{0011}\ket{0,0}\bra{1,1} + n^{(i)}_{0010}\ket{0,0}\bra{1,0}\right.\nonumber\\
    &\left.+ n^{(i)}_{001-1}\ket{0,0}\bra{1,-1}\right),\\
    n_{1,i}=&\left(n^{(i)}_{1120}\ket{1,1}\bra{2,0}+n^{(i)}_{1020}\ket{1,0}\bra{2,0}\right.\nonumber\\ &\left.+n^{(i)}_{1-120}\ket{1,-1}\bra{2,0}\right).
\end{align}
\end{subequations}
The coefficients $n^{(i)}_{\ell m\ell'm'}$ are given in Eq.~\eqref{eq:plmlm}. 

Inserting \eqref{eq:c1RotationalDecoherence} into \eqref{eq:dissipatorQuasistatic} yields the dissipator
\begin{align}\label{eq:DissipatorRotationalDecoherence}
    \mathcal{L}\rho=&\frac{2p^2}{\hbar}\sum_{\ell=0}^1\sum_{i,j=1}^3h_{ij}(\bls{R}_{\rm cm},\omega_\ell)\left(n_{\ell,i}\rho n_{\ell,j}^\dagger-\frac{1}{2}n^\dagger_{\ell,j} n_{\ell,i}\rho\right.\nonumber\\
    &\left.-\frac{1}{2}\rho n_{\ell,j}^\dagger n_{\ell,i}\right)
\end{align}
with the surface-fluctuation kernel
\begin{align}
    h_{ij}(\bls{R}_{\rm cm},\omega_\ell)=&-\left(\blg{\epsilon}_j\cdot\frac{\partial}{\partial\bls{r}}\right)\left(\blg{\epsilon}_i\cdot\frac{\partial}{\partial\bls{r}'}\right)\nonumber\\
    &\left.\vphantom{\frac{1}{2}}\times{\rm Im}\left[g(\bls{r},\bls{r}',\omega_\ell)\right]\right|_{\bls{r}=\bls{r}'=\bls{R}_{\rm cm}}.
\end{align}

\begin{figure}
\centering
\includegraphics[width = \columnwidth]{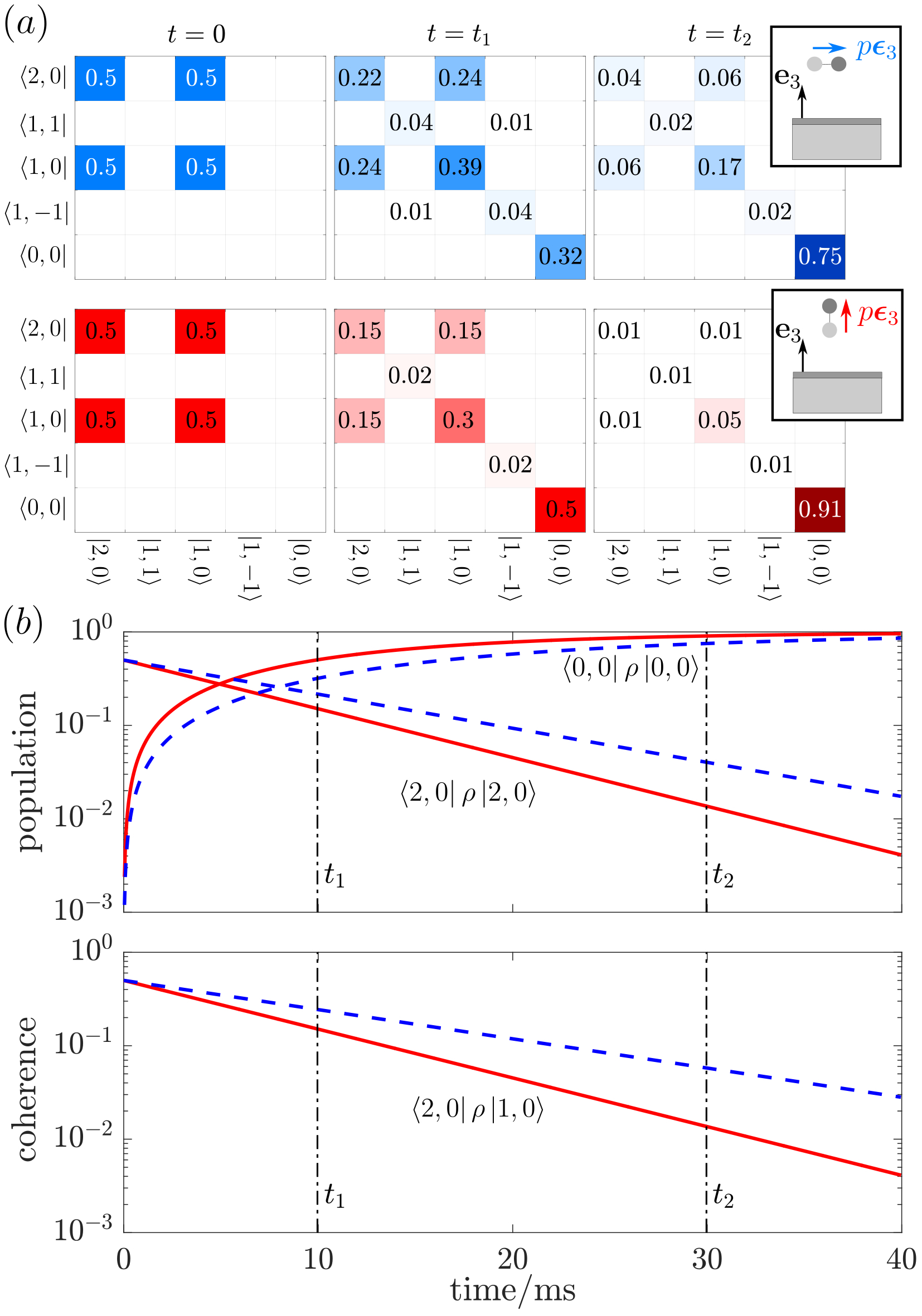}
\caption{Rotational decoherence of a point dipole near a superconducting surface. (a) The particle is initially prepared in an angular momentum superposition state with initial dipole $p\blg{\epsilon}_3$ pointing perpendicular (blue) or parallel (red) to the surface normal vector $\bls{e}_3$. The noise originating from the surface is dominated by a thin layer covering the superconducting device (see Sec.~\ref{Sec:decoherenceAndHeating}). The dynamics governed by \eqref{eq:DissipatorRotationalDecoherence} are displayed via the density operator in the basis of the lowest angular momentum states $|\ell m\rangle$ (quantization axis $\blg{\epsilon}_3$) for the times $t=0$, $t_1=10\,{\rm ms}$ and $t_2=30\,{\rm ms}$. Surface-induced dissipation relaxes the particle to its rotational ground state. (b)(top) While the groundstate population  approaches unity for large times, the excited state population decreases at a constant rate. The latter differs in the two scenarios, marked in dashed blue and solid red, respectively. (b)(bottom) The coherence decay in (a) occurs also at constant rate. Decoherence in the red case is stronger, since the field fluctuations are stronger in the direction of the surface normal.} \label{figure3}
\end{figure}

The particle is located at a distance of $\bls{R}_{\rm cm}\cdot\bls{e}_3=100\,{\rm nm}$ \cite{andre2006coherent} from the superconducting surface, with normal vector $\bls{e}_3$, which is covered by a thin dielectric layer. We assume a layer thickness of $d_{\rm s}=4\,{\rm nm}$ and $\varepsilon_{\rm s}(\omega_\ell)=3\times(1+0.001i)$, as is realistic for e.g.\ an oxide layer \cite{westphal1977dielectric,kumph2016electric}. This permittivity is of the same order of magnetitude as the response shown in Fig.~\ref{figure2}. At these frequencies, its value is dominated by the high frequency peak rather than the low-frequency features. The quasistatic approximation of the electric fields is justified since (i) retardation between the surface and the particle is negligible and since (ii) the noise from the layer dominates over blackbody radiation and noise from a realistic metal or superconductor bulk. As discussed in the previous example, the surface-fluctuation correlations of the spectrum shown in Fig.~\ref{figure2} decay at a rate of $10^5{\rm s}^{-1}$. Consequently, the Markov approximation is justified because the particle decoheres much slower than surface correlations decay and than the particle rotates, as we will show next.

In Fig.~\ref{figure3}, we solve the master equation \eqref{eq:MasterEquationUnspecifiedSlowParticle} with the dissipator \eqref{eq:DissipatorRotationalDecoherence}, neglecting the coherent surface interaction. Using Eq.~\eqref{eq:thinLayerReflectingMetal} we can approximate
\begin{align}\label{eq:fluctuationKernelExample2}
    h_{ij}(\bls{R}_{\rm cm},\omega_\ell)\simeq& \frac{1}{4\pi\varepsilon_0}\frac{{\rm Im}\left[\varepsilon_{\rm s}(\omega_\ell)\right]}{|\varepsilon_{\rm s}(\omega_\ell)|^2}\frac{3d_{\rm s}}{8|\bls{R}_{\rm cm}\cdot\bls{e}_3|^4}\nonumber\\
    &\times\left[\delta_{ij}+(\blg{\epsilon}_j\cdot\bls{e}_3)(\blg{\epsilon}_i\cdot\bls{e}_3)\right],
\end{align}
expressing that the noise is predominantly due to the surface layer. The initial state $\rho(0) = |\Psi\rangle \langle \Psi|$ generates a mean dipole moment that points along the $\blg{\epsilon}_3$ axis. In contrast, pure angular momentum states, or an incoherent mixture thereof, lead to a vanishing mean dipole. As displayed in Fig.~\ref{figure3}, the decoherence rate crucially depends on the alignment between the dipole and the surface normal. The order of magnitude of the decoherence rate can be roughly estimated as $p^2\sum_i h_{ii}(\bls{R}_{\rm cm},\omega_\ell)/\hbar\approx 4\times 10^2/{\rm s}$. This rate will increase for higher temperatures and a dielectric with a larger loss tangent, becoming relevant for hybrid quantum systems with polar molecules and superconducting devices \cite{xiang2013hybrid, andre2006coherent}.

\subsection{Relation to previous results}

{\it Electron beams near a metal surface---}In the slow-particle limit the decoherence rate of a monopole reduces to that calculated in Ref.~\cite{scheel2012path}. Specifically, the decoherence rate follows from Tab.~\ref{tab:MasterEquations}a. and \eqref{eq:definingEquationDecoherenceRate} together with the surface-fluctuation kernel of a Drude metal \eqref{eq:fFunctDrudeMetal} as
\begin{align}
    \Gamma(\bls{R},\bls{R}')=&\frac{1}{4\pi\varepsilon_0}\frac{q^2k_{\rm B}T}{\hbar^2}\frac{2\gamma_{\rm m} }{\omega_{\rm pm}^2}\left[\frac{1}{|\bls{R}-\mathsf{M}\bls{R}|}\right.\nonumber\\
    &\left.+\frac{1}{|\bls{R}'-\mathsf{M}\bls{R}'|}-2\frac{1}{|\bls{R}-\mathsf{M}\bls{R}'|}\right].
\end{align}
For superposed electron beams traveling parallel to a metal surface with a fixed distance between the superposition branches, this is identical to the path decoherence predicted in Ref.~\cite{scheel2012path} and measured in \cite{kerker2020quantum}.

{\it Motional decoherence of ions in quadrupole traps---}For a point monopole, the resonant master equation \eqref{eq:dissipatorQuasistatic} reduces to the one used for describing heating of ions levitated in quadrupole traps. Specifically, using row {\it d.}\ in Tab.~\ref{tab:MasterEquations} at a trapping frequency $\omega_k=\omega_0$ and oscillation direction $\blg{\epsilon}_k=\blg{\epsilon}_0$, yields the groundstate heating rate $\Gamma_{\rm h}(\omega_0)=q^2n(\omega_0)h_0(\bls{R}_{\rm eq})/m\omega_0$ \cite{henkel1999loss,brownnutt2015ion}. For $k_{\rm B}T/\hbar\omega_0\gg 1$, as is the case for typical trapping conditions \cite{brownnutt2015ion}, the heating rate can be approximated by means of the electric-field PSD \eqref{eq:52}
\begin{align}\label{eq:heatingRateFieldPSD}
    \Gamma_{\rm h}(\omega_0)&\simeq\frac{q^2}{2m\hbar\omega_0}\blg{\epsilon}_0\cdot\mathsf{S}_{EE}(\bls{R}_{\rm eq},\omega_0)\blg{\epsilon}_0,
\end{align}
which exhibits the characteristic distance and frequency scaling discussed in Sec.~\ref{sec:specific_dielectrics}.

{\it Rotational heating of polar molecules---}Groundstate heating of dipolar rotors near surfaces has been studied in Ref.~\cite{buhmann2008surface}. In order to verify that the two results agree, we evaluate the master equation \eqref{eq:dissipatorQuasistatic} for a point dipole, neglecting all other multipole moments, at a fixed center-of-mass position with non-degenerate rotational states. Considering the quasistatic approximation \eqref{eq:quasistaticGDerivation} in Ref.~\cite{buhmann2008surface} shows that both theories coincide.

\section{Discussion}\label{sec:conclusion}

Given the dielectric properties and geometry of the surface, the presented toolbox serves to predict quantitatively the induced heating and decoherence for a multitude of particle shapes and charge distributions. By disentangling the role of the particle from that of the surface, it provides novel tools for systematically studying the problem of surface-induced noise and decoherence in electric traps. In the following, we discuss a possible roadmap for future investigations, then comment on potential future extensions of the framework, and finally provide our conclusions.

\subsection{Taming the surface noise}

{\it Characterizing the surface---}As a main feature of the presented toolbox, the material properties of the surface enter only via the dielectric response functions of bulk and surface layer. Ex-situ measurements of the low-frequency behavior of the dielectric material can therefore be used to predict its impact on the particle dynamics. Performing such measurements for different materials and surface contaminants will serve to identify the origin and strength of broad, low-frequency features in the dielectric permittivity.

{\it Sensing field fluctuations---}The levitated particle acts as a highly sensitive probe of electric surface noise  since the heating force felt by a monopole and the torque experienced by a dipole are fully characterized by the electric field PSD \eqref{eq:52}. Specifically, the decoherence rate of a slow monopole at position ${\bf R}_0$ in a superposition of size $\Delta {\bf R} = {\bf R} - {\bf R}'$ (with $|{\bf R}_0|\gg |\Delta {\bf R}|$) reads
\begin{equation}\label{eq:decoherenceRateSmallSuperposition}
    \Gamma(\bls{R},\bls{R}')\simeq\frac{q^2}{2\hbar^2}\Delta\bls{R}\cdot \mathsf{S}_{EE}(\bls{R}_0,0)\Delta\bls{R}.
\end{equation} 
Such a quadratic  dependence on $\Delta {\bf R}$ is accompanied by momentum diffusion, and thus heating, with diffusion tensor $q^2 \mathsf{S}_{EE}({\bf R}_0,0)/2$. Likewise, for orientational superpositions of a slow point dipole in a superposition of orientations $\Omega$ and $\Omega'$, the decoherence rate reads
\begin{align}\label{eq:decoherenceDipoleSamePosition}
    \Gamma_{\Omega\,\Omega'}(\bls{R}_{\rm cm},\bls{R}_{\rm cm})=&\frac{1}{2\hbar^2}\Delta{\bf p}\cdot \mathsf{S}_{EE}(\bls{R}_{\rm cm},0)\Delta \bls{p},
\end{align}
with $\Delta {\bf p} = \bls{p}_\Omega-\bls{p}_{\Omega'}$. Again, this decoherence is associated with angular momentum diffusion and thus heating \cite{papendell2017}.

{\it Sensing gradient fluctuations---}In more general situations, the electric field PSD does not suffice to describe the induced heating. For instance, the orientational decoherence and heating of a point quadrupole probes the PSD of the field gradient, a fourth-rank tensor, at its center-of-mass position. Likewise, the center-of-mass decoherence of a point dipole is determined by field gradient fluctuations characterized by the same tensor. For instance, the orientational decoherence of a point quadrupole at position $\bls{R}_{\rm cm}$ reads
\begin{align}\label{eq:DecoRateSlowQuadrupoleOrientational}
    \Gamma_{\Omega\,\Omega'}&(\bls{R}_{\rm cm},\bls{R}_{\rm cm})=\frac{1}{36\hbar}\left[\frac{\partial}{\partial \bls{R}}\cdot (\mathsf{Q}_\Omega-\mathsf{Q}_{\Omega'})\frac{\partial}{\partial \bls{R}}\right]\nonumber\\
    &\times\left.\left[\frac{\partial}{\partial \bls{R}'}\cdot (\mathsf{Q}_\Omega-\mathsf{Q}_{\Omega'})\frac{\partial}{\partial \bls{R}'}\right]h(\bls{R},\bls{R}')\right|_{\bls{R}=\bls{R}'=\bls{R}_{\rm cm}}.
\end{align}
Note that this equation exhibits a surface distance dependence distinct from the scaling in Eq.~\eqref{eq:electricFieldNoiseThinLayer}, showing that large-quadrupole particles might yield further insight into the origins of anomalous heating \cite{brown2021materials}.

{\it Probing non-local field correlations---}It follows from Eq.~\eqref{eq:DecoRateSlowParticle} that the center-of-mass decoherence rate of a spatially delocalized wavepacket depends on non-local noise correlations, rather than on the local electric-field PSD. In particular, the decoherence rate remains bounded for large distances in the superposition, as observed in experiments with electron beams \cite{kerker2020quantum}. Creating and monitoring spatially extended superposition states thus enables deducing information on the surface noise that is not encoded in the heating rates, e.g. by utilizing modern parameter estimation techniques \cite{papivc2021neural}.

{\it Modelling the dielectric response---}Identifying the microscopic origins of surface noise will require input from atomistic theoretical models. This encompasses accounting for the role of microscopic processes as sources and as mediators of electromagnetic noise. Ab-initio calculations \cite{botti2007time,ping2013electronic} and molecular-dynamics simulations \cite{foulon2022omega} link microscopic models to the macroscopic dielectric response, which in turn serves as one input for the presented toolbox and can thus be tested experimentally.

{\it Tailoring the surface response---}Once the mechanism of surface noise is understood, surface-induced decoherence may be suppressed by choosing materials with beneficial dielectric properties. Specific surface coatings may help suppressing such noise at the relevant particle frequencies or prevent adsorption of surface contaminants. Plasma cleaning \cite{mcconnell2015reduction} and ion milling \cite{hite2012100} of electrode surfaces in ion traps have shown that even slight surface modifications can have great impact on the noise level. 

{\it Geometry design---}As a further handle to reduce the impact of unavoidable surface fluctuations, one may adjust the surface geometry to modify the electric fields at the particle position. For instance, when working with dipolar particles, one can minimize field noise, as compared to field gradient noise, by a mirror-symmetric arrangement of short-circuited electrodes.

\subsection{Extending the toolbox}

The presented toolbox is based on a number of assumptions, namely the quasistatic approximation of the surface-induced field, the Born-Markov approximation for the particle dynamics, the rigid description of the particle charge distribution, and the neglect of magnetic surface effects, all of which is justified for typical ion and nanoparticle trapping experiments. That said, the toolbox could be extended in following ways:

{\it Transverse electric field---}Taking the transverse field into account requires two steps: First, the Green tensor, which describes both the longitudinal and the transverse field propagation, must be calculated instead of the scalar Green function. Second, the coupling between the particle velocity and the transverse fields must be accounted for. Note that the impact of Thomson scattering, which describes scattering from the purely transverse field at vanishing particle velocity, is discussed in App.~\ref{app:ThomsonScattering}, where it is shown that its impact is negligibly small compared to decoherence due to the longitudinal field and other experimentally relevant  decoherence processes.

{\it Anisotropic and non-local dielectrics---}To incorporate the possible anisotropy and non-locality of surface materials requires using the tensorial and non-local dielectric response in the definition of the surface polarization operators \cite{scheel2008macroscopic} and the scalar Green function. While the subsequent calculation is presumably tedious, we expect the derivation presented in this paper to carry over to the generalized case.

{\it Non-rigid charge distributions---}The fact that the bound charges inside the particle can fluctuate and react to fields emanating from the surface can be included in terms of the particle polarization. This adds to the dynamics (i) the Casimir-Polder interaction between the particle and the surface \cite{buhmann1} as well as (ii) the associated decoherence channel \cite{sinha2020quantum}. This effect is expected to become relevant for weakly-charged micron-sized dielectrics.

{\it Magnetic effects---}Magnetic contributions to the particle-surface interaction can be taken into account by adding to the Hamiltonian the coupling of fluctuating surface magnetization currents to the charge distribution (see App.~\ref{sec:macroscopicQED}). Likewise, if the particle exhibits permanent or fluctuating magnetic moments, their coupling to the surface polarization and magnetization fields can be included \cite{pino2018chip,buhmann1}. These extensions may serve to model the ro-translational heating of levitated magnets or superconductors close to surfaces \cite{gieseler2020single,vinante2020ultralow,streltsov2021ground,latorre2022chip}, which are an alternative to electrically levitated  particles \cite{gonzalez2021levitodynamics,stickler2021}.

{\it Ro-translational thermalization---}The dynamics of ro-translational damping are fully accounted for in the resonant limit. In the slow-particle limit, they can be described by including to first order the particle motion during the decay of surface-fluctuation correlations \cite{breuer2002}. This adds to the master equation terms that are linear in the center-of-mass momentum and angular momentum operators. The resulting master equation describes ro-translational thermalization if it is of a specific form \cite{stickler2018rotational}.

{\it Non-Markovian effects---}If the surface-fluctuation correlations decay on a timescale comparable to that of the particle motion, the weak-coupling Born-Markov approximation loses justification, even though it may still provide a good description for most situations. The impact of colored noise and related non-Markovian effects can be taken into account via the method of influence functionals \cite{breuer2002} or by using quantum Langevin equations \cite{gardinerzoller2004}. The latter feature an operator-valued noise to describe the dynamics of selected observables. If necessitated by empirical evidence, such a treatment is possible in principle, though significantly more involved and less versatile.

\subsection{Conclusions}

The framework described in this article  predicts quantitatively surface-induced rotational and translational decoherence and heating of extended objects with arbitrary shape and charge distribution. The quantum master equations \eqref{eq:MasterEquationSlowParticleIntro} and \eqref{eq:dissipatorQuasistatic} incorporate the surface properties in terms of (i) the dielectric permittivity, which is empirically accessible and microscopically computable, and (ii) the scalar Green function, which describes the field propagation in arbitrary geometries. Our work applies to a variety of state-of-the-art experimental setups, ranging from trapped ion setups \cite{brown2021materials}, to hybrid quantum devices with charged atoms and molecules \cite{xiang2013hybrid}, to nanoparticles in electric and optical traps \cite{stickler2021}. We expect the presented framework and the worked-out toolbox Tab.~\ref{tab:MasterEquations} to be instrumental for a wide class of future quantum experiments with charged objects of finite extension.

\acknowledgements
This research was funded by the Deutsche Forschungsgemeinschaft (DFG, German Research Foundation) -- 411042854.

\appendix
\section{Derivation of the coupling Hamiltonian}\label{sec:derivation}

This appendix derives the coupling Hamiltonian between the charged particle and the surface. We first derive the minimal coupling Hamiltonian \eqref{eq:couplingHamiltonian}  from the Lorentz force and torque laws for a moving and revolving rigid charge distribution. Inserting into this Hamiltonian the electromagnetic fields in terms of surface excitations then yields the surface-particle interaction Hamiltonian, which serves as a starting point for the derivation of the master equations.

\subsection{Rotor-field coupling}\label{sec:RotorFieldCoupling}

To describe the classical dynamics of a particle in the electromagnetic field, we introduce the position $\bls{R}$ and the orientation via Euler angles $\Omega=\{\alpha,\beta,\gamma\}$ in the $z$-$y'$-$z''$ convention. The particle charge distribution at a given point $\bls{r}$ reads $\varrho(\bls{r})=\varrho_0\!\left[\mathsf{R}_\Omega^{\rm T}(\bls{r}-\bls{R})\right]$. Here, $\varrho_0$ describes the charge distribution in the reference position and the rotation tensor $\mathsf{R}_\Omega$ relates the space-fixed axes $\bls{e}_i$ to the principal axes $\bls{n}_i=\mathsf{R}_\Omega\bls{e}_i$. The angular velocity is defined via the kinematic relation $\dot{\sf R}_\Omega = {\bf \omega}\times {\sf R}_\Omega$. The coupling Lagrangian,
\begin{align}\label{eq:LagrangeWithKineticEnergy}
    L=E_{\rm kin}-\int d^3r\,\varrho(\bls{r})\phi(\bls{r},t)+\int d^3r\,\bls{j}(\bls{r})\cdot\bls{A}(\bls{r},t),
\end{align}
involves the charge distribution and the current density,
\begin{align}
    \bls{j}(\bls{r})=\varrho_0\left[\mathsf{R}_\Omega^{\rm T}(\bls{r}-\bls{R})\right]\left[\dot{\bls{R}}+\blg{\omega}\times(\bls{r}-\bls{R})\right],
\end{align}
due to the rigid motion of the body. 

We first show that this Lagrangian is consistent with the Lorentz force and torque laws. To demonstrate this equivalence, we use that
\begin{equation}\label{eq:angularSpeedAngles}
    \blg{\omega}=\dot{\alpha}\bls{e}_3+\dot{\beta}\bls{e}_\xi+\dot{\gamma}\bls{n}_3
\end{equation}
with $\bls{e}_\xi=-\sin\alpha\bls{e}_1+\cos\alpha\bls{e}_2$. We employ \eqref{eq:angularSpeedAngles} to express the kinetic energy $E_{\rm kin}$ and the current density in terms of the Euler angles. We then transform the integrals in \eqref{eq:LagrangeWithKineticEnergy} to the body-fixed frame yielding
\begin{align}
    L=&\frac{1}{2}\blg{\omega}\cdot{\rm I}_\Omega\blg{\omega}+\frac{1}{2}m\dot{\bls{R}}^2-\int d^3r\,\varrho_0(\bls{r})\phi(\bls{R}+\mathsf{R}_\Omega\bls{r},t)\nonumber\\
    &+\int d^3r\,\varrho_0(\bls{r})\left(\dot{\bls{R}}+\blg{\omega}\times\mathsf{R}_\Omega\bls{r}\right)\cdot\bls{A}(\bls{R}+\mathsf{R}_\Omega\bls{r},t).
\end{align}

The Lagrange equation for the center of mass 
\begin{equation}
    \frac{d}{dt}\frac{\partial L}{\partial\dot{\bls{R}}}-\frac{\partial L}{\partial\bls{R}}=0
\end{equation}
involves the total time derivative of the vector potential 
\begin{align}\label{eq:totalTimeDerivativeA}
    \frac{d}{dt}\bls{A}(\bls{R}+\mathsf{R}_\Omega\bls{r},t)=&\left(\dot{\bls{R}}+\blg{\omega}\times\mathsf{R}_\Omega\bls{r}\right)\cdot\frac{\partial}{\partial \bls{R}_0}\bls{A}(\bls{R}_0,t)\nonumber\\
    &+\frac{\partial}{\partial t}\bls{A}(\bls{R}+\mathsf{R}_\Omega\bls{r},t)
\end{align}
with $\bls{R}_0=\bls{R}+\mathsf{R}_\Omega\bls{r}$. Using \eqref{eq:totalTimeDerivativeA} the Lagrange equation yields
\begin{align}
    m\ddot{\bls{R}}=&\int d^3r\,\varrho_0(\bls{r})\left(\dot{\bls{R}}+\blg{\omega}\times\mathsf{R}_\Omega\bls{r}\right)\times\bls{B}(\bls{R}+\mathsf{R}_\Omega\bls{r},t)\nonumber\\
    &+\int d^3r\, \varrho_0(\bls{r})\bls{E}(\bls{R}+\mathsf{R}_\Omega\bls{r},t),
\end{align}
which describes the center-of-mass dynamics of an extended charge distribution in presence of the electric and magnetic fields
\begin{align}
    \bls{E}(\bls{r},t)&=-\frac{\partial}{\partial \bls{r}}\phi(\bls{r},t)-\frac{\partial}{\partial t}\bls{A}(\bls{r},t),\\
    \bls{B}(\bls{r},t)&=\frac{\partial}{\partial \bls{r}}\times\bls{A}(\bls{r},t).
\end{align}

The Lagrange equations for the Euler angles are more involved. First, we deduce some useful relations from the definition of the rotation matrix and \eqref{eq:angularSpeedAngles},
\begin{subequations}
\begin{align}
    &\frac{\partial}{\partial \dot{\mu}}=\bls{e}_\mu\cdot \frac{\partial}{\partial \blg{\omega}},\label{eq:omegaDerivatives}\\
    &\frac{\partial \mathsf{R}_\Omega}{\partial \mu}=\bls{e}_\mu\times\mathsf{R}_\Omega,\label{eq:angularDerivativesRot}\\
   &\frac{d\bls{e}_\mu}{dt}=\mathsf{R}_\Omega\frac{\partial}{\partial \mu}\mathsf{R}_\Omega^{\rm T}\blg{\omega},\label{eq:timeDerivativesBasisvectors}
\end{align}
\end{subequations}
with $\mu \in \{\alpha,\beta,\gamma\}$ and $\bls{e}_\mu\in\{\bls{e}_3,\bls{e}_\xi,\bls{n}_3\}$.
Note that $d\bls{e}_3/dt =0$ since the space-fixed axis does not move. Let us start with the angle $\alpha$,
\begin{equation}
    \frac{d}{dt}\frac{\partial L}{\partial\dot{\alpha}}-\frac{\partial L}{\partial\alpha}=0,
\end{equation}
and use \eqref{eq:omegaDerivatives} to write 
\begin{align}
    \frac{d}{dt}\frac{\partial}{\partial \dot{\alpha}}\frac{1}{2}\blg{\omega}\cdot{\rm I}_\Omega\blg{\omega}&= \frac{d}{dt}\bls{e}_3\cdot{\rm I}_\Omega\blg{\omega}\nonumber\\
    &=\bls{e}_3\cdot\frac{d}{dt}{\rm I}_\Omega\blg{\omega}+\frac{d\bls{e}_3}{dt}\cdot{\rm I}_\Omega\blg{\omega}\label{eq:freeRotorLagrangeAlpha1}.
\end{align}
Since ${\rm I}_0=\mathsf{R}_\Omega^{\rm T}{\rm I}_\Omega\mathsf{R}_\Omega$ is independent of the Euler angles, we find
\begin{align}
    \frac{\partial}{\partial \alpha}\frac{1}{2}\blg{\omega}\cdot{\rm I}_\Omega\blg{\omega}=&\frac{\partial}{\partial \alpha}\frac{1}{2}(\mathsf{R}_\Omega^{\rm T}\blg{\omega})\cdot{\rm I}_0(\mathsf{R}_\Omega^{\rm T}\blg{\omega})\nonumber\\
    =&\frac{d\bls{e}_3}{dt}\cdot{\rm I}_\Omega\blg{\omega}\label{eq:freeRotorLagrangeAlpha2},
\end{align}
where we used \eqref{eq:timeDerivativesBasisvectors}. Combining \eqref{eq:freeRotorLagrangeAlpha1} and \eqref{eq:freeRotorLagrangeAlpha2}, one thus obtains the $\bls{e}_3$-component of the angular-momentum time-derivative
\begin{equation}\label{eq:torquePart1}
    \left(\frac{d}{dt}\frac{\partial}{\partial\dot{\alpha}}-\frac{\partial}{\partial \alpha}\right)\frac{1}{2}\blg{\omega}\cdot{\rm I}_\Omega\blg{\omega}=\bls{e}_3\cdot\frac{d}{dt}{\rm I}_\Omega\blg{\omega}.
\end{equation}

The term involving the scalar potential can be rewritten using \eqref{eq:angularDerivativesRot},
\begin{align}
    \left(\frac{d}{dt}\frac{\partial}{\partial\dot{\alpha}}-\frac{\partial}{\partial \alpha}\right)&\phi(\bls{R}+\mathsf{R}_\Omega\bls{r},t)\nonumber\\
    &=-\bls{e}_3\cdot(\mathsf{R}_\Omega\bls{r})\times\frac{\partial}{\partial \bls{R}_0}\phi(\bls{R}_0,t),\label{eq:torquePart2}
\end{align}
again with $\bls{R}_0=\bls{R}+\mathsf{R}_\Omega\bls{r}$. Likewise, we evaluate the term involving the vector potential. In a first step, we compute
\begin{align}
\frac{d}{dt}\frac{\partial}{\partial\dot{\alpha}}&\left[\left(\dot{\bls{R}}+\blg{\omega}\times\mathsf{R}_\Omega\bls{r}\right)\cdot\bls{A}(\bls{R}+\mathsf{R}_\Omega\bls{r},t)\right]\nonumber\\
=&\frac{d}{dt}\left[\bls{e}_3\cdot\left(\mathsf{R}_\Omega\bls{r}\right)\times\bls{A}(\bls{R}+\mathsf{R}_\Omega\bls{r},t)\right]\nonumber\\\label{eq:timeAngularDerivativeA}
=&\bls{e}_3\cdot\left(\mathsf{R}_\Omega\bls{r}\right)\times\frac{d}{dt}\bls{A}(\bls{R}+\mathsf{R}_\Omega\bls{r},t)\nonumber\\
&+\bls{A}(\bls{R}+\mathsf{R}_\Omega\bls{r},t)\cdot\frac{d}{dt} \bls{e}_3\times\left(\mathsf{R}_\Omega\bls{r}\right),
\end{align}
where we used \eqref{eq:omegaDerivatives}. Noting 
\begin{align}
    \frac{d}{dt}\bls{e}_3\times\left(\mathsf{R}_\Omega\bls{r}\right)=\frac{d}{dt} \frac{\partial}{\partial \alpha}\mathsf{R}_\Omega\bls{r}=\frac{\partial}{\partial \alpha}\blg{\omega}\times\mathsf{R}_\Omega\bls{r},
\end{align}
Eq.~\eqref{eq:timeAngularDerivativeA} yields
\begin{align}
   \frac{d}{dt}\frac{\partial}{\partial\dot{\alpha}}&\left[\left(\dot{\bls{R}}+\blg{\omega}\times\mathsf{R}_\Omega\bls{r}\right)\cdot\bls{A}(\bls{R}+\mathsf{R}_\Omega\bls{r},t)\right]\nonumber\\
=&\bls{e}_3\cdot\left(\mathsf{R}_\Omega\bls{r}\right)\times\frac{d}{dt}\bls{A}(\bls{R}+\mathsf{R}_\Omega\bls{r},t)\nonumber\\
&+\bls{A}(\bls{R}+\mathsf{R}_\Omega\bls{r},t)\cdot\frac{\partial}{\partial \alpha}\blg{\omega}\times(\mathsf{R}_\Omega\bls{r}).
\end{align}
Subtracting the $\alpha$-derivative gives
\begin{align}
   \left(\frac{d}{dt}\frac{\partial}{\partial\dot{\alpha}}-\frac{\partial}{\partial\alpha}\right)&\left[\left(\dot{\bls{R}}+\blg{\omega}\times\mathsf{R}_\Omega\bls{r}\right)\cdot\bls{A}(\bls{R}+\mathsf{R}_\Omega\bls{r},t)\right]\nonumber\\
   =&\bls{e}_3\cdot\left(\mathsf{R}_\Omega\bls{r}\right)\times\frac{d}{dt}\bls{A}(\bls{R}+\mathsf{R}_\Omega\bls{r},t)\nonumber\\
   &-\left(\dot{\bls{R}}+\blg{\omega}\times\mathsf{R}_\Omega\bls{r}\right)\cdot\frac{\partial}{\partial\alpha}\bls{A}(\bls{R}+\mathsf{R}_\Omega\bls{r},t).
\end{align}
With the help of \eqref{eq:angularDerivativesRot} one finds that
\begin{align}
   &\left(\frac{d}{dt}\frac{\partial}{\partial\dot{\alpha}}-\frac{\partial}{\partial\alpha}\right)\left[\left(\dot{\bls{R}}+\blg{\omega}\times\mathsf{R}_\Omega\bls{r}\right)\cdot\bls{A}(\bls{R}+\mathsf{R}_\Omega\bls{r},t)\right]\nonumber\\
   &=\bls{e}_3\cdot\left(\mathsf{R}_\Omega\bls{r}\right)\times\frac{d}{dt}\bls{A}(\bls{R}+\mathsf{R}_\Omega\bls{r},t)\nonumber\\
   &-\left(\bls{e}_3\cdot(\mathsf{R}_\Omega\bls{r})\times\frac{\partial}{\partial \bls{R}_0}\right)\left(\dot{\bls{R}}+\blg{\omega}\times\mathsf{R}_\Omega\bls{r}\right)\cdot\bls{A}(\bls{R}_0,t).\\
   &=\bls{e}_3\cdot\left(\mathsf{R}_\Omega\bls{r}\right)\times\frac{\partial}{\partial t}\bls{A}(\bls{R}+\mathsf{R}_\Omega\bls{r},t)\nonumber\\
   &-\bls{e}_3\cdot(\mathsf{R}_\Omega\bls{r})\times\left\{\left(\dot{\bls{R}}+\blg{\omega}\times\mathsf{R}_\Omega\bls{r}\right)\times\left[\frac{\partial}{\partial \bls{R}_0}\times\bls{A}(\bls{R}_0,t)\right]\right\}\label{eq:torquePart3}
\end{align}
where \eqref{eq:totalTimeDerivativeA} and the Graßmann identity have been used. We combine \eqref{eq:torquePart1}, \eqref{eq:torquePart2} and \eqref{eq:torquePart3} with the Lagrange equation for $\alpha$ to finally obtain
\begin{align}\label{eq:RotEquationMotionAlpha}
    \bls{e}_3\cdot\frac{d}{dt} {\rm I}_\Omega\blg{\omega}=\bls{e}_3\cdot\int d^3r\,\varrho_0(\bls{r})\left(\mathsf{R}_\Omega\bls{r}\right)\times\Big[\bls{E}(\bls{R}+\mathsf{R}_\Omega\bls{r},t)\nonumber\\
    +\left(\dot{\bls{R}}+\blg{\omega}\times\mathsf{R}_\Omega\bls{r}\right)\times\bls{B}(\bls{R}+\mathsf{R}_\Omega\bls{r},t)\Big].
\end{align}
This is the $\bls{e}_3$-component of the Lorentz torque acting on an extended charge distribution. 

Since \eqref{eq:omegaDerivatives}-\eqref{eq:timeDerivativesBasisvectors} equally hold for $\beta$ and $\gamma$, the corresponding Lagrange equations yield the analogue of \eqref{eq:RotEquationMotionAlpha}, but with $\bls{e}_3$ replaced by $\bls{e}_\xi$ and $\bls{n}_3$, respectively. Thus, we demonstrated that the Lagrangian \eqref{eq:LagrangeWithKineticEnergy} is indeed consistent with the Lorentz torque acting on an extended charge distribution,
\begin{align}
    \frac{d}{dt} {\rm I}_\Omega\blg{\omega}=\int d^3r\,\varrho_0(\bls{r})\left(\mathsf{R}_\Omega\bls{r}\right)\times\Big[\bls{E}(\bls{R}+\mathsf{R}_\Omega\bls{r},t)\nonumber\\
    +\left(\dot{\bls{R}}+\blg{\omega}\times\mathsf{R}_\Omega\bls{r}\right)\times\bls{B}(\bls{R}+\mathsf{R}_\Omega\bls{r},t)\Big].
\end{align}

In order to obtain the Hamiltonian we first define the canonical linear and angular momenta
\begin{align}
    \bls{P}&=\frac{\partial L}{\partial \dot{\bls{R}}}=m\dot{\bls{R}}+\int d^3r\,\varrho_0(\bls{r})\bls{A}(\bls{R}+\mathsf{R}_\Omega\bls{r},t),\\
    \bls{J}&=\frac{\partial L}{\partial \blg{\omega}}={\rm I}_\Omega\blg{\omega}+\int d^3r\,\varrho_0(\bls{r})(\mathsf{R}_\Omega\bls{r})\times\bls{A}(\bls{R}+\mathsf{R}_\Omega\bls{r},t),
\end{align}
and then perform a Legendre transformation. Here we used \eqref{eq:omegaDerivatives} and related the angular momentum vector to the canonical Euler momenta via $p_\alpha=\bls{e}_3\cdot\bls{J}$, $p_\beta=\bls{e}_\xi\cdot\bls{J}$ and $p_\gamma=\bls{n}_3\cdot\bls{J}$.
The Legendre transformation
\begin{subequations}
\begin{align}
    H&=\dot{\bls{R}}\cdot\bls{P}+\dot{\alpha}p_\alpha+\dot{\beta}p_\beta+\dot{\gamma}p_\gamma-L\\
   &=\dot{\bls{R}}\cdot\bls{P}+\blg{\omega}\cdot\bls{J}-L
\end{align}
\end{subequations}
yields the Hamiltonian \eqref{eq:couplingHamiltonian}.

\subsection{Macroscopic quantum electrodynamics}\label{sec:macroscopicQED}
One central ingredient of macroscopic quantum electrodynamics is the Green tensor $\mathsf{G}(\bls{r},\bls{r}',\omega)$, giving the electromagnetic field at the position $\bls{r}$ due to a source current located at $\bls{r}'$. In the presence of dispersive media with permittivity $\varepsilon_{\rm r}(\bls{r},\omega)$ and permeability $\mu_{\rm r}(\bls{r},\omega)$, it fulfills the defining equation
\begin{align}\label{eq:DGLFullGreenTensor}
    &\frac{\partial}{\partial \bls{r}}\times\frac{1}{\mu_{\rm r}(\bls{r},\omega)}\left[\frac{\partial}{\partial \bls{r}}\times\mathsf{G}(\bls{r},\bls{r'},\omega)\right]-\frac{\omega^2}{c^2}\varepsilon_{\rm r}(\bls{r},\omega)\mathsf{G}(\bls{r},\bls{r'},\omega)\nonumber\\
    &=\mathbb{1}\delta(\bls{r}-\bls{r'}),
\end{align}
along with the boundary condition $\mathsf{G}(\bls{r},\bls{r'},\omega)=\mathbb{0}$ for $|\bls{r}-\bls{r'}|\to \infty$. The electric field 
\begin{equation}
    \bls{E}(\bls{r},t)=\frac{1}{2\pi}\int_0^\infty d\omega\, \bls{E}(\bls{r},\omega)e^{-i\omega t}+{\rm c.c.}
\end{equation}
due to the current density $\bls{j}(\bls{r},\omega)$ is then given by
\begin{equation}\label{eq:electricFieldCurrent}
    \bls{E}(\bls{r},\omega)=i\mu_0\omega\int d^3r'\,\mathsf{G}(\bls{r},\bls{r}',\omega)\bls{j}(\bls{r}',\omega).
\end{equation}
If the current density obeys the fluctuation-dissipation theorem, one obtains \cite{buhmann1},
\begin{align}\label{eq:classicalCorrelationElectricFieldOperators}
    &\braket{\bls{E}^*(\bls{r},\omega)\otimes\bls{E}(\bls{r},\omega')}\nonumber\\
    &=4\pi k_{\rm B}T\mu_0\omega\delta(\omega-\omega'){\rm Im}\left[\mathsf{G}(\bls{r},\bls{r},\omega)\right].
\end{align}

The theory is quantized by expressing the current density in terms of quantum harmonic oscillators \cite{buhmann1,scheel2008macroscopic}, the so-called polarization and magnetization oscillators $\bls{f}_\lambda(\bls{r},\omega)$, $\lambda\in\{{\rm e,m}\}$. They obey the canonical commutation relations $[\bls{f}_{\lambda}(\bls{r},\omega),\bls{f}_{\lambda'}(\bls{r'},\omega')]=\mathbb{0}$ and $[\bls{f}_{\lambda}(\bls{r},\omega),\bls{f}^\dagger_{\lambda'}(\bls{r'},\omega')]=\mathbb{1}\delta_{\lambda,\lambda'}\delta(\bls{r}-\bls{r'})\delta(\omega-\omega')$. These oscillators are related to the (noise) polarization and magnetization of the medium through 
\begin{align}
    \bls{P}_{\rm N}(\bls{r},\omega)&=i\sqrt{4\pi\hbar\varepsilon_0{\rm Im}\left[\varepsilon_{\rm r}(\bls{r},\omega)\right]}\bls{f}_{\rm e}(\bls{r},\omega),\\
    \bls{M}_{\rm N}(\bls{r},\omega)&=\sqrt{\frac{4\pi\hbar}{\mu_0}\frac{{\rm Im}\left[\mu_{\rm r}(\bls{r},\omega)\right]}{|\mu_{\rm r}(\bls{r},\omega)|^2}}\bls{f}_{\rm m}(\bls{r},\omega),
\end{align}
so that the associated noise current density 
\begin{equation}
\bls{j}_{\rm N}(\bls{r},\omega)=-i\omega\bls{P}_{\rm N}(\bls{r},\omega)+\frac{\partial}{\partial\bls{r}}\times\bls{M}_{\rm N}(\bls{r},\omega)
\end{equation}
acts as a source for the electromagnetic field. The free field dynamics are governed by the Hamiltonian
\begin{equation}\label{eq:freeFieldEnergy}
 H_{\rm f}=\sum_{\lambda={\rm e,m}}\int d^3r\,\int_0^\infty d\omega \,\hbar \omega\,\bls{f}^\dagger_{\lambda}(\bls{r},\omega)\cdot\bls{f}_{\lambda}(\bls{r},\omega). 
\end{equation}

In Coulomb gauge the electromagnetic potentials originating from the dielectric can be expressed through the (Schrödinger-picture) polarization and magnetization oscillators as \cite{buhmann1,scheel2008macroscopic}
\begin{align}\label{eq:MQE_vectorpotential}
\bsf{A}(\bls{r})=&\sum_{\lambda={\rm e,m}}\int d^3r'\,\int_0^\infty d\omega\, \frac{1}{i\omega} \vspace{0mm}^\perp \mathsf{G}_{\lambda}(\bls{r},\bls{r'},\omega)\bls{f}_{\lambda}(\bls{r}',\omega)\nonumber\\
&+{\rm h.c.},\\
    \phi(\bls{r})=&-\frac{1}{4\pi}\sum_{\lambda={\rm e,m}}\int d^3r'\,\int_0^\infty d\omega\, \int d^3s\,\frac{\bls{r}-\bls{s}}{|\bls{r}-\bls{s}|^3}\nonumber\\
    &\cdot \mathsf{G}_{\lambda}(\bls{s},\bls{r'},\omega)\bls{f}_{\lambda}(\bls{r}',\omega)+{\rm h.c.}.\label{eq:MQE_scalarpotential}
\end{align}
Here 
\begin{equation}
^\perp\mathsf{G}_\lambda(\bls{r},\bls{r}',\omega)=\int d^3s\, \delta^\perp(\bls{r}-\bls{s})\mathsf{G}_\lambda(\bls{s},\bls{r}',\omega)
\end{equation}
is the left-transverse part of the tensors $\mathsf{G}_\lambda$, which are related to the Green tensor,
\begin{align}
    \mathsf{G}_{\rm e}(\bls{r},\bls{r'},\omega)=&i\frac{\omega^2}{c^2}\sqrt{\frac{\hbar}{\pi\varepsilon_0}{\rm Im}\left[\varepsilon_{\rm r}(\bls{r'},\omega)\right]}\mathsf{G}(\bls{r},\bls{r'},\omega),\\
    \mathsf{G}_{\rm m}(\bls{r},\bls{r'},\omega)=&i\frac{\omega}{c}\sqrt{\frac{\hbar}{\pi\varepsilon_0}\frac{{\rm Im}\left[\mu_{\rm r}(\bls{r'},\omega)\right]}{|\mu_{\rm r}(\bls{r'},\omega)|^2}}\left[\frac{\partial}{\partial \bls{r}'}\!\times\!\mathsf{G}(\bls{r}',\bls{r},\omega)\right]^{\rm T}.
\end{align}
The transverse delta function is defined as
\begin{equation}\label{eq:definitionTransverseDelta}
    \delta^\perp(\bls{r})=\frac{\partial}{\partial \bls{r}}\times\left(\frac{\partial}{\partial \bls{r}}\times \mathbb{1}\right)\frac{1}{4\pi r}.
\end{equation}

Expressing the Green tensor in terms of the scalar Green function $g(\bls{r},\bls{r}',\omega)$ (see App.~\ref{sec:quasi_static}) shows that the vector potential vanishes, while the scalar potential is of the form Eq.~\eqref{eq:phiviag}. In the following, we will formulate the theory in terms of the Green tensor to keep the discussion general. Simplifications associated with using the scalar Green function $g$ are discussed along the way.

The quantized electric field \eqref{eq:electricFieldCurrent} thus reads
\begin{equation}\label{eq:electricFieldOperator}
    \bls{E}(\bls{r})=\sum_{\lambda={\rm e,m}}\int d^3r'\,\int_0^\infty d\omega\, \mathsf{G}_{\lambda}(\bls{r},\bls{r'},\omega)\bls{f}_{\lambda}(\bls{r}',\omega)+{\rm h.c.}.
\end{equation}
We note that in presence of a particle, the field \eqref{eq:electricFieldOperator} is not yet the physical electric field, since the Coulomb field of the particle charge in absence of the dielectric medium has still to be added \cite{buhmann1}. The dynamics generated by the minimal coupling Hamiltonian \eqref{eq:couplingHamiltonian} and the free-field dynamics \eqref{eq:freeFieldEnergy} are consistent with the Maxwell equations, the Lorentz force and torque, and thermal field fluctuations in absorbing media.

In the following, we neglect the vector potential to get the total Hamiltonian
\begin{align}
 H&= H_0+H_{\rm f}+ H_{\rm int},
\end{align}
where the free Hamiltonian 
\begin{equation}\label{eq:H0_MQE}
     H_0=\frac{1}{2m}\bsf{P}^2+\frac{1}{2}\bsf{J}\cdot {\rm I}_\Omega^{-1}\bsf{J}+V_{\rm ext}(\bls{R},\Omega),
\end{equation}
involves the external potential $V_{\rm ext}(\bls{R},\Omega)$ and the particle-field interaction reads as
\begin{align}\label{eq:hint}
     H_{\rm int}=&-\frac{1}{4\pi}\int d^3r\,\varrho_0(\bls{r})
    \sum_{\lambda={\rm e,m}}\int d^3r'\,\int_0^\infty d\omega\, \int d^3s\nonumber\\
    &\times\frac{\bls{R}+\mathsf{R}_\Omega\bls{r}-\bls{s}}{|\bls{R}+\mathsf{R}_\Omega\bls{r}-\bls{s}|^3}\cdot \mathsf{G}_{\lambda}(\bls{s},\bls{r'},\omega)\bls{f}_{\lambda}(\bls{r}',\omega)+{\rm h.c.}.
\end{align}
Here $\bls{R}$, $\bls{P}$, $\Omega=\{\alpha,\beta,\gamma\}$, $p_\alpha=\bls{e}_3\cdot\bls{J}$, $p_\beta=\bls{e}_\xi\cdot\bls{J}$ and $p_\gamma=\bls{n}_3\cdot\bls{J}$ are operators satisfying the canonical commutation relations of translation and rotation. 

\section{Born-Markov approximation}\label{sec:derivation_markovian_master_equation}

This section derives the two main master equations from the above discussed interaction Hamiltonian. In a first step, we will perform the weak-coupling approximation, and then evaluate the resulting Markovian master equation in the slow-particle limit and in the resonant limit.

\subsection{Weak-coupling approximation}

In the weak-coupling approximation, the time evolution of the interaction-picture density operator $\tilde{\rho}(t)=U_0^\dagger(t)\rho(t)U_0(t)$ with $U_0(t)=\exp\left[-it (H_0+ H_{\rm f})/\hbar\right]$ can be written as \cite{breuer2002}
\begin{align}\label{eq:weakCoupling}
    \frac{\partial \tilde{\rho}}{\partial t}\simeq \frac{1}{\hbar^2}\int_0^\infty d\tau {\rm tr_B}\left\{\left[\tilde{ H}_{\rm int}(t),\left[\tilde{\rho}(t)\otimes\rho_{\rm B},\tilde{ H}_{\rm int}(t-\tau)\right]\right]\right\}
\end{align}
where the bath density operator $\rho_{\rm B}$ determines the state of the field degrees-of-freedom, which are assumed to be uncorrelated with the particle state, while the bath correlation (surface-fluctuation correlation) is assumed to decay much faster than the particle-relaxation time. The interaction Hamiltonian \eqref{eq:hint} in the interaction picture decomposes as
\begin{align}\label{eq:InteractionHamiltonianCb}
    \tilde{ H}_{\rm int}(t)=U_0^\dagger(t) H_{\rm int}U_0(t)=\int d^3s\, \bls{C}(\bls{s},t)\cdot \bls{b}(\bls{s},t)
\end{align}
where we defined the Coulomb field operator 
\begin{align}
    \bsf{C}(\bls{s},t)=&U_0^\dagger(t)\int d^3r\,\varrho_0(\bls{r})\frac{\bls{s}-\bsf{R}-\mathsf{R}_\Omega\bls{r}}{|\bls{s}-\bsf{R}-\mathsf{R}_\Omega\bls{r}|^3}U_0(t),\label{eq:CoulombFieldOperator}
\end{align}
and the bath operator
\begin{align}
    \bls{b}(\bls{s},t)=&\sum_{\lambda={\rm e,m}}\int d^3r'\,\int_0^\infty d\omega\, \mathsf{G}_{\lambda}(\bls{s},\bls{r'},\omega)\bls{f}_{\lambda}(\bls{r}',\omega)\frac{e^{-i\omega t}}{4\pi}\nonumber\\
    &+{\rm h.c.}.
\end{align}

We assume the bath state $\rho_{\rm B}$ to be thermal with respect to the free field Hamiltonian, implying that $\braket{ \bls{f}_{\lambda}(\bls{r}',\omega)}_{\rm B}=0$, where $\braket{\cdot}_{\rm B}={\rm tr}_{\rm B}\left\{\cdot\rho_{\rm B}\right\}$. The second moments of the bath oscillators are finite,
\begin{align}\label{eq:thermalSecondMomentsF}
\braket{\bls{f}^\dagger_{\lambda}(\bls{r},\omega)\otimes\bls{f}_{\lambda'}(\bls{r}',\omega')}_{\rm B}=n(\omega)\delta(\bls{r}-\bls{r}')\delta(\omega-\omega')\delta_{\lambda,\lambda'}\mathbb{1},
\end{align}
where we introduced the Bose-Einstein distribution $n(\omega)=1/[{\rm exp}(\hbar\omega/k_{\rm B}T)-1]$. The second moments determine the bath-operator autocorrelations,
\begin{align}
     \braket{\bls{b}(\bls{s},t)\otimes\bls{b}(\bls{s}',t-\tau)}_{\rm B}=\frac{\mathsf{N}(\bls{s},\bls{s}',\tau)-i\mathsf{D}(\bls{s},\bls{s}',\tau)}{2},\\
     \braket{\bls{b}(\bls{s}',t-\tau)\otimes\bls{b}(\bls{s},t)}_{\rm B}=\frac{\mathsf{N}(\bls{s}',\bls{s},\tau)+i\mathsf{D}(\bls{s}',\bls{s},\tau)}{2},
\end{align}
These autocorrelations contain the tensorial dissipation kernel \cite{breuer2002}
\begin{align}\label{eq:dissipationKernel}
    \mathsf{D}(\bls{s},\bls{s}',\tau)=\frac{2\hbar}{(4\pi)^2}\int_0^\infty d\omega\, \mathsf{J}(\bls{s},\bls{s}',\omega){\rm sin}(\omega \tau)
\end{align}
and the tensorial noise kernel
\begin{align}\label{eq:noiseKernel}
    \mathsf{N}(\bls{s},\bls{s}',\tau)=&\frac{2\hbar}{(4\pi)^2}\int_0^\infty d\omega\, \mathsf{J}(\bls{s},\bls{s}',\omega){\rm coth}\left(\frac{\hbar\omega}{2k_{\rm B}T}\right)\nonumber\\
    &\times{\rm cos}(\omega \tau).
\end{align}
Both are determined by the spectral density
\begin{equation}\label{eq:spectralDensityDefinition}
    \mathsf{J}(\bls{s},\bls{s}',\omega)=\frac{\mu_0}{\pi}\omega^2{\rm Im}\left[\mathsf{G}(\bls{s},\bls{s}',\omega)\right],
\end{equation}
as follows by using the relation \cite{buhmann1},
\begin{align}\label{eq:relationGGImGreenTensor}
    \sum_{\lambda={e,m}}\int d^3r'\,\mathsf{G}_{\lambda}(\bls{s},\bls{r}',\omega)&{\mathsf{G}^*_{\lambda}}^{\rm T}(\bls{s}',\bls{r}',\omega)\nonumber\\
    =&\frac{\hbar\mu_0}{\pi}\omega^2{\rm Im}\left[\mathsf{G}(\bls{s},\bls{s}',\omega)\right].
\end{align}
In addition, the Onsager reciprocity \cite{buhmann1} of the Green tensor,
\begin{equation}
    \mathsf{G}^{\rm T}(\bls{s},\bls{s}',\omega)=\mathsf{G}(\bls{s}',\bls{s},\omega),
\end{equation}
also applies to $\mathsf{D}$, $\mathsf{N}$ and $\mathsf{J}$, while the Schwarz reflection principle of $\mathsf{G}$ leads to
\begin{equation}
    \mathsf{J}(\bls{s},\bls{s}',-\omega)=-\mathsf{J}(\bls{s},\bls{s}',\omega).
\end{equation} 
In summary, Eq.~\eqref{eq:weakCoupling} can be rewritten as
\begin{align}\label{eq:nonMarkovianMasterEquation}
    \frac{\partial \tilde{\rho}}{\partial t}=&\frac{1}{\hbar^2}\int_0^\infty d\tau \int d^3sd^3s'\nonumber\\
    &\times\left(\frac{i}{2}\Big[\bls{C}(\bls{s},t),\big\{\mathsf{D}(\bls{s},\bls{s}',\tau)\bls{C}(\bls{s}',t-\tau),\tilde{\rho}(t)\big\}\Big]\right.\nonumber\\
    &\left.-\frac{1}{2}\Big[\bls{C}(\bls{s},t),\big[\mathsf{N}(\bls{s},\bls{s}',\tau)\bls{C}(\bls{s}',t-\tau),\tilde{\rho}(t)\big]\Big]\right).
\end{align}
This non-Markovian evolution equation serves as a starting point to derive the Markovian master equations in two different limiting cases: the slow-particle limit and the resonant limit.

\subsection{Slow-particle limit}\label{sec:derivationSlowParticle}
If the free particle motion is slow, so that the autocorrelations $\mathsf{D}$ and $\mathsf{N}$ drop on a much shorter timescale than the timescale over which the Coulomb field operator $\bls{C}(\bls{s}',t)$ changes, one can approximate
\begin{equation}\label{eq:slowParticleApproximationC}
    \bls{C}(\bls{s}',t-\tau)\simeq\bls{C}(\bls{s}',t).
\end{equation}
In this case, the time-integration in \eqref{eq:nonMarkovianMasterEquation} can be carried out, yielding
\begin{align}
    \int_0^\infty d\tau \mathsf{D}(\bls{s},\bls{s}',\tau)
    =&\frac{2\hbar}{(4\pi)^2}\frac{\mu_0}{\pi}\int_0^\infty d\omega\, \omega\,{\rm Im}\left[\mathsf{G}(\bls{s},\bls{s}',\omega)\right],\\
    \int_0^\infty d\tau \mathsf{N}(\bls{s},\bls{s}',\tau)=& \frac{2k_{\rm B}T\mu_0}{(4\pi)^2}\lim_{\omega\downarrow 0}\omega\,{\rm Im}\left[\mathsf{G}(\bls{s},\bls{s}',\omega)\right].
\end{align}
Transforming equation \eqref{eq:nonMarkovianMasterEquation} back to the Schrödinger picture one obtains
\begin{align}\label{eq:slowParticleMasterEqNonLindblad}
    \frac{\partial \rho}{\partial t}=&-\frac{i}{\hbar}\left[ H_0+ H_{\rm si},\rho\right]+\frac{2k_{\rm B}T\mu_0}{(4\pi)^2\hbar^2}\int d^3sd^3s'\nonumber\\
    &\times\left(\bls{C}(\bls{s},0)\cdot\rho\lim_{\omega\downarrow 0}\omega{\rm Im}\left[\mathsf{G}(\bls{s},\bls{s}',\omega)\right]\bls{C}(\bls{s}',0)\right.\nonumber\\
    &-\frac{1}{2}\left.\left\{\bls{C}(\bls{s},0)\cdot\lim_{\omega\downarrow 0}\omega{\rm Im}\left[\mathsf{G}(\bls{s},\bls{s}',\omega)\right]\bls{C}(\bls{s}',0),\rho\right\}\right),
\end{align}
where we introduced the coherent surface interaction (Lamb shift)
\begin{align}\label{eq:LambShiftSlowParticle}
     H_{\rm si}=&-\frac{1}{(4\pi)^2}\frac{\mu_0}{\pi}\int d^3sd^3s'\,\bls{C}(\bls{s},0)\nonumber\\
    &\cdot\int_0^\infty d\omega\, \omega{\rm Im}\left[\mathsf{G}(\bls{s},\bls{s}',\omega)\right]\bls{C}(\bls{s}',0).
\end{align}
The frequency integral included in \eqref{eq:LambShiftSlowParticle} can be rewritten as an integral over the whole frequency axis, 
\begin{equation}
    \int_0^\infty d\omega\, \omega{\rm Im}\left[\mathsf{G}(\bls{s},\bls{s}',\omega)\right]=\frac{1}{2i}\mathsf{P}\int_{-\infty}^\infty d\omega\, \omega\mathsf{G}(\bls{s},\bls{s}',\omega),
\end{equation}
where ${\sf P}$ denotes the Cauchy principal value. Using contour integration, exploiting that the Green tensor is holomorphic in the upper complex frequency plane, that it approaches $\mathsf{G}(\bls{s},\bls{s}',\omega)\sim-c^2\mathbb{1}\delta(\bls{s}-\bls{s}')/\omega^2$ for large $\omega$ \cite{buhmann1}, and that it is longitudinal with \eqref{eq:quasistaticGDerivation} for $\omega$ towards zero, the integral can be evaluated
\begin{align}
    \frac{1}{2i}\mathsf{P}\int_{-\infty}^\infty d\omega\, \omega\mathsf{G}(\bls{s},\bls{s}',\omega)=&-\frac{\pi}{2\mu_0}\frac{\partial}{\partial \bls{s}}\otimes\frac{\partial}{\partial \bls{s}'}g(\bls{s},\bls{s}',0)\nonumber\\
    &+ \frac{\pi c^2}{2}\mathbb{1}\delta(\bls{s}-\bls{s}').
\end{align}
Consequently, the surface-interaction Hamiltonian contains the negative electrostatic self energy of the rigid particle charge distribution and an additional contribution due to the electrostatic interaction of particle and the dielectric,
\begin{align}\label{eq:LambShiftSlowParticleWithg}
    H_{\rm si}=&\frac{1}{2}\int d^3sd^3s'\,\left[-\frac{\varrho_0(\bls{s})\varrho_0(\bls{s}')}{4\pi\varepsilon_0|\bls{s}-\bls{s}'|}+\varrho_0\!\left[\mathsf{R}_\Omega^{\rm T}(\bls{s}-\bls{R})\right]\right.\nonumber\\
    &\left.\times\varrho_0\!\left[\mathsf{R}_\Omega^{\rm T}(\bls{s}'-\bls{R})\right]g(\bls{s},\bls{s}',0)\vphantom{\frac{1}{2}}\right].
\end{align}
The self energy, given by the first term, is independent of the particle operators $\bls{R}$ and $\mathsf{R}_\Omega$ and thus does not have an impact on the particle dynamics in the slow-particle limit.

The second term gives the image-charge interaction. It is the potential energy of a classical charge distribution due to the presence of a dielectric surface. For instance, a monopole with $\varrho_0(\bls{r})=q\delta(\bls{r}-\bls{R})$ in front of a Drude-metal half space (see Tab.~\ref{tab:GreenfunctionHalfSpace}) with permittivity \eqref{eq:responseDrudeMetal}, experiences the image potential 
\begin{equation}
    H_{\rm si}=-\frac{1}{2}\frac{q^2}{4\pi\epsilon_0}\frac{1}{|2\bls{R}\cdot\bls{e}_3|}.
\end{equation}

The master equation \eqref{eq:slowParticleMasterEqNonLindblad} can be brought into Lindblad form by using the relation \eqref{eq:relationGGImGreenTensor},
\begin{align}
    &\frac{\partial \rho}{\partial t}=-\frac{i}{\hbar}\left[ H_0+ H_{\rm si},\rho\right]+\frac{\varepsilon_0}{\hbar}\lim_{\omega\downarrow 0}\frac{2k_{\rm B}T}{\hbar \omega}\int d^3r\nonumber\\
    &\times\left[\frac{{\rm Im}[\varepsilon_{\rm r}(\bls{r},\omega)]}{|\varepsilon_{\rm r}(\bls{r},\omega)|^2}\Big(\bls{L}_{\rm e}(\bls{r})\cdot \rho \bls{L}^\dagger_{\rm e}(\bls{r})-\frac{1}{2}\left\{\bls{L}^\dagger_{\rm e}(\bls{r})\cdot \bls{L}_{\rm e}(\bls{r}),\rho\right\}\Big)\right.\nonumber\\
    &\left.+\frac{{\rm Im}[\mu_{\rm r}(\bls{r},\omega)]}{|\mu_{\rm r}(\bls{r},\omega)|^2}\Big(\bls{L}_{\rm m}(\bls{r})\cdot \rho \bls{L}^\dagger_{\rm m}(\bls{r})
    -\frac{1}{2}\left\{\bls{L}^\dagger_{\rm m}(\bls{r})\cdot \bls{L}_{\rm m}(\bls{r}),\rho\right\}\Big)\right]
\end{align}
with the Lindblad operators
\begin{align}\label{eq:LindbladOperatorSlowParticle}
    \bls{L}_{\rm e}(\bls{r})&=\lim_{\omega\downarrow 0}\frac{\mu_0\omega^2}{4\pi}\varepsilon_{\rm r}(\bls{r},\omega)\int d^3s\, \mathsf{G}(\bls{r},\bls{s},\omega)\bls{C}(\bls{s},0),\\
    \bls{L}_{\rm m}(\bls{r})&=\lim_{\omega\downarrow 0}\frac{\mu_0\omega c}{4\pi}\int d^3s\, \frac{\partial}{\partial \bls{r}}\times\mathsf{G}(\bls{r},\bls{s},\omega)\bls{C}(\bls{s},0).
\end{align}
In the quasistatic approximation, the Green tensor is longitudinal and takes the form \eqref{eq:quasistaticGDerivation}. As a result, $\bls{L}_{\rm m}(\bls{r})= 0$, so that the master equation simplifies to
\begin{align}\label{eq:masterEquationSlowParticleLindblad}
    \frac{\partial \rho}{\partial t} &=-\frac{i}{\hbar}\left[ H_0+ H_{\rm si},\rho\right]+\frac{\varepsilon_0}{\hbar}\lim_{\omega\downarrow 0}\frac{2k_{\rm B}T}{\hbar \omega}\int d^3r\,\frac{{\rm Im}[\varepsilon_{\rm r}(\bls{r},\omega)]}{|\varepsilon_{\rm r}(\bls{r},\omega)|^2}\nonumber\\
    &\times\Big(\bls{L}_{\rm e}(\bls{r})\cdot \rho \bls{L}^\dagger_{\rm e}(\bls{r})-\frac{1}{2}\left\{\bls{L}^\dagger_{\rm e}(\bls{r})\cdot \bls{L}_{\rm e}(\bls{r}),\rho\right\}\Big),
\end{align}
with
\begin{align}
    \bls{L}_{\rm e}(\bls{r})&=\int d^3s \,\varrho_0\!\left[\mathsf{R}_\Omega^{\rm T}\left(\bls{s}-\bls{R}\right)\right] \lim_{\omega\downarrow 0}\varepsilon_{\rm r}(\bls{r},\omega)\frac{\partial}{\partial \bls{r}} g(\bls{r},\bls{s},\omega),
\end{align}
where we performed a partial integration. The charge distribution enters through
\begin{align}
\frac{\partial}{\partial \bls{s}}\cdot\bls{C}(\bls{s},0)=4\pi\varrho_0\!\left[\mathsf{R}_\Omega^{\rm T}\left(\bls{s}-\bls{R}\right)\right],
\end{align}
as a direct consequence of the definition \eqref{eq:CoulombFieldOperator}.

Having shown that the master equation has Lindblad form, we can again use \eqref{eq:relationScalarGreenfunction} to simplify the master equation in the quasistatic limit, yielding Eq.~\eqref{eq:MasterEquationSlowParticleIntro}. This equation may also be obtained directly by inserting the quasistatic Green tensor \eqref{eq:quasistaticGDerivation} into \eqref{eq:slowParticleMasterEqNonLindblad} and subsequently carrying out partial integrations.

\subsection{Resonant limit}\label{sec:DerivationResonantLimit}
A Markovian master equation can also be derived in the limit that the free particle dynamics are much faster than the bath-induced particle-relaxation time. We start from the non-Markovian evolution equation \eqref{eq:nonMarkovianMasterEquation} and expand the Coulomb field operator \eqref{eq:CoulombFieldOperator} in eigenoperators $\bls{C}_\ell(\bls{s})$ of the free evolution,
\begin{equation}
\bls{C}(\bls{s},t)=\bls{C}_0(\bls{s})+\left(\sum_{\omega_\ell > 0}\bsf{C}_\ell(\bls{s})e^{-i\omega_\ell t}+{\rm h.c.}\right).\label{eq:expansionCoulombFieldOperator}
\end{equation}
The index $\ell$ covers all positive energy differences $\hbar\omega_\ell$. We express
\begin{align}
    \bls{C}_\ell(\bls{s})=&\sum_{\substack{n,m \\E_m-E_n=\hbar\omega_\ell>0}}\bls{C}_{nm}(\bls{s})\ket{\Psi_n}\bra{\Psi_m},\\
    \bls{C}_0(\bls{s})=&\sum_{\substack{n,m \\E_m-E_n=0}}\bls{C}_{nm}(\bls{s})\ket{\Psi_n}\bra{\Psi_m}
\end{align}
through the eigenstates $\ket{\Psi_n}$ of $H_0$ with energies $E_n$ and coefficients
\begin{equation}
    \bls{C}_{nm}(\bls{s})=\bra{\Psi_n}\int d^3r\,\varrho_0(\bls{r})\frac{\bls{s}-\bsf{R}-\mathsf{R}_\Omega\bls{r}}{|\bls{s}-\bsf{R}-\mathsf{R}_\Omega\bls{r}|^3}\ket{\Psi_m}.
\end{equation}

The operators $c_\ell(\bls{s})$ in \eqref{eq:eigenoperatorsExplicitly1} expand the charge distribution,
\begin{align}\label{eq:expansion_ChargeDistribution}
    U_0^\dagger(t)\varrho_0\left[\mathsf{R}_\Omega^{\rm T}(\bls{r}-\bls{R})\right]U_0(t)=&c_0(\bls{r}) + \left(\sum_{\omega_\ell > 0} c_\ell(\bls{r})e^{-i\omega_\ell t}\right.\nonumber\\
    &\left.\vphantom{\sum_\ell}+{\rm h.c.}\right),
\end{align}
and are thus related to \eqref{eq:expansionCoulombFieldOperator} via
\begin{subequations}\label{eq:eigenoperatorsExplicitly}
\begin{align}
    c_\ell(\bls{r})=&\frac{1}{4\pi}\frac{\partial}{\partial \bls{r}}\cdot\bls{C}_\ell(\bls{r})=\sum_{\substack{n,m \\E_m-E_n=\hbar\omega_\ell>0}}c_{nm}(\bls{r})\ket{\Psi_n}\bra{\Psi_m},\\
    c_0(\bls{r})=&\frac{1}{4\pi}\frac{\partial}{\partial \bls{r}}\cdot\bls{C}_0(\bls{r})=\sum_{\substack{n,m \\E_m-E_n=0}}c_{nm}(\bls{r})\ket{\Psi_n}\bra{\Psi_m},\label{eq:eigenoperatorsExplicitly2}
\end{align}
\end{subequations}
where the coefficients read
\begin{equation}
    c_{nm}(\bls{r})=\bra{\Psi_n}\varrho_0\!\left[\mathsf{R}_\Omega^{\rm T}(\bls{r}-\bls{R})\right]\ket{\Psi_m}.
\end{equation}  

We define the Fourier integrals of the damping and noise kernel,
\begin{align}\label{eq:FTdampingKernel}
    \mathsf{\Gamma}_{\mathsf{D}}(\bls{s},\bls{s}',\omega)=&\int_0^\infty d\tau \mathsf{D}(\bls{s},\bls{s}',\tau)e^{i\omega \tau}\nonumber\\
    =&\frac{i\hbar \pi}{(4\pi)^2}\mathsf{J}(\bls{s},\bls{s}',\omega)+\frac{\hbar}{(4\pi)^2}\mathsf{P}\int_{-\infty}^\infty d\omega'\, \frac{\mathsf{J}(\bls{s},\bls{s}',\omega')}{\omega'-\omega},
\end{align}
and
\begin{align}\label{eq:FTnoiseKernel}
    \mathsf{\Gamma}_{\mathsf{N}}(\bls{s},\bls{s}',\omega)=&\int_0^\infty d\tau \mathsf{N}(\bls{s},\bls{s}',\tau)e^{i\omega \tau}\nonumber\\
    =&\frac{\hbar \pi}{(4\pi)^2}\mathsf{J}(\bls{s},\bls{s}',\omega){\rm coth}\left(\frac{\hbar\omega}{2k_{\rm B}T}\right)\nonumber\\
    -&\frac{i\hbar}{(4\pi)^2}\mathsf{P}\int_{-\infty}^\infty d\omega'\, \frac{\mathsf{J}(\bls{s},\bls{s}',\omega')}{\omega'-\omega}{\rm coth}\left(\frac{\hbar\omega'}{2k_{\rm B}T}\right),
\end{align}
in order to write
\begin{align}\label{eq:commutatorD1CRho}
    \Bigg[\int_0^\infty d\tau \mathsf{N}&(\bls{s},\bls{s}',\tau)\bls{C}(\bls{s}',t-\tau),\tilde{\rho}(t)\Bigg]\nonumber\\
    =\sum_{\omega_\ell > 0}\Bigg[&\mathsf{\Gamma}_{\mathsf{N}}(\bls{s},\bls{s}',\omega_\ell)\bls{C}_\ell(\bls{s}')e^{-i\omega_\ell t}\nonumber\\
    &+\mathsf{\Gamma}_{\mathsf{N}}(\bls{s},\bls{s}',-\omega_\ell)\bls{C}^\dagger_\ell(\bls{s}')e^{i\omega_\ell t},\tilde{\rho}(t)\Bigg].
\end{align}
Here we assumed that the time-independent term in \eqref{eq:expansionCoulombFieldOperator},  $\bls{C}_0(\bls{s})$, is a c-number, as it is the case in all situations discussed in the main text. The case of an operator-valued $\bls{C}_0(\bls{s})$ could be included along the lines of the derivation in the slow-particle limit in Sec.~\ref{sec:derivationSlowParticle}. 

With \eqref{eq:commutatorD1CRho} we find
\begin{align}
    &\left[\bls{C}(\bls{s},t), \left[\int_0^\infty d\tau\mathsf{N}(\bls{s},\bls{s}',\tau)\bls{C}(\bls{s}',t-\tau),\tilde{\rho}(t)\right]\right]\nonumber\\
    &=\left[\sum_{\omega_n > 0}\bls{C}_n(\bls{s})e^{-i\omega_n t}+{\rm h.c.},\right.\nonumber\left[\sum_{\omega_\ell > 0}\mathsf{\Gamma}_{\mathsf{N}}(\bls{s},\bls{s}',\omega_\ell)\bls{C}_\ell(\bls{s}')\right.\nonumber\\
    &\left.\left.\times e^{-i\omega_\ell t}+\sum_{\omega_\ell > 0}\mathsf{\Gamma}_{\mathsf{N}}(\bls{s},\bls{s}',-\omega_\ell)\bls{C}^\dagger_\ell(\bls{s}')e^{i\omega_\ell t},\tilde{\rho}(t)\right]\right].
\end{align}
If the free particle dynamics are fast, so that all frequencies and frequency differences are much larger than the rate at which the density operator $\tilde{\rho}(t)$ relaxes, we can carry out the rotating wave approximation and cancel all rapidly oscillating terms. This yields
\begin{align}\label{eq:RWACommutator}
\int_0^\infty d\tau \Big[\bls{C}(\bls{s},t)&,\big[\mathsf{N}(\bls{s},\bls{s}',\tau)\bls{C}(\bls{s}',t-\tau),\tilde{\rho}(t)\big]\Big]\nonumber\\
\simeq&\sum_{\omega_\ell > 0}\left[\bls{C}^\dagger_\ell(\bls{s}),\left[\mathsf{\Gamma}_{\mathsf{N}}(\bls{s},\bls{s}',\omega_\ell)\bls{C}_\ell(\bls{s}'),\tilde{\rho}(t)\right]\right]\nonumber\\
    +&\sum_{\omega_\ell > 0}\left[\bls{C}_\ell(\bls{s}),\left[\mathsf{\Gamma}_{\mathsf{N}}(\bls{s},\bls{s}',-\omega_\ell)\bls{C}^\dagger_\ell(\bls{s}'),\tilde{\rho}(t)\right]\right]
\end{align}
Along the same lines one finds that
\begin{align} \label{eq:RWAAnticommutator}
\int_0^\infty d\tau &\Big[\bls{C}(\bls{s},t),\big\{\mathsf{D}(\bls{s},\bls{s}',\tau)\bls{C}(\bls{s}',t-\tau),\tilde{\rho}(t)\big\}\Big]\nonumber\\
\simeq&\sum_{\omega_\ell > 0}\left[\bls{C}^\dagger_\ell(\bls{s}),\left\{\mathsf{\Gamma}_{\mathsf{D}}(\bls{s},\bls{s}',\omega_\ell)\bls{C}_\ell(\bls{s}'),\tilde{\rho}(t)\right\}\right]\nonumber\\
    &+\sum_{\omega_\ell > 0}\left[\bls{C}_\ell(\bls{s}),\left\{\mathsf{\Gamma}_{\mathsf{D}}(\bls{s},\bls{s}',-\omega_\ell)\bls{C}^\dagger_\ell(\bls{s}'),\tilde{\rho}(t)\right\}\right].
\end{align}
Inserting \eqref{eq:RWACommutator} and \eqref{eq:RWAAnticommutator} with \eqref{eq:FTdampingKernel} and \eqref{eq:FTnoiseKernel} into \eqref{eq:nonMarkovianMasterEquation} and transforming back to the Schrödinger picture yields a Markovian master equation,
\begin{align}\label{eq:markovianMasterEquationResonantLimitFullG}
  \frac{\partial \rho}{\partial t}=&-\frac{i}{\hbar}\left[ H_0+ H_{\rm si},\rho\right]+\frac{1}{8\pi\hbar}\int d^3sd^3s'\,\sum_{\omega_\ell > 0}\nonumber\\
  &\times\left(\left[n(\omega_\ell)+1\right]\left[\bls{C}_\ell(\bls{s})\cdot\rho\mathsf{J}(\bls{s},\bls{s}',\omega_\ell)\bls{C}^\dagger_\ell(\bls{s}')\right.\right.\nonumber\\
  &\left.\left.-\frac{1}{2}\left\{\bls{C}^\dagger_\ell(\bls{s})\cdot\mathsf{J}(\bls{s},\bls{s}',\omega_\ell)\bls{C}_\ell(\bls{s}'),\rho\right\}\right]\right.\nonumber\\
  &+n(\omega_\ell)\left[\bls{C}^\dagger_\ell(\bls{s})\cdot\rho\mathsf{J}(\bls{s},\bls{s}',\omega_\ell)\bls{C}_\ell(\bls{s}')\right.\nonumber\\
  &\left.\left.-\frac{1}{2}\left\{\bls{C}_\ell(\bls{s})\cdot\mathsf{J}(\bls{s},\bls{s}',\omega_\ell)\bls{C}^\dagger_\ell(\bls{s}'),\rho\right\}\right]\right),
\end{align}
after a rearrangement that uses the symmetries in $\omega$, $\bls{s}$ and $\bls{s}'$. Here
\begin{align}\label{eq:LambdShiftResonantGeneral}
     H_{\rm si}=&\frac{1}{(4\pi)^2}\int d^3sd^3s'\,\sum_{\omega_\ell > 0}\mathsf{P}\int_{-\infty}^\infty d\omega\,\frac{1}{\omega_\ell-\omega}\nonumber\\
    &\Big(\left[n(\omega)+1\right]\bsf{C}^\dagger_\ell(\bls{s})\cdot \mathsf{J}(\bls{s},\bls{s}',\omega)\bsf{C}_\ell(\bls{s}')\nonumber\\
    &-n(\omega)\bsf{C}_\ell(\bls{s})\cdot \mathsf{J}(\bls{s},\bls{s}',\omega)\bsf{C}^\dagger_\ell(\bls{s}')\Big)
\end{align}
describes the coherent surface interaction in the resonant limit. The two principal-value integrals can be evaluated using contour integration. We express the imaginary part of the Green tensor through the total Green tensor and use the Schwarz reflection principle to then exploit that the Green tensor is holomorphic in the positive imaginary plane. Accounting for the pole at $\omega=\omega_\ell$, all the poles of $n(\omega)$ in the positive imaginary plane and at $\omega=0$, and for the behaviour of the integrand at infinity, the two integrals yield
\begin{subequations}
\begin{align}\label{eq:IntegralSpectralDensityResonant1}
    \mathsf{P}\int_{-\infty}^\infty d\omega\,\frac{1}{\omega_\ell-\omega}\mathsf{J}(\bls{s},\bls{s}',\omega)=&-\mu_0c^2\mathbb{1}\delta(\bls{s}-\bls{s}')\nonumber\\
    &-\mu_0\omega_\ell^2{\rm Re}\left[\mathsf{G}(\bls{s},\bls{s}',\omega_\ell)\right],
\end{align}
and 
\begin{align}\label{eq:IntegralSpectralDensityResonant2}
    &\mathsf{P}\int_{-\infty}^\infty d\omega\,\frac{1}{\omega_\ell-\omega}n(\omega)\mathsf{J}(\bls{s},\bls{s}',\omega)=\frac{\mu_0c^2}{2}\mathbb{1}\delta(\bls{s}-\bls{s}')\nonumber\\
    &-\mu_0\omega_\ell^2n(\omega_\ell){\rm Re}\left[\mathsf{G}(\bls{s},\bls{s}',\omega_\ell)\right]+\frac{\mu_0\omega_{\rm th}}{2\pi\omega_\ell}\lim_{\omega\to 0}\omega^2\mathsf{G}(\bls{s},\bls{s}',\omega)\nonumber\\
    &-\frac{\mu_0\omega_{\rm th}\omega_\ell}{\pi}\sum_{m=1}^\infty \frac{(\omega_{\rm th}m)^2}{\omega_\ell^2+(\omega_{\rm th}m)^2}\mathsf{G}(\bls{s},\bls{s}',i\omega_{\rm th}m),
\end{align}
\end{subequations}
Here $\omega_{\rm th}=2\pi k_{\rm B}T/\hbar$ denotes the thermal Matsubara frequency. In the limit $T\to 0$ the Matsubara frequency summation transforms into an ordinary integral that is independent of temperature. In the quasistatic limit \eqref{eq:quasistaticGDerivation} one then obtains \eqref{eq:LambShiftResonantQuasistaticTzero}.

Similar to the the slow-particle limit we now briefly give the result for a particle in front of a perfectly conducting half space. The quasistatic Green tensor \eqref{eq:quasistaticGDerivation} in front of a perfectly conducting half space (see Tab.~\ref{tab:GreenfunctionHalfSpace}) reads
\begin{align}
    \mathsf{G}(\bls{s},\bls{s}',\omega)=&-\frac{c^2}{\omega^2}\frac{1}{4\pi}\frac{\partial}{\partial\bls{s}}\otimes\frac{\partial}{\partial\bls{s}'}\frac{1}{|\bls{s}-\bls{s}'|}\nonumber\\
    &+\frac{c^2}{\omega^2}\frac{1}{4\pi}\frac{\partial}{\partial\bls{s}}\otimes\frac{\partial}{\partial\bls{s}'}\frac{1}{|\bls{s}-\mathsf{M}\bls{s}'|}
\end{align}
yielding for \eqref{eq:IntegralSpectralDensityResonant1} and \eqref{eq:IntegralSpectralDensityResonant2} temperature-independent expressions,
\begin{subequations}
\begin{align}
\mathsf{P}\int_{-\infty}^\infty d\omega\,\frac{1}{\omega_\ell-\omega}\mathsf{J}(\bls{s},\bls{s}',\omega)=-\mu_0c^2\delta^{\perp}(\bls{s}-\bls{s}')\nonumber\\
-\mu_0c^2\frac{1}{4\pi}\frac{\partial}{\partial\bls{s}}\otimes\frac{\partial}{\partial\bls{s}'}\frac{1}{|\bls{s}-\mathsf{M}\bls{s}'|},\\
\mathsf{P}\int_{-\infty}^\infty d\omega\,\frac{1}{\omega_\ell-\omega}n(\omega)\mathsf{J}(\bls{s},\bls{s}',\omega)=\frac{\mu_0c^2}{2}\delta^{\perp}(\bls{s}-\bls{s}')\nonumber\\
+\frac{\mu_0c^2}{2}\frac{1}{4\pi}\frac{\partial}{\partial\bls{s}}\otimes\frac{\partial}{\partial\bls{s}'}\frac{1}{|\bls{s}-\mathsf{M}\bls{s}'|}.
\end{align}
\end{subequations}
The surface-interaction Hamiltonian \eqref{eq:LambdShiftResonantGeneral} can now be evaluated. The transverse delta-function does not contribute and one finds
\begin{align}
    H_{\rm si}=&-\frac{1}{2}\frac{1}{4\pi\varepsilon_0}\sum_{\omega_\ell > 0}\int d^3sd^3s'\,\frac{1}{|\bls{s}-\mathsf{M}\bls{s}'|}\left\{c_\ell^\dagger(\bls{s}),c_\ell(\bls{s}')\right\},
\end{align}
after a partial integration by use of \eqref{eq:eigenoperatorsExplicitly}. A small-amplitude oscillating monopole with \eqref{eq:resonantMonopolecl} and a single frequency $\omega_0$, oscillation direction $\blg{\epsilon}_0$ and annihilation operator $a_0$ is consequently exposed to the potential
\begin{align}
    H_{\rm si}=-\frac{q^2}{4\pi\varepsilon_0}\frac{\hbar}{2m\omega_0}\frac{1}{(2\bls{R}_{\rm eq}\cdot\bls{e}_3)^3}\left[1+(\blg{\epsilon}_0\cdot\bls{e}_3)^2\right]a_0^\dagger a_0.
\end{align}
Here $\bls{e}_3$ is the surface normal. This quantum potential could also be derived by taking the classical image-charge potential of an extended charge distribution, promoting the charge distribution to an operator, accounting for the small-amplitude oscillations of the monopole and carrying out a rotating wave approximation.

To demonstrate that \eqref{eq:markovianMasterEquationResonantLimitFullG} has Lindblad form we use \eqref{eq:relationGGImGreenTensor} to obtain
\begin{align}
&\frac{\partial \rho}{\partial t}=-\frac{i}{\hbar}\left[ H_0+ H_{\rm si},\rho\right]+\frac{\varepsilon_0}{\hbar}\sum_{\omega_\ell > 0}\int d^3r\,\frac{{\rm Im}\left[\varepsilon_{\rm r}(\bls{r},\omega_\ell)\right]}{|\varepsilon_{\rm r}(\bls{r},\omega_\ell)|^2}\nonumber\\
    &\times\left[ \left[n(\omega_\ell)+1\right]\left(\bsf{L}_\ell^{\rm e}(\bls{r})\cdot\rho{\bsf{L}_\ell^{\rm e}}^\dagger(\bls{r})-\frac{1}{2}\left\{{\bsf{L}_\ell^{\rm e}}^\dagger(\bls{r})\cdot\bsf{L}^{\rm e}_\ell(\bls{r}),\rho\right\}\right)\right.\nonumber\\
    &+\left.n(\omega_\ell)\left({\bsf{L}^{\rm e}_\ell}^\dagger(\bls{r})\cdot\rho\bsf{L}^{\rm e}_\ell(\bls{r})-\frac{1}{2}\left\{\bsf{L}^{\rm e}_\ell(\bls{r})\cdot{\bsf{L}_\ell^{\rm e}}^\dagger(\bls{r}),\rho\right\}\right)\right]\nonumber\\
    &+\frac{\varepsilon_0}{\hbar}\sum_{\omega_\ell > 0}\int d^3r\,\frac{{\rm Im}\left[\mu_{\rm r}(\bls{r},\omega_\ell)\right]}{|\mu_{\rm r}(\bls{r},\omega_\ell)|^2}\left[ \left[n(\omega_\ell)+1\right]\left(\bsf{L}^{\rm m}_\ell(\bls{r})\vphantom{\frac{1}{2}}\right.\right.\nonumber\\
    &\left.\cdot\rho{\bsf{L}_\ell^{\rm m}}^\dagger(\bls{r})-\frac{1}{2}\left\{{\bsf{L}_\ell^{\rm m}}^\dagger(\bls{r})\cdot\bsf{L}^{\rm m}_\ell(\bls{r}),\rho\right\}\right)\nonumber\\
    &+\left.n(\omega_\ell)\left({\bsf{L}^{\rm m}_\ell}^\dagger(\bls{r})\cdot\rho\bsf{L}^{\rm m}_\ell(\bls{r})-\frac{1}{2}\left\{\bsf{L}^{\rm m}_\ell(\bls{r})\cdot{\bsf{L}_\ell^{\rm m}}^\dagger(\bls{r}),\rho\right\}\right)\right]
\end{align}
with Lindblad operators 
\begin{align}
    \bsf{L}^{\rm e}_\ell(\bls{r})=\frac{\sqrt{2}\mu_0\omega_\ell^2}{4\pi}\varepsilon_{\rm r}(\bls{r},\omega_\ell)\int d^3s\, \mathsf{G}(\bls{r},\bls{s},\omega_\ell)\bsf{C}_\ell(\bls{s}),\\
    \bsf{L}^{\rm m}_\ell(\bls{r})=\frac{\sqrt{2}\mu_0\omega_\ell c}{4\pi}\int d^3s\, \frac{\partial}{\partial\bls{r}}\times\mathsf{G}(\bls{r},\bls{s},\omega_\ell)\bsf{C}_\ell(\bls{s}).
\end{align}
In case of a longitudinal Green tensor of the form \eqref{eq:quasistaticGDerivation}, $\bsf{L}^{\rm m}_\ell(\bls{r})=0$, and the Lindblad master equation reduces to 
\begin{align}\label{eq:LindbladMasterEquation}
    &\frac{\partial \rho}{\partial t}=-\frac{i}{\hbar}\left[ H_0+ H_{\rm si},\rho\right]+\frac{\varepsilon_0}{\hbar}\sum_{\omega_\ell > 0}\int d^3r\,\frac{{\rm Im}\left[\varepsilon_{\rm r}(\bls{r},\omega_\ell)\right]}{|\varepsilon_{\rm r}(\bls{r},\omega_\ell)|^2}\nonumber\\
    &\times\left[ \left[n(\omega_\ell)+1\right]\left(\bsf{L}^{\rm e}_\ell(\bls{r})\cdot\rho{\bsf{L}_\ell^{\rm e}}^\dagger(\bls{r})-\frac{1}{2}\left\{{\bsf{L}_\ell^{\rm e}}^\dagger(\bls{r})\cdot\bsf{L}^{\rm e}_\ell(\bls{r}),\rho\right\}\right)\right.\nonumber\\
    &+\left.n(\omega_\ell)\left({\bsf{L}_\ell^{\rm e}}^\dagger(\bls{r})\cdot\rho\bsf{L}^{\rm e}_\ell(\bls{r})-\frac{1}{2}\left\{\bsf{L}^{\rm e}_\ell(\bls{r})\cdot{\bsf{L}_\ell^{\rm e}}^\dagger(\bls{r}),\rho\right\}\right)\right].
\end{align}
The rates are positive since ${\rm Im}\left[\varepsilon_{\rm r}(\bls{r},\omega_\ell)\right]\geq 0$.

To simplify the Lindblad master equation \eqref{eq:LindbladMasterEquation} to a scalar form we can insert the quasistatic Green tensor \eqref{eq:quasistaticGDerivation},
\begin{align}
    \bsf{L}^{\rm e}_\ell(\bls{r})=\sqrt{2}\int d^3s\, c_\ell(\bls{s}) \varepsilon_{\rm r}(\bls{r},\omega_\ell)\frac{\partial}{\partial\bls{r}}g(\bls{r},\bls{s},\omega_\ell).
\end{align}
Here the Lindblad operators do not depend on the $\bls{C}_\ell(\bls{s})$, but instead involve the operators $c_\ell(\bls{s})$ used to expand the charge distribution. We use relation \eqref{eq:relationScalarGreenfunction} to finally arrive at \eqref{eq:dissipatorQuasistatic}. Similar to the slow-particle limit, this equation may also be obtained directly by inserting the quasistatic Green tensor \eqref{eq:quasistaticGDerivation} into \eqref{eq:markovianMasterEquationResonantLimitFullG} and subsequently carrying out partial integrations.

\section{Quasistatic approximation of the Green tensor}\label{sec:quasi_static}
Depending on the material response $\varepsilon_{\rm r}(\bls{r},\omega)$, the frequency, and the distance $|\bls{r}-\bls{r}'|$, retardation effects in the propagation of the electromagnetic field to the particle can be neglected. In this quasistatic limit, the Fourier transformed electric field $\bls{E}(\bls{r},\omega)$,
\begin{align}
    \bls{E}(\bls{r},\omega)=\int_{-\infty}^\infty dt\bls{E}(\bls{r},t)e^{i\omega t},
\end{align}
generated by an arbitrary charge distribution $\rho(\bls{r},\omega)$ follows as the solution of the quasistatic Maxwell equations
\begin{subequations}\label{eq:quasistaticME}
\begin{align}\label{eq:quasistaticME1}
    \frac{\partial}{\partial \bls{r}}\cdot \varepsilon_0\varepsilon_{\rm r}(\bls{r},\omega)\bls{E}(\bls{r},\omega)&=\varrho(\bls{r},\omega),\\\label{eq:quasistaticME2}
    \frac{\partial}{\partial \bls{r}}\times\bls{E}(\bls{r},\omega)&=0.
\end{align}
\end{subequations}
The charge distribution is related to the current density by the continuity equation
\begin{equation}
    -i\omega\varrho(\bls{r},\omega)+\frac{\partial}{\partial \bls{r}}\cdot\bls{j}(\bls{r},\omega)=0.
\end{equation}

Expressing the electric field in terms of the Green tensor and using \eqref{eq:electricFieldCurrent}, the quasistatic Maxwell equations \eqref{eq:quasistaticME} can be rewritten as
\begin{subequations}
\begin{align}
    \frac{\partial}{\partial \bls{r}}\cdot\varepsilon_{\rm r}(\bls{r},\omega)\mathsf{G}(\bls{r},\bls{r}',\omega)&=-\frac{c^2}{\omega^2}\frac{\partial}{\partial \bls{r}}\delta(\bls{r}-\bls{r}'),\\
    \frac{\partial}{\partial \bls{r}}\times\mathsf{G}(\bls{r},\bls{r}',\omega)&=\bls{0}.
\end{align}
\end{subequations}
They are solved by
\begin{equation}\label{eq:quasistaticGDerivation}
    \mathsf{G}(\bls{r},\bls{r}',\omega)=-\frac{1}{\mu_0\omega^2}\frac{\partial}{\partial\bls{r}}\otimes \frac{\partial}{\partial\bls{r}'}g(\bls{r},\bls{r}',\omega),
\end{equation}
where the scalar Green function $g(\bls{r},\bls{r}',\omega)$ is the solution of 
\begin{equation}\label{eq:DGLforScalarg}
   \frac{\partial}{\partial\bls{r}}\cdot\left[ \varepsilon_0 \varepsilon_{\rm r}(\bls{r},\omega)\frac{\partial}{\partial\bls{r}}g(\bls{r},\bls{r}',\omega)\right]=-\delta(\bls{r}-\bls{r}').
\end{equation}
Comparing Eqs.~\eqref{eq:DGLforScalarg} and \eqref{eq:DGLFullGreenTensor} shows that the complexity of the problem was reduced considerably. From the defining equation \eqref{eq:DGLforScalarg} and the Schwarz reflection principle for the dielectric response follows the Schwarz reflection principle for the Green function $g^*(\bls{r},\bls{r}',\omega)=g(\bls{r},\bls{r}',-\omega^*)$.

The boundary conditions of the problem can be deduced from inserting \eqref{eq:quasistaticGDerivation} into the electric field \eqref{eq:electricFieldCurrent} and carrying out a partial integration,
\begin{align}\label{eq:electricFieldScalarGreensFunction}
    \bls{E}(\bls{r},\omega)=&\frac{i}{\omega}\int d^3r'\,\left[\frac{\partial}{\partial \bls{r}}g(\bls{r},\bls{r}',\omega)\right]\frac{\partial}{\partial \bls{r}'}\cdot\bls{j}(\bls{r}',\omega)\nonumber\\
    &=-\int d^3r'\,\left[\frac{\partial}{\partial \bls{r}}g(\bls{r},\bls{r}',\omega)\right]\varrho(\bls{r}',\omega).
\end{align}
To ensure that the  electric field vanishes  infinitely far from the source, one requires that $\partial g(\bls{r},\bls{r}',\omega)/\partial\bls{r}= 0$ for $|\bls{r}-\bls{r}'|\to\infty$. Equation \eqref{eq:DGLforScalarg} determines $ g(\bls{r},\bls{r}',\omega)$ up to a constant, so that we can set $ g(\bls{r},\bls{r}',\omega) = 0$ for $|\bls{r}-\bls{r}'|\to\infty$, as is beneficial for partial integration. From Eq.~\eqref{eq:electricFieldScalarGreensFunction} we can see that $g(\bls{r},\bls{r}',\omega)$ yields the quasi-electrostatic potential at $\bls{r}$ of a test charge located at $\bls{r}'$ oscillating in magnitude with $\omega$.
Inserting \eqref{eq:quasistaticGDerivation} into the scalar potential \eqref{eq:MQE_scalarpotential} and a subsequent partial integration yields, after transforming to the interaction picture, the simplified form \eqref{eq:phiviag}.

\subsection{Continuity at an interface}

At an interface $\varepsilon_{\rm r}(\bls{r},\omega)$ changes discontinuously, so that the defining equation \eqref{eq:DGLforScalarg} implies continuity conditions. These conditions will be used in Sec.~\ref{sec:derivationGHalfSpace} and Sec.~\ref{sec:derivationGLayer} to derive the scalar Green function of the dielectric half-space and the dielectric layer configuration.

For example, a flat interface between two homogeneous dielectrics with $\varepsilon_{\rm r}(\bls{r},\omega)=\varepsilon_<$ for $\bls{r}\cdot \bls{e}_3<0$ and $\varepsilon_{\rm r}(\bls{r},\omega)=\varepsilon_>$ for $\bls{r}\cdot \bls{e}_3>0$, volume integration at the boundary gives the condition
\begin{subequations}\label{eq:continuity}
\begin{equation}\label{eq:continuity1}
    \varepsilon_>\bls{e}_3\cdot\frac{\partial}{\partial \bls{r}}g(\bls{r}_>,\bls{r}',\omega)=\varepsilon_<\bls{e}_3\cdot\frac{\partial}{\partial \bls{r}}g(\bls{r}_<,\bls{r}',\omega),
\end{equation}
where $\bls{r}_{>}$ and $\bls{r}_{<}$  approach the surface from above and below. This expresses the continuity of the displacement field's normal component. 

Likewise, by integrating over a loop through the interface we find the continuity for the electric field's parallel component
\begin{equation}\label{eq:continuity2}
   \bls{e}_\parallel\cdot\frac{\partial}{\partial \bls{r}}g(\bls{r}_>,\bls{r}',\omega)=\bls{e}_\parallel\cdot\frac{\partial}{\partial \bls{r}}g(\bls{r}_<,\bls{r}',\omega),
\end{equation}
\end{subequations}
where $\bls{e}_\parallel$ is perpendicular to $\bls{e}_3$.

\subsection{Symmetry of the Green function}\label{sec:symmetryOfGreenFunction}
To prove the relation \eqref{eq:relationScalarGreenfunction}, we first show that
\begin{equation}\label{eq:symmetryScalarg}
    g(\bls{r},\bls{r}',\omega)=g(\bls{r}',\bls{r},\omega).
\end{equation}
This symmetry follows directly from multiplying \eqref{eq:DGLforScalarg} by $g(\bls{s},\bls{r},\omega)$ and integrating over $\bls{r}$ to obtain
\begin{align}
&\varepsilon_0\int d^3r\, g(\bls{s},\bls{r},\omega)\frac{\partial}{\partial \bls{r}}\cdot \left[\varepsilon_{\rm r}(\bls{r},\omega)\frac{\partial}{\partial \bls{r}}g(\bls{r},\bls{r}',\omega)\right]\nonumber\\
&=-g(\bls{s},\bls{r}',\omega).
\end{align}
Two partial integrations then yield
  \begin{align}
&\varepsilon_0\int d^3r\, \left(\frac{\partial}{\partial \bls{r}}\cdot \left[\varepsilon_{\rm r}(\bls{r},\omega)\frac{\partial}{\partial \bls{r}}g(\bls{s},\bls{r},\omega)\right]\right)g(\bls{r},\bls{r}',\omega)\nonumber\\
&=-g(\bls{s},\bls{r}',\omega).
\end{align} 
This equation is valid for all $\bls{s}$, $\bls{r}'$ and $\omega$, so that
\begin{equation}
    \varepsilon_0\frac{\partial}{\partial \bls{r}}\cdot \left[\varepsilon_{\rm r}(\bls{r},\omega)\frac{\partial}{\partial \bls{r}}g(\bls{s},\bls{r},\omega)\right]=-\delta(\bls{r}-\bls{s})
\end{equation}
must hold. Renaming ${\bf s}$ by ${\bf r}'$ and comparing the equation with  \eqref{eq:DGLforScalarg} shows that $g(\bls{r},\bls{r}',\omega)$ and $g(\bls{r}',\bls{r},\omega)$ solve the same defining equation with the same boundary conditions and, thus, that they are equal.

\subsection{A useful relation}
We are now in the position to prove \eqref{eq:relationScalarGreenfunction}. We start with the defining equation
\begin{equation}
    \varepsilon_0\frac{\partial}{\partial \bls{s}}\cdot\left[\varepsilon_{\rm r}(\bls{s},\omega)\frac{\partial}{\partial \bls{s}}g(\bls{s},\bls{r}',\omega)\right]=-\delta(\bls{s}-\bls{r}'),
\end{equation}
multiply it by $g^*(\bls{r},\bls{s},\omega)$, integrate over $\bls{s}$, and perform a partial integration to find
\begin{align}\label{eq:derivationRelationg1}
    \varepsilon_0\int d^3s\,&\varepsilon_{\rm r}(\bls{s},\omega)\left(\frac{\partial}{\partial \bls{s}}g^*(\bls{r},\bls{s},\omega)\right)\cdot \left(\frac{\partial}{\partial \bls{s}}g(\bls{s},\bls{r}',\omega)\right)\nonumber\\
    &=g^*(\bls{r},\bls{r}',\omega).
\end{align}
Then taking the complex conjugate of the defining equation
\begin{equation}
    \varepsilon_0\frac{\partial}{\partial \bls{s}}\cdot\left[\varepsilon_{\rm r}^*(\bls{s},\omega)\frac{\partial}{\partial \bls{s}}g^*(\bls{s},\bls{r},\omega)\right]=-\delta(\bls{s}-\bls{r}),
\end{equation}
multiplying  by $g(\bls{s},\bls{r}',\omega)$, integrating over $\bls{s}$, and performing a partial integration yields
\begin{align}
     \varepsilon_0\int d^3s\,&\varepsilon_{\rm r}^*(\bls{s},\omega)\left(\frac{\partial}{\partial \bls{s}}g^*(\bls{s},\bls{r},\omega)\right)\cdot \left(\frac{\partial}{\partial \bls{s}}g(\bls{s},\bls{r}',\omega)\right)\nonumber\\
    &=g(\bls{r},\bls{r}',\omega).
\end{align}
The symmetry \eqref{eq:symmetryScalarg} leads to
\begin{align}\label{eq:derivationRelationg2}
     \varepsilon_0\int d^3s\,&\varepsilon_{\rm r}^*(\bls{s},\omega)\left(\frac{\partial}{\partial \bls{s}}g^*(\bls{r},\bls{s},\omega)\right)\cdot \left(\frac{\partial}{\partial \bls{s}}g(\bls{s},\bls{r}',\omega)\right)\nonumber\\
    &=g(\bls{r},\bls{r}',\omega).
\end{align}
We then subtract \eqref{eq:derivationRelationg1} and \eqref{eq:derivationRelationg2} to find that
\begin{align}
    &{\rm Im}\left[g(\bls{r},\bls{r}',\omega)\right]=-\varepsilon_0\int d^3s \,{\rm Im}\left[\varepsilon_{\rm r}(\bls{s},\omega)\right]\nonumber\\
    & \times\left(\frac{\partial}{\partial \bls{s}}g^*(\bls{r},\bls{s},\omega)\right)\cdot \left(\frac{\partial}{\partial \bls{s}}g(\bls{s},\bls{r}',\omega)\right).
\end{align}
Subsequent complex conjugation yields \eqref{eq:relationScalarGreenfunction}.

\begin{center}
\begin{table*}[htbp]
\caption{The Green function for a dielectric half-space with $\varepsilon_{\rm r}(\omega)$ for $\bls{r}\cdot\bls{e}_3<0$ and vacuum for $\bls{r}\cdot\bls{e}_3>0$.}
\bgroup\normalsize
\setlength{\tabcolsep}{6pt} 
\renewcommand{\arraystretch}{2} 
\newcolumntype{?}{!{\vrule width 1pt}}
\begin{tabular}{c?c|c}
$g(\bls{r},\bls{r}',\omega)$ & $\bls{r}\cdot\bls{e}_3<0$ & $0<\displaystyle\bls{r}\cdot\bls{e}_3$ \\ \noalign{\hrule height 1.0pt}
$\bls{r}'\cdot\bls{e}_3<0$ & $\begin{aligned}\vphantom{\frac{1}{\big[}}\frac{1}{4\pi\varepsilon_0\varepsilon_{\rm r}(\omega)}\left(\frac{1}{|\bls{r}-\bls{r}'|}-\frac{1-\varepsilon_{\rm r}(\omega)}{1+\varepsilon_{\rm r}(\omega)}\frac{1}{|\bls{r}-\mathsf{M}\bls{r}'|}\right)\end{aligned}$ &   $\begin{aligned}\frac{1}{4\pi\varepsilon_0}\frac{2}{1+\varepsilon_{\rm r}(\omega)}\frac{1}{|\bls{r}-\bls{r}'|}\end{aligned}$\\ \noalign{\hrule height 1.0pt}
$0<\bls{r}'\cdot\bls{e}_3$ & $\begin{aligned}\frac{1}{4\pi\varepsilon_0}\frac{2}{1+\varepsilon_{\rm r}(\omega)}\frac{1}{|\bls{r}-\bls{r}'|}\end{aligned}$ & $\begin{aligned}\frac{1}{4\pi\varepsilon_0}\left(\frac{1}{|\bls{r}-\bls{r}'|}+\frac{1-\varepsilon_{\rm r}(\omega)}{1+\varepsilon_{\rm r}(\omega)}\frac{1}{|\bls{r}-\mathsf{M}\bls{r}'|}\right)\end{aligned}$ \\
\end{tabular}
\egroup
\label{tab:GreenfunctionHalfSpace}
\end{table*}
\end{center}

\subsection{Homogeneous half-space}\label{sec:derivationGHalfSpace}
We now use these results to derive the Green function $g(\bls{r},\bls{r}',\omega)$ for the dielectric half-space. For a space-filling dielectric $\varepsilon_{\rm r}(\omega)$, the defining equation \eqref{eq:DGLscalarGreenfunction} is solved by the particular solution
\begin{equation}\label{eq:particularSolutionHalfSpace}
    g(\bls{r},\bls{r}',\omega)=\frac{1}{4\pi\varepsilon_0\varepsilon_{\rm r}(\omega)}\frac{1}{|\bls{r}-\bls{r}'|},
\end{equation}
which fulfills the boundary condition at infinity. 

For the half-space geometry depicted in Tab.~\ref{tab:dielectrics}, where we assume vacuum for $\bls{r}\cdot\bls{e}_3>0$, while for $\bls{r}\cdot\bls{e}_3<0$ we set $\varepsilon_{\rm r}(\bls{r},\omega)=\varepsilon_{\rm r}(\omega)$ to be the response of an arbitrary homogeneous dielectric. With $\bls{r}'\cdot\bls{e}_3>0$ in vacuum the solution can be obtained from the ansatz
\begin{subequations}
\begin{align}
    g(\bls{r},\bls{r}',\omega)&=\frac{1}{4\pi\varepsilon_0}\frac{1}{|\bls{r}-\bls{r}'|}+\frac{b_1}{4\pi\varepsilon_0}\frac{1}{|\bls{r}-\mathsf{M}\bls{r}'|},&\,\bls{r}\cdot\bls{e}_3>0,\label{eq:gscalar1}\\
    g(\bls{r},\bls{r}',\omega)&=\frac{b_2}{4\pi\varepsilon_0}\frac{1}{|\bls{r}-\bls{r}'|},&\,\bls{r}\cdot\bls{e}_3<0,
\end{align}
\end{subequations}
with the mirror tensor $\mathsf{M}=\mathbb{1}-2\bls{e}_3\otimes\bls{e}_3$. The second term in Eq.\,\eqref{eq:gscalar1} is a homogeneous solution, since ${\bf e}_3\cdot \mathsf{M}\bls{r}'<0$. Continuity at the interface \eqref{eq:continuity} implies
\begin{subequations}
\begin{align}
    b_1&=\frac{1-\varepsilon_{\rm r}(\omega)}{1+\varepsilon_{\rm r}(\omega)},\\
    b_2&=\frac{2}{1+\varepsilon_{\rm r}(\omega)}.
\end{align}
\end{subequations}
The homogeneous solution can be interpreted as the potential due to the presence of an image charge. The same calculation for $\bls{r}'\cdot\bls{e}_3<0$ in the dielectric yields the entire Green function  given in Tab.~\ref{tab:GreenfunctionHalfSpace}.

\subsection{Half-space covered by a surface layer}\label{sec:derivationGLayer}
We next derive the Green function $g(\bls{r},\bls{r}',\omega)$ in a geometry consisting of a vacuum half-space with $\varepsilon_{\rm r}(\bls{r},\omega)=1$ for $\bls{r}\cdot\bls{e}_3>0$, a dielectric layer with $\varepsilon_{\rm r}(\bls{r},\omega)=\varepsilon_{\rm s}(\omega)$ for $-d_{\rm s}<\bls{r}\cdot\bls{e}_3<0$, and a dielectric half-space  with $\varepsilon_{\rm r}(\bls{r},\omega)=\varepsilon_{\rm b}(\omega)$ for $\bls{r}\cdot\bls{e}_3<-d_{\rm s}$, see Tab.~\ref{tab:dielectrics}. 

In a geometry with two interfaces, an infinite number of image charges appears. We will now use a compact formulation by expanding the particular solution \eqref{eq:particularSolutionHalfSpace} in exponential functions \cite{kleefstra1980influence},
\begin{align}
    \frac{1}{|\bls{r}-\bls{r}'|}=\int_0^\infty dk \,J_0\left(k\Delta r\right)e^{-k|(\bls{r}-\bls{r}')\cdot\bls{e}_3|},
\end{align}
with the in-plane distance $\Delta r^2=\left[(\bls{r}-\bls{r}')\cdot\bls{e}_1\right]^2+\left[(\bls{r}-\bls{r}')\cdot\bls{e}_2\right]^2$. One can show that
\begin{equation}
    \frac{\partial}{\partial \bls{r}}\cdot\frac{\partial}{\partial \bls{r}}e^{\pm k\bls{r}\cdot\bls{e}_3}J_0\left(k\Delta r\right)= 0,
\end{equation}
so that we can make the ansatz 
\begin{equation}\label{eq:AnsatzDielectricLayer}
g(\bls{r},\bls{r}',\omega)=\frac{1}{4\pi\varepsilon_0}\int_0^\infty dk\, g_k(\bls{r},\bls{r}',\omega) J_0\left(k\Delta r\right),
\end{equation}
with the $g_k(\bls{r},\bls{r}',\omega)$ given in Tab.~\ref{tab:GreenfunctionLayer}. Then \eqref{eq:AnsatzDielectricLayer} solves the defining differential equation \eqref{eq:DGLforScalarg} in all regions of constant $\varepsilon_{\rm r}(\bls{r},\omega)$ and accounts for the boundary condition at infinity and the symmetry \eqref{eq:symmetryScalarg}. The coefficients $c_1$ to $c_{10}$ appearing in Tab.~\ref{tab:GreenfunctionLayer} can be determined by using the continuity at the interface \eqref{eq:continuity1} and \eqref{eq:continuity2}. Defining
\begin{subequations}
\begin{align}
    \xi_{\rm b}(\omega)=& \frac{\varepsilon_{\rm s}(\omega)-\varepsilon_{\rm b}(\omega)}{\varepsilon_{\rm s}(\omega)+\varepsilon_{\rm b}(\omega)},\\
 \xi_{\rm v}(\omega)=&\frac{\varepsilon_{\rm s}(\omega)-1}{\varepsilon_{\rm s}(\omega)+1},   
\end{align}
\end{subequations}
the coefficients read
\begin{subequations}\label{eq:coefficientsDielectricLayer}
\begin{align}
    c_1&=\frac{\xi_{\rm b}(\omega)e^{-2kd_{\rm s}}-\xi_{\rm v}(\omega)}{1-\xi_{\rm b}(\omega)\xi_{\rm v}(\omega)e^{-2kd_{\rm s}}},\\
    c_2&=\frac{1-\xi_{\rm v}(\omega)}{1-\xi_{\rm b}(\omega)\xi_{\rm v}(\omega)e^{-2kd_{\rm s}}},\\
    c_3&=\frac{[1-\xi_{\rm v}(\omega)]\xi_{\rm b}(\omega)e^{-2kd_{\rm s}}}{1-\xi_{\rm b}(\omega)\xi_{\rm v}(\omega)e^{-2kd_{\rm s}}},\\
    c_4&=\frac{[1-\xi_{\rm v}(\omega)][1+\xi_{\rm b}(\omega)]}{1-\xi_{\rm b}(\omega)\xi_{\rm v}(\omega)e^{-2kd_{\rm s}}},\\
    c_5&=\frac{\xi_{\rm v}(\omega)\xi_{\rm b}(\omega)e^{-2kd_{\rm s}}}{1-\xi_{\rm b}(\omega)\xi_{\rm v}(\omega)e^{-2kd_{\rm s}}},\\
    c_6&=\frac{\xi_{\rm v}(\omega)}{1-\xi_{\rm b}(\omega)\xi_{\rm v}(\omega)e^{-2kd_{\rm s}}},\\
    c_7&=\frac{\xi_{\rm b}(\omega)e^{-2kd_{\rm s}}}{1-\xi_{\rm b}(\omega)\xi_{\rm v}(\omega)e^{-2kd_{\rm s}}},\\
    c_{8}&=\frac{1+\xi_{\rm b}(\omega)}{1-\xi_{\rm b}(\omega)\xi_{\rm v}(\omega)e^{-2kd_{\rm s}}},\\
    c_{9}&=\frac{[1+\xi_{\rm b}(\omega)]\xi_{\rm v}(\omega)}{1-\xi_{\rm b}(\omega)\xi_{\rm v}(\omega)e^{-2kd_{\rm s}}},\\
    c_{10}&=\frac{\xi_{\rm v}(\omega)-\xi_{\rm b}(\omega)e^{2kd_{\rm s}}}{1-\xi_{\rm b}(\omega)\xi_{\rm v}(\omega)e^{-2kd_{\rm s}}},
\end{align}
\end{subequations}
some of which can also be found in \cite{kleefstra1980influence,kumph2016electric}. Expanding the denominator in \eqref{eq:coefficientsDielectricLayer} using the geometric series yields an infinite sum of image potentials for the layer configuration.

\begin{center}
\begin{table*}[htbp]
\caption{The Green function for a geometry consisting of a vacuum half-space with $\varepsilon_{\rm r}(\bls{r},\omega)=1$ for $0<\bls{r}\cdot\bls{e}_3$, a dielectric layer with $\varepsilon_{\rm r}(\bls{r},\omega)=\varepsilon_{\rm s}(\omega)$ for $-d_{\rm s}<\bls{r}\cdot\bls{e}_3<0$ and a dielectric half-space  with $\varepsilon_{\rm r}(\bls{r},\omega)=\varepsilon_{\rm b}(\omega)$ for $\bls{r}\cdot\bls{e}_3<-d_{\rm s}$. For the table we defined $z=\bls{r}\cdot\bls{e}_3$ and $z'=\bls{r}'\cdot\bls{e}_3$. For the dependence of the Green function on $g_k(\bls{r},\bls{r}',\omega)$, see Eq.~\eqref{eq:AnsatzDielectricLayer}.}
\bgroup\normalsize
\setlength{\tabcolsep}{6pt} 
\renewcommand{\arraystretch}{2} 
\newcolumntype{?}{!{\vrule width 1pt}}
\begin{tabular}{c?c|c|c}
$g_k(\bls{r},\bls{r}',\omega)$ &$z<-d_{\rm s}$ & $-d_{\rm s}<z<0$ & $0<z$ \\ \noalign{\hrule height 1.0pt}

$z'<-d_{\rm s}$ & $\begin{aligned}\vphantom{\frac{1}{\big[}}\frac{1}{\varepsilon_{\rm b}(\omega)}\left(e^{-k|z-z'|}+c_{10}e^{k(z+z')}\right)\end{aligned}$ & $\begin{aligned}\vphantom{\frac{1}{\big[}}\frac{1}{\varepsilon_{\rm s}(\omega)}\left(c_{8}e^{-k(z-z')}+c_{9}e^{k(z+z')}\right)\end{aligned}$ &  $c_4e^{-k(z-z')}$\\ \noalign{\hrule height 1.0pt}

$-d_{\rm s}<z'<0$ & $\begin{aligned}\frac{1}{\varepsilon_{\rm s}(\omega)}\left(c_{8}e^{k(z-z')}+c_{9}e^{k(z+z')}\right)\end{aligned}$ &$\begin{aligned}\vphantom{\frac{1}{\big[}}&\left(e^{-k|z-z'|}+c_5e^{k(z-z')}+c_6e^{k(z+z')}\right.\\&\left.+c_5e^{-k(z-z')}+c_7e^{-k(z+z')}\right)\frac{1}{\varepsilon_{\rm s}(\omega)}\vphantom{\frac{1}{\big[}}\end{aligned}$ &  $c_2e^{-k(z-z')}+c_3e^{-k(z+z')}$\\ \noalign{\hrule height 1.0pt}

$0<z'$ & $c_4e^{k(z-z')}$  & $c_2e^{k(z-z')}+c_3e^{-k(z+z')}$ &  $e^{-k|z-z'|}+c_1e^{-k(z+z')}$
\end{tabular}
\egroup
\label{tab:GreenfunctionLayer}
\end{table*}
\end{center}

\section{Decoherence due to Thomson scattering}\label{app:ThomsonScattering}

\subsection{Thomson scattering in presence of dielectrics}

In this section, we calculate decoherence due to the $\bls{A}^2$-terms in Hamiltonian \eqref{eq:couplingHamiltonian}. We show that these terms can be associated with scattering decoherence due to Thomson scattering, i.e.\ the radiation due to the acceleration of a charge in an external electromagnetic field. Since the scattering cross section is small for massive particles, this decoherence will turn out to be negligible.

We start by taking the coupling Hamiltonian \eqref{eq:couplingHamiltonian} and the free-field Hamiltonian $H_{\rm f}$ and only keeping coupling terms that involve the square of the vector potential,
\begin{align}
    H=H_0+ H_{\rm f}+H_{\rm int}
\end{align}
with 
\begin{align}\label{eq:interactionHalimtonianThomsonScattering}
    H_{\rm int}=&\frac{1}{2}\int d^3rd^3r'\, \varrho_0\left[\mathsf{R}^{\rm T}_\Omega(\bls{r}-\bls{R})\right]\varrho_0\left[\mathsf{R}^{\rm T}_\Omega(\bls{r}'-\bls{R})\right]\nonumber\\
    &\times\bls{A}(\bls{r})\cdot \mathsf{T}(\bls{r},\bls{r}')\bls{A}(\bls{r}').
\end{align}
The free particle Hamiltonian $H_0$ is given by \eqref{eq:H0_MQE} and $\bls{A}(\bls{r})$ can be found in \eqref{eq:MQE_vectorpotential}. Here, the tensor
\begin{equation}\label{eq:definitionTTensor}
    \mathsf{T}(\bls{r},\bls{r}')=\frac{1}{m}\mathbb{1}-(\bls{r}-\bls{R})\times{\rm I}_\Omega^{-1}\times(\bls{r}'-\bls{R})
\end{equation}
depends on the the mass and the inertia tensor. The interaction Hamiltonian describes a coupling of the field operators $\bls{f}_\lambda(\bls{r},\omega)$ due to the presence of the particle.

As above, we define the bath operator
\begin{align}
    \bls{b}(\bls{s},t)=&\sum_{\lambda={\rm e,m}}\int d^3r'\,\int_0^\infty d\omega\,\frac{1}{i\omega} \mathsf{G}_{\lambda}(\bls{s},\bls{r'},\omega)\bls{f}_{\lambda}(\bls{r}',\omega)e^{-i\omega t}\nonumber\\
    &+{\rm h.c.},
\end{align}
and the tensor-valued particle operator that depends on the particle charge,
\begin{align}\label{eq:CTensorOperatorThomson}
    \mathsf{C}&(\bls{s},\bls{s}',t)=U_0^\dagger(t)\int d^3rd^3r'\,\varrho_0\left[\mathsf{R}^{\rm T}_\Omega(\bls{r}-\bls{R})\right]\nonumber\\
    &\times\varrho_0\left[\mathsf{R}^{\rm T}_\Omega(\bls{r}'-\bls{R})\right]\delta^\perp(\bls{r}-\bls{s})\mathsf{T}(\bls{r},\bls{r}')\delta^\perp(\bls{r}'-\bls{s}')U_0(t).
\end{align}
Here, the transverse delta function, given by \eqref{eq:definitionTransverseDelta}, arises from the transverse part of the Green tensor included in the vector potential. 

The interaction Hamiltonian in the interaction picture reads
\begin{align}
    \tilde{ H}_{\rm int}(t)&=\frac{1}{2}\int d^3sd^3s'\,\bls{b}(\bls{s},t)\cdot \mathsf{C}(\bls{s},\bls{s}',t)\bls{b}(\bls{s}',t).
\end{align}
In the weak coupling limit, one has to evaluate
\begin{widetext}
\begin{align}\label{eq:doubleCommutatorThomson}
    \int_0^\infty d\tau {\rm tr_B}&\left\{\left[\tilde{ H}_{\rm int}(t),\left[\tilde{ H}_{\rm int}(t-\tau),\tilde{\rho}(t)\otimes\rho_{\rm B}\right]\right]\right\}=\frac{1}{4}\int d^3s_1d^3s_1'd^3s_2d^3s_2'\,\sum_{i_1i_1'i_2i_2'}\int_0^\infty d\tau\nonumber\\
    &\times\left(-\frac{i}{2}\left[C_{i_1i_1'}(\bls{s}_1,\bls{s}_1',t),\left\{D_{i_1i_1'i_2i_2'}(\bls{s}_1,\bls{s}_1',\bls{s}_2,\bls{s}_2',\tau)C_{i_2i_2'}(\bls{s}_2,\bls{s}_2',t-\tau),\tilde\rho(t)\right\}\right]\right.\nonumber\\
    &\left.+\frac{1}{2}\left[C_{i_1i_1'}(\bls{s}_1,\bls{s}_1',t),\left[N_{i_1i_1'i_2i_2'}(\bls{s}_1,\bls{s}_1',\bls{s}_2,\bls{s}_2',\tau)C_{i_2i_2'}(\bls{s}_2,\bls{s}_2',t-\tau),\tilde\rho(t)\right]\right]\right)
\end{align}
involving the dissipation kernel \cite{breuer2002}
\begin{align}
    D_{i_1i_1'i_2i_2'}(\bls{s}_1,\bls{s}_1',\bls{s}_2,\bls{s}_2',\tau)=i\braket{b_{i_1}(\bls{s}_1,t)b_{i_1'}(\bls{s}_1',t)b_{i_2}(\bls{s}_2,t-\tau)b_{i_2'}(\bls{s}_2',t-\tau)-b_{i_2}(\bls{s}_2,t-\tau)b_{i_2'}(\bls{s}_2',t-\tau)b_{i_1}(\bls{s}_1,t)b_{i_1'}(\bls{s}_1',t)}_{\rm B}
\end{align}
and the noise kernel 
\begin{align}\label{eq:dampingKernelThomson}
    N_{i_1i_1'i_2i_2'}(\bls{s}_1,\bls{s}_1',\bls{s}_2,\bls{s}_2',\tau)=\braket{b_{i_1}(\bls{s}_1,t)b_{i_1'}(\bls{s}_1',t)b_{i_2}(\bls{s}_2,t-\tau)b_{i_2'}(\bls{s}_2',t-\tau)+b_{i_2}(\bls{s}_2,t-\tau)b_{i_2'}(\bls{s}_2',t-\tau)b_{i_1}(\bls{s}_1,t)b_{i_1'}(\bls{s}_1',t)}_{\rm B}.
\end{align}
Here, $\braket{\cdot}_B={\rm tr}_{\rm B}\left\{\cdot\rho_{\rm B}\right\}$ is the expectation value of bath operators. The indices refer to the components of the vectors $\bls{b}(\bls{s},t)$ and tensors $\mathsf{C}(\bls{s},\bls{s}',t)$.

In order to calculate the decoherence rate, one has to determine the dissipator  $\mathcal{L}_{\rm Th}$ from the double commutator in \eqref{eq:doubleCommutatorThomson}. In the slow particle limit
\begin{align}
    C_{i_2i_2'}(\bls{s}_2,\bls{s}_2',t-\tau)\approx C_{i_2i_2'}(\bls{s}_2,\bls{s}_2',t),
\end{align}
and after transforming back to the Schrödinger picture we find
\begin{align}\label{eq:DissipatorDoubleCommutator}
    \mathcal{L}_{\rm Th}\rho=-\frac{1}{8\hbar^2}\int d^3s_1d^3s_1'd^3s_2d^3s_2'\,\sum_{i_1i_1'i_2i_2'}\left[C_{i_1i_1'}(\bls{s}_1,\bls{s}_1'),\left[\int_0^\infty d\tau N_{i_1i_1'i_2i_2'}(\bls{s}_1,\bls{s}_1',\bls{s}_2,\bls{s}_2',\tau)C_{i_2i_2'}(\bls{s}_2,\bls{s}_2'),\rho\right]\right]
\end{align}
where $C_{i_1i_1'}(\bls{s},\bls{s}'):=C_{i_1i_1'}(\bls{s},\bls{s}',0)$. The Born-Markov approximation assumed here is justified if the particle operator $\mathsf{C}(\bls{s},\bls{s}',t)$ evolves slowly in comparison to the timescale on which the dissipation kernel decays. 

To calculate the noise kernel we first evaluate the fourth-order thermal oscillator correlation function. Correlation functions involving unequal numbers of daggered and undaggered operators vanish. As the only non-trivial correlation function, we find
\begin{align}\label{eq:fourthOrderCorrelation}
    &\braket{
    \big[\bls{f}^\dagger_{\lambda_1}(\bls{r}_1,\omega_1)\big]_{j_1}
    \big[\bls{f}^\dagger_{\lambda_1'}(\bls{r}_1',\omega_1')\big]_{j_1'}
    \big[\bls{f}_{\lambda_2}(\bls{r}_2,\omega_2)\big]_{j_2}
    \big[\bls{f}_{\lambda_2'}(\bls{r}_2',\omega_2')\big]_{j_2'}}=n(\omega_1)n(\omega_1')\left[\delta(\bls{r}_1-\bls{r}_2)\delta(\bls{r}_1'-\bls{r}_2')\delta(\omega_1-\omega_2)\right.\nonumber\\
    &\left.\times\delta(\omega_1'-\omega_2')\delta_{\lambda_1\lambda_2}\delta_{\lambda_1'\lambda_2'}\delta_{j_1j_2}\delta_{j_1'j_2'}+\delta(\bls{r}_1-\bls{r}_2')\delta(\bls{r}_1'-\bls{r}_2)\delta(\omega_1-\omega_2')\delta(\omega_1'-\omega_2)\delta_{\lambda_1\lambda_2'}\delta_{\lambda_1'\lambda_2}\delta_{j_1j_2'}\delta_{j_1'j_2}\right],
\end{align}
while all other correlation functions follow from the commutation relations, see above \eqref{eq:freeFieldEnergy}. 

A tedious but straight-forward calculation yields
\begin{align}\label{eq:dampingKernelTimeDependentThomson}
    &N_{i_1i_1'i_2i_2'}(\bls{s}_1,\bls{s}_1',\bls{s}_2,\bls{s}_2',\tau)=2\hbar^2\int_0^\infty d\omega_1d\omega_1'\frac{1}{\omega_1^2\omega_1'^2}J_{i_1'i_1}(\bls{s}'_1,\bls{s}_1,\omega_1)J_{i_2'i_2}(\bls{s}_2',\bls{s}_2,\omega_1')\left[2n(\omega_1)+1\right]\left[2n(\omega_1')+1\right]\nonumber\\
    &+\hbar^2\int_0^\infty d\omega_1d\omega_1'\frac{1}{\omega_1^2\omega_1'^2}\left[J_{i_2i_1}(\bls{s}_2,\bls{s}_1,\omega_1)J_{i_2'i_1'}(\bls{s}_2',\bls{s}_1',\omega_1')+J_{i_2'i_1}(\bls{s}_2',\bls{s}_1,\omega_1)J_{i_2i_1'}(\bls{s}_2,\bls{s}_1',\omega_1')\right]\nonumber\\
    &\times\left\{\big[2n(\omega_1)+2n(\omega_1')+4n(\omega_1)n(\omega_1')\big]\cos[(\omega_1-\omega_1')\tau]+\big[2n(\omega_1)+2n(\omega_1')+4n(\omega_1)n(\omega_1')+2\big]\cos[(\omega_1+\omega_1')\tau]\right\}
\end{align}
using relation \eqref{eq:relationGGImGreenTensor} and the definition of the spectral density \eqref{eq:spectralDensityDefinition} and its components $J_{ii'}(\bls{s},\bls{s}',\omega)$.

The first term in \eqref{eq:dampingKernelTimeDependentThomson} is independent of time and arises from those contributions $\big[\bls{f}^\dagger_\lambda(\bls{r},\omega)\big]_j\big[\bls{f}_\lambda(\bls{r},\omega)\big]_j$ and $\big[\bls{f}_\lambda(\bls{r},\omega)\big]_j\big[\bls{f}^\dagger_\lambda(\bls{r},\omega)\big]_j$ in the interaction Hamiltonian \eqref{eq:interactionHalimtonianThomsonScattering} which are equal in all indices and arguments. To avoid the time integral in the slow-particle limit from diverging, we drop the time-independent terms in \eqref{eq:dampingKernelTimeDependentThomson}. This procedure does not affect the particle-induced coupling in \eqref{eq:interactionHalimtonianThomsonScattering} of field degrees of freedom $\bls{f}_\lambda(\bls{r},\omega)$ with different indices and arguments. It is justified by the fact that in vacuum the resulting master equation coincides with the scattering master equation \cite{stickler2016b} obtained by means of the Thompson scattering amplitudes, see Eq.~\eqref{eq:scatteringAmplitudeThomson}.

The integral of the noise kernel without the time-independent contribution yields
\begin{align}\label{eq:IntegralNoiseKernelThomson}
    \int_0^\infty d\tau N_{i_1i_1'i_2i_2'}(\bls{s}_1,\bls{s}_1',\bls{s}_2,\bls{s}_2',\tau)=&\int_0^\infty d\omega\,\frac{4\pi\hbar^2}{\omega^4}\left[J_{i_2i_1}(\bls{s}_2,\bls{s}_1,\omega)J_{i_2'i_1'}(\bls{s}_2',\bls{s}_1',\omega)+J_{i_2'i_1}(\bls{s}_2',\bls{s}_1,\omega)J_{i_2i_1'}(\bls{s}_2,\bls{s}_1',\omega)\right]\nonumber\\
    &\times n(\omega)\left[n(\omega)+1\right].
\end{align}
Inserting \eqref{eq:IntegralNoiseKernelThomson} into \eqref{eq:DissipatorDoubleCommutator},
using
\begin{align}\label{eq:COperatorThomsonScatteringSlowParticle}
    \mathsf{C}(\bls{s},\bls{s}')=&\int d^3rd^3r'\,\varrho_0\left[\mathsf{R}^{\rm T}_\Omega(\bls{r}-\bls{R})\right]\varrho_0\left[\mathsf{R}^{\rm T}_\Omega(\bls{r}'-\bls{R})\right]
    \delta^\perp(\bls{r}-\bls{s})\mathsf{T}(\bls{r},\bls{r}')\delta^\perp(\bls{r}'-\bls{s}'),
\end{align}
and the symmetries $\mathsf{J}^{\rm T}(\bls{s},\bls{s}',\omega)=\mathsf{J}(\bls{s}',\bls{s},\omega)$, $\mathsf{C}^{\rm T}(\bls{s},\bls{s}')=\mathsf{C}(\bls{s}',\bls{s})$,
one arrives at
\begin{align}\label{eq:dissipatorArbitraryGreentensor}
    \mathcal{L}_{\rm Th}\rho=\int_0^\infty d\omega\, \frac{2\pi}{\omega^4}n(\omega)\left[n(\omega)+1\right]\int d^3s_1 d^3s_2\,{\rm Tr}\left[\mathsf{L}(\bls{s}_1,\bls{s}_2,\omega)\rho\mathsf{L}(\bls{s}_2,\bls{s}_1,\omega)-\frac{1}{2}\left\{\mathsf{L}(\bls{s}_1,\bls{s}_2,\omega)\mathsf{L}(\bls{s}_2,\bls{s}_1,\omega),\rho\right\}\right]
\end{align}
with 
\begin{equation}\label{eq:definitionLTensorThomson}
    \mathsf{L}(\bls{s}_1,\bls{s}_2,\omega)=\int d^3s\, \mathsf{J}(\bls{s}_1,\bls{s},\omega)\mathsf{C}(\bls{s},\bls{s}_2).
\end{equation}
Here ${\rm Tr}\left[\cdot\right]$ is the tensor trace. The dissipator $\mathcal{L}_{\rm Th}$ describes decoherence of a particle with arbitrary charge distribution near a dielectric medium due to the $\bls{A}^2$-term in the coupling Hamiltonian. The properties of the dielectric enter through the spectral density $\mathsf{J}(\bls{s},\bls{s}',\omega)$, while the particle properties enter through the operator $\mathsf{C}(\bls{s},\bls{s}')$.

\subsection{Free-space Thomson scattering}
We are now in a position to describe decoherence due to free-space Thompson scattering by inserting into \eqref{eq:dissipatorArbitraryGreentensor} the free-space Green tensor
\begin{align}\label{eq:freeSpaceGreentensor}
    \mathsf{G}(\bls{r},\bls{r}',\omega)=\frac{c^2}{(2\pi)^3}\lim_{\epsilon\downarrow 0}\int d^3k\left(\mathbb{1}-\frac{c\bls{k}}{\omega+i\epsilon}\otimes\frac{c\bls{k}}{\omega+i\epsilon}\right)\frac{e^{i\bls{k}\cdot(\bls{r}-\bls{r}')}}{c^2k^2-(\omega+i\epsilon)^2}.
\end{align}
Carrying out a cyclic permutation inside the trace, we obtain
\begin{align}\label{eq:DissipatorFirstTerm1}
    &\int d^3s_1 d^3s_2\,{\rm Tr}\left[\mathsf{L}(\bls{s}_1,\bls{s}_2,\omega)\,\rho\,\mathsf{L}(\bls{s}_2,\bls{s}_1,\omega)\right]=\int d^3r_1d^3r_1'd^3r_2d^3r_2'\,\varrho_0\!\left[\mathsf{R}^{\rm T}_\Omega(\bls{r}_1-\bls{R})\right]\varrho_0\!\left[\mathsf{R}^{\rm T}_\Omega(\bls{r}_1'-\bls{R})\right]\nonumber\\
    &\times {\rm Tr}\left[\vspace{0mm}^\perp\!\mathsf{J}^\perp(\bls{r}_2',\bls{r}_1,\omega)\mathsf{T}(\bls{r}_1,\bls{r}_1')\,\rho\,\vspace{0mm}^\perp\!\mathsf{J}^\perp(\bls{r}_1',\bls{r}_2,\omega)\mathsf{T}(\bls{r}_2,\bls{r}_2')\right]\varrho_0\!\left[\mathsf{R}^{\rm T}_\Omega(\bls{r}_2-\bls{R})\right]\varrho_0\!\left[\mathsf{R}^{\rm T}_\Omega(\bls{r}_2'-\bls{R})\right].
\end{align}
Here, $\vspace{0mm}^\perp\!\mathsf{J}^\perp(\bls{r},\bls{r}',\omega)\equiv\mu_0\omega^2{\rm Im}\vspace{0mm}^\perp\mathsf{G}^\perp(\bls{r},\bls{r}',\omega)/\pi$ is the left- and right-transverse of the spectral density  \eqref{eq:spectralDensityDefinition}. It can be obtained  from
\begin{align}
    \vspace{0mm}^\perp\mathsf{G}^\perp(\bls{r},\bls{r}',\omega)\equiv \int d^3sd^3s'\, \delta^\perp(\bls{r}-\bls{s})\mathsf{G}(\bls{s},\bls{s}',\omega)\delta^\perp(\bls{r}'-\bls{s}')=-\lim_{\epsilon\downarrow 0}\frac{c^2}{(2\pi)^3}\int d^3k\frac{e^{i\bls{k}\cdot(\bls{r}-\bls{r}')}}{c^2k^2-(\omega+i\epsilon)^2}\frac{\bls{k}}{k}\times\left(\frac{\bls{k}}{k}\times \mathbb{1}\right)
\end{align}
where we used \eqref{eq:freeSpaceGreentensor} with the Fourier-representation of the transverse $\delta$-distribution \eqref{eq:definitionTransverseDelta},
\begin{equation}
    \delta^\perp(\bls{r}-\bls{r}')=-\frac{1}{(2\pi)^3}\int d^3k \frac{\bls{k}}{k}\times\left(\frac{\bls{k}}{k}\times \mathbb{1}\right)e^{i\bls{k}\cdot(\bls{r}-\bls{r}')}.
\end{equation}
Noting
\begin{equation}
    \lim_{\epsilon\downarrow 0}\left(\frac{c^2k^2}{c^2k^2-(\omega+i\epsilon)^2}-\frac{c^2k^2}{c^2k^2-{(\omega-i\epsilon)}^2}\right)=\frac{i\pi\omega}{c}\delta\!\left(k-\frac{\omega}{c}\right)
\end{equation}
and $d^3k= dk\,k^2\,d^2n$ with ${\bls k}=k \bls n$ this yields
\begin{align}\label{eq:spectralDensityFreeSpaceThomson}
\vspace{0mm}^\perp\!\mathsf{J}^\perp(\bls{r},\bls{r}',\omega)=-\frac{\mu_0\omega^3}{2c(2\pi)^3}\int d^2n\,\bls{n}\times\left(\bls{n}\times \mathbb{1}\right)e^{i\frac{\omega}{c}\bls{n}\cdot(\bls{r}-\bls{r}')},
\end{align}

Next, one inserts \eqref{eq:spectralDensityFreeSpaceThomson} into \eqref{eq:DissipatorFirstTerm1}, defines the orientation dependent form-factor operators
\begin{subequations}
\begin{align}
F(\bls{k})&=\int d^3r\, \varrho_0(\bls{r})e^{i\bls{k}\cdot \mathsf{R}_\Omega\bls{r}}=F^\dagger(-\bls{k}),\\
\bls{F}(\bls{k})&=\int d^3r\, \varrho_0(\bls{r})e^{i\bls{k}\cdot \mathsf{R}_\Omega\bls{r}}\mathsf{R}_\Omega \bls{r}=\bls{F}^\dagger(-\bls{k}),
\end{align}
\end{subequations}
and uses that 
\begin{align}
    \bls{n}\times(\bls{n}\times\mathbb{1})=&\bls{n}\otimes\bls{n}-\mathbb{1}=-\left[\bls{n}\otimes\bls{n}-\mathbb{1}\right]\left[\bls{n}\otimes\bls{n}-\mathbb{1}\right],\\
    \bls{n}'\times(\bls{n}'\times\mathbb{1})=&-\sum_{j=1}^2\blg{\epsilon}_j'\otimes\blg{\epsilon}_j',
\end{align}
with $\blg{\epsilon}_1'$, $\blg{\epsilon}_2'$ forming an orthonormal basis with $\bls{n}'$ (like the two polarization vectors of a plane wave traveling in direction $\bls{n}'$). With these definitions and \eqref{eq:definitionTTensor} one can write
\begin{align}\label{eq:DissipatorFirstTerm3}
    &\int d^3s_1 d^3s_2\,{\rm Tr}\left[\mathsf{L}(\bls{s}_1,\bls{s}_2,\omega)\,\rho\,\mathsf{L}(\bls{s}_2,\bls{s}_1,\omega)\right]=\left(\frac{\mu_0\omega^3}{2c(2\pi)^3}\right)^2\int d^2nd^2n'\sum_{j=1}^2{\rm Tr}\left\{\left[\bls{n}\otimes\bls{n}-\mathbb{1}\right]\left[\mathbb{1}\frac{F(-\omega\bls{n}/c)F(\omega\bls{n}'/c)}{m}\right.\right.\nonumber\\
    &\left.\left.\vphantom{\frac{1}{2}}-\bls{F}(-\omega\bls{n}/c)\times{\bf I}^{-1}_\Omega\times\bls{F}(\omega\bls{n}'/c)\right]\blg{\epsilon}'_j\otimes\blg{\epsilon}'_je^{i\frac{\omega}{c}(\bls{n}'-\bls{n})\cdot\bls{R}}\,\rho\,e^{-i\frac{\omega}{c}(\bls{n}'-\bls{n})\cdot\bls{R}}\left[\mathbb{1}\frac{F^\dagger(\omega\bls{n}'/c)F^\dagger(-\omega\bls{n}/c)}{m}\right.\right.\nonumber\\
    &\left.\left.\vphantom{\frac{1}{2}}-\bls{F}^\dagger(\omega\bls{n}'/c)\times{\bf I}^{-1}_\Omega\times\bls{F}^\dagger(-\omega\bls{n}/c)\right]\left[\bls{n}\otimes\bls{n}-\mathbb{1}\right]\right\}.
\end{align}
Here, one can identify the orientation-dependent scatting amplitude of Thomson scattering
\begin{align}\label{eq:scatteringAmplitudeThomson}
    \bls{F}_j(\bls{k},\bls{k}')=\frac{1}{4\pi\epsilon_0c^2}\left[\frac{\bls{k}}{k}\otimes\frac{\bls{k}}{k}-\mathbb{1}\right]\left[\mathbb{1}\frac{F(-\bls{k})F(\bls{k}')}{m}-\bls{F}(-\bls{k})\times{\bf I}^{-1}_\Omega\times\bls{F}(\bls{k}')\right]\blg{\epsilon}'_j
\end{align}
of an electromagnetic wave with incoming wave vector $\bls{k}'$ and polarization $\blg{\epsilon}_j'$, elastically scattered at a rigid charge distribution to an outgoing direction with wave vector $\bls{k}$, as can be derived from classical electrodynamics \cite{martinetzDiss}. With this  \eqref{eq:DissipatorFirstTerm3} can be simplified to
\begin{align}
    &\int d^3s_1 d^3s_2\,{\rm Tr}\left[\mathsf{L}(\bls{s}_1,\bls{s}_2,\omega)\,\rho\,\mathsf{L}(\bls{s}_2,\bls{s}_1,\omega)\right]\nonumber\\
    &=\left(\frac{\omega^3}{c(2\pi)^2}\right)^2\int d^2nd^2n'\sum_{j=1}^2{\rm Tr}\left\{\bls{F}_j\left(\frac{\omega}{c}\bls{n},\frac{\omega}{c}\bls{n}'\right)e^{i\frac{\omega}{c}(\bls{n}'-\bls{n})\cdot\bls{R}}\,\rho\,\otimes\bls{F}^\dagger_j\left(\frac{\omega}{c}\bls{n},\frac{\omega}{c}\bls{n}'\right)e^{-i\frac{\omega}{c}(\bls{n}'-\bls{n})\cdot\bls{R}}\right\}\\
    &=\left(\frac{\omega^3}{c(2\pi)^2}\right)^2\int d^2nd^2n'\sum_{j=1}^2\bls{F}_j\left(\frac{\omega}{c}\bls{n},\frac{\omega}{c}\bls{n}'\right)e^{i\frac{\omega}{c}(\bls{n}'-\bls{n})\cdot\bls{R}}\,\rho\,\cdot\bls{F}^\dagger_j\left(\frac{\omega}{c}\bls{n},\frac{\omega}{c}\bls{n}'\right)e^{-i\frac{\omega}{c}(\bls{n}'-\bls{n})\cdot\bls{R}}.
\end{align}
Defining the Lindblad operators 
\begin{align}
    \bls{L}_j(\bls{k},\bls{k}')=e^{i(\bls{k}'-\bls{k})\cdot\bls{R}}\bls{F}_j(\bls{k},\bls{k}')
\end{align}
with symmetry $\bls{L}_j(\bls{k},\bls{k}')=\bls{L}^\dagger_j(-\bls{k},-\bls{k}')$ we can write the total dissipator \eqref{eq:dissipatorArbitraryGreentensor} as
\begin{align}\label{eq:FinalDissipatorThomsonScattering}
    \mathcal{L}_{\rm Th}\rho=\int_0^\infty dk\frac{ck^2}{(2\pi)^3} n(ck)\left[n(ck)+1\right]\int d^2n d^2n'\sum_{j=1}^2 \left[\bls{L}_j(k\bls{n},k\bls{n}')\cdot \,\rho\,\bls{L}^\dagger_j(k\bls{n},k\bls{n}')-\frac{1}{2}\left\{\bls{L}^\dagger_j(k\bls{n},k\bls{n}')\cdot \bls{L}_j(k\bls{n},k\bls{n}'),\,\rho\,\right\}\right].
\end{align}
The dissipator \eqref{eq:FinalDissipatorThomsonScattering} is of the typical form to describe scattering decoherence. The Lindblad operators are given by plane waves multiplied by the scattering amplitude, while the pre-factor accounts for the statistics of the scattering incidents. The dissipator is consistent with the dissipator derived in the Monitoring Approach \cite{hornberger2007,stickler2016b}. 
\end{widetext}

\subsection{Thomson-scattering decoherence of a point monopole}
To estimate decoherence of a point monopole  $\varrho_0(\bls{r})=q\delta(\bls{r})$, we use that the transverse electromagnetic field, and thus the vector potential, dominate the particle-surface interaction for distances larger than the wavelength, e.g. the thermal wavelength $2\pi\hbar c/k_{\rm B}T$. For simplicity, we thus assume the particle to be located in free space, where \eqref{eq:FinalDissipatorThomsonScattering} applies. For the point particle, the scattering amplitude \eqref{eq:scatteringAmplitudeThomson} reduces to
\begin{align}
    \bls{F}_j(k\bls{n},k\bls{n}')&=\frac{q^2}{4\pi\epsilon_0c^2m}\left[\bls{n}\otimes\bls{n}-\mathbb{1}\right]\blg{\epsilon}'_j,
\end{align}
which fulfills
\begin{align}
    \sum_{j=1}^2\bls{F}_j(k\bls{n},k\bls{n}')&\cdot \bls{F}^\dagger_j(k\bls{n},k\bls{n}')\nonumber\\
    &=\left(\frac{q^2}{4\pi\epsilon_0c^2m}\right)^2\left[1+(\bls{n}\cdot\bls{n}')^2\right].
\end{align}
Thus, the coherences in the position basis decrease according to $\bra{\bls{R}}\mathcal{L}_{\rm Th}\rho\ket{\bls{R}'}=-\Gamma_{\rm Th}(\bls{R},\bls{R}')\bra{\bls{R}}\rho\ket{\bls{R}'}$ with the decoherence rate
\begin{align}
    &\Gamma_{\rm Th}(\bls{R},\bls{R}')=\left(\frac{q^2}{4\pi\epsilon_0c^2m}\right)^2\int_0^\infty dk\frac{ck^2}{(2\pi)^3} n(ck)\left[n(ck)+1\right]\nonumber\\
    &\times\int d^2n d^2n'\left[1+(\bls{n}\cdot\bls{n}')^2\right]\left(1-e^{ik(\bls{n}'-\bls{n})\cdot(\bls{R}-\bls{R}')}\right).
\end{align}
We express the spherical integrals through the spherical Bessel functions of the first kind $j_n(x)$,
\begin{align}
    &\Gamma_{\rm Th}(\bls{R},\bls{R}')=\left(\frac{q^2}{4\pi\epsilon_0c^2m}\right)^2\int_0^\infty dk\frac{2ck^2}{\pi} n(ck)\left[n(ck)+1\right]\nonumber\\
    &\times\left\{\frac{4}{3}-2\left[\left(j_0(x)-\frac{j_1(x)}{x}\right)^2+2\frac{j^2_1(x)}{x^2}\right]\right\},
\end{align}
where $x=k|\bls{R}-\bls{R}'|$. The rate vanishes for $|\bls{R}-\bls{R}'|=0$, while for  separations $|\bls{R}-\bls{R}'|$ much larger than the thermal wavelength it saturates at the value
\begin{align}
    \Gamma_\infty=
    \left(\frac{q^2}{4\pi\epsilon_0c^3m}\right)^2\frac{8\pi }{9}\left(\frac{k_{\rm B}T}{\hbar }\right)^3.
\end{align}

The impact of decoherence due to the $\bls{A}^2$-term can now be evaluated explicitly.
For example a single electron in vacuum at temperature $T=300\,{\rm K}$ is subject to a decoherence rate of $$\Gamma_\infty\approx1.5\times10^{-5}{\rm s}^{-1},$$ which is negligible in comparison to the decoherence due to the longitudinal field. This is a consequence of  the tiny scattering cross section of Thomson scattering, which is even smaller for nanoparticles with their reduced ratio of $q^2/m$.


\begin{thebibliography}{128}%
\makeatletter
\providecommand \@ifxundefined [1]{%
 \@ifx{#1\undefined}
}%
\providecommand \@ifnum [1]{%
 \ifnum #1\expandafter \@firstoftwo
 \else \expandafter \@secondoftwo
 \fi
}%
\providecommand \@ifx [1]{%
 \ifx #1\expandafter \@firstoftwo
 \else \expandafter \@secondoftwo
 \fi
}%
\providecommand \natexlab [1]{#1}%
\providecommand \enquote  [1]{``#1''}%
\providecommand \bibnamefont  [1]{#1}%
\providecommand \bibfnamefont [1]{#1}%
\providecommand \citenamefont [1]{#1}%
\providecommand \href@noop [0]{\@secondoftwo}%
\providecommand \href [0]{\begingroup \@sanitize@url \@href}%
\providecommand \@href[1]{\@@startlink{#1}\@@href}%
\providecommand \@@href[1]{\endgroup#1\@@endlink}%
\providecommand \@sanitize@url [0]{\catcode `\\12\catcode `\$12\catcode
  `\&12\catcode `\#12\catcode `\^12\catcode `\_12\catcode `\%12\relax}%
\providecommand \@@startlink[1]{}%
\providecommand \@@endlink[0]{}%
\providecommand \url  [0]{\begingroup\@sanitize@url \@url }%
\providecommand \@url [1]{\endgroup\@href {#1}{\urlprefix }}%
\providecommand \urlprefix  [0]{URL }%
\providecommand \Eprint [0]{\href }%
\providecommand \doibase [0]{https://doi.org/}%
\providecommand \selectlanguage [0]{\@gobble}%
\providecommand \bibinfo  [0]{\@secondoftwo}%
\providecommand \bibfield  [0]{\@secondoftwo}%
\providecommand \translation [1]{[#1]}%
\providecommand \BibitemOpen [0]{}%
\providecommand \bibitemStop [0]{}%
\providecommand \bibitemNoStop [0]{.\EOS\space}%
\providecommand \EOS [0]{\spacefactor3000\relax}%
\providecommand \BibitemShut  [1]{\csname bibitem#1\endcsname}%
\let\auto@bib@innerbib\@empty
\bibitem [{\citenamefont {Monroe}\ \emph {et~al.}(2021)\citenamefont {Monroe},
  \citenamefont {Campbell}, \citenamefont {Duan}, \citenamefont {Gong},
  \citenamefont {Gorshkov}, \citenamefont {Hess}, \citenamefont {Islam},
  \citenamefont {Kim}, \citenamefont {Linke}, \citenamefont {Pagano},
  \citenamefont {Richerme}, \citenamefont {Senko},\ and\ \citenamefont
  {Yao}}]{monroe2021}%
  \BibitemOpen
  \bibfield  {author} {\bibinfo {author} {\bibfnamefont {C.}~\bibnamefont
  {Monroe}}, \bibinfo {author} {\bibfnamefont {W.~C.}\ \bibnamefont
  {Campbell}}, \bibinfo {author} {\bibfnamefont {L.-M.}\ \bibnamefont {Duan}},
  \bibinfo {author} {\bibfnamefont {Z.-X.}\ \bibnamefont {Gong}}, \bibinfo
  {author} {\bibfnamefont {A.~V.}\ \bibnamefont {Gorshkov}}, \bibinfo {author}
  {\bibfnamefont {P.~W.}\ \bibnamefont {Hess}}, \bibinfo {author}
  {\bibfnamefont {R.}~\bibnamefont {Islam}}, \bibinfo {author} {\bibfnamefont
  {K.}~\bibnamefont {Kim}}, \bibinfo {author} {\bibfnamefont {N.~M.}\
  \bibnamefont {Linke}}, \bibinfo {author} {\bibfnamefont {G.}~\bibnamefont
  {Pagano}}, \bibinfo {author} {\bibfnamefont {P.}~\bibnamefont {Richerme}},
  \bibinfo {author} {\bibfnamefont {C.}~\bibnamefont {Senko}},\ and\ \bibinfo
  {author} {\bibfnamefont {N.~Y.}\ \bibnamefont {Yao}},\ }\bibfield  {title}
  {\bibinfo {title} {Programmable quantum simulations of spin systems with
  trapped ions},\ }\href {https://doi.org/10.1103/RevModPhys.93.025001}
  {\bibfield  {journal} {\bibinfo  {journal} {Rev. Mod. Phys.}\ }\textbf
  {\bibinfo {volume} {93}},\ \bibinfo {pages} {025001} (\bibinfo {year}
  {2021})}\BibitemShut {NoStop}%
\bibitem [{\citenamefont {de~Leon}\ \emph {et~al.}(2021)\citenamefont
  {de~Leon}, \citenamefont {Itoh}, \citenamefont {Kim}, \citenamefont {Mehta},
  \citenamefont {Northup}, \citenamefont {Paik}, \citenamefont {Palmer},
  \citenamefont {Samarth}, \citenamefont {Sangtawesin},\ and\ \citenamefont
  {Steuerman}}]{de2021materials}%
  \BibitemOpen
  \bibfield  {author} {\bibinfo {author} {\bibfnamefont {N.~P.}\ \bibnamefont
  {de~Leon}}, \bibinfo {author} {\bibfnamefont {K.~M.}\ \bibnamefont {Itoh}},
  \bibinfo {author} {\bibfnamefont {D.}~\bibnamefont {Kim}}, \bibinfo {author}
  {\bibfnamefont {K.~K.}\ \bibnamefont {Mehta}}, \bibinfo {author}
  {\bibfnamefont {T.~E.}\ \bibnamefont {Northup}}, \bibinfo {author}
  {\bibfnamefont {H.}~\bibnamefont {Paik}}, \bibinfo {author} {\bibfnamefont
  {B.}~\bibnamefont {Palmer}}, \bibinfo {author} {\bibfnamefont
  {N.}~\bibnamefont {Samarth}}, \bibinfo {author} {\bibfnamefont
  {S.}~\bibnamefont {Sangtawesin}},\ and\ \bibinfo {author} {\bibfnamefont
  {D.}~\bibnamefont {Steuerman}},\ }\bibfield  {title} {\bibinfo {title}
  {Materials challenges and opportunities for quantum computing hardware},\
  }\href@noop {} {\bibfield  {journal} {\bibinfo  {journal} {Science}\ }\textbf
  {\bibinfo {volume} {372}},\ \bibinfo {pages} {eabb2823} (\bibinfo {year}
  {2021})}\BibitemShut {NoStop}%
\bibitem [{\citenamefont {Gilmore}\ \emph {et~al.}(2021)\citenamefont
  {Gilmore}, \citenamefont {Affolter}, \citenamefont {Lewis-Swan},
  \citenamefont {Barberena}, \citenamefont {Jordan}, \citenamefont {Rey},\ and\
  \citenamefont {Bollinger}}]{gilmore2021quantum}%
  \BibitemOpen
  \bibfield  {author} {\bibinfo {author} {\bibfnamefont {K.~A.}\ \bibnamefont
  {Gilmore}}, \bibinfo {author} {\bibfnamefont {M.}~\bibnamefont {Affolter}},
  \bibinfo {author} {\bibfnamefont {R.~J.}\ \bibnamefont {Lewis-Swan}},
  \bibinfo {author} {\bibfnamefont {D.}~\bibnamefont {Barberena}}, \bibinfo
  {author} {\bibfnamefont {E.}~\bibnamefont {Jordan}}, \bibinfo {author}
  {\bibfnamefont {A.~M.}\ \bibnamefont {Rey}},\ and\ \bibinfo {author}
  {\bibfnamefont {J.~J.}\ \bibnamefont {Bollinger}},\ }\bibfield  {title}
  {\bibinfo {title} {Quantum-enhanced sensing of displacements and electric
  fields with two-dimensional trapped-ion crystals},\ }\href@noop {} {\bibfield
   {journal} {\bibinfo  {journal} {Science}\ }\textbf {\bibinfo {volume}
  {373}},\ \bibinfo {pages} {673} (\bibinfo {year} {2021})}\BibitemShut
  {NoStop}%
\bibitem [{\citenamefont {Cetina}\ \emph {et~al.}(2022)\citenamefont {Cetina},
  \citenamefont {Egan}, \citenamefont {Noel}, \citenamefont {Goldman},
  \citenamefont {Biswas}, \citenamefont {Risinger}, \citenamefont {Zhu},\ and\
  \citenamefont {Monroe}}]{cetina2022quantum}%
  \BibitemOpen
  \bibfield  {author} {\bibinfo {author} {\bibfnamefont {M.}~\bibnamefont
  {Cetina}}, \bibinfo {author} {\bibfnamefont {L.}~\bibnamefont {Egan}},
  \bibinfo {author} {\bibfnamefont {C.}~\bibnamefont {Noel}}, \bibinfo {author}
  {\bibfnamefont {M.}~\bibnamefont {Goldman}}, \bibinfo {author} {\bibfnamefont
  {D.}~\bibnamefont {Biswas}}, \bibinfo {author} {\bibfnamefont
  {A.}~\bibnamefont {Risinger}}, \bibinfo {author} {\bibfnamefont
  {D.}~\bibnamefont {Zhu}},\ and\ \bibinfo {author} {\bibfnamefont
  {C.}~\bibnamefont {Monroe}},\ }\bibfield  {title} {\bibinfo {title} {Quantum
  gates on individually-addressed atomic qubits subject to noisy transverse
  motion},\ }\href@noop {} {\bibfield  {journal} {\bibinfo  {journal} {PRX
  Quantum}\ }\textbf {\bibinfo {volume} {3}},\ \bibinfo {pages} {010334}
  (\bibinfo {year} {2022})}\BibitemShut {NoStop}%
\bibitem [{\citenamefont {Zhang}\ \emph {et~al.}(2017)\citenamefont {Zhang},
  \citenamefont {Hess}, \citenamefont {Kyprianidis}, \citenamefont {Becker},
  \citenamefont {Lee}, \citenamefont {Smith}, \citenamefont {Pagano},
  \citenamefont {Potirniche}, \citenamefont {Potter}, \citenamefont
  {Vishwanath} \emph {et~al.}}]{zhang2017observation1}%
  \BibitemOpen
  \bibfield  {author} {\bibinfo {author} {\bibfnamefont {J.}~\bibnamefont
  {Zhang}}, \bibinfo {author} {\bibfnamefont {P.}~\bibnamefont {Hess}},
  \bibinfo {author} {\bibfnamefont {A.}~\bibnamefont {Kyprianidis}}, \bibinfo
  {author} {\bibfnamefont {P.}~\bibnamefont {Becker}}, \bibinfo {author}
  {\bibfnamefont {A.}~\bibnamefont {Lee}}, \bibinfo {author} {\bibfnamefont
  {J.}~\bibnamefont {Smith}}, \bibinfo {author} {\bibfnamefont
  {G.}~\bibnamefont {Pagano}}, \bibinfo {author} {\bibfnamefont {I.-D.}\
  \bibnamefont {Potirniche}}, \bibinfo {author} {\bibfnamefont {A.~C.}\
  \bibnamefont {Potter}}, \bibinfo {author} {\bibfnamefont {A.}~\bibnamefont
  {Vishwanath}}, \emph {et~al.},\ }\bibfield  {title} {\bibinfo {title}
  {Observation of a discrete time crystal},\ }\href@noop {} {\bibfield
  {journal} {\bibinfo  {journal} {Nature}\ }\textbf {\bibinfo {volume} {543}},\
  \bibinfo {pages} {217} (\bibinfo {year} {2017})}\BibitemShut {NoStop}%
\bibitem [{\citenamefont {Kyprianidis}\ \emph {et~al.}(2021)\citenamefont
  {Kyprianidis}, \citenamefont {Machado}, \citenamefont {Morong}, \citenamefont
  {Becker}, \citenamefont {Collins}, \citenamefont {Else}, \citenamefont
  {Feng}, \citenamefont {Hess}, \citenamefont {Nayak}, \citenamefont {Pagano}
  \emph {et~al.}}]{kyprianidis2021observation}%
  \BibitemOpen
  \bibfield  {author} {\bibinfo {author} {\bibfnamefont {A.}~\bibnamefont
  {Kyprianidis}}, \bibinfo {author} {\bibfnamefont {F.}~\bibnamefont
  {Machado}}, \bibinfo {author} {\bibfnamefont {W.}~\bibnamefont {Morong}},
  \bibinfo {author} {\bibfnamefont {P.}~\bibnamefont {Becker}}, \bibinfo
  {author} {\bibfnamefont {K.~S.}\ \bibnamefont {Collins}}, \bibinfo {author}
  {\bibfnamefont {D.~V.}\ \bibnamefont {Else}}, \bibinfo {author}
  {\bibfnamefont {L.}~\bibnamefont {Feng}}, \bibinfo {author} {\bibfnamefont
  {P.~W.}\ \bibnamefont {Hess}}, \bibinfo {author} {\bibfnamefont
  {C.}~\bibnamefont {Nayak}}, \bibinfo {author} {\bibfnamefont
  {G.}~\bibnamefont {Pagano}}, \emph {et~al.},\ }\bibfield  {title} {\bibinfo
  {title} {Observation of a prethermal discrete time crystal},\ }\href@noop {}
  {\bibfield  {journal} {\bibinfo  {journal} {Science}\ }\textbf {\bibinfo
  {volume} {372}},\ \bibinfo {pages} {1192} (\bibinfo {year}
  {2021})}\BibitemShut {NoStop}%
\bibitem [{\citenamefont {Andr{\'e}}\ \emph {et~al.}(2006)\citenamefont
  {Andr{\'e}}, \citenamefont {DeMille}, \citenamefont {Doyle}, \citenamefont
  {Lukin}, \citenamefont {Maxwell}, \citenamefont {Rabl}, \citenamefont
  {Schoelkopf},\ and\ \citenamefont {Zoller}}]{andre2006coherent}%
  \BibitemOpen
  \bibfield  {author} {\bibinfo {author} {\bibfnamefont {A.}~\bibnamefont
  {Andr{\'e}}}, \bibinfo {author} {\bibfnamefont {D.}~\bibnamefont {DeMille}},
  \bibinfo {author} {\bibfnamefont {J.~M.}\ \bibnamefont {Doyle}}, \bibinfo
  {author} {\bibfnamefont {M.~D.}\ \bibnamefont {Lukin}}, \bibinfo {author}
  {\bibfnamefont {S.~E.}\ \bibnamefont {Maxwell}}, \bibinfo {author}
  {\bibfnamefont {P.}~\bibnamefont {Rabl}}, \bibinfo {author} {\bibfnamefont
  {R.~J.}\ \bibnamefont {Schoelkopf}},\ and\ \bibinfo {author} {\bibfnamefont
  {P.}~\bibnamefont {Zoller}},\ }\bibfield  {title} {\bibinfo {title} {A
  coherent all-electrical interface between polar molecules and mesoscopic
  superconducting resonators},\ }\href@noop {} {\bibfield  {journal} {\bibinfo
  {journal} {Nat. Phys.}\ }\textbf {\bibinfo {volume} {2}},\ \bibinfo {pages}
  {636} (\bibinfo {year} {2006})}\BibitemShut {NoStop}%
\bibitem [{\citenamefont {Rabl}\ \emph {et~al.}(2006)\citenamefont {Rabl},
  \citenamefont {DeMille}, \citenamefont {Doyle}, \citenamefont {Lukin},
  \citenamefont {Schoelkopf},\ and\ \citenamefont {Zoller}}]{rabl2006hybrid}%
  \BibitemOpen
  \bibfield  {author} {\bibinfo {author} {\bibfnamefont {P.}~\bibnamefont
  {Rabl}}, \bibinfo {author} {\bibfnamefont {D.}~\bibnamefont {DeMille}},
  \bibinfo {author} {\bibfnamefont {J.~M.}\ \bibnamefont {Doyle}}, \bibinfo
  {author} {\bibfnamefont {M.~D.}\ \bibnamefont {Lukin}}, \bibinfo {author}
  {\bibfnamefont {R.}~\bibnamefont {Schoelkopf}},\ and\ \bibinfo {author}
  {\bibfnamefont {P.}~\bibnamefont {Zoller}},\ }\bibfield  {title} {\bibinfo
  {title} {Hybrid quantum processors: molecular ensembles as quantum memory for
  solid state circuits},\ }\href@noop {} {\bibfield  {journal} {\bibinfo
  {journal} {Phys. Rev. Lett.}\ }\textbf {\bibinfo {volume} {97}},\ \bibinfo
  {pages} {033003} (\bibinfo {year} {2006})}\BibitemShut {NoStop}%
\bibitem [{\citenamefont {Rabl}\ and\ \citenamefont
  {Zoller}(2007)}]{rabl2007molecular}%
  \BibitemOpen
  \bibfield  {author} {\bibinfo {author} {\bibfnamefont {P.}~\bibnamefont
  {Rabl}}\ and\ \bibinfo {author} {\bibfnamefont {P.}~\bibnamefont {Zoller}},\
  }\bibfield  {title} {\bibinfo {title} {Molecular dipolar crystals as
  high-fidelity quantum memory for hybrid quantum computing},\ }\href@noop {}
  {\bibfield  {journal} {\bibinfo  {journal} {Phys. Rev. A}\ }\textbf {\bibinfo
  {volume} {76}},\ \bibinfo {pages} {042308} (\bibinfo {year}
  {2007})}\BibitemShut {NoStop}%
\bibitem [{\citenamefont {Xiang}\ \emph {et~al.}(2013)\citenamefont {Xiang},
  \citenamefont {Ashhab}, \citenamefont {You},\ and\ \citenamefont
  {Nori}}]{xiang2013hybrid}%
  \BibitemOpen
  \bibfield  {author} {\bibinfo {author} {\bibfnamefont {Z.-L.}\ \bibnamefont
  {Xiang}}, \bibinfo {author} {\bibfnamefont {S.}~\bibnamefont {Ashhab}},
  \bibinfo {author} {\bibfnamefont {J.}~\bibnamefont {You}},\ and\ \bibinfo
  {author} {\bibfnamefont {F.}~\bibnamefont {Nori}},\ }\bibfield  {title}
  {\bibinfo {title} {Hybrid quantum circuits: Superconducting circuits
  interacting with other quantum systems},\ }\href@noop {} {\bibfield
  {journal} {\bibinfo  {journal} {Rev. Mod. Phys.}\ }\textbf {\bibinfo {volume}
  {85}},\ \bibinfo {pages} {623} (\bibinfo {year} {2013})}\BibitemShut
  {NoStop}%
\bibitem [{\citenamefont {Albert}\ \emph {et~al.}(2020)\citenamefont {Albert},
  \citenamefont {Covey},\ and\ \citenamefont {Preskill}}]{albert2020robust}%
  \BibitemOpen
  \bibfield  {author} {\bibinfo {author} {\bibfnamefont {V.~V.}\ \bibnamefont
  {Albert}}, \bibinfo {author} {\bibfnamefont {J.~P.}\ \bibnamefont {Covey}},\
  and\ \bibinfo {author} {\bibfnamefont {J.}~\bibnamefont {Preskill}},\
  }\bibfield  {title} {\bibinfo {title} {Robust encoding of a qubit in a
  molecule},\ }\href@noop {} {\bibfield  {journal} {\bibinfo  {journal} {Phys.
  Rev. X}\ }\textbf {\bibinfo {volume} {10}},\ \bibinfo {pages} {031050}
  (\bibinfo {year} {2020})}\BibitemShut {NoStop}%
\bibitem [{\citenamefont {Sawant}\ \emph {et~al.}(2020)\citenamefont {Sawant},
  \citenamefont {Blackmore}, \citenamefont {Gregory}, \citenamefont
  {Mur-Petit}, \citenamefont {Jaksch}, \citenamefont {Aldegunde}, \citenamefont
  {Hutson}, \citenamefont {Tarbutt},\ and\ \citenamefont
  {Cornish}}]{sawant2020ultracold}%
  \BibitemOpen
  \bibfield  {author} {\bibinfo {author} {\bibfnamefont {R.}~\bibnamefont
  {Sawant}}, \bibinfo {author} {\bibfnamefont {J.~A.}\ \bibnamefont
  {Blackmore}}, \bibinfo {author} {\bibfnamefont {P.~D.}\ \bibnamefont
  {Gregory}}, \bibinfo {author} {\bibfnamefont {J.}~\bibnamefont {Mur-Petit}},
  \bibinfo {author} {\bibfnamefont {D.}~\bibnamefont {Jaksch}}, \bibinfo
  {author} {\bibfnamefont {J.}~\bibnamefont {Aldegunde}}, \bibinfo {author}
  {\bibfnamefont {J.~M.}\ \bibnamefont {Hutson}}, \bibinfo {author}
  {\bibfnamefont {M.}~\bibnamefont {Tarbutt}},\ and\ \bibinfo {author}
  {\bibfnamefont {S.~L.}\ \bibnamefont {Cornish}},\ }\bibfield  {title}
  {\bibinfo {title} {Ultracold polar molecules as qudits},\ }\href@noop {}
  {\bibfield  {journal} {\bibinfo  {journal} {New J. Phys.}\ }\textbf {\bibinfo
  {volume} {22}},\ \bibinfo {pages} {013027} (\bibinfo {year}
  {2020})}\BibitemShut {NoStop}%
\bibitem [{\citenamefont {Campbell}\ and\ \citenamefont
  {Hudson}(2020)}]{campbell2020dipole}%
  \BibitemOpen
  \bibfield  {author} {\bibinfo {author} {\bibfnamefont {W.~C.}\ \bibnamefont
  {Campbell}}\ and\ \bibinfo {author} {\bibfnamefont {E.~R.}\ \bibnamefont
  {Hudson}},\ }\bibfield  {title} {\bibinfo {title} {Dipole-phonon quantum
  logic with trapped polar molecular ions},\ }\href@noop {} {\bibfield
  {journal} {\bibinfo  {journal} {Phys. Rev. Lett.}\ }\textbf {\bibinfo
  {volume} {125}},\ \bibinfo {pages} {120501} (\bibinfo {year}
  {2020})}\BibitemShut {NoStop}%
\bibitem [{\citenamefont {Gregory}\ \emph {et~al.}(2021)\citenamefont
  {Gregory}, \citenamefont {Blackmore}, \citenamefont {Bromley}, \citenamefont
  {Hutson},\ and\ \citenamefont {Cornish}}]{gregory2021robust}%
  \BibitemOpen
  \bibfield  {author} {\bibinfo {author} {\bibfnamefont {P.~D.}\ \bibnamefont
  {Gregory}}, \bibinfo {author} {\bibfnamefont {J.~A.}\ \bibnamefont
  {Blackmore}}, \bibinfo {author} {\bibfnamefont {S.~L.}\ \bibnamefont
  {Bromley}}, \bibinfo {author} {\bibfnamefont {J.~M.}\ \bibnamefont
  {Hutson}},\ and\ \bibinfo {author} {\bibfnamefont {S.~L.}\ \bibnamefont
  {Cornish}},\ }\bibfield  {title} {\bibinfo {title} {Robust storage qubits in
  ultracold polar molecules},\ }\href@noop {} {\bibfield  {journal} {\bibinfo
  {journal} {Nat. Phys.}\ }\textbf {\bibinfo {volume} {17}},\ \bibinfo {pages}
  {1149} (\bibinfo {year} {2021})}\BibitemShut {NoStop}%
\bibitem [{\citenamefont {Krems}(2008)}]{krems2008cold}%
  \BibitemOpen
  \bibfield  {author} {\bibinfo {author} {\bibfnamefont {R.~V.}\ \bibnamefont
  {Krems}},\ }\bibfield  {title} {\bibinfo {title} {Cold controlled
  chemistry},\ }\href@noop {} {\bibfield  {journal} {\bibinfo  {journal} {Phys.
  Chem. Chem. Phys.}\ }\textbf {\bibinfo {volume} {10}},\ \bibinfo {pages}
  {4079} (\bibinfo {year} {2008})}\BibitemShut {NoStop}%
\bibitem [{\citenamefont {Ospelkaus}\ \emph {et~al.}(2010)\citenamefont
  {Ospelkaus}, \citenamefont {Ni}, \citenamefont {Wang}, \citenamefont
  {De~Miranda}, \citenamefont {Neyenhuis}, \citenamefont {Qu{\'e}m{\'e}ner},
  \citenamefont {Julienne}, \citenamefont {Bohn}, \citenamefont {Jin},\ and\
  \citenamefont {Ye}}]{ospelkaus2010quantum}%
  \BibitemOpen
  \bibfield  {author} {\bibinfo {author} {\bibfnamefont {S.}~\bibnamefont
  {Ospelkaus}}, \bibinfo {author} {\bibfnamefont {K.-K.}\ \bibnamefont {Ni}},
  \bibinfo {author} {\bibfnamefont {D.}~\bibnamefont {Wang}}, \bibinfo {author}
  {\bibfnamefont {M.}~\bibnamefont {De~Miranda}}, \bibinfo {author}
  {\bibfnamefont {B.}~\bibnamefont {Neyenhuis}}, \bibinfo {author}
  {\bibfnamefont {.~G.}\ \bibnamefont {Qu{\'e}m{\'e}ner}}, \bibinfo {author}
  {\bibfnamefont {P.}~\bibnamefont {Julienne}}, \bibinfo {author}
  {\bibfnamefont {J.}~\bibnamefont {Bohn}}, \bibinfo {author} {\bibfnamefont
  {D.}~\bibnamefont {Jin}},\ and\ \bibinfo {author} {\bibfnamefont
  {J.}~\bibnamefont {Ye}},\ }\bibfield  {title} {\bibinfo {title}
  {Quantum-state controlled chemical reactions of ultracold potassium-rubidium
  molecules},\ }\href@noop {} {\bibfield  {journal} {\bibinfo  {journal}
  {Science}\ }\textbf {\bibinfo {volume} {327}},\ \bibinfo {pages} {853}
  (\bibinfo {year} {2010})}\BibitemShut {NoStop}%
\bibitem [{\citenamefont {Quemener}\ and\ \citenamefont
  {Julienne}(2012)}]{quemener2012ultracold}%
  \BibitemOpen
  \bibfield  {author} {\bibinfo {author} {\bibfnamefont {G.}~\bibnamefont
  {Quemener}}\ and\ \bibinfo {author} {\bibfnamefont {P.~S.}\ \bibnamefont
  {Julienne}},\ }\bibfield  {title} {\bibinfo {title} {Ultracold molecules
  under control!},\ }\href@noop {} {\bibfield  {journal} {\bibinfo  {journal}
  {Chem. Rev.}\ }\textbf {\bibinfo {volume} {112}},\ \bibinfo {pages} {4949}
  (\bibinfo {year} {2012})}\BibitemShut {NoStop}%
\bibitem [{\citenamefont {Balakrishnan}(2016)}]{balakrishnan2016perspective}%
  \BibitemOpen
  \bibfield  {author} {\bibinfo {author} {\bibfnamefont {N.}~\bibnamefont
  {Balakrishnan}},\ }\bibfield  {title} {\bibinfo {title} {Perspective:
  Ultracold molecules and the dawn of cold controlled chemistry},\ }\href@noop
  {} {\bibfield  {journal} {\bibinfo  {journal} {J. Chem. Phys.}\ }\textbf
  {\bibinfo {volume} {145}},\ \bibinfo {pages} {150901} (\bibinfo {year}
  {2016})}\BibitemShut {NoStop}%
\bibitem [{\citenamefont {Hu}\ \emph {et~al.}(2021)\citenamefont {Hu},
  \citenamefont {Liu}, \citenamefont {Nichols}, \citenamefont {Zhu},
  \citenamefont {Qu{\'e}m{\'e}ner}, \citenamefont {Dulieu},\ and\ \citenamefont
  {Ni}}]{hu2021nuclear}%
  \BibitemOpen
  \bibfield  {author} {\bibinfo {author} {\bibfnamefont {M.-G.}\ \bibnamefont
  {Hu}}, \bibinfo {author} {\bibfnamefont {Y.}~\bibnamefont {Liu}}, \bibinfo
  {author} {\bibfnamefont {M.~A.}\ \bibnamefont {Nichols}}, \bibinfo {author}
  {\bibfnamefont {L.}~\bibnamefont {Zhu}}, \bibinfo {author} {\bibfnamefont
  {G.}~\bibnamefont {Qu{\'e}m{\'e}ner}}, \bibinfo {author} {\bibfnamefont
  {O.}~\bibnamefont {Dulieu}},\ and\ \bibinfo {author} {\bibfnamefont {K.-K.}\
  \bibnamefont {Ni}},\ }\bibfield  {title} {\bibinfo {title} {Nuclear spin
  conservation enables state-to-state control of ultracold molecular
  reactions},\ }\href@noop {} {\bibfield  {journal} {\bibinfo  {journal} {Nat.
  Chem.}\ }\textbf {\bibinfo {volume} {13}},\ \bibinfo {pages} {435} (\bibinfo
  {year} {2021})}\BibitemShut {NoStop}%
\bibitem [{\citenamefont {Hirzler}\ \emph {et~al.}(2022)\citenamefont
  {Hirzler}, \citenamefont {Lous}, \citenamefont {Trimby}, \citenamefont
  {P{\'e}rez-R{\'\i}os}, \citenamefont {Safavi-Naini},\ and\ \citenamefont
  {Gerritsma}}]{hirzler2022observation}%
  \BibitemOpen
  \bibfield  {author} {\bibinfo {author} {\bibfnamefont {H.}~\bibnamefont
  {Hirzler}}, \bibinfo {author} {\bibfnamefont {R.}~\bibnamefont {Lous}},
  \bibinfo {author} {\bibfnamefont {E.}~\bibnamefont {Trimby}}, \bibinfo
  {author} {\bibfnamefont {J.}~\bibnamefont {P{\'e}rez-R{\'\i}os}}, \bibinfo
  {author} {\bibfnamefont {A.}~\bibnamefont {Safavi-Naini}},\ and\ \bibinfo
  {author} {\bibfnamefont {R.}~\bibnamefont {Gerritsma}},\ }\bibfield  {title}
  {\bibinfo {title} {Observation of chemical reactions between a trapped ion
  and ultracold {F}eshbach dimers},\ }\href@noop {} {\bibfield  {journal}
  {\bibinfo  {journal} {Phys. Rev. Lett.}\ }\textbf {\bibinfo {volume} {128}},\
  \bibinfo {pages} {103401} (\bibinfo {year} {2022})}\BibitemShut {NoStop}%
\bibitem [{\citenamefont {Zelevinsky}\ \emph {et~al.}(2008)\citenamefont
  {Zelevinsky}, \citenamefont {Kotochigova},\ and\ \citenamefont
  {Ye}}]{zelevinsky2008precision}%
  \BibitemOpen
  \bibfield  {author} {\bibinfo {author} {\bibfnamefont {T.}~\bibnamefont
  {Zelevinsky}}, \bibinfo {author} {\bibfnamefont {S.}~\bibnamefont
  {Kotochigova}},\ and\ \bibinfo {author} {\bibfnamefont {J.}~\bibnamefont
  {Ye}},\ }\bibfield  {title} {\bibinfo {title} {Precision test of mass-ratio
  variations with lattice-confined ultracold molecules},\ }\href@noop {}
  {\bibfield  {journal} {\bibinfo  {journal} {Phys. Rev. Lett.}\ }\textbf
  {\bibinfo {volume} {100}},\ \bibinfo {pages} {043201} (\bibinfo {year}
  {2008})}\BibitemShut {NoStop}%
\bibitem [{\citenamefont {Hudson}\ \emph {et~al.}(2011)\citenamefont {Hudson},
  \citenamefont {Kara}, \citenamefont {Smallman}, \citenamefont {Sauer},
  \citenamefont {Tarbutt},\ and\ \citenamefont {Hinds}}]{hudson2011improved}%
  \BibitemOpen
  \bibfield  {author} {\bibinfo {author} {\bibfnamefont {J.~J.}\ \bibnamefont
  {Hudson}}, \bibinfo {author} {\bibfnamefont {D.~M.}\ \bibnamefont {Kara}},
  \bibinfo {author} {\bibfnamefont {I.}~\bibnamefont {Smallman}}, \bibinfo
  {author} {\bibfnamefont {B.~E.}\ \bibnamefont {Sauer}}, \bibinfo {author}
  {\bibfnamefont {M.~R.}\ \bibnamefont {Tarbutt}},\ and\ \bibinfo {author}
  {\bibfnamefont {E.~A.}\ \bibnamefont {Hinds}},\ }\bibfield  {title} {\bibinfo
  {title} {Improved measurement of the shape of the electron},\ }\href@noop {}
  {\bibfield  {journal} {\bibinfo  {journal} {Nature}\ }\textbf {\bibinfo
  {volume} {473}},\ \bibinfo {pages} {493} (\bibinfo {year}
  {2011})}\BibitemShut {NoStop}%
\bibitem [{\citenamefont {DeMille}\ \emph {et~al.}(2017)\citenamefont
  {DeMille}, \citenamefont {Doyle},\ and\ \citenamefont
  {Sushkov}}]{demille2017probing}%
  \BibitemOpen
  \bibfield  {author} {\bibinfo {author} {\bibfnamefont {D.}~\bibnamefont
  {DeMille}}, \bibinfo {author} {\bibfnamefont {J.~M.}\ \bibnamefont {Doyle}},\
  and\ \bibinfo {author} {\bibfnamefont {A.~O.}\ \bibnamefont {Sushkov}},\
  }\bibfield  {title} {\bibinfo {title} {Probing the frontiers of particle
  physics with tabletop-scale experiments},\ }\href@noop {} {\bibfield
  {journal} {\bibinfo  {journal} {Science}\ }\textbf {\bibinfo {volume}
  {357}},\ \bibinfo {pages} {990} (\bibinfo {year} {2017})}\BibitemShut
  {NoStop}%
\bibitem [{\citenamefont {Safronova}\ \emph {et~al.}(2018)\citenamefont
  {Safronova}, \citenamefont {Budker}, \citenamefont {DeMille}, \citenamefont
  {Kimball}, \citenamefont {Derevianko},\ and\ \citenamefont
  {Clark}}]{safronova2018search}%
  \BibitemOpen
  \bibfield  {author} {\bibinfo {author} {\bibfnamefont {M.}~\bibnamefont
  {Safronova}}, \bibinfo {author} {\bibfnamefont {D.}~\bibnamefont {Budker}},
  \bibinfo {author} {\bibfnamefont {D.}~\bibnamefont {DeMille}}, \bibinfo
  {author} {\bibfnamefont {D.~F.~J.}\ \bibnamefont {Kimball}}, \bibinfo
  {author} {\bibfnamefont {A.}~\bibnamefont {Derevianko}},\ and\ \bibinfo
  {author} {\bibfnamefont {C.~W.}\ \bibnamefont {Clark}},\ }\bibfield  {title}
  {\bibinfo {title} {Search for new physics with atoms and molecules},\
  }\href@noop {} {\bibfield  {journal} {\bibinfo  {journal} {Rev. Mod. Phys.}\
  }\textbf {\bibinfo {volume} {90}},\ \bibinfo {pages} {025008} (\bibinfo
  {year} {2018})}\BibitemShut {NoStop}%
\bibitem [{\citenamefont {Urban}\ \emph {et~al.}(2019)\citenamefont {Urban},
  \citenamefont {Glikin}, \citenamefont {Mouradian}, \citenamefont {Krimmel},
  \citenamefont {Hemmerling},\ and\ \citenamefont
  {Haeffner}}]{urban2019coherent}%
  \BibitemOpen
  \bibfield  {author} {\bibinfo {author} {\bibfnamefont {E.}~\bibnamefont
  {Urban}}, \bibinfo {author} {\bibfnamefont {N.}~\bibnamefont {Glikin}},
  \bibinfo {author} {\bibfnamefont {S.}~\bibnamefont {Mouradian}}, \bibinfo
  {author} {\bibfnamefont {K.}~\bibnamefont {Krimmel}}, \bibinfo {author}
  {\bibfnamefont {B.}~\bibnamefont {Hemmerling}},\ and\ \bibinfo {author}
  {\bibfnamefont {H.}~\bibnamefont {Haeffner}},\ }\bibfield  {title} {\bibinfo
  {title} {Coherent control of the rotational degree of freedom of a two-ion
  {C}oulomb crystal},\ }\href@noop {} {\bibfield  {journal} {\bibinfo
  {journal} {Phys. Rev. Lett.}\ }\textbf {\bibinfo {volume} {123}},\ \bibinfo
  {pages} {133202} (\bibinfo {year} {2019})}\BibitemShut {NoStop}%
\bibitem [{\citenamefont {Budker}\ \emph {et~al.}(2022)\citenamefont {Budker},
  \citenamefont {Graham}, \citenamefont {Ramani}, \citenamefont
  {Schmidt-Kaler}, \citenamefont {Smorra},\ and\ \citenamefont
  {Ulmer}}]{budker2022millicharged}%
  \BibitemOpen
  \bibfield  {author} {\bibinfo {author} {\bibfnamefont {D.}~\bibnamefont
  {Budker}}, \bibinfo {author} {\bibfnamefont {P.~W.}\ \bibnamefont {Graham}},
  \bibinfo {author} {\bibfnamefont {H.}~\bibnamefont {Ramani}}, \bibinfo
  {author} {\bibfnamefont {F.}~\bibnamefont {Schmidt-Kaler}}, \bibinfo {author}
  {\bibfnamefont {C.}~\bibnamefont {Smorra}},\ and\ \bibinfo {author}
  {\bibfnamefont {S.}~\bibnamefont {Ulmer}},\ }\bibfield  {title} {\bibinfo
  {title} {Millicharged dark matter detection with ion traps},\ }\href@noop {}
  {\bibfield  {journal} {\bibinfo  {journal} {PRX Quantum}\ }\textbf {\bibinfo
  {volume} {3}},\ \bibinfo {pages} {010330} (\bibinfo {year}
  {2022})}\BibitemShut {NoStop}%
\bibitem [{\citenamefont {Gonzalez-Ballestero}\ \emph
  {et~al.}(2021)\citenamefont {Gonzalez-Ballestero}, \citenamefont
  {Aspelmeyer}, \citenamefont {Novotny}, \citenamefont {Quidant},\ and\
  \citenamefont {Romero-Isart}}]{gonzalez2021levitodynamics}%
  \BibitemOpen
  \bibfield  {author} {\bibinfo {author} {\bibfnamefont {C.}~\bibnamefont
  {Gonzalez-Ballestero}}, \bibinfo {author} {\bibfnamefont {M.}~\bibnamefont
  {Aspelmeyer}}, \bibinfo {author} {\bibfnamefont {L.}~\bibnamefont {Novotny}},
  \bibinfo {author} {\bibfnamefont {R.}~\bibnamefont {Quidant}},\ and\ \bibinfo
  {author} {\bibfnamefont {O.}~\bibnamefont {Romero-Isart}},\ }\bibfield
  {title} {\bibinfo {title} {Levitodynamics: Levitation and control of
  microscopic objects in vacuum},\ }\href@noop {} {\bibfield  {journal}
  {\bibinfo  {journal} {Science}\ }\textbf {\bibinfo {volume} {374}},\ \bibinfo
  {pages} {eabg3027} (\bibinfo {year} {2021})}\BibitemShut {NoStop}%
\bibitem [{\citenamefont {Perdriat}\ \emph {et~al.}(2021)\citenamefont
  {Perdriat}, \citenamefont {Pellet-Mary}, \citenamefont {Huillery},
  \citenamefont {Rondin},\ and\ \citenamefont {H{\'e}tet}}]{perdriat2021spin}%
  \BibitemOpen
  \bibfield  {author} {\bibinfo {author} {\bibfnamefont {M.}~\bibnamefont
  {Perdriat}}, \bibinfo {author} {\bibfnamefont {C.}~\bibnamefont
  {Pellet-Mary}}, \bibinfo {author} {\bibfnamefont {P.}~\bibnamefont
  {Huillery}}, \bibinfo {author} {\bibfnamefont {L.}~\bibnamefont {Rondin}},\
  and\ \bibinfo {author} {\bibfnamefont {G.}~\bibnamefont {H{\'e}tet}},\
  }\bibfield  {title} {\bibinfo {title} {Spin-mechanics with nitrogen-vacancy
  centers and trapped particles},\ }\href@noop {} {\bibfield  {journal}
  {\bibinfo  {journal} {Micromachines}\ }\textbf {\bibinfo {volume} {12}},\
  \bibinfo {pages} {651} (\bibinfo {year} {2021})}\BibitemShut {NoStop}%
\bibitem [{\citenamefont {Goldwater}\ \emph {et~al.}(2019)\citenamefont
  {Goldwater}, \citenamefont {Stickler}, \citenamefont {Martinetz},
  \citenamefont {Northup}, \citenamefont {Hornberger},\ and\ \citenamefont
  {Millen}}]{goldwater2018levitated}%
  \BibitemOpen
  \bibfield  {author} {\bibinfo {author} {\bibfnamefont {D.}~\bibnamefont
  {Goldwater}}, \bibinfo {author} {\bibfnamefont {B.~A.}\ \bibnamefont
  {Stickler}}, \bibinfo {author} {\bibfnamefont {L.}~\bibnamefont {Martinetz}},
  \bibinfo {author} {\bibfnamefont {T.~E.}\ \bibnamefont {Northup}}, \bibinfo
  {author} {\bibfnamefont {K.}~\bibnamefont {Hornberger}},\ and\ \bibinfo
  {author} {\bibfnamefont {J.}~\bibnamefont {Millen}},\ }\bibfield  {title}
  {\bibinfo {title} {Levitated electromechanics: all-electrical cooling of
  charged nano-and micro-particles},\ }\href@noop {} {\bibfield  {journal}
  {\bibinfo  {journal} {Quantum Sci. Technol.}\ }\textbf {\bibinfo {volume}
  {4}},\ \bibinfo {pages} {024003} (\bibinfo {year} {2019})}\BibitemShut
  {NoStop}%
\bibitem [{\citenamefont {Magrini}\ \emph {et~al.}(2021)\citenamefont
  {Magrini}, \citenamefont {Rosenzweig}, \citenamefont {Bach}, \citenamefont
  {Deutschmann-Olek}, \citenamefont {Hofer}, \citenamefont {Hong},
  \citenamefont {Kiesel}, \citenamefont {Kugi},\ and\ \citenamefont
  {Aspelmeyer}}]{magrini2020optimal}%
  \BibitemOpen
  \bibfield  {author} {\bibinfo {author} {\bibfnamefont {L.}~\bibnamefont
  {Magrini}}, \bibinfo {author} {\bibfnamefont {P.}~\bibnamefont {Rosenzweig}},
  \bibinfo {author} {\bibfnamefont {C.}~\bibnamefont {Bach}}, \bibinfo {author}
  {\bibfnamefont {A.}~\bibnamefont {Deutschmann-Olek}}, \bibinfo {author}
  {\bibfnamefont {S.~G.}\ \bibnamefont {Hofer}}, \bibinfo {author}
  {\bibfnamefont {S.}~\bibnamefont {Hong}}, \bibinfo {author} {\bibfnamefont
  {N.}~\bibnamefont {Kiesel}}, \bibinfo {author} {\bibfnamefont
  {A.}~\bibnamefont {Kugi}},\ and\ \bibinfo {author} {\bibfnamefont
  {M.}~\bibnamefont {Aspelmeyer}},\ }\bibfield  {title} {\bibinfo {title}
  {Optimal quantum control of mechanical motion at room temperature:
  ground-state cooling},\ }\href@noop {} {\bibfield  {journal} {\bibinfo
  {journal} {Nature}\ }\textbf {\bibinfo {volume} {595}},\ \bibinfo {pages}
  {373} (\bibinfo {year} {2021})}\BibitemShut {NoStop}%
\bibitem [{\citenamefont {Tebbenjohanns}\ \emph {et~al.}(2021)\citenamefont
  {Tebbenjohanns}, \citenamefont {Mattana}, \citenamefont {Rossi},
  \citenamefont {Frimmer},\ and\ \citenamefont
  {Novotny}}]{tebbenjohanns2021quantum}%
  \BibitemOpen
  \bibfield  {author} {\bibinfo {author} {\bibfnamefont {F.}~\bibnamefont
  {Tebbenjohanns}}, \bibinfo {author} {\bibfnamefont {M.~L.}\ \bibnamefont
  {Mattana}}, \bibinfo {author} {\bibfnamefont {M.}~\bibnamefont {Rossi}},
  \bibinfo {author} {\bibfnamefont {M.}~\bibnamefont {Frimmer}},\ and\ \bibinfo
  {author} {\bibfnamefont {L.}~\bibnamefont {Novotny}},\ }\bibfield  {title}
  {\bibinfo {title} {{Quantum control of a nanoparticle optically levitated in
  cryogenic free space}},\ }\href@noop {} {\bibfield  {journal} {\bibinfo
  {journal} {Nature}\ }\textbf {\bibinfo {volume} {595}},\ \bibinfo {pages}
  {378} (\bibinfo {year} {2021})}\BibitemShut {NoStop}%
\bibitem [{\citenamefont {Arndt}\ and\ \citenamefont
  {Hornberger}(2014)}]{arndt2014a}%
  \BibitemOpen
  \bibfield  {author} {\bibinfo {author} {\bibfnamefont {M.}~\bibnamefont
  {Arndt}}\ and\ \bibinfo {author} {\bibfnamefont {K.}~\bibnamefont
  {Hornberger}},\ }\bibfield  {title} {\bibinfo {title} {{Testing the limits of
  quantum mechanical superpositions}},\ }\href@noop {} {\bibfield  {journal}
  {\bibinfo  {journal} {Nat. Phys.}\ }\textbf {\bibinfo {volume} {10}},\
  \bibinfo {pages} {271} (\bibinfo {year} {2014})}\BibitemShut {NoStop}%
\bibitem [{\citenamefont {Millen}\ and\ \citenamefont
  {Stickler}(2020)}]{millen2020quantum}%
  \BibitemOpen
  \bibfield  {author} {\bibinfo {author} {\bibfnamefont {J.}~\bibnamefont
  {Millen}}\ and\ \bibinfo {author} {\bibfnamefont {B.~A.}\ \bibnamefont
  {Stickler}},\ }\bibfield  {title} {\bibinfo {title} {Quantum experiments with
  microscale particles},\ }\href
  {https://doi.org/10.1080/00107514.2020.1854497} {\bibfield  {journal}
  {\bibinfo  {journal} {Contemp. Phys.}\ }\textbf {\bibinfo {volume} {61}},\
  \bibinfo {pages} {155} (\bibinfo {year} {2020})}\BibitemShut {NoStop}%
\bibitem [{\citenamefont {Martinetz}\ \emph {et~al.}(2020)\citenamefont
  {Martinetz}, \citenamefont {Hornberger}, \citenamefont {Millen},
  \citenamefont {Kim},\ and\ \citenamefont {Stickler}}]{martinetz2020}%
  \BibitemOpen
  \bibfield  {author} {\bibinfo {author} {\bibfnamefont {L.}~\bibnamefont
  {Martinetz}}, \bibinfo {author} {\bibfnamefont {K.}~\bibnamefont
  {Hornberger}}, \bibinfo {author} {\bibfnamefont {J.}~\bibnamefont {Millen}},
  \bibinfo {author} {\bibfnamefont {M.}~\bibnamefont {Kim}},\ and\ \bibinfo
  {author} {\bibfnamefont {B.~A.}\ \bibnamefont {Stickler}},\ }\bibfield
  {title} {\bibinfo {title} {Quantum electromechanics with levitated
  nanoparticles},\ }\href@noop {} {\bibfield  {journal} {\bibinfo  {journal}
  {npj Quantum Inf.}\ }\textbf {\bibinfo {volume} {6}},\ \bibinfo {pages} {101}
  (\bibinfo {year} {2020})}\BibitemShut {NoStop}%
\bibitem [{\citenamefont {Carney}\ \emph {et~al.}(2021)\citenamefont {Carney},
  \citenamefont {H{\"a}ffner}, \citenamefont {Moore},\ and\ \citenamefont
  {Taylor}}]{carney2021trapped}%
  \BibitemOpen
  \bibfield  {author} {\bibinfo {author} {\bibfnamefont {D.}~\bibnamefont
  {Carney}}, \bibinfo {author} {\bibfnamefont {H.}~\bibnamefont {H{\"a}ffner}},
  \bibinfo {author} {\bibfnamefont {D.~C.}\ \bibnamefont {Moore}},\ and\
  \bibinfo {author} {\bibfnamefont {J.~M.}\ \bibnamefont {Taylor}},\ }\bibfield
   {title} {\bibinfo {title} {Trapped electrons and ions as particle
  detectors},\ }\href@noop {} {\bibfield  {journal} {\bibinfo  {journal} {Phys.
  Rev. Lett.}\ }\textbf {\bibinfo {volume} {127}},\ \bibinfo {pages} {061804}
  (\bibinfo {year} {2021})}\BibitemShut {NoStop}%
\bibitem [{\citenamefont {Lemeshko}\ \emph {et~al.}(2013)\citenamefont
  {Lemeshko}, \citenamefont {Krems}, \citenamefont {Doyle},\ and\ \citenamefont
  {Kais}}]{lemeshko2013manipulation}%
  \BibitemOpen
  \bibfield  {author} {\bibinfo {author} {\bibfnamefont {M.}~\bibnamefont
  {Lemeshko}}, \bibinfo {author} {\bibfnamefont {R.~V.}\ \bibnamefont {Krems}},
  \bibinfo {author} {\bibfnamefont {J.~M.}\ \bibnamefont {Doyle}},\ and\
  \bibinfo {author} {\bibfnamefont {S.}~\bibnamefont {Kais}},\ }\bibfield
  {title} {\bibinfo {title} {Manipulation of molecules with electromagnetic
  fields},\ }\href@noop {} {\bibfield  {journal} {\bibinfo  {journal} {Mol.
  Phys.}\ }\textbf {\bibinfo {volume} {111}},\ \bibinfo {pages} {1648}
  (\bibinfo {year} {2013})}\BibitemShut {NoStop}%
\bibitem [{\citenamefont {Koch}\ \emph {et~al.}(2019)\citenamefont {Koch},
  \citenamefont {Lemeshko},\ and\ \citenamefont {Sugny}}]{koch2019quantum}%
  \BibitemOpen
  \bibfield  {author} {\bibinfo {author} {\bibfnamefont {C.~P.}\ \bibnamefont
  {Koch}}, \bibinfo {author} {\bibfnamefont {M.}~\bibnamefont {Lemeshko}},\
  and\ \bibinfo {author} {\bibfnamefont {D.}~\bibnamefont {Sugny}},\ }\bibfield
   {title} {\bibinfo {title} {Quantum control of molecular rotation},\
  }\href@noop {} {\bibfield  {journal} {\bibinfo  {journal} {Rev. Mod. Phys.}\
  }\textbf {\bibinfo {volume} {91}},\ \bibinfo {pages} {035005} (\bibinfo
  {year} {2019})}\BibitemShut {NoStop}%
\bibitem [{\citenamefont {Stickler}\ \emph {et~al.}(2021)\citenamefont
  {Stickler}, \citenamefont {Hornberger},\ and\ \citenamefont
  {Kim}}]{stickler2021}%
  \BibitemOpen
  \bibfield  {author} {\bibinfo {author} {\bibfnamefont {B.~A.}\ \bibnamefont
  {Stickler}}, \bibinfo {author} {\bibfnamefont {K.}~\bibnamefont
  {Hornberger}},\ and\ \bibinfo {author} {\bibfnamefont {M.}~\bibnamefont
  {Kim}},\ }\bibfield  {title} {\bibinfo {title} {Quantum rotations of
  nanoparticles},\ }\href@noop {} {\bibfield  {journal} {\bibinfo  {journal}
  {Nat. Rev. Phys.}\ }\textbf {\bibinfo {volume} {3}},\ \bibinfo {pages} {589}
  (\bibinfo {year} {2021})}\BibitemShut {NoStop}%
\bibitem [{\citenamefont {DeMille}(2002)}]{demille2002quantum}%
  \BibitemOpen
  \bibfield  {author} {\bibinfo {author} {\bibfnamefont {D.}~\bibnamefont
  {DeMille}},\ }\bibfield  {title} {\bibinfo {title} {Quantum computation with
  trapped polar molecules},\ }\href@noop {} {\bibfield  {journal} {\bibinfo
  {journal} {Phys. Rev. Lett.}\ }\textbf {\bibinfo {volume} {88}},\ \bibinfo
  {pages} {067901} (\bibinfo {year} {2002})}\BibitemShut {NoStop}%
\bibitem [{\citenamefont {Hudson}\ and\ \citenamefont
  {Campbell}(2018)}]{hudson2018dipolar}%
  \BibitemOpen
  \bibfield  {author} {\bibinfo {author} {\bibfnamefont {E.~R.}\ \bibnamefont
  {Hudson}}\ and\ \bibinfo {author} {\bibfnamefont {W.~C.}\ \bibnamefont
  {Campbell}},\ }\bibfield  {title} {\bibinfo {title} {Dipolar quantum logic
  for freely rotating trapped molecular ions},\ }\href@noop {} {\bibfield
  {journal} {\bibinfo  {journal} {Phys. Rev. A}\ }\textbf {\bibinfo {volume}
  {98}},\ \bibinfo {pages} {040302} (\bibinfo {year} {2018})}\BibitemShut
  {NoStop}%
\bibitem [{\citenamefont {Hoang}\ \emph {et~al.}(2016)\citenamefont {Hoang},
  \citenamefont {Ma}, \citenamefont {Ahn}, \citenamefont {Bang}, \citenamefont
  {Robicheaux}, \citenamefont {Yin},\ and\ \citenamefont {Li}}]{hoang2016}%
  \BibitemOpen
  \bibfield  {author} {\bibinfo {author} {\bibfnamefont {T.~M.}\ \bibnamefont
  {Hoang}}, \bibinfo {author} {\bibfnamefont {Y.}~\bibnamefont {Ma}}, \bibinfo
  {author} {\bibfnamefont {J.}~\bibnamefont {Ahn}}, \bibinfo {author}
  {\bibfnamefont {J.}~\bibnamefont {Bang}}, \bibinfo {author} {\bibfnamefont
  {F.}~\bibnamefont {Robicheaux}}, \bibinfo {author} {\bibfnamefont {Z.-Q.}\
  \bibnamefont {Yin}},\ and\ \bibinfo {author} {\bibfnamefont {T.}~\bibnamefont
  {Li}},\ }\bibfield  {title} {\bibinfo {title} {Torsional optomechanics of a
  levitated nonspherical nanoparticle},\ }\href@noop {} {\bibfield  {journal}
  {\bibinfo  {journal} {Phys. Rev. Lett.}\ }\textbf {\bibinfo {volume} {117}},\
  \bibinfo {pages} {123604} (\bibinfo {year} {2016})}\BibitemShut {NoStop}%
\bibitem [{\citenamefont {Kuhn}\ \emph
  {et~al.}(2017{\natexlab{a}})\citenamefont {Kuhn}, \citenamefont {Kosloff},
  \citenamefont {Stickler}, \citenamefont {Patolsky}, \citenamefont
  {Hornberger}, \citenamefont {Arndt},\ and\ \citenamefont
  {Millen}}]{kuhn2017a}%
  \BibitemOpen
  \bibfield  {author} {\bibinfo {author} {\bibfnamefont {S.}~\bibnamefont
  {Kuhn}}, \bibinfo {author} {\bibfnamefont {A.}~\bibnamefont {Kosloff}},
  \bibinfo {author} {\bibfnamefont {B.~A.}\ \bibnamefont {Stickler}}, \bibinfo
  {author} {\bibfnamefont {F.}~\bibnamefont {Patolsky}}, \bibinfo {author}
  {\bibfnamefont {K.}~\bibnamefont {Hornberger}}, \bibinfo {author}
  {\bibfnamefont {M.}~\bibnamefont {Arndt}},\ and\ \bibinfo {author}
  {\bibfnamefont {J.}~\bibnamefont {Millen}},\ }\bibfield  {title} {\bibinfo
  {title} {Full rotational control of levitated silicon nanorods},\ }\href@noop
  {} {\bibfield  {journal} {\bibinfo  {journal} {Optica}\ }\textbf {\bibinfo
  {volume} {4}},\ \bibinfo {pages} {356} (\bibinfo {year}
  {2017}{\natexlab{a}})}\BibitemShut {NoStop}%
\bibitem [{\citenamefont {Kuhn}\ \emph
  {et~al.}(2017{\natexlab{b}})\citenamefont {Kuhn}, \citenamefont {Stickler},
  \citenamefont {Kosloff}, \citenamefont {Patolsky}, \citenamefont
  {Hornberger}, \citenamefont {Arndt},\ and\ \citenamefont
  {Millen}}]{kuhn2017b}%
  \BibitemOpen
  \bibfield  {author} {\bibinfo {author} {\bibfnamefont {S.}~\bibnamefont
  {Kuhn}}, \bibinfo {author} {\bibfnamefont {B.~A.}\ \bibnamefont {Stickler}},
  \bibinfo {author} {\bibfnamefont {A.}~\bibnamefont {Kosloff}}, \bibinfo
  {author} {\bibfnamefont {F.}~\bibnamefont {Patolsky}}, \bibinfo {author}
  {\bibfnamefont {K.}~\bibnamefont {Hornberger}}, \bibinfo {author}
  {\bibfnamefont {M.}~\bibnamefont {Arndt}},\ and\ \bibinfo {author}
  {\bibfnamefont {J.}~\bibnamefont {Millen}},\ }\bibfield  {title} {\bibinfo
  {title} {Optically driven ultra-stable nanomechanical rotor},\ }\href@noop {}
  {\bibfield  {journal} {\bibinfo  {journal} {Nat. Commun.}\ }\textbf {\bibinfo
  {volume} {8}},\ \bibinfo {pages} {1670} (\bibinfo {year}
  {2017}{\natexlab{b}})}\BibitemShut {NoStop}%
\bibitem [{\citenamefont {Rashid}\ \emph {et~al.}(2018)\citenamefont {Rashid},
  \citenamefont {Toro{\v{s}}}, \citenamefont {Setter},\ and\ \citenamefont
  {Ulbricht}}]{rashid2018precession}%
  \BibitemOpen
  \bibfield  {author} {\bibinfo {author} {\bibfnamefont {M.}~\bibnamefont
  {Rashid}}, \bibinfo {author} {\bibfnamefont {M.}~\bibnamefont {Toro{\v{s}}}},
  \bibinfo {author} {\bibfnamefont {A.}~\bibnamefont {Setter}},\ and\ \bibinfo
  {author} {\bibfnamefont {H.}~\bibnamefont {Ulbricht}},\ }\bibfield  {title}
  {\bibinfo {title} {Precession motion in levitated optomechanics},\
  }\href@noop {} {\bibfield  {journal} {\bibinfo  {journal} {Phys. Rev. Lett.}\
  }\textbf {\bibinfo {volume} {121}},\ \bibinfo {pages} {253601} (\bibinfo
  {year} {2018})}\BibitemShut {NoStop}%
\bibitem [{\citenamefont {Reimann}\ \emph {et~al.}(2018)\citenamefont
  {Reimann}, \citenamefont {Doderer}, \citenamefont {Hebestreit}, \citenamefont
  {Diehl}, \citenamefont {Frimmer}, \citenamefont {Windey}, \citenamefont
  {Tebbenjohanns},\ and\ \citenamefont {Novotny}}]{reimann2018ghz}%
  \BibitemOpen
  \bibfield  {author} {\bibinfo {author} {\bibfnamefont {R.}~\bibnamefont
  {Reimann}}, \bibinfo {author} {\bibfnamefont {M.}~\bibnamefont {Doderer}},
  \bibinfo {author} {\bibfnamefont {E.}~\bibnamefont {Hebestreit}}, \bibinfo
  {author} {\bibfnamefont {R.}~\bibnamefont {Diehl}}, \bibinfo {author}
  {\bibfnamefont {M.}~\bibnamefont {Frimmer}}, \bibinfo {author} {\bibfnamefont
  {D.}~\bibnamefont {Windey}}, \bibinfo {author} {\bibfnamefont
  {F.}~\bibnamefont {Tebbenjohanns}},\ and\ \bibinfo {author} {\bibfnamefont
  {L.}~\bibnamefont {Novotny}},\ }\bibfield  {title} {\bibinfo {title} {{GHz}
  rotation of an optically trapped nanoparticle in vacuum},\ }\href@noop {}
  {\bibfield  {journal} {\bibinfo  {journal} {Phys. Rev. Lett.}\ }\textbf
  {\bibinfo {volume} {121}},\ \bibinfo {pages} {033602} (\bibinfo {year}
  {2018})}\BibitemShut {NoStop}%
\bibitem [{\citenamefont {Ahn}\ \emph {et~al.}(2018)\citenamefont {Ahn},
  \citenamefont {Xu}, \citenamefont {Bang}, \citenamefont {Deng}, \citenamefont
  {Hoang}, \citenamefont {Han}, \citenamefont {Ma},\ and\ \citenamefont
  {Li}}]{ahn2018optically}%
  \BibitemOpen
  \bibfield  {author} {\bibinfo {author} {\bibfnamefont {J.}~\bibnamefont
  {Ahn}}, \bibinfo {author} {\bibfnamefont {Z.}~\bibnamefont {Xu}}, \bibinfo
  {author} {\bibfnamefont {J.}~\bibnamefont {Bang}}, \bibinfo {author}
  {\bibfnamefont {Y.-H.}\ \bibnamefont {Deng}}, \bibinfo {author}
  {\bibfnamefont {T.~M.}\ \bibnamefont {Hoang}}, \bibinfo {author}
  {\bibfnamefont {Q.}~\bibnamefont {Han}}, \bibinfo {author} {\bibfnamefont
  {R.-M.}\ \bibnamefont {Ma}},\ and\ \bibinfo {author} {\bibfnamefont
  {T.}~\bibnamefont {Li}},\ }\bibfield  {title} {\bibinfo {title} {Optically
  levitated nanodumbbell torsion balance and {GHz} nanomechanical rotor},\
  }\href@noop {} {\bibfield  {journal} {\bibinfo  {journal} {Phys. Rev. Lett.}\
  }\textbf {\bibinfo {volume} {121}},\ \bibinfo {pages} {033603} (\bibinfo
  {year} {2018})}\BibitemShut {NoStop}%
\bibitem [{\citenamefont {Ahn}\ \emph {et~al.}(2020)\citenamefont {Ahn},
  \citenamefont {Xu}, \citenamefont {Bang}, \citenamefont {Ju}, \citenamefont
  {Gao},\ and\ \citenamefont {Li}}]{ahn2020ultrasensitive}%
  \BibitemOpen
  \bibfield  {author} {\bibinfo {author} {\bibfnamefont {J.}~\bibnamefont
  {Ahn}}, \bibinfo {author} {\bibfnamefont {Z.}~\bibnamefont {Xu}}, \bibinfo
  {author} {\bibfnamefont {J.}~\bibnamefont {Bang}}, \bibinfo {author}
  {\bibfnamefont {P.}~\bibnamefont {Ju}}, \bibinfo {author} {\bibfnamefont
  {X.}~\bibnamefont {Gao}},\ and\ \bibinfo {author} {\bibfnamefont
  {T.}~\bibnamefont {Li}},\ }\bibfield  {title} {\bibinfo {title}
  {Ultrasensitive torque detection with an optically levitated nanorotor},\
  }\href@noop {} {\bibfield  {journal} {\bibinfo  {journal} {Nat.
  Nanotechnol.}\ }\textbf {\bibinfo {volume} {15}},\ \bibinfo {pages} {89}
  (\bibinfo {year} {2020})}\BibitemShut {NoStop}%
\bibitem [{\citenamefont {Bang}\ \emph {et~al.}(2020)\citenamefont {Bang},
  \citenamefont {Seberson}, \citenamefont {Ju}, \citenamefont {Ahn},
  \citenamefont {Xu}, \citenamefont {Gao}, \citenamefont {Robicheaux},\ and\
  \citenamefont {Li}}]{bang2020}%
  \BibitemOpen
  \bibfield  {author} {\bibinfo {author} {\bibfnamefont {J.}~\bibnamefont
  {Bang}}, \bibinfo {author} {\bibfnamefont {T.}~\bibnamefont {Seberson}},
  \bibinfo {author} {\bibfnamefont {P.}~\bibnamefont {Ju}}, \bibinfo {author}
  {\bibfnamefont {J.}~\bibnamefont {Ahn}}, \bibinfo {author} {\bibfnamefont
  {Z.}~\bibnamefont {Xu}}, \bibinfo {author} {\bibfnamefont {X.}~\bibnamefont
  {Gao}}, \bibinfo {author} {\bibfnamefont {F.}~\bibnamefont {Robicheaux}},\
  and\ \bibinfo {author} {\bibfnamefont {T.}~\bibnamefont {Li}},\ }\bibfield
  {title} {\bibinfo {title} {Five-dimensional cooling and nonlinear dynamics of
  an optically levitated nanodumbbell},\ }\href
  {https://doi.org/10.1103/PhysRevResearch.2.043054} {\bibfield  {journal}
  {\bibinfo  {journal} {Phys. Rev. Research}\ }\textbf {\bibinfo {volume}
  {2}},\ \bibinfo {pages} {043054} (\bibinfo {year} {2020})}\BibitemShut
  {NoStop}%
\bibitem [{\citenamefont {Sch{\"a}fer}\ \emph {et~al.}(2021)\citenamefont
  {Sch{\"a}fer}, \citenamefont {Rudolph}, \citenamefont {Hornberger},\ and\
  \citenamefont {Stickler}}]{schafer2020}%
  \BibitemOpen
  \bibfield  {author} {\bibinfo {author} {\bibfnamefont {J.}~\bibnamefont
  {Sch{\"a}fer}}, \bibinfo {author} {\bibfnamefont {H.}~\bibnamefont
  {Rudolph}}, \bibinfo {author} {\bibfnamefont {K.}~\bibnamefont
  {Hornberger}},\ and\ \bibinfo {author} {\bibfnamefont {B.~A.}\ \bibnamefont
  {Stickler}},\ }\bibfield  {title} {\bibinfo {title} {Cooling nanorotors by
  elliptic coherent scattering},\ }\href@noop {} {\bibfield  {journal}
  {\bibinfo  {journal} {Phys. Rev. Lett.}\ }\textbf {\bibinfo {volume} {126}},\
  \bibinfo {pages} {163603} (\bibinfo {year} {2021})}\BibitemShut {NoStop}%
\bibitem [{\citenamefont {van~der Laan}\ \emph {et~al.}(2021)\citenamefont
  {van~der Laan}, \citenamefont {Tebbenjohanns}, \citenamefont {Reimann},
  \citenamefont {Vijayan}, \citenamefont {Novotny},\ and\ \citenamefont
  {Frimmer}}]{vanderlaan2020observation}%
  \BibitemOpen
  \bibfield  {author} {\bibinfo {author} {\bibfnamefont {F.}~\bibnamefont
  {van~der Laan}}, \bibinfo {author} {\bibfnamefont {F.}~\bibnamefont
  {Tebbenjohanns}}, \bibinfo {author} {\bibfnamefont {R.}~\bibnamefont
  {Reimann}}, \bibinfo {author} {\bibfnamefont {J.}~\bibnamefont {Vijayan}},
  \bibinfo {author} {\bibfnamefont {L.}~\bibnamefont {Novotny}},\ and\ \bibinfo
  {author} {\bibfnamefont {M.}~\bibnamefont {Frimmer}},\ }\bibfield  {title}
  {\bibinfo {title} {Sub-{K}elvin feedback cooling and heating dynamics of an
  optically levitated librator},\ }\href@noop {} {\bibfield  {journal}
  {\bibinfo  {journal} {Phys. Rev. Lett.}\ }\textbf {\bibinfo {volume} {127}},\
  \bibinfo {pages} {123605} (\bibinfo {year} {2021})}\BibitemShut {NoStop}%
\bibitem [{\citenamefont {Jin}\ \emph {et~al.}(2021)\citenamefont {Jin},
  \citenamefont {Yan}, \citenamefont {Rahman}, \citenamefont {Li},
  \citenamefont {Yu},\ and\ \citenamefont {Zhang}}]{jin20216}%
  \BibitemOpen
  \bibfield  {author} {\bibinfo {author} {\bibfnamefont {Y.}~\bibnamefont
  {Jin}}, \bibinfo {author} {\bibfnamefont {J.}~\bibnamefont {Yan}}, \bibinfo
  {author} {\bibfnamefont {S.~J.}\ \bibnamefont {Rahman}}, \bibinfo {author}
  {\bibfnamefont {J.}~\bibnamefont {Li}}, \bibinfo {author} {\bibfnamefont
  {X.}~\bibnamefont {Yu}},\ and\ \bibinfo {author} {\bibfnamefont
  {J.}~\bibnamefont {Zhang}},\ }\bibfield  {title} {\bibinfo {title} {6 {GHz}
  hyperfast rotation of an optically levitated nanoparticle in vacuum},\
  }\href@noop {} {\bibfield  {journal} {\bibinfo  {journal} {Photonics Res.}\
  }\textbf {\bibinfo {volume} {9}},\ \bibinfo {pages} {1344} (\bibinfo {year}
  {2021})}\BibitemShut {NoStop}%
\bibitem [{\citenamefont {Delord}\ \emph {et~al.}(2020)\citenamefont {Delord},
  \citenamefont {Huillery}, \citenamefont {Nicolas},\ and\ \citenamefont
  {H{\'e}tet}}]{delord2020spincoupling}%
  \BibitemOpen
  \bibfield  {author} {\bibinfo {author} {\bibfnamefont {T.}~\bibnamefont
  {Delord}}, \bibinfo {author} {\bibfnamefont {P.}~\bibnamefont {Huillery}},
  \bibinfo {author} {\bibfnamefont {L.}~\bibnamefont {Nicolas}},\ and\ \bibinfo
  {author} {\bibfnamefont {G.}~\bibnamefont {H{\'e}tet}},\ }\bibfield  {title}
  {\bibinfo {title} {Spin-cooling of the motion of a trapped diamond},\
  }\href@noop {} {\bibfield  {journal} {\bibinfo  {journal} {Nature}\ }\textbf
  {\bibinfo {volume} {580}},\ \bibinfo {pages} {56} (\bibinfo {year}
  {2020})}\BibitemShut {NoStop}%
\bibitem [{\citenamefont {Delord}\ \emph {et~al.}(2018)\citenamefont {Delord},
  \citenamefont {Huillery}, \citenamefont {Schwab}, \citenamefont {Nicolas},
  \citenamefont {Lecordier},\ and\ \citenamefont {H\'etet}}]{delord2018}%
  \BibitemOpen
  \bibfield  {author} {\bibinfo {author} {\bibfnamefont {T.}~\bibnamefont
  {Delord}}, \bibinfo {author} {\bibfnamefont {P.}~\bibnamefont {Huillery}},
  \bibinfo {author} {\bibfnamefont {L.}~\bibnamefont {Schwab}}, \bibinfo
  {author} {\bibfnamefont {L.}~\bibnamefont {Nicolas}}, \bibinfo {author}
  {\bibfnamefont {L.}~\bibnamefont {Lecordier}},\ and\ \bibinfo {author}
  {\bibfnamefont {G.}~\bibnamefont {H\'etet}},\ }\bibfield  {title} {\bibinfo
  {title} {Ramsey interferences and spin echoes from electron spins inside a
  levitating macroscopic particle},\ }\href
  {https://doi.org/10.1103/PhysRevLett.121.053602} {\bibfield  {journal}
  {\bibinfo  {journal} {Phys. Rev. Lett.}\ }\textbf {\bibinfo {volume} {121}},\
  \bibinfo {pages} {053602} (\bibinfo {year} {2018})}\BibitemShut {NoStop}%
\bibitem [{\citenamefont {Moore}\ and\ \citenamefont
  {Geraci}(2021)}]{moore2021searching}%
  \BibitemOpen
  \bibfield  {author} {\bibinfo {author} {\bibfnamefont {D.~C.}\ \bibnamefont
  {Moore}}\ and\ \bibinfo {author} {\bibfnamefont {A.~A.}\ \bibnamefont
  {Geraci}},\ }\bibfield  {title} {\bibinfo {title} {Searching for new physics
  using optically levitated sensors},\ }\href@noop {} {\bibfield  {journal}
  {\bibinfo  {journal} {Quantum Sci. Technol.}\ }\textbf {\bibinfo {volume}
  {6}},\ \bibinfo {pages} {014008} (\bibinfo {year} {2021})}\BibitemShut
  {NoStop}%
\bibitem [{\citenamefont {Rusconi}\ \emph {et~al.}(2017)\citenamefont
  {Rusconi}, \citenamefont {P{\"o}chhacker}, \citenamefont {Kustura},
  \citenamefont {Cirac},\ and\ \citenamefont
  {Romero-Isart}}]{rusconi2017quantum}%
  \BibitemOpen
  \bibfield  {author} {\bibinfo {author} {\bibfnamefont {C.~C.}\ \bibnamefont
  {Rusconi}}, \bibinfo {author} {\bibfnamefont {V.}~\bibnamefont
  {P{\"o}chhacker}}, \bibinfo {author} {\bibfnamefont {K.}~\bibnamefont
  {Kustura}}, \bibinfo {author} {\bibfnamefont {J.~I.}\ \bibnamefont {Cirac}},\
  and\ \bibinfo {author} {\bibfnamefont {O.}~\bibnamefont {Romero-Isart}},\
  }\bibfield  {title} {\bibinfo {title} {Quantum spin stabilized magnetic
  levitation},\ }\href@noop {} {\bibfield  {journal} {\bibinfo  {journal}
  {Phys. Rev. Lett.}\ }\textbf {\bibinfo {volume} {119}},\ \bibinfo {pages}
  {167202} (\bibinfo {year} {2017})}\BibitemShut {NoStop}%
\bibitem [{\citenamefont {Stickler}\ \emph
  {et~al.}(2018{\natexlab{a}})\citenamefont {Stickler}, \citenamefont
  {Papendell}, \citenamefont {Kuhn}, \citenamefont {Schrinski}, \citenamefont
  {Millen}, \citenamefont {Arndt},\ and\ \citenamefont
  {Hornberger}}]{stickler2018probing}%
  \BibitemOpen
  \bibfield  {author} {\bibinfo {author} {\bibfnamefont {B.~A.}\ \bibnamefont
  {Stickler}}, \bibinfo {author} {\bibfnamefont {B.}~\bibnamefont {Papendell}},
  \bibinfo {author} {\bibfnamefont {S.}~\bibnamefont {Kuhn}}, \bibinfo {author}
  {\bibfnamefont {B.}~\bibnamefont {Schrinski}}, \bibinfo {author}
  {\bibfnamefont {J.}~\bibnamefont {Millen}}, \bibinfo {author} {\bibfnamefont
  {M.}~\bibnamefont {Arndt}},\ and\ \bibinfo {author} {\bibfnamefont
  {K.}~\bibnamefont {Hornberger}},\ }\bibfield  {title} {\bibinfo {title}
  {Probing macroscopic quantum superpositions with nanorotors},\ }\href@noop {}
  {\bibfield  {journal} {\bibinfo  {journal} {New J. Phys.}\ }\textbf {\bibinfo
  {volume} {20}},\ \bibinfo {pages} {122001} (\bibinfo {year}
  {2018}{\natexlab{a}})}\BibitemShut {NoStop}%
\bibitem [{\citenamefont {Ma}\ \emph {et~al.}(2020)\citenamefont {Ma},
  \citenamefont {Khosla}, \citenamefont {Stickler},\ and\ \citenamefont
  {Kim}}]{ma2020quantum}%
  \BibitemOpen
  \bibfield  {author} {\bibinfo {author} {\bibfnamefont {Y.}~\bibnamefont
  {Ma}}, \bibinfo {author} {\bibfnamefont {K.~E.}\ \bibnamefont {Khosla}},
  \bibinfo {author} {\bibfnamefont {B.~A.}\ \bibnamefont {Stickler}},\ and\
  \bibinfo {author} {\bibfnamefont {M.}~\bibnamefont {Kim}},\ }\bibfield
  {title} {\bibinfo {title} {Quantum persistent tennis racket dynamics of
  nanorotors},\ }\href@noop {} {\bibfield  {journal} {\bibinfo  {journal}
  {Phys. Rev. Lett.}\ }\textbf {\bibinfo {volume} {125}},\ \bibinfo {pages}
  {053604} (\bibinfo {year} {2020})}\BibitemShut {NoStop}%
\bibitem [{\citenamefont {Kaltenbaek}\ \emph {et~al.}(2022)\citenamefont
  {Kaltenbaek}, \citenamefont {Arndt}, \citenamefont {Aspelmeyer},
  \citenamefont {Barker}, \citenamefont {Bassi}, \citenamefont {Bateman},
  \citenamefont {Belenchia}, \citenamefont {Berg{\'e}}, \citenamefont {Bose},
  \citenamefont {Braxmaier} \emph {et~al.}}]{kaltenbaek2022maqro}%
  \BibitemOpen
  \bibfield  {author} {\bibinfo {author} {\bibfnamefont {R.}~\bibnamefont
  {Kaltenbaek}}, \bibinfo {author} {\bibfnamefont {M.}~\bibnamefont {Arndt}},
  \bibinfo {author} {\bibfnamefont {M.}~\bibnamefont {Aspelmeyer}}, \bibinfo
  {author} {\bibfnamefont {P.~F.}\ \bibnamefont {Barker}}, \bibinfo {author}
  {\bibfnamefont {A.}~\bibnamefont {Bassi}}, \bibinfo {author} {\bibfnamefont
  {J.}~\bibnamefont {Bateman}}, \bibinfo {author} {\bibfnamefont
  {A.}~\bibnamefont {Belenchia}}, \bibinfo {author} {\bibfnamefont
  {J.}~\bibnamefont {Berg{\'e}}}, \bibinfo {author} {\bibfnamefont
  {S.}~\bibnamefont {Bose}}, \bibinfo {author} {\bibfnamefont {C.}~\bibnamefont
  {Braxmaier}}, \emph {et~al.},\ }\bibfield  {title} {\bibinfo {title}
  {{MAQRO--BPS 2023} research campaign whitepaper},\ }\href@noop {} {\bibfield
  {journal} {\bibinfo  {journal} {arXiv:2202.01535}\ } (\bibinfo {year}
  {2022})}\BibitemShut {NoStop}%
\bibitem [{\citenamefont {Tian}\ \emph {et~al.}(2004)\citenamefont {Tian},
  \citenamefont {Rabl}, \citenamefont {Blatt},\ and\ \citenamefont
  {Zoller}}]{tian2004interfacing}%
  \BibitemOpen
  \bibfield  {author} {\bibinfo {author} {\bibfnamefont {L.}~\bibnamefont
  {Tian}}, \bibinfo {author} {\bibfnamefont {P.}~\bibnamefont {Rabl}}, \bibinfo
  {author} {\bibfnamefont {R.}~\bibnamefont {Blatt}},\ and\ \bibinfo {author}
  {\bibfnamefont {P.}~\bibnamefont {Zoller}},\ }\bibfield  {title} {\bibinfo
  {title} {Interfacing quantum-optical and solid-state qubits},\ }\href@noop {}
  {\bibfield  {journal} {\bibinfo  {journal} {Phys. Rev. Lett.}\ }\textbf
  {\bibinfo {volume} {92}},\ \bibinfo {pages} {247902} (\bibinfo {year}
  {2004})}\BibitemShut {NoStop}%
\bibitem [{\citenamefont {S{\o}rensen}\ \emph {et~al.}(2004)\citenamefont
  {S{\o}rensen}, \citenamefont {van~der Wal}, \citenamefont {Childress},\ and\
  \citenamefont {Lukin}}]{sorensen2004capacitive}%
  \BibitemOpen
  \bibfield  {author} {\bibinfo {author} {\bibfnamefont {A.~S.}\ \bibnamefont
  {S{\o}rensen}}, \bibinfo {author} {\bibfnamefont {C.~H.}\ \bibnamefont
  {van~der Wal}}, \bibinfo {author} {\bibfnamefont {L.~I.}\ \bibnamefont
  {Childress}},\ and\ \bibinfo {author} {\bibfnamefont {M.~D.}\ \bibnamefont
  {Lukin}},\ }\bibfield  {title} {\bibinfo {title} {Capacitive coupling of
  atomic systems to mesoscopic conductors},\ }\href@noop {} {\bibfield
  {journal} {\bibinfo  {journal} {Phys. Rev. Lett.}\ }\textbf {\bibinfo
  {volume} {92}},\ \bibinfo {pages} {063601} (\bibinfo {year}
  {2004})}\BibitemShut {NoStop}%
\bibitem [{\citenamefont {Martinetz}\ \emph {et~al.}(2021)\citenamefont
  {Martinetz}, \citenamefont {Hornberger},\ and\ \citenamefont
  {Stickler}}]{martinetz2021electric}%
  \BibitemOpen
  \bibfield  {author} {\bibinfo {author} {\bibfnamefont {L.}~\bibnamefont
  {Martinetz}}, \bibinfo {author} {\bibfnamefont {K.}~\bibnamefont
  {Hornberger}},\ and\ \bibinfo {author} {\bibfnamefont {B.~A.}\ \bibnamefont
  {Stickler}},\ }\bibfield  {title} {\bibinfo {title} {Electric trapping and
  circuit cooling of charged nanorotors},\ }\href@noop {} {\bibfield  {journal}
  {\bibinfo  {journal} {New J. Phys.}\ }\textbf {\bibinfo {volume} {23}},\
  \bibinfo {pages} {093001} (\bibinfo {year} {2021})}\BibitemShut {NoStop}%
\bibitem [{\citenamefont {An}\ \emph {et~al.}(2022)\citenamefont {An},
  \citenamefont {Alonso}, \citenamefont {Matthiesen},\ and\ \citenamefont
  {H{\"a}ffner}}]{an2021coupling}%
  \BibitemOpen
  \bibfield  {author} {\bibinfo {author} {\bibfnamefont {D.}~\bibnamefont
  {An}}, \bibinfo {author} {\bibfnamefont {A.~M.}\ \bibnamefont {Alonso}},
  \bibinfo {author} {\bibfnamefont {C.}~\bibnamefont {Matthiesen}},\ and\
  \bibinfo {author} {\bibfnamefont {H.}~\bibnamefont {H{\"a}ffner}},\
  }\bibfield  {title} {\bibinfo {title} {Coupling two laser-cooled ions via a
  room-temperature conductor},\ }\href@noop {} {\bibfield  {journal} {\bibinfo
  {journal} {Phys. Rev. Lett.}\ }\textbf {\bibinfo {volume} {128}},\ \bibinfo
  {pages} {063201} (\bibinfo {year} {2022})}\BibitemShut {NoStop}%
\bibitem [{\citenamefont {Negretti}\ \emph {et~al.}(2011)\citenamefont
  {Negretti}, \citenamefont {Treutlein},\ and\ \citenamefont
  {Calarco}}]{negretti2011quantum}%
  \BibitemOpen
  \bibfield  {author} {\bibinfo {author} {\bibfnamefont {A.}~\bibnamefont
  {Negretti}}, \bibinfo {author} {\bibfnamefont {P.}~\bibnamefont
  {Treutlein}},\ and\ \bibinfo {author} {\bibfnamefont {T.}~\bibnamefont
  {Calarco}},\ }\bibfield  {title} {\bibinfo {title} {Quantum computing
  implementations with neutral particles},\ }\href@noop {} {\bibfield
  {journal} {\bibinfo  {journal} {Quantum Inf. Process.}\ }\textbf {\bibinfo
  {volume} {10}},\ \bibinfo {pages} {721} (\bibinfo {year} {2011})}\BibitemShut
  {NoStop}%
\bibitem [{\citenamefont {Kurizki}\ \emph {et~al.}(2015)\citenamefont
  {Kurizki}, \citenamefont {Bertet}, \citenamefont {Kubo}, \citenamefont
  {M{\o}lmer}, \citenamefont {Petrosyan}, \citenamefont {Rabl},\ and\
  \citenamefont {Schmiedmayer}}]{kurizki2015quantum}%
  \BibitemOpen
  \bibfield  {author} {\bibinfo {author} {\bibfnamefont {G.}~\bibnamefont
  {Kurizki}}, \bibinfo {author} {\bibfnamefont {P.}~\bibnamefont {Bertet}},
  \bibinfo {author} {\bibfnamefont {Y.}~\bibnamefont {Kubo}}, \bibinfo {author}
  {\bibfnamefont {K.}~\bibnamefont {M{\o}lmer}}, \bibinfo {author}
  {\bibfnamefont {D.}~\bibnamefont {Petrosyan}}, \bibinfo {author}
  {\bibfnamefont {P.}~\bibnamefont {Rabl}},\ and\ \bibinfo {author}
  {\bibfnamefont {J.}~\bibnamefont {Schmiedmayer}},\ }\bibfield  {title}
  {\bibinfo {title} {Quantum technologies with hybrid systems},\ }\href@noop {}
  {\bibfield  {journal} {\bibinfo  {journal} {PNAS}\ }\textbf {\bibinfo
  {volume} {112}},\ \bibinfo {pages} {3866} (\bibinfo {year}
  {2015})}\BibitemShut {NoStop}%
\bibitem [{\citenamefont {Bruzewicz}\ \emph {et~al.}(2019)\citenamefont
  {Bruzewicz}, \citenamefont {Chiaverini}, \citenamefont {McConnell},\ and\
  \citenamefont {Sage}}]{bruzewicz2019trapped}%
  \BibitemOpen
  \bibfield  {author} {\bibinfo {author} {\bibfnamefont {C.~D.}\ \bibnamefont
  {Bruzewicz}}, \bibinfo {author} {\bibfnamefont {J.}~\bibnamefont
  {Chiaverini}}, \bibinfo {author} {\bibfnamefont {R.}~\bibnamefont
  {McConnell}},\ and\ \bibinfo {author} {\bibfnamefont {J.~M.}\ \bibnamefont
  {Sage}},\ }\bibfield  {title} {\bibinfo {title} {Trapped-ion quantum
  computing: Progress and challenges},\ }\href@noop {} {\bibfield  {journal}
  {\bibinfo  {journal} {Appl. Phys. Rev.}\ }\textbf {\bibinfo {volume} {6}},\
  \bibinfo {pages} {021314} (\bibinfo {year} {2019})}\BibitemShut {NoStop}%
\bibitem [{\citenamefont {Brown}\ \emph {et~al.}(2021)\citenamefont {Brown},
  \citenamefont {Chiaverini}, \citenamefont {Sage},\ and\ \citenamefont
  {H{\"a}ffner}}]{brown2021materials}%
  \BibitemOpen
  \bibfield  {author} {\bibinfo {author} {\bibfnamefont {K.~R.}\ \bibnamefont
  {Brown}}, \bibinfo {author} {\bibfnamefont {J.}~\bibnamefont {Chiaverini}},
  \bibinfo {author} {\bibfnamefont {J.~M.}\ \bibnamefont {Sage}},\ and\
  \bibinfo {author} {\bibfnamefont {H.}~\bibnamefont {H{\"a}ffner}},\
  }\bibfield  {title} {\bibinfo {title} {Materials challenges for trapped-ion
  quantum computers},\ }\href@noop {} {\bibfield  {journal} {\bibinfo
  {journal} {Nat. Rev. Mater.}\ }\textbf {\bibinfo {volume} {6}},\ \bibinfo
  {pages} {892} (\bibinfo {year} {2021})}\BibitemShut {NoStop}%
\bibitem [{\citenamefont {Wineland}\ \emph {et~al.}(1998)\citenamefont
  {Wineland}, \citenamefont {Monroe}, \citenamefont {Itano}, \citenamefont
  {Leibfried}, \citenamefont {King},\ and\ \citenamefont
  {Meekhof}}]{wineland1998experimental}%
  \BibitemOpen
  \bibfield  {author} {\bibinfo {author} {\bibfnamefont {D.~J.}\ \bibnamefont
  {Wineland}}, \bibinfo {author} {\bibfnamefont {C.}~\bibnamefont {Monroe}},
  \bibinfo {author} {\bibfnamefont {W.~M.}\ \bibnamefont {Itano}}, \bibinfo
  {author} {\bibfnamefont {D.}~\bibnamefont {Leibfried}}, \bibinfo {author}
  {\bibfnamefont {B.~E.}\ \bibnamefont {King}},\ and\ \bibinfo {author}
  {\bibfnamefont {D.~M.}\ \bibnamefont {Meekhof}},\ }\bibfield  {title}
  {\bibinfo {title} {Experimental issues in coherent quantum-state manipulation
  of trapped atomic ions},\ }\href@noop {} {\bibfield  {journal} {\bibinfo
  {journal} {J. Res. Natl. Inst. Stand. Technol.}\ }\textbf {\bibinfo {volume}
  {103}},\ \bibinfo {pages} {259} (\bibinfo {year} {1998})}\BibitemShut
  {NoStop}%
\bibitem [{\citenamefont {Leibfried}\ \emph {et~al.}(2003)\citenamefont
  {Leibfried}, \citenamefont {Blatt}, \citenamefont {Monroe},\ and\
  \citenamefont {Wineland}}]{leibfried2003quantum}%
  \BibitemOpen
  \bibfield  {author} {\bibinfo {author} {\bibfnamefont {D.}~\bibnamefont
  {Leibfried}}, \bibinfo {author} {\bibfnamefont {R.}~\bibnamefont {Blatt}},
  \bibinfo {author} {\bibfnamefont {C.}~\bibnamefont {Monroe}},\ and\ \bibinfo
  {author} {\bibfnamefont {D.}~\bibnamefont {Wineland}},\ }\bibfield  {title}
  {\bibinfo {title} {Quantum dynamics of single trapped ions},\ }\href@noop {}
  {\bibfield  {journal} {\bibinfo  {journal} {Rev. Mod. Phys.}\ }\textbf
  {\bibinfo {volume} {75}},\ \bibinfo {pages} {281} (\bibinfo {year}
  {2003})}\BibitemShut {NoStop}%
\bibitem [{\citenamefont {Hite}\ \emph {et~al.}(2013)\citenamefont {Hite},
  \citenamefont {Colombe}, \citenamefont {Wilson}, \citenamefont {Allcock},
  \citenamefont {Leibfried}, \citenamefont {Wineland},\ and\ \citenamefont
  {Pappas}}]{hite2013surface}%
  \BibitemOpen
  \bibfield  {author} {\bibinfo {author} {\bibfnamefont {D.}~\bibnamefont
  {Hite}}, \bibinfo {author} {\bibfnamefont {Y.}~\bibnamefont {Colombe}},
  \bibinfo {author} {\bibfnamefont {A.~C.}\ \bibnamefont {Wilson}}, \bibinfo
  {author} {\bibfnamefont {D.}~\bibnamefont {Allcock}}, \bibinfo {author}
  {\bibfnamefont {D.}~\bibnamefont {Leibfried}}, \bibinfo {author}
  {\bibfnamefont {D.}~\bibnamefont {Wineland}},\ and\ \bibinfo {author}
  {\bibfnamefont {D.}~\bibnamefont {Pappas}},\ }\bibfield  {title} {\bibinfo
  {title} {Surface science for improved ion traps},\ }\href@noop {} {\bibfield
  {journal} {\bibinfo  {journal} {MRS Bull.}\ }\textbf {\bibinfo {volume}
  {38}},\ \bibinfo {pages} {826} (\bibinfo {year} {2013})}\BibitemShut
  {NoStop}%
\bibitem [{\citenamefont {Monroe}\ and\ \citenamefont
  {Kim}(2013)}]{monroe2013scaling}%
  \BibitemOpen
  \bibfield  {author} {\bibinfo {author} {\bibfnamefont {C.}~\bibnamefont
  {Monroe}}\ and\ \bibinfo {author} {\bibfnamefont {J.}~\bibnamefont {Kim}},\
  }\bibfield  {title} {\bibinfo {title} {Scaling the ion trap quantum
  processor},\ }\href@noop {} {\bibfield  {journal} {\bibinfo  {journal}
  {Science}\ }\textbf {\bibinfo {volume} {339}},\ \bibinfo {pages} {1164}
  (\bibinfo {year} {2013})}\BibitemShut {NoStop}%
\bibitem [{\citenamefont {Brownnutt}\ \emph {et~al.}(2015)\citenamefont
  {Brownnutt}, \citenamefont {Kumph}, \citenamefont {Rabl},\ and\ \citenamefont
  {Blatt}}]{brownnutt2015ion}%
  \BibitemOpen
  \bibfield  {author} {\bibinfo {author} {\bibfnamefont {M.}~\bibnamefont
  {Brownnutt}}, \bibinfo {author} {\bibfnamefont {M.}~\bibnamefont {Kumph}},
  \bibinfo {author} {\bibfnamefont {P.}~\bibnamefont {Rabl}},\ and\ \bibinfo
  {author} {\bibfnamefont {R.}~\bibnamefont {Blatt}},\ }\bibfield  {title}
  {\bibinfo {title} {Ion-trap measurements of electric-field noise near
  surfaces},\ }\href@noop {} {\bibfield  {journal} {\bibinfo  {journal} {Rev.
  Mod. Phys.}\ }\textbf {\bibinfo {volume} {87}},\ \bibinfo {pages} {1419}
  (\bibinfo {year} {2015})}\BibitemShut {NoStop}%
\bibitem [{\citenamefont {Sonnentag}\ and\ \citenamefont
  {Hasselbach}(2007)}]{sonnentag2007measurement}%
  \BibitemOpen
  \bibfield  {author} {\bibinfo {author} {\bibfnamefont {P.}~\bibnamefont
  {Sonnentag}}\ and\ \bibinfo {author} {\bibfnamefont {F.}~\bibnamefont
  {Hasselbach}},\ }\bibfield  {title} {\bibinfo {title} {Measurement of
  decoherence of electron waves and visualization of the quantum-classical
  transition},\ }\href@noop {} {\bibfield  {journal} {\bibinfo  {journal}
  {Phys. Rev. Lett.}\ }\textbf {\bibinfo {volume} {98}},\ \bibinfo {pages}
  {200402} (\bibinfo {year} {2007})}\BibitemShut {NoStop}%
\bibitem [{\citenamefont {Beierle}\ \emph {et~al.}(2018)\citenamefont
  {Beierle}, \citenamefont {Zhang},\ and\ \citenamefont
  {Batelaan}}]{beierle2018experimental}%
  \BibitemOpen
  \bibfield  {author} {\bibinfo {author} {\bibfnamefont {P.~J.}\ \bibnamefont
  {Beierle}}, \bibinfo {author} {\bibfnamefont {L.}~\bibnamefont {Zhang}},\
  and\ \bibinfo {author} {\bibfnamefont {H.}~\bibnamefont {Batelaan}},\
  }\bibfield  {title} {\bibinfo {title} {Experimental test of decoherence
  theory using electron matter waves},\ }\href@noop {} {\bibfield  {journal}
  {\bibinfo  {journal} {New J. Phys.}\ }\textbf {\bibinfo {volume} {20}},\
  \bibinfo {pages} {113030} (\bibinfo {year} {2018})}\BibitemShut {NoStop}%
\bibitem [{\citenamefont {Kerker}\ \emph {et~al.}(2020)\citenamefont {Kerker},
  \citenamefont {R{\"o}pke}, \citenamefont {Steinert}, \citenamefont {Pooch},\
  and\ \citenamefont {Stibor}}]{kerker2020quantum}%
  \BibitemOpen
  \bibfield  {author} {\bibinfo {author} {\bibfnamefont {N.}~\bibnamefont
  {Kerker}}, \bibinfo {author} {\bibfnamefont {R.}~\bibnamefont {R{\"o}pke}},
  \bibinfo {author} {\bibfnamefont {L.-M.}\ \bibnamefont {Steinert}}, \bibinfo
  {author} {\bibfnamefont {A.}~\bibnamefont {Pooch}},\ and\ \bibinfo {author}
  {\bibfnamefont {A.}~\bibnamefont {Stibor}},\ }\bibfield  {title} {\bibinfo
  {title} {Quantum decoherence by {C}oulomb interaction},\ }\href@noop {}
  {\bibfield  {journal} {\bibinfo  {journal} {New J. Phys.}\ }\textbf {\bibinfo
  {volume} {22}},\ \bibinfo {pages} {063039} (\bibinfo {year}
  {2020})}\BibitemShut {NoStop}%
\bibitem [{\citenamefont {Anglin}\ \emph {et~al.}(1997)\citenamefont {Anglin},
  \citenamefont {Paz},\ and\ \citenamefont {Zurek}}]{anglin1997deconstructing}%
  \BibitemOpen
  \bibfield  {author} {\bibinfo {author} {\bibfnamefont {J.}~\bibnamefont
  {Anglin}}, \bibinfo {author} {\bibfnamefont {J.}~\bibnamefont {Paz}},\ and\
  \bibinfo {author} {\bibfnamefont {W.}~\bibnamefont {Zurek}},\ }\bibfield
  {title} {\bibinfo {title} {Deconstructing decoherence},\ }\href@noop {}
  {\bibfield  {journal} {\bibinfo  {journal} {Phys. Rev. A}\ }\textbf {\bibinfo
  {volume} {55}},\ \bibinfo {pages} {4041} (\bibinfo {year}
  {1997})}\BibitemShut {NoStop}%
\bibitem [{\citenamefont {Machnikowski}(2006)}]{machnikowski2006theory}%
  \BibitemOpen
  \bibfield  {author} {\bibinfo {author} {\bibfnamefont {P.}~\bibnamefont
  {Machnikowski}},\ }\bibfield  {title} {\bibinfo {title} {Theory of which path
  dephasing in single electron interference due to trace in conductive
  environment},\ }\href@noop {} {\bibfield  {journal} {\bibinfo  {journal}
  {Phys. Rev. B}\ }\textbf {\bibinfo {volume} {73}},\ \bibinfo {pages} {155109}
  (\bibinfo {year} {2006})}\BibitemShut {NoStop}%
\bibitem [{\citenamefont {Howie}(2011)}]{howie2011mechanisms}%
  \BibitemOpen
  \bibfield  {author} {\bibinfo {author} {\bibfnamefont {A.}~\bibnamefont
  {Howie}},\ }\bibfield  {title} {\bibinfo {title} {Mechanisms of decoherence
  in electron microscopy},\ }\href@noop {} {\bibfield  {journal} {\bibinfo
  {journal} {Ultramicroscopy}\ }\textbf {\bibinfo {volume} {111}},\ \bibinfo
  {pages} {761} (\bibinfo {year} {2011})}\BibitemShut {NoStop}%
\bibitem [{\citenamefont {Safavi-Naini}\ \emph {et~al.}(2011)\citenamefont
  {Safavi-Naini}, \citenamefont {Rabl}, \citenamefont {Weck},\ and\
  \citenamefont {Sadeghpour}}]{safavi2011microscopic}%
  \BibitemOpen
  \bibfield  {author} {\bibinfo {author} {\bibfnamefont {A.}~\bibnamefont
  {Safavi-Naini}}, \bibinfo {author} {\bibfnamefont {P.}~\bibnamefont {Rabl}},
  \bibinfo {author} {\bibfnamefont {P.}~\bibnamefont {Weck}},\ and\ \bibinfo
  {author} {\bibfnamefont {H.}~\bibnamefont {Sadeghpour}},\ }\bibfield  {title}
  {\bibinfo {title} {Microscopic model of electric-field-noise heating in ion
  traps},\ }\href@noop {} {\bibfield  {journal} {\bibinfo  {journal} {Phys.
  Rev. A}\ }\textbf {\bibinfo {volume} {84}},\ \bibinfo {pages} {023412}
  (\bibinfo {year} {2011})}\BibitemShut {NoStop}%
\bibitem [{\citenamefont {Safavi-Naini}\ \emph {et~al.}(2013)\citenamefont
  {Safavi-Naini}, \citenamefont {Kim}, \citenamefont {Weck}, \citenamefont
  {Rabl},\ and\ \citenamefont {Sadeghpour}}]{safavi2013influence}%
  \BibitemOpen
  \bibfield  {author} {\bibinfo {author} {\bibfnamefont {A.}~\bibnamefont
  {Safavi-Naini}}, \bibinfo {author} {\bibfnamefont {E.}~\bibnamefont {Kim}},
  \bibinfo {author} {\bibfnamefont {P.}~\bibnamefont {Weck}}, \bibinfo {author}
  {\bibfnamefont {P.}~\bibnamefont {Rabl}},\ and\ \bibinfo {author}
  {\bibfnamefont {H.}~\bibnamefont {Sadeghpour}},\ }\bibfield  {title}
  {\bibinfo {title} {Influence of monolayer contamination on
  electric-field-noise heating in ion traps},\ }\href@noop {} {\bibfield
  {journal} {\bibinfo  {journal} {Phys. Rev. A}\ }\textbf {\bibinfo {volume}
  {87}},\ \bibinfo {pages} {023421} (\bibinfo {year} {2013})}\BibitemShut
  {NoStop}%
\bibitem [{\citenamefont {Kim}\ \emph {et~al.}(2017)\citenamefont {Kim},
  \citenamefont {Safavi-Naini}, \citenamefont {Hite}, \citenamefont {McKay},
  \citenamefont {Pappas}, \citenamefont {Weck},\ and\ \citenamefont
  {Sadeghpour}}]{kim2017electric}%
  \BibitemOpen
  \bibfield  {author} {\bibinfo {author} {\bibfnamefont {E.}~\bibnamefont
  {Kim}}, \bibinfo {author} {\bibfnamefont {A.}~\bibnamefont {Safavi-Naini}},
  \bibinfo {author} {\bibfnamefont {D.}~\bibnamefont {Hite}}, \bibinfo {author}
  {\bibfnamefont {K.}~\bibnamefont {McKay}}, \bibinfo {author} {\bibfnamefont
  {D.}~\bibnamefont {Pappas}}, \bibinfo {author} {\bibfnamefont
  {P.}~\bibnamefont {Weck}},\ and\ \bibinfo {author} {\bibfnamefont
  {H.}~\bibnamefont {Sadeghpour}},\ }\bibfield  {title} {\bibinfo {title}
  {Electric-field noise from carbon-adatom diffusion on a {Au} (110) surface:
  First-principles calculations and experiments},\ }\href@noop {} {\bibfield
  {journal} {\bibinfo  {journal} {Phys. Rev. A}\ }\textbf {\bibinfo {volume}
  {95}},\ \bibinfo {pages} {033407} (\bibinfo {year} {2017})}\BibitemShut
  {NoStop}%
\bibitem [{\citenamefont {Sandoghdar}\ \emph {et~al.}(1992)\citenamefont
  {Sandoghdar}, \citenamefont {Sukenik}, \citenamefont {Hinds},\ and\
  \citenamefont {Haroche}}]{sandoghdar1992direct}%
  \BibitemOpen
  \bibfield  {author} {\bibinfo {author} {\bibfnamefont {V.}~\bibnamefont
  {Sandoghdar}}, \bibinfo {author} {\bibfnamefont {C.}~\bibnamefont {Sukenik}},
  \bibinfo {author} {\bibfnamefont {E.}~\bibnamefont {Hinds}},\ and\ \bibinfo
  {author} {\bibfnamefont {S.}~\bibnamefont {Haroche}},\ }\bibfield  {title}
  {\bibinfo {title} {Direct measurement of the {van der Waals} interaction
  between an atom and its images in a micron-sized cavity},\ }\href@noop {}
  {\bibfield  {journal} {\bibinfo  {journal} {Phys. Rev. Lett.}\ }\textbf
  {\bibinfo {volume} {68}},\ \bibinfo {pages} {3432} (\bibinfo {year}
  {1992})}\BibitemShut {NoStop}%
\bibitem [{\citenamefont {Turchette}\ \emph {et~al.}(2000)\citenamefont
  {Turchette}, \citenamefont {King}, \citenamefont {Leibfried}, \citenamefont
  {Meekhof}, \citenamefont {Myatt}, \citenamefont {Rowe}, \citenamefont
  {Sackett}, \citenamefont {Wood}, \citenamefont {Itano}, \citenamefont
  {Monroe} \emph {et~al.}}]{turchette2000heating}%
  \BibitemOpen
  \bibfield  {author} {\bibinfo {author} {\bibfnamefont {Q.~A.}\ \bibnamefont
  {Turchette}}, \bibinfo {author} {\bibfnamefont {B.}~\bibnamefont {King}},
  \bibinfo {author} {\bibfnamefont {D.}~\bibnamefont {Leibfried}}, \bibinfo
  {author} {\bibfnamefont {D.}~\bibnamefont {Meekhof}}, \bibinfo {author}
  {\bibfnamefont {C.}~\bibnamefont {Myatt}}, \bibinfo {author} {\bibfnamefont
  {M.}~\bibnamefont {Rowe}}, \bibinfo {author} {\bibfnamefont {C.}~\bibnamefont
  {Sackett}}, \bibinfo {author} {\bibfnamefont {C.}~\bibnamefont {Wood}},
  \bibinfo {author} {\bibfnamefont {W.}~\bibnamefont {Itano}}, \bibinfo
  {author} {\bibfnamefont {C.}~\bibnamefont {Monroe}}, \emph {et~al.},\
  }\bibfield  {title} {\bibinfo {title} {Heating of trapped ions from the
  quantum ground state},\ }\href@noop {} {\bibfield  {journal} {\bibinfo
  {journal} {Phys. Rev. A}\ }\textbf {\bibinfo {volume} {61}},\ \bibinfo
  {pages} {063418} (\bibinfo {year} {2000})}\BibitemShut {NoStop}%
\bibitem [{\citenamefont {Dubessy}\ \emph {et~al.}(2009)\citenamefont
  {Dubessy}, \citenamefont {Coudreau},\ and\ \citenamefont
  {Guidoni}}]{dubessy2009electric}%
  \BibitemOpen
  \bibfield  {author} {\bibinfo {author} {\bibfnamefont {R.}~\bibnamefont
  {Dubessy}}, \bibinfo {author} {\bibfnamefont {T.}~\bibnamefont {Coudreau}},\
  and\ \bibinfo {author} {\bibfnamefont {L.}~\bibnamefont {Guidoni}},\
  }\bibfield  {title} {\bibinfo {title} {Electric field noise above surfaces: A
  model for heating-rate scaling law in ion traps},\ }\href@noop {} {\bibfield
  {journal} {\bibinfo  {journal} {Phys. Rev. A}\ }\textbf {\bibinfo {volume}
  {80}},\ \bibinfo {pages} {031402} (\bibinfo {year} {2009})}\BibitemShut
  {NoStop}%
\bibitem [{\citenamefont {Low}\ \emph {et~al.}(2011)\citenamefont {Low},
  \citenamefont {Herskind},\ and\ \citenamefont {Chuang}}]{low2011finite}%
  \BibitemOpen
  \bibfield  {author} {\bibinfo {author} {\bibfnamefont {G.~H.}\ \bibnamefont
  {Low}}, \bibinfo {author} {\bibfnamefont {P.~F.}\ \bibnamefont {Herskind}},\
  and\ \bibinfo {author} {\bibfnamefont {I.~L.}\ \bibnamefont {Chuang}},\
  }\bibfield  {title} {\bibinfo {title} {Finite-geometry models of electric
  field noise from patch potentials in ion traps},\ }\href@noop {} {\bibfield
  {journal} {\bibinfo  {journal} {Phys. Rev. A}\ }\textbf {\bibinfo {volume}
  {84}},\ \bibinfo {pages} {053425} (\bibinfo {year} {2011})}\BibitemShut
  {NoStop}%
\bibitem [{\citenamefont {Noel}\ \emph {et~al.}(2019)\citenamefont {Noel},
  \citenamefont {Berlin-Udi}, \citenamefont {Matthiesen}, \citenamefont {Yu},
  \citenamefont {Zhou}, \citenamefont {Lordi},\ and\ \citenamefont
  {H{\"a}ffner}}]{noel2019electric}%
  \BibitemOpen
  \bibfield  {author} {\bibinfo {author} {\bibfnamefont {C.}~\bibnamefont
  {Noel}}, \bibinfo {author} {\bibfnamefont {M.}~\bibnamefont {Berlin-Udi}},
  \bibinfo {author} {\bibfnamefont {C.}~\bibnamefont {Matthiesen}}, \bibinfo
  {author} {\bibfnamefont {J.}~\bibnamefont {Yu}}, \bibinfo {author}
  {\bibfnamefont {Y.}~\bibnamefont {Zhou}}, \bibinfo {author} {\bibfnamefont
  {V.}~\bibnamefont {Lordi}},\ and\ \bibinfo {author} {\bibfnamefont
  {H.}~\bibnamefont {H{\"a}ffner}},\ }\bibfield  {title} {\bibinfo {title}
  {Electric-field noise from thermally activated fluctuators in a surface ion
  trap},\ }\href@noop {} {\bibfield  {journal} {\bibinfo  {journal} {Phys. Rev.
  A}\ }\textbf {\bibinfo {volume} {99}},\ \bibinfo {pages} {063427} (\bibinfo
  {year} {2019})}\BibitemShut {NoStop}%
\bibitem [{\citenamefont {Wylie}\ and\ \citenamefont
  {Sipe}(1984)}]{wylie1984quantum}%
  \BibitemOpen
  \bibfield  {author} {\bibinfo {author} {\bibfnamefont {J.~M.}\ \bibnamefont
  {Wylie}}\ and\ \bibinfo {author} {\bibfnamefont {J.}~\bibnamefont {Sipe}},\
  }\bibfield  {title} {\bibinfo {title} {Quantum electrodynamics near an
  interface},\ }\href@noop {} {\bibfield  {journal} {\bibinfo  {journal} {Phys.
  Rev. A}\ }\textbf {\bibinfo {volume} {30}},\ \bibinfo {pages} {1185}
  (\bibinfo {year} {1984})}\BibitemShut {NoStop}%
\bibitem [{\citenamefont {Buhmann}\ \emph {et~al.}(2008)\citenamefont
  {Buhmann}, \citenamefont {Tarbutt}, \citenamefont {Scheel},\ and\
  \citenamefont {Hinds}}]{buhmann2008surface}%
  \BibitemOpen
  \bibfield  {author} {\bibinfo {author} {\bibfnamefont {S.~Y.}\ \bibnamefont
  {Buhmann}}, \bibinfo {author} {\bibfnamefont {M.}~\bibnamefont {Tarbutt}},
  \bibinfo {author} {\bibfnamefont {S.}~\bibnamefont {Scheel}},\ and\ \bibinfo
  {author} {\bibfnamefont {E.}~\bibnamefont {Hinds}},\ }\bibfield  {title}
  {\bibinfo {title} {Surface-induced heating of cold polar molecules},\
  }\href@noop {} {\bibfield  {journal} {\bibinfo  {journal} {Phys. Rev. A}\
  }\textbf {\bibinfo {volume} {78}},\ \bibinfo {pages} {052901} (\bibinfo
  {year} {2008})}\BibitemShut {NoStop}%
\bibitem [{\citenamefont {Scheel}\ and\ \citenamefont
  {Buhmann}(2012)}]{scheel2012path}%
  \BibitemOpen
  \bibfield  {author} {\bibinfo {author} {\bibfnamefont {S.}~\bibnamefont
  {Scheel}}\ and\ \bibinfo {author} {\bibfnamefont {S.~Y.}\ \bibnamefont
  {Buhmann}},\ }\bibfield  {title} {\bibinfo {title} {Path decoherence of
  charged and neutral particles near surfaces},\ }\href@noop {} {\bibfield
  {journal} {\bibinfo  {journal} {Phys. Rev. A}\ }\textbf {\bibinfo {volume}
  {85}},\ \bibinfo {pages} {030101} (\bibinfo {year} {2012})}\BibitemShut
  {NoStop}%
\bibitem [{\citenamefont {Kumph}\ \emph {et~al.}(2016)\citenamefont {Kumph},
  \citenamefont {Henkel}, \citenamefont {Rabl}, \citenamefont {Brownnutt},\
  and\ \citenamefont {Blatt}}]{kumph2016electric}%
  \BibitemOpen
  \bibfield  {author} {\bibinfo {author} {\bibfnamefont {M.}~\bibnamefont
  {Kumph}}, \bibinfo {author} {\bibfnamefont {C.}~\bibnamefont {Henkel}},
  \bibinfo {author} {\bibfnamefont {P.}~\bibnamefont {Rabl}}, \bibinfo {author}
  {\bibfnamefont {M.}~\bibnamefont {Brownnutt}},\ and\ \bibinfo {author}
  {\bibfnamefont {R.}~\bibnamefont {Blatt}},\ }\bibfield  {title} {\bibinfo
  {title} {Electric-field noise above a thin dielectric layer on metal
  electrodes},\ }\href@noop {} {\bibfield  {journal} {\bibinfo  {journal} {New
  J. Phys.}\ }\textbf {\bibinfo {volume} {18}},\ \bibinfo {pages} {023020}
  (\bibinfo {year} {2016})}\BibitemShut {NoStop}%
\bibitem [{\citenamefont {Teller}\ \emph {et~al.}(2021)\citenamefont {Teller},
  \citenamefont {Fioretto}, \citenamefont {Holz}, \citenamefont {Schindler},
  \citenamefont {Messerer}, \citenamefont {Sch{\"u}ppert}, \citenamefont {Zou},
  \citenamefont {Blatt}, \citenamefont {Chiaverini}, \citenamefont {Sage} \emph
  {et~al.}}]{teller2021heating}%
  \BibitemOpen
  \bibfield  {author} {\bibinfo {author} {\bibfnamefont {M.}~\bibnamefont
  {Teller}}, \bibinfo {author} {\bibfnamefont {D.~A.}\ \bibnamefont
  {Fioretto}}, \bibinfo {author} {\bibfnamefont {P.~C.}\ \bibnamefont {Holz}},
  \bibinfo {author} {\bibfnamefont {P.}~\bibnamefont {Schindler}}, \bibinfo
  {author} {\bibfnamefont {V.}~\bibnamefont {Messerer}}, \bibinfo {author}
  {\bibfnamefont {K.}~\bibnamefont {Sch{\"u}ppert}}, \bibinfo {author}
  {\bibfnamefont {Y.}~\bibnamefont {Zou}}, \bibinfo {author} {\bibfnamefont
  {R.}~\bibnamefont {Blatt}}, \bibinfo {author} {\bibfnamefont
  {J.}~\bibnamefont {Chiaverini}}, \bibinfo {author} {\bibfnamefont
  {J.}~\bibnamefont {Sage}}, \emph {et~al.},\ }\bibfield  {title} {\bibinfo
  {title} {Heating of a trapped ion induced by dielectric materials},\
  }\href@noop {} {\bibfield  {journal} {\bibinfo  {journal} {Phys. Rev. Lett.}\
  }\textbf {\bibinfo {volume} {126}},\ \bibinfo {pages} {230505} (\bibinfo
  {year} {2021})}\BibitemShut {NoStop}%
\bibitem [{\citenamefont {Buhmann}(2012{\natexlab{a}})}]{buhmann1}%
  \BibitemOpen
  \bibfield  {author} {\bibinfo {author} {\bibfnamefont {S.~Y.}\ \bibnamefont
  {Buhmann}},\ }\href@noop {} {\emph {\bibinfo {title} {{Dispersion Forces I -
  Macroscopic Quantum Electrodynamics and Ground-State Casimir, Casimir-Polder
  and van der Waals Forces}}}}\ (\bibinfo  {publisher} {Springer - Berlin},\
  \bibinfo {year} {2012})\BibitemShut {NoStop}%
\bibitem [{\citenamefont {Buhmann}(2012{\natexlab{b}})}]{buhmann2}%
  \BibitemOpen
  \bibfield  {author} {\bibinfo {author} {\bibfnamefont {S.~Y.}\ \bibnamefont
  {Buhmann}},\ }\href@noop {} {\emph {\bibinfo {title} {{Dispersion Forces II -
  Many-Body Effects, Excited Atoms, Finite Temperature and Quantum
  Friction}}}}\ (\bibinfo  {publisher} {Springer - Berlin},\ \bibinfo {year}
  {2012})\BibitemShut {NoStop}%
\bibitem [{\citenamefont {Holz}\ \emph {et~al.}(2021)\citenamefont {Holz},
  \citenamefont {Lakhmanskiy}, \citenamefont {Rathje}, \citenamefont
  {Schindler}, \citenamefont {Colombe},\ and\ \citenamefont
  {Blatt}}]{holz2021electric}%
  \BibitemOpen
  \bibfield  {author} {\bibinfo {author} {\bibfnamefont {P.~C.}\ \bibnamefont
  {Holz}}, \bibinfo {author} {\bibfnamefont {K.}~\bibnamefont {Lakhmanskiy}},
  \bibinfo {author} {\bibfnamefont {D.}~\bibnamefont {Rathje}}, \bibinfo
  {author} {\bibfnamefont {P.}~\bibnamefont {Schindler}}, \bibinfo {author}
  {\bibfnamefont {Y.}~\bibnamefont {Colombe}},\ and\ \bibinfo {author}
  {\bibfnamefont {R.}~\bibnamefont {Blatt}},\ }\bibfield  {title} {\bibinfo
  {title} {Electric field noise in a high-temperature superconducting surface
  ion trap},\ }\href@noop {} {\bibfield  {journal} {\bibinfo  {journal} {Phys.
  Rev. B}\ }\textbf {\bibinfo {volume} {104}} (\bibinfo {year}
  {2021})}\BibitemShut {NoStop}%
\bibitem [{\citenamefont {Skagerstam}\ \emph {et~al.}(2006)\citenamefont
  {Skagerstam}, \citenamefont {Hohenester}, \citenamefont {Eiguren},\ and\
  \citenamefont {Rekdal}}]{skagerstam2006spin}%
  \BibitemOpen
  \bibfield  {author} {\bibinfo {author} {\bibfnamefont {B.-S.~K.}\
  \bibnamefont {Skagerstam}}, \bibinfo {author} {\bibfnamefont
  {U.}~\bibnamefont {Hohenester}}, \bibinfo {author} {\bibfnamefont
  {A.}~\bibnamefont {Eiguren}},\ and\ \bibinfo {author} {\bibfnamefont {P.~K.}\
  \bibnamefont {Rekdal}},\ }\bibfield  {title} {\bibinfo {title} {Spin
  decoherence in superconducting atom chips},\ }\href@noop {} {\bibfield
  {journal} {\bibinfo  {journal} {Phys. Rev. Lett.}\ }\textbf {\bibinfo
  {volume} {97}},\ \bibinfo {pages} {070401} (\bibinfo {year}
  {2006})}\BibitemShut {NoStop}%
\bibitem [{\citenamefont {Pino}\ \emph {et~al.}(2018)\citenamefont {Pino},
  \citenamefont {Prat-Camps}, \citenamefont {Sinha}, \citenamefont
  {Venkatesh},\ and\ \citenamefont {Romero-Isart}}]{pino2018chip}%
  \BibitemOpen
  \bibfield  {author} {\bibinfo {author} {\bibfnamefont {H.}~\bibnamefont
  {Pino}}, \bibinfo {author} {\bibfnamefont {J.}~\bibnamefont {Prat-Camps}},
  \bibinfo {author} {\bibfnamefont {K.}~\bibnamefont {Sinha}}, \bibinfo
  {author} {\bibfnamefont {B.~P.}\ \bibnamefont {Venkatesh}},\ and\ \bibinfo
  {author} {\bibfnamefont {O.}~\bibnamefont {Romero-Isart}},\ }\bibfield
  {title} {\bibinfo {title} {On-chip quantum interference of a superconducting
  microsphere},\ }\href@noop {} {\bibfield  {journal} {\bibinfo  {journal}
  {Quantum Sci. Technol.}\ }\textbf {\bibinfo {volume} {3}},\ \bibinfo {pages}
  {025001} (\bibinfo {year} {2018})}\BibitemShut {NoStop}%
\bibitem [{\citenamefont {Sinha}\ and\ \citenamefont
  {Suba{\c{s}}{\i}}(2020)}]{sinha2020quantum}%
  \BibitemOpen
  \bibfield  {author} {\bibinfo {author} {\bibfnamefont {K.}~\bibnamefont
  {Sinha}}\ and\ \bibinfo {author} {\bibfnamefont {Y.}~\bibnamefont
  {Suba{\c{s}}{\i}}},\ }\bibfield  {title} {\bibinfo {title} {Quantum
  {Brownian} motion of a particle from {Casimir-Polder} interactions},\
  }\href@noop {} {\bibfield  {journal} {\bibinfo  {journal} {Phys. Rev. A}\
  }\textbf {\bibinfo {volume} {101}},\ \bibinfo {pages} {032507} (\bibinfo
  {year} {2020})}\BibitemShut {NoStop}%
\bibitem [{\citenamefont {Botti}\ \emph {et~al.}(2007)\citenamefont {Botti},
  \citenamefont {Schindlmayr}, \citenamefont {Del~Sole},\ and\ \citenamefont
  {Reining}}]{botti2007time}%
  \BibitemOpen
  \bibfield  {author} {\bibinfo {author} {\bibfnamefont {S.}~\bibnamefont
  {Botti}}, \bibinfo {author} {\bibfnamefont {A.}~\bibnamefont {Schindlmayr}},
  \bibinfo {author} {\bibfnamefont {R.}~\bibnamefont {Del~Sole}},\ and\
  \bibinfo {author} {\bibfnamefont {L.}~\bibnamefont {Reining}},\ }\bibfield
  {title} {\bibinfo {title} {Time-dependent density-functional theory for
  extended systems},\ }\href@noop {} {\bibfield  {journal} {\bibinfo  {journal}
  {Rep. Prog. Phys.}\ }\textbf {\bibinfo {volume} {70}},\ \bibinfo {pages}
  {357} (\bibinfo {year} {2007})}\BibitemShut {NoStop}%
\bibitem [{\citenamefont {Ping}\ \emph {et~al.}(2013)\citenamefont {Ping},
  \citenamefont {Rocca},\ and\ \citenamefont {Galli}}]{ping2013electronic}%
  \BibitemOpen
  \bibfield  {author} {\bibinfo {author} {\bibfnamefont {Y.}~\bibnamefont
  {Ping}}, \bibinfo {author} {\bibfnamefont {D.}~\bibnamefont {Rocca}},\ and\
  \bibinfo {author} {\bibfnamefont {G.}~\bibnamefont {Galli}},\ }\bibfield
  {title} {\bibinfo {title} {Electronic excitations in light absorbers for
  photoelectrochemical energy conversion: first principles calculations based
  on many body perturbation theory},\ }\href@noop {} {\bibfield  {journal}
  {\bibinfo  {journal} {Chem. Soc. Rev.}\ }\textbf {\bibinfo {volume} {42}},\
  \bibinfo {pages} {2437} (\bibinfo {year} {2013})}\BibitemShut {NoStop}%
\bibitem [{\citenamefont {Foulon}\ \emph {et~al.}(2022)\citenamefont {Foulon},
  \citenamefont {Ray}, \citenamefont {Kim}, \citenamefont {Liu}, \citenamefont
  {Rubenstein},\ and\ \citenamefont {Lordi}}]{foulon2022omega}%
  \BibitemOpen
  \bibfield  {author} {\bibinfo {author} {\bibfnamefont {B.~L.}\ \bibnamefont
  {Foulon}}, \bibinfo {author} {\bibfnamefont {K.~G.}\ \bibnamefont {Ray}},
  \bibinfo {author} {\bibfnamefont {C.-E.}\ \bibnamefont {Kim}}, \bibinfo
  {author} {\bibfnamefont {Y.}~\bibnamefont {Liu}}, \bibinfo {author}
  {\bibfnamefont {B.~M.}\ \bibnamefont {Rubenstein}},\ and\ \bibinfo {author}
  {\bibfnamefont {V.}~\bibnamefont {Lordi}},\ }\bibfield  {title} {\bibinfo
  {title} {1/$\omega$ electric-field noise in surface ion traps from correlated
  adsorbate dynamics},\ }\href@noop {} {\bibfield  {journal} {\bibinfo
  {journal} {Phys. Rev. A}\ }\textbf {\bibinfo {volume} {105}},\ \bibinfo
  {pages} {013107} (\bibinfo {year} {2022})}\BibitemShut {NoStop}%
\bibitem [{\citenamefont {Scheel}\ and\ \citenamefont
  {Buhmann}(2008)}]{scheel2008macroscopic}%
  \BibitemOpen
  \bibfield  {author} {\bibinfo {author} {\bibfnamefont {S.}~\bibnamefont
  {Scheel}}\ and\ \bibinfo {author} {\bibfnamefont {S.~Y.}\ \bibnamefont
  {Buhmann}},\ }\bibfield  {title} {\bibinfo {title} {Macroscopic quantum
  electrodynamics - concepts and applications},\ }\href@noop {} {\bibfield
  {journal} {\bibinfo  {journal} {Acta Phys. Slovaca}\ }\textbf {\bibinfo
  {volume} {58}},\ \bibinfo {pages} {675} (\bibinfo {year} {2008})}\BibitemShut
  {NoStop}%
\bibitem [{\citenamefont {Breuer}\ and\ \citenamefont
  {Petruccione}(2002)}]{breuer2002}%
  \BibitemOpen
  \bibfield  {author} {\bibinfo {author} {\bibfnamefont {H.-P.}\ \bibnamefont
  {Breuer}}\ and\ \bibinfo {author} {\bibfnamefont {F.}~\bibnamefont
  {Petruccione}},\ }\href@noop {} {\emph {\bibinfo {title} {{The theory of open
  quantum systems}}}}\ (\bibinfo  {publisher} {Oxford University Press},\
  \bibinfo {year} {2002})\BibitemShut {NoStop}%
\bibitem [{\citenamefont {Jackson}(1999)}]{jackson1999}%
  \BibitemOpen
  \bibfield  {author} {\bibinfo {author} {\bibfnamefont {J.~D.}\ \bibnamefont
  {Jackson}},\ }\href@noop {} {\emph {\bibinfo {title} {{Classical
  Electrodynamics}}}}\ (\bibinfo  {publisher} {Wiley},\ \bibinfo {address} {New
  York},\ \bibinfo {year} {1999})\BibitemShut {NoStop}%
\bibitem [{\citenamefont {London}\ and\ \citenamefont
  {London}(1935)}]{london1935electromagnetic}%
  \BibitemOpen
  \bibfield  {author} {\bibinfo {author} {\bibfnamefont {F.}~\bibnamefont
  {London}}\ and\ \bibinfo {author} {\bibfnamefont {H.}~\bibnamefont
  {London}},\ }\bibfield  {title} {\bibinfo {title} {The electromagnetic
  equations of the supraconductor},\ }\href@noop {} {\bibfield  {journal}
  {\bibinfo  {journal} {Proc. R. Soc. A: Math. Phys. Eng. Sci.}\ }\textbf
  {\bibinfo {volume} {149}},\ \bibinfo {pages} {71} (\bibinfo {year}
  {1935})}\BibitemShut {NoStop}%
\bibitem [{\citenamefont {Hohenester}\ \emph {et~al.}(2007)\citenamefont
  {Hohenester}, \citenamefont {Eiguren}, \citenamefont {Scheel},\ and\
  \citenamefont {Hinds}}]{hohenester2007spin}%
  \BibitemOpen
  \bibfield  {author} {\bibinfo {author} {\bibfnamefont {U.}~\bibnamefont
  {Hohenester}}, \bibinfo {author} {\bibfnamefont {A.}~\bibnamefont {Eiguren}},
  \bibinfo {author} {\bibfnamefont {S.}~\bibnamefont {Scheel}},\ and\ \bibinfo
  {author} {\bibfnamefont {E.}~\bibnamefont {Hinds}},\ }\bibfield  {title}
  {\bibinfo {title} {Spin-flip lifetimes in superconducting atom chips:
  {Bardeen-Cooper-Schrieffer} versus {Eliashberg} theory},\ }\href@noop {}
  {\bibfield  {journal} {\bibinfo  {journal} {Phys. Rev. A}\ }\textbf {\bibinfo
  {volume} {76}},\ \bibinfo {pages} {033618} (\bibinfo {year}
  {2007})}\BibitemShut {NoStop}%
\bibitem [{\citenamefont {Henkel}\ \emph {et~al.}(1999)\citenamefont {Henkel},
  \citenamefont {P{\"o}tting},\ and\ \citenamefont {Wilkens}}]{henkel1999loss}%
  \BibitemOpen
  \bibfield  {author} {\bibinfo {author} {\bibfnamefont {C.}~\bibnamefont
  {Henkel}}, \bibinfo {author} {\bibfnamefont {S.}~\bibnamefont
  {P{\"o}tting}},\ and\ \bibinfo {author} {\bibfnamefont {M.}~\bibnamefont
  {Wilkens}},\ }\bibfield  {title} {\bibinfo {title} {Loss and heating of
  particles in small and noisy traps},\ }\href@noop {} {\bibfield  {journal}
  {\bibinfo  {journal} {Appl. Phys. B}\ }\textbf {\bibinfo {volume} {69}},\
  \bibinfo {pages} {379} (\bibinfo {year} {1999})}\BibitemShut {NoStop}%
\bibitem [{Note1()}]{Note1}%
  \BibitemOpen
  \bibinfo {note} {Note however that the quasistatic approximation of the
  electric field emanating from the metal ceases to be valid for distances
  larger than the skin depth $\delta $, with $\delta ^2=c^2\gamma _{\protect
  \rm m}/\omega \omega _{\protect \rm pm}^2$ \cite {kumph2016electric,
  henkel1999loss}.}\BibitemShut {Stop}%
\bibitem [{\citenamefont {Jonscher}(1977)}]{jonscher1977universal}%
  \BibitemOpen
  \bibfield  {author} {\bibinfo {author} {\bibfnamefont {A.~K.}\ \bibnamefont
  {Jonscher}},\ }\bibfield  {title} {\bibinfo {title} {The ‘universal’
  dielectric response},\ }\href@noop {} {\bibfield  {journal} {\bibinfo
  {journal} {Nature}\ }\textbf {\bibinfo {volume} {267}},\ \bibinfo {pages}
  {673} (\bibinfo {year} {1977})}\BibitemShut {NoStop}%
\bibitem [{\citenamefont {Delord}\ \emph {et~al.}(2017)\citenamefont {Delord},
  \citenamefont {Nicolas}, \citenamefont {Schwab},\ and\ \citenamefont
  {H{\'e}tet}}]{delord2017b}%
  \BibitemOpen
  \bibfield  {author} {\bibinfo {author} {\bibfnamefont {T.}~\bibnamefont
  {Delord}}, \bibinfo {author} {\bibfnamefont {L.}~\bibnamefont {Nicolas}},
  \bibinfo {author} {\bibfnamefont {L.}~\bibnamefont {Schwab}},\ and\ \bibinfo
  {author} {\bibfnamefont {G.}~\bibnamefont {H{\'e}tet}},\ }\bibfield  {title}
  {\bibinfo {title} {Electron spin resonance from {NV} centers in diamonds
  levitating in an ion trap},\ }\href@noop {} {\bibfield  {journal} {\bibinfo
  {journal} {New J. Phys.}\ }\textbf {\bibinfo {volume} {19}},\ \bibinfo
  {pages} {033031} (\bibinfo {year} {2017})}\BibitemShut {NoStop}%
\bibitem [{\citenamefont {Nagornykh}\ \emph {et~al.}(2017)\citenamefont
  {Nagornykh}, \citenamefont {Coppock}, \citenamefont {Murphy},\ and\
  \citenamefont {Kane}}]{nagornykh2017optical}%
  \BibitemOpen
  \bibfield  {author} {\bibinfo {author} {\bibfnamefont {P.}~\bibnamefont
  {Nagornykh}}, \bibinfo {author} {\bibfnamefont {J.~E.}\ \bibnamefont
  {Coppock}}, \bibinfo {author} {\bibfnamefont {J.~P.}\ \bibnamefont
  {Murphy}},\ and\ \bibinfo {author} {\bibfnamefont {B.}~\bibnamefont {Kane}},\
  }\bibfield  {title} {\bibinfo {title} {Optical and magnetic measurements of
  gyroscopically stabilized graphene nanoplatelets levitated in an ion trap},\
  }\href@noop {} {\bibfield  {journal} {\bibinfo  {journal} {Phys. Rev. B}\
  }\textbf {\bibinfo {volume} {96}},\ \bibinfo {pages} {035402} (\bibinfo
  {year} {2017})}\BibitemShut {NoStop}%
\bibitem [{\citenamefont {Bykov}\ \emph {et~al.}(2019)\citenamefont {Bykov},
  \citenamefont {Mestres}, \citenamefont {Dania}, \citenamefont
  {Schm{\"o}ger},\ and\ \citenamefont {Northup}}]{bykov2019}%
  \BibitemOpen
  \bibfield  {author} {\bibinfo {author} {\bibfnamefont {D.~S.}\ \bibnamefont
  {Bykov}}, \bibinfo {author} {\bibfnamefont {P.}~\bibnamefont {Mestres}},
  \bibinfo {author} {\bibfnamefont {L.}~\bibnamefont {Dania}}, \bibinfo
  {author} {\bibfnamefont {L.}~\bibnamefont {Schm{\"o}ger}},\ and\ \bibinfo
  {author} {\bibfnamefont {T.~E.}\ \bibnamefont {Northup}},\ }\bibfield
  {title} {\bibinfo {title} {Direct loading of nanoparticles under high vacuum
  into a {Paul} trap for levitodynamical experiments},\ }\href@noop {}
  {\bibfield  {journal} {\bibinfo  {journal} {Appl. Phys. Lett.}\ }\textbf
  {\bibinfo {volume} {115}},\ \bibinfo {pages} {034101} (\bibinfo {year}
  {2019})}\BibitemShut {NoStop}%
\bibitem [{\citenamefont {Tebbenjohanns}\ \emph {et~al.}(2019)\citenamefont
  {Tebbenjohanns}, \citenamefont {Frimmer}, \citenamefont {Militaru},
  \citenamefont {Jain},\ and\ \citenamefont {Novotny}}]{tebbenjohanns2019cold}%
  \BibitemOpen
  \bibfield  {author} {\bibinfo {author} {\bibfnamefont {F.}~\bibnamefont
  {Tebbenjohanns}}, \bibinfo {author} {\bibfnamefont {M.}~\bibnamefont
  {Frimmer}}, \bibinfo {author} {\bibfnamefont {A.}~\bibnamefont {Militaru}},
  \bibinfo {author} {\bibfnamefont {V.}~\bibnamefont {Jain}},\ and\ \bibinfo
  {author} {\bibfnamefont {L.}~\bibnamefont {Novotny}},\ }\bibfield  {title}
  {\bibinfo {title} {Cold damping of an optically levitated nanoparticle to
  microkelvin temperatures},\ }\href@noop {} {\bibfield  {journal} {\bibinfo
  {journal} {Phys. Rev. Lett.}\ }\textbf {\bibinfo {volume} {122}},\ \bibinfo
  {pages} {223601} (\bibinfo {year} {2019})}\BibitemShut {NoStop}%
\bibitem [{\citenamefont {Conangla}\ \emph {et~al.}(2019)\citenamefont
  {Conangla}, \citenamefont {Ricci}, \citenamefont {Cuairan}, \citenamefont
  {Schell}, \citenamefont {Meyer},\ and\ \citenamefont
  {Quidant}}]{conangla2019optimal}%
  \BibitemOpen
  \bibfield  {author} {\bibinfo {author} {\bibfnamefont {G.~P.}\ \bibnamefont
  {Conangla}}, \bibinfo {author} {\bibfnamefont {F.}~\bibnamefont {Ricci}},
  \bibinfo {author} {\bibfnamefont {M.~T.}\ \bibnamefont {Cuairan}}, \bibinfo
  {author} {\bibfnamefont {A.~W.}\ \bibnamefont {Schell}}, \bibinfo {author}
  {\bibfnamefont {N.}~\bibnamefont {Meyer}},\ and\ \bibinfo {author}
  {\bibfnamefont {R.}~\bibnamefont {Quidant}},\ }\bibfield  {title} {\bibinfo
  {title} {Optimal feedback cooling of a charged levitated nanoparticle with
  adaptive control},\ }\href@noop {} {\bibfield  {journal} {\bibinfo  {journal}
  {Phys. Rev. Lett.}\ }\textbf {\bibinfo {volume} {122}},\ \bibinfo {pages}
  {223602} (\bibinfo {year} {2019})}\BibitemShut {NoStop}%
\bibitem [{\citenamefont {Breuer}\ \emph {et~al.}(2016)\citenamefont {Breuer},
  \citenamefont {Laine}, \citenamefont {Piilo},\ and\ \citenamefont
  {Vacchini}}]{breuer2016colloquium}%
  \BibitemOpen
  \bibfield  {author} {\bibinfo {author} {\bibfnamefont {H.-P.}\ \bibnamefont
  {Breuer}}, \bibinfo {author} {\bibfnamefont {E.-M.}\ \bibnamefont {Laine}},
  \bibinfo {author} {\bibfnamefont {J.}~\bibnamefont {Piilo}},\ and\ \bibinfo
  {author} {\bibfnamefont {B.}~\bibnamefont {Vacchini}},\ }\bibfield  {title}
  {\bibinfo {title} {Colloquium: {Non-Markovian} dynamics in open quantum
  systems},\ }\href@noop {} {\bibfield  {journal} {\bibinfo  {journal} {Rev.
  Mod. Phys.}\ }\textbf {\bibinfo {volume} {88}},\ \bibinfo {pages} {021002}
  (\bibinfo {year} {2016})}\BibitemShut {NoStop}%
\bibitem [{\citenamefont {Westphal}\ and\ \citenamefont
  {Sils}(1977)}]{westphal1977dielectric}%
  \BibitemOpen
  \bibfield  {author} {\bibinfo {author} {\bibfnamefont {W.~B.}\ \bibnamefont
  {Westphal}}\ and\ \bibinfo {author} {\bibfnamefont {A.}~\bibnamefont
  {Sils}},\ }\href@noop {} {\emph {\bibinfo {title} {Dielectric constant and
  loss data}}}\ (\bibinfo  {publisher} {MIT Technical Report AFML-74-250, Part
  III},\ \bibinfo {year} {1977})\BibitemShut {NoStop}%
\bibitem [{\citenamefont {Papendell}\ \emph {et~al.}(2017)\citenamefont
  {Papendell}, \citenamefont {Stickler},\ and\ \citenamefont
  {Hornberger}}]{papendell2017}%
  \BibitemOpen
  \bibfield  {author} {\bibinfo {author} {\bibfnamefont {B.}~\bibnamefont
  {Papendell}}, \bibinfo {author} {\bibfnamefont {B.~A.}\ \bibnamefont
  {Stickler}},\ and\ \bibinfo {author} {\bibfnamefont {K.}~\bibnamefont
  {Hornberger}},\ }\bibfield  {title} {\bibinfo {title} {Quantum angular
  momentum diffusion of rigid bodies},\ }\href
  {http://stacks.iop.org/1367-2630/19/i=12/a=122001} {\bibfield  {journal}
  {\bibinfo  {journal} {New J. Phys.}\ }\textbf {\bibinfo {volume} {19}},\
  \bibinfo {pages} {122001} (\bibinfo {year} {2017})}\BibitemShut {NoStop}%
\bibitem [{\citenamefont {Papi\ifmmode~\check{c}\else \v{c}\fi{}}\ and\
  \citenamefont {de~Vega}(2022)}]{papivc2021neural}%
  \BibitemOpen
  \bibfield  {author} {\bibinfo {author} {\bibfnamefont {M.}~\bibnamefont
  {Papi\ifmmode~\check{c}\else \v{c}\fi{}}}\ and\ \bibinfo {author}
  {\bibfnamefont {I.}~\bibnamefont {de~Vega}},\ }\bibfield  {title} {\bibinfo
  {title} {Neural-network-based qubit-environment characterization},\ }\href
  {https://doi.org/10.1103/PhysRevA.105.022605} {\bibfield  {journal} {\bibinfo
   {journal} {Phys. Rev. A}\ }\textbf {\bibinfo {volume} {105}},\ \bibinfo
  {pages} {022605} (\bibinfo {year} {2022})}\BibitemShut {NoStop}%
\bibitem [{\citenamefont {McConnell}\ \emph {et~al.}(2015)\citenamefont
  {McConnell}, \citenamefont {Bruzewicz}, \citenamefont {Chiaverini},\ and\
  \citenamefont {Sage}}]{mcconnell2015reduction}%
  \BibitemOpen
  \bibfield  {author} {\bibinfo {author} {\bibfnamefont {R.}~\bibnamefont
  {McConnell}}, \bibinfo {author} {\bibfnamefont {C.}~\bibnamefont
  {Bruzewicz}}, \bibinfo {author} {\bibfnamefont {J.}~\bibnamefont
  {Chiaverini}},\ and\ \bibinfo {author} {\bibfnamefont {J.}~\bibnamefont
  {Sage}},\ }\bibfield  {title} {\bibinfo {title} {Reduction of trapped-ion
  anomalous heating by in situ surface plasma cleaning},\ }\href@noop {}
  {\bibfield  {journal} {\bibinfo  {journal} {Phys. Rev. A}\ }\textbf {\bibinfo
  {volume} {92}},\ \bibinfo {pages} {020302} (\bibinfo {year}
  {2015})}\BibitemShut {NoStop}%
\bibitem [{\citenamefont {Hite}\ \emph {et~al.}(2012)\citenamefont {Hite},
  \citenamefont {Colombe}, \citenamefont {Wilson}, \citenamefont {Brown},
  \citenamefont {Warring}, \citenamefont {J{\"o}rdens}, \citenamefont {Jost},
  \citenamefont {McKay}, \citenamefont {Pappas}, \citenamefont {Leibfried}
  \emph {et~al.}}]{hite2012100}%
  \BibitemOpen
  \bibfield  {author} {\bibinfo {author} {\bibfnamefont {D.~A.}\ \bibnamefont
  {Hite}}, \bibinfo {author} {\bibfnamefont {Y.}~\bibnamefont {Colombe}},
  \bibinfo {author} {\bibfnamefont {A.~C.}\ \bibnamefont {Wilson}}, \bibinfo
  {author} {\bibfnamefont {K.~R.}\ \bibnamefont {Brown}}, \bibinfo {author}
  {\bibfnamefont {U.}~\bibnamefont {Warring}}, \bibinfo {author} {\bibfnamefont
  {R.}~\bibnamefont {J{\"o}rdens}}, \bibinfo {author} {\bibfnamefont {J.~D.}\
  \bibnamefont {Jost}}, \bibinfo {author} {\bibfnamefont {K.}~\bibnamefont
  {McKay}}, \bibinfo {author} {\bibfnamefont {D.}~\bibnamefont {Pappas}},
  \bibinfo {author} {\bibfnamefont {D.}~\bibnamefont {Leibfried}}, \emph
  {et~al.},\ }\bibfield  {title} {\bibinfo {title} {100-fold reduction of
  electric-field noise in an ion trap cleaned with in situ argon-ion-beam
  bombardment},\ }\href@noop {} {\bibfield  {journal} {\bibinfo  {journal}
  {Phys. Rev. Lett.}\ }\textbf {\bibinfo {volume} {109}},\ \bibinfo {pages}
  {103001} (\bibinfo {year} {2012})}\BibitemShut {NoStop}%
\bibitem [{\citenamefont {Gieseler}\ \emph {et~al.}(2020)\citenamefont
  {Gieseler}, \citenamefont {Kabcenell}, \citenamefont {Rosenfeld},
  \citenamefont {Schaefer}, \citenamefont {Safira}, \citenamefont {Schuetz},
  \citenamefont {Gonzalez-Ballestero}, \citenamefont {Rusconi}, \citenamefont
  {Romero-Isart},\ and\ \citenamefont {Lukin}}]{gieseler2020single}%
  \BibitemOpen
  \bibfield  {author} {\bibinfo {author} {\bibfnamefont {J.}~\bibnamefont
  {Gieseler}}, \bibinfo {author} {\bibfnamefont {A.}~\bibnamefont {Kabcenell}},
  \bibinfo {author} {\bibfnamefont {E.}~\bibnamefont {Rosenfeld}}, \bibinfo
  {author} {\bibfnamefont {J.}~\bibnamefont {Schaefer}}, \bibinfo {author}
  {\bibfnamefont {A.}~\bibnamefont {Safira}}, \bibinfo {author} {\bibfnamefont
  {M.~J.}\ \bibnamefont {Schuetz}}, \bibinfo {author} {\bibfnamefont
  {C.}~\bibnamefont {Gonzalez-Ballestero}}, \bibinfo {author} {\bibfnamefont
  {C.~C.}\ \bibnamefont {Rusconi}}, \bibinfo {author} {\bibfnamefont
  {O.}~\bibnamefont {Romero-Isart}},\ and\ \bibinfo {author} {\bibfnamefont
  {M.~D.}\ \bibnamefont {Lukin}},\ }\bibfield  {title} {\bibinfo {title}
  {Single-spin magnetomechanics with levitated micromagnets},\ }\href@noop {}
  {\bibfield  {journal} {\bibinfo  {journal} {Phys. Rev. Lett.}\ }\textbf
  {\bibinfo {volume} {124}},\ \bibinfo {pages} {163604} (\bibinfo {year}
  {2020})}\BibitemShut {NoStop}%
\bibitem [{\citenamefont {Vinante}\ \emph {et~al.}(2020)\citenamefont
  {Vinante}, \citenamefont {Falferi}, \citenamefont {Gasbarri}, \citenamefont
  {Setter}, \citenamefont {Timberlake},\ and\ \citenamefont
  {Ulbricht}}]{vinante2020ultralow}%
  \BibitemOpen
  \bibfield  {author} {\bibinfo {author} {\bibfnamefont {A.}~\bibnamefont
  {Vinante}}, \bibinfo {author} {\bibfnamefont {P.}~\bibnamefont {Falferi}},
  \bibinfo {author} {\bibfnamefont {G.}~\bibnamefont {Gasbarri}}, \bibinfo
  {author} {\bibfnamefont {A.}~\bibnamefont {Setter}}, \bibinfo {author}
  {\bibfnamefont {C.}~\bibnamefont {Timberlake}},\ and\ \bibinfo {author}
  {\bibfnamefont {H.}~\bibnamefont {Ulbricht}},\ }\bibfield  {title} {\bibinfo
  {title} {Ultralow mechanical damping with {M}eissner-levitated ferromagnetic
  microparticles},\ }\href@noop {} {\bibfield  {journal} {\bibinfo  {journal}
  {Phys. Rev. Appl.}\ }\textbf {\bibinfo {volume} {13}},\ \bibinfo {pages}
  {064027} (\bibinfo {year} {2020})}\BibitemShut {NoStop}%
\bibitem [{\citenamefont {Streltsov}\ \emph {et~al.}(2021)\citenamefont
  {Streltsov}, \citenamefont {Pedernales},\ and\ \citenamefont
  {Plenio}}]{streltsov2021ground}%
  \BibitemOpen
  \bibfield  {author} {\bibinfo {author} {\bibfnamefont {K.}~\bibnamefont
  {Streltsov}}, \bibinfo {author} {\bibfnamefont {J.~S.}\ \bibnamefont
  {Pedernales}},\ and\ \bibinfo {author} {\bibfnamefont {M.~B.}\ \bibnamefont
  {Plenio}},\ }\bibfield  {title} {\bibinfo {title} {Ground-state cooling of
  levitated magnets in low-frequency traps},\ }\href@noop {} {\bibfield
  {journal} {\bibinfo  {journal} {Phys. Rev. Lett.}\ }\textbf {\bibinfo
  {volume} {126}},\ \bibinfo {pages} {193602} (\bibinfo {year}
  {2021})}\BibitemShut {NoStop}%
\bibitem [{\citenamefont {Latorre}\ \emph {et~al.}(2022)\citenamefont
  {Latorre}, \citenamefont {Paradkar}, \citenamefont {Hambraeus}, \citenamefont
  {Higgins},\ and\ \citenamefont {Wieczorek}}]{latorre2022chip}%
  \BibitemOpen
  \bibfield  {author} {\bibinfo {author} {\bibfnamefont {M.~G.}\ \bibnamefont
  {Latorre}}, \bibinfo {author} {\bibfnamefont {A.}~\bibnamefont {Paradkar}},
  \bibinfo {author} {\bibfnamefont {D.}~\bibnamefont {Hambraeus}}, \bibinfo
  {author} {\bibfnamefont {G.}~\bibnamefont {Higgins}},\ and\ \bibinfo {author}
  {\bibfnamefont {W.}~\bibnamefont {Wieczorek}},\ }\bibfield  {title} {\bibinfo
  {title} {A chip-based superconducting magnetic trap for levitating
  superconducting microparticles},\ }\href@noop {} {\bibfield  {journal}
  {\bibinfo  {journal} {IEEE Trans. Appl. Supercond.}\ }\textbf {\bibinfo
  {volume} {32}},\ \bibinfo {pages} {1} (\bibinfo {year} {2022})}\BibitemShut
  {NoStop}%
\bibitem [{\citenamefont {Stickler}\ \emph
  {et~al.}(2018{\natexlab{b}})\citenamefont {Stickler}, \citenamefont
  {Schrinski},\ and\ \citenamefont {Hornberger}}]{stickler2018rotational}%
  \BibitemOpen
  \bibfield  {author} {\bibinfo {author} {\bibfnamefont {B.~A.}\ \bibnamefont
  {Stickler}}, \bibinfo {author} {\bibfnamefont {B.}~\bibnamefont
  {Schrinski}},\ and\ \bibinfo {author} {\bibfnamefont {K.}~\bibnamefont
  {Hornberger}},\ }\bibfield  {title} {\bibinfo {title} {Rotational friction
  and diffusion of quantum rotors},\ }\href@noop {} {\bibfield  {journal}
  {\bibinfo  {journal} {Phys. Rev. Lett.}\ }\textbf {\bibinfo {volume} {121}},\
  \bibinfo {pages} {040401} (\bibinfo {year} {2018}{\natexlab{b}})}\BibitemShut
  {NoStop}%
\bibitem [{\citenamefont {Gardiner}\ \emph {et~al.}(2004)\citenamefont
  {Gardiner}, \citenamefont {Zoller},\ and\ \citenamefont
  {Zoller}}]{gardinerzoller2004}%
  \BibitemOpen
  \bibfield  {author} {\bibinfo {author} {\bibfnamefont {C.}~\bibnamefont
  {Gardiner}}, \bibinfo {author} {\bibfnamefont {P.}~\bibnamefont {Zoller}},\
  and\ \bibinfo {author} {\bibfnamefont {P.}~\bibnamefont {Zoller}},\
  }\href@noop {} {\emph {\bibinfo {title} {Quantum noise: a handbook of
  Markovian and non-Markovian quantum stochastic methods with applications to
  quantum optics}}}\ (\bibinfo  {publisher} {Springer Science \& Business
  Media},\ \bibinfo {year} {2004})\BibitemShut {NoStop}%
\bibitem [{\citenamefont {Kleefstra}\ and\ \citenamefont
  {Herman}(1980)}]{kleefstra1980influence}%
  \BibitemOpen
  \bibfield  {author} {\bibinfo {author} {\bibfnamefont {M.}~\bibnamefont
  {Kleefstra}}\ and\ \bibinfo {author} {\bibfnamefont {G.}~\bibnamefont
  {Herman}},\ }\bibfield  {title} {\bibinfo {title} {Influence of the image
  force on the band gap in semiconductors and insulators},\ }\href@noop {}
  {\bibfield  {journal} {\bibinfo  {journal} {J. Appl. Phys.}\ }\textbf
  {\bibinfo {volume} {51}},\ \bibinfo {pages} {4923} (\bibinfo {year}
  {1980})}\BibitemShut {NoStop}%
\bibitem [{\citenamefont {Stickler}\ \emph {et~al.}(2016)\citenamefont
  {Stickler}, \citenamefont {Papendell},\ and\ \citenamefont
  {Hornberger}}]{stickler2016b}%
  \BibitemOpen
  \bibfield  {author} {\bibinfo {author} {\bibfnamefont {B.~A.}\ \bibnamefont
  {Stickler}}, \bibinfo {author} {\bibfnamefont {B.}~\bibnamefont
  {Papendell}},\ and\ \bibinfo {author} {\bibfnamefont {K.}~\bibnamefont
  {Hornberger}},\ }\bibfield  {title} {\bibinfo {title} {Spatio-orientational
  decoherence of nanoparticles},\ }\href@noop {} {\bibfield  {journal}
  {\bibinfo  {journal} {Phys. Rev. A}\ }\textbf {\bibinfo {volume} {94}},\
  \bibinfo {pages} {033828} (\bibinfo {year} {2016})}\BibitemShut {NoStop}%
\bibitem [{\citenamefont {Martinetz}(2022)}]{martinetzDiss}%
  \BibitemOpen
  \bibfield  {author} {\bibinfo {author} {\bibfnamefont {L.}~\bibnamefont
  {Martinetz}},\ }\emph {\bibinfo {title} {in preparation}},\ \href@noop {}
  {Ph.D. thesis},\ \bibinfo  {school} {University of Duisburg-Essen} (\bibinfo
  {year} {2022})\BibitemShut {NoStop}%
\bibitem [{\citenamefont {Hornberger}(2007)}]{hornberger2007}%
  \BibitemOpen
  \bibfield  {author} {\bibinfo {author} {\bibfnamefont {K.}~\bibnamefont
  {Hornberger}},\ }\bibfield  {title} {\bibinfo {title} {{Monitoring approach
  to open quantum dynamics using scattering theory}},\ }\href@noop {}
  {\bibfield  {journal} {\bibinfo  {journal} {Europhys. Lett.}\ }\textbf
  {\bibinfo {volume} {77}},\ \bibinfo {pages} {50007} (\bibinfo {year}
  {2007})}\BibitemShut {NoStop}%
\end{thebibliography}

%

\end{document}